\documentclass[utf8,loar]{jydiss2}

\usepackage{csquotes}
\usepackage{enumitem}
\usepackage{graphicx}
\usepackage{subcaption}
\usepackage{tikz}
\usetikzlibrary{arrows.meta,calc,backgrounds}
\pgfdeclarelayer{background}
\pgfdeclarelayer{foreground}
\pgfsetlayers{background,main,foreground}

\usepackage{epstopdf}
\usepackage{slashed}
\usepackage{dsfont}
\usepackage{bbm}
\usepackage{amsbsy}
\usepackage{amssymb}
\usepackage{mathtools}
\usepackage{empheq}
\usepackage{booktabs}
\usepackage{makecell}
\usepackage{float}

\newcommand*\diff{\mathop{}\!\mathrm{d}}
\newcommand*\dd{\mathop{}\!\mathrm{d}}

\newcommand{\tr}[1]{\mathop{\mathrm{tr}}\left[#1\right]}

\usepackage[dvipsnames]{xcolor}
\definecolor{myred}{RGB}{195, 50, 60}
\definecolor{myblue}{RGB}{60, 110, 195}
\definecolor{mypink}{RGB}{220, 20, 120}
\usepackage[most]{tcolorbox}
\tcbuselibrary{theorems}

\newtcolorbox{emphbox}[2]{
  enhanced, sharp corners, boxrule=0.8pt,
  colback=white, colframe=black,
  coltitle=black, fonttitle=\bfseries\itshape,
  fontupper=\itshape,
  left=4mm, right=4mm, top=2mm, bottom=4mm,
  attach boxed title to top left={yshift*=-\tcboxedtitleheight/2, xshift=5mm},
  boxed title style={sharp corners, frame hidden, colback=white, left=1mm, right=2mm},
  title={#2},
}

\graphicspath{{images/}{diagrams/}}

\renewcommand{\Large}{\fontsize{16}{20pt plus 0.15pt}\selectfont}
\renewcommand{\large}{\fontsize{14}{16pt plus 0.15pt}\selectfont}

\usepackage{hyperref}
\hypersetup{
    colorlinks=true,
    linkcolor=myred,
    citecolor=myblue,
    urlcolor=myblue,
}
\usepackage[capitalise]{cleveref}
\crefname{equation}{Eq.}{Eqs.}
\crefname{figure}{Fig.}{Figs.}
\crefname{appendix}{Appendix}{Appendices}
\Crefname{appendix}{Appendix}{Appendices}
\crefname{section}{Section}{Sections}
\crefname{subsection}{Section}{Sections}
\Crefname{section}{Section}{Sections}
\Crefname{subsection}{Section}{Sections}
\crefname{chapter}{Chapter}{Chapters}
\Crefname{chapter}{Chapter}{Chapters}

\usepackage{subfiles}
\usepackage{accsupp}

\usepackage[style=numeric-comp,sorting=none,isbn=false,url=false,
  doi=false,giveninits=true,maxbibnames=99]{biblatex}
\usepackage{xpatch}
\renewbibmacro{in:}{}

\DeclareFieldFormat[article,inbook,incollection,inproceedings,thesis,unpublished]{title}{#1\isdot}
\DeclareFieldFormat{pages}{#1}

\DeclareNameAlias{sortname}{first-last}

\DefineBibliographyExtras{english}{%

}

\DeclareFieldFormat[article,periodical]{volume}{\mkbibbold{#1}}
\DeclareFieldFormat[article,periodical]{journaltitle}{#1}
\DeclareFieldFormat[article,periodical]{pages}{#1}

\renewcommand*{\mkbibnamefamily}[1]{%
  \ifstrequal{#1}{Avramescu}{\textit{#1}}{#1}}
\renewcommand*{\mkbibnamegiven}[1]{%
  \ifstrequal{\namepartfamily}{Avramescu}{\textit{#1}}{#1}}

\DeclareFieldFormat{note}{\textit{#1}}

\DeclareFieldFormat{doilink}{%
  \iffieldundef{doi}{#1}{\href{https://doi.org/\thefield{doi}}{#1}}}

\xpatchbibdriver{article}
  {\usebibmacro{journal+issuetitle}%
   \newunit
   \usebibmacro{byeditor+others}%
   \newunit
   \usebibmacro{note+pages}}
  {\printtext[doilink]{%
     \usebibmacro{journal+issuetitle}%
     \newunit
     \usebibmacro{note+pages}}}
  {}{}

\newbool{titlefirst}

\newbibmacro*{titlefirst}{%
  \ifboolexpr{
    test {\iffieldundef{title}}
    and
    test {\iffieldundef{subtitle}}
  }{}
  {\printtext[title]{%
      \bfseries\itshape\printfield[titlecase]{title}%
      \setunit{\subtitlepunct}%
      \printfield[titlecase]{subtitle}}%
   \newunit}%
  \printfield{titleaddon}}

\AtEveryBibitem{%
  \ifbool{titlefirst}{%
    \renewbibmacro*{begentry}{%
      \usebibmacro{titlefirst}\newline\nopunct\newunit}%
    \renewbibmacro*{title}{}
  }{}}

\DeclareBibliographyCategory{ownpapers}
\addtocategory{ownpapers}{ar1,ar2,ar3,ar4,ar5}

\DeclareBibliographyCategory{authoredother}
\addtocategory{authoredother}{Pooja:2026hjg}

\defbibfilter{notauthored}{not category=ownpapers and not category=authoredother}

\defbibnote{lopdescr}{%
  This thesis consists of an introductory part and of the following publications:
}
\defbibnote{lopcontr}{%
  The author wrote the numerical code, performed the simulations and drew the figures for Articles~\cite{ar1,ar2,ar3,ar4}, and contributed to the classical simulations for Article~\cite{ar5}. The author wrote the original draft for Articles~\cite{ar1,ar2}, and contributed to writing and editing the manuscripts for Articles~\cite{ar3,ar4,ar5}. The glasma solver is based on the code written by collaborator D. M\"{u}ller, while the particle solver was developed by the author. All diagrams and figures in this thesis, which are not taken from the included articles, were drawn by the author. Claude (Anthropic) was used to assist with grammar checking, proofreading, and language editing.
}

\renewcommand{\references}{%
  \unnumberedchapter{References}%
  \addcontentsline{toc}{chapter}{REFERENCES}%
  \printbibliography[filter=notauthored,heading=none]%
}

\title{Transport of hard probes in the glasma}
\setauthor{Dana}{Avramescu}

\abstract{
The quark gluon plasma (QGP), formed in relativistic heavy-ion collisions at RHIC and LHC, thermalizes from an out-of-equilibrium state of strong classical gluon fields known as the glasma. Jets and heavy quarks are produced in the initial scatterings and interact with these pre-equilibrium color fields. The glasma stage is neglected in phenomenological studies of jets and heavy quarks. This simplification may affect the extraction of their QGP transport coefficients from experimental data.

This thesis investigates the impact of the glasma fields on the transport of jets and heavy quarks. The glasma is formulated using the Color Glass Condensate (CGC) effective theory and obtained by solving the classical Yang-Mills equations on a real-time lattice. Initially, the fields consist of strong longitudinal electric and magnetic flux tubes, with the saturation scale $Q_s$ setting the transverse correlation length. Jets and heavy quarks in the glasma are studied using both classical transport equations and quantum evolution based on the light-front Hamiltonian approach.

The transport coefficients $\hat{q}$ for jets and $\kappa$ for heavy quarks are significantly larger than in the QGP and exhibit anisotropy between the transverse and longitudinal directions. On the phenomenological side, the azimuthal decorrelation of back-to-back $c\overline{c}$ pairs in the glasma is comparable to that induced by the QGP. Thus, $D\bar{D}$ angular correlations provide a potentially measurable signature of the glasma stage. The glasma also modifies the nuclear modification factor $R_{AA}$, but the effect is moderate compared to the nuclear parton distribution function (nPDF) contribution.

A systematic analysis of gauge invariance shows that the distinction between kinetic and canonical momenta is essential for comparing the classical and quantum calculations, and that Coulomb gauge fixing largely reduces the gauge artifacts. The quantum evolution of a high-energy quark in the eikonal limit is studied using the time-dependent Basis Light-Front Quantization (tBLFQ) technique. The quantum results for momentum broadening in the glasma are in agreement with the classical estimates and lay the groundwork for including quantum radiation effects.
}

\fititle{Jettien ja raskaiden kvarkkien kuljetus glasmassa}

\abstractfin{
Relativistisissa raskasionit\"{o}rm\"{a}yksiss\"{a} RHIC:ss\"{a} ja LHC:ss\"{a} muodostuva kvarkkigluoniplasma (QGP) saavuttaa termisen tasapainon vahvasti vuorovaikuttavien klassisten gluonikenttien ep\"{a}tasapainotilasta, joka tunnetaan nimell\"{a} glasma. T\"{o}rm\"{a}yksen ensimm\"{a}isiss\"{a} sironnoissa muodostuu jettej\"{a} ja raskaita kvarkkeja, jotka vuorovaikuttavat n\"{a}iden termist\"{a} tasapainoa edelt\"{a}vien v\"{a}rikenttien kanssa. Glasmavaihe j\"{a}tet\"{a}\"{a}n huomiotta jettien ja raskaiden kvarkkien fenomenologisissa tutkimuksissa. T\"{a}m\"{a} yksinkertaistus saattaa vaikuttaa kokeelliseen dataan perustuvan QGP:n kuljetuskertoimien m\"{a}\"{a}ritykseen.

T\"{a}m\"{a}n v\"{a}it\"{o}skirjan aiheena on glasmakenttien vaikutus jettien ja raskaiden kvarkkien kuljetukseen. Glasmaa kuvataan v\"{a}rilasikondensaatiksi (CGC) kutsutun efektiivisen teorian avulla, jossa kentt\"{a}t johdetaan ratkaisemalla klassiset Yang-Mills -yht\"{a}l\"{o}t reaaliaikahilalla. Aluksi glasma koostuu voimakkaista pitkit\"{a}issuuntaisista s\"{a}hk\"{o}- ja magneettivuoputkista, ja saturaatioskaalan $Q_s$ m\"{a}\"{a}r\"{a}\"{a} poikittaissuuntainen korrelaatiopituus. Jettej\"{a} ja raskaita kvarkkeja tutkitaan glasmassa sek\"{a} klassisten kuljetusyht\"{a}l\"{o}iden ett\"{a} valokartiohamiltonilaiseen l\"{a}hestymistapaan perustuvaan kvanttievoluution avulla.

Kuljetuskertoimet $\hat{q}$ jeteille ja $\kappa$ raskaille kvarkeille ovat merkitt\"{a}v\"{a}sti suuremmat kuin QGP:ss\"{a}, ja niiss\"{a} esiintyy ep\"{a}isotropisuutta pituus- ja poikittaissuuntien v\"{a}lill\"{a}. Tutkimuksen fenomenologisella puolella vastakkaissuuntaisten $c\overline{c}$-parien atsimutaalinen dekorrelaatio on verrattavissa samaan QGP:n indusoimaan vaikutukseen. N\"{a}in ollen $D\bar{D}$ kulmakorrelaatiot tarjoavat mahdollisesti mitattavan tunnusmerkin glasmatilasta. Glasma my\"{o}s muuttaa ydinmodifikaatiotekij\"{a}\"{a} $R_{AA}$, mutta glasman vaikutus on maltillinen verrattuna ydinpartonijakauman (nPDF) kontribuutioon.

Mittainvarianssin j\"{a}rjestelm\"{a}llinen tutkimus osoittaa, ett\"{a} erottelu kineettisten ja kanonisten liikem\"{a}\"{a}rien v\"{a}lill\"{a} on olennaista klassisten ja kvanttilaskujen vertailuun, ja ett\"{a} Coulombin mitan kiinnitt\"{a}minen huomattavasti v\"{a}hent\"{a}\"{a} mitta-artefakteja. Korkeaenergisen kvarkin kvanttievoluutiota tutkitaan eikonaalisella rajalla k\"{a}ytt\"{a}en valokartiokvantisaation aikariippuvan kannan (tBLFQ) menetelm\"{a}\"{a}. Kvanttimekaaniset tulokset liikem\"{a}\"{a}r\"{a}n levenemiselle glasmassa yhtyv\"{a}t klassisiin arvioihin ja luovat pohjaa kvanttimekaanisten s\"{a}teilyefektien sis\"{a}llytt\"{a}miselle.
}
\avainsanat{raskasionit\"{o}rm\"{a}ykset, v\"{a}rilasikondensaatti, glasma, jetit, raskaat kvarkit, liikem\"{a}\"{a}r\"{a}n leveneminen, valokartiokvantisointi\\[1.5em]\hspace*{\fill}\textit{Translated by Pyry Runko}}

\people{
  \item[Author] Dana Avramescu\\
    Department of Physics\\
    University of Jyväskylä\\
    Finland
  \item[Supervisors] Prof. Tuomas Lappi\\
    Department of Physics\\
    University of Jyväskylä\\
    Finland

    Assoc. Prof. Heikki M\"{a}ntysaari\\
    Department of Physics\\
    University of Jyväskylä\\
    Finland

  \item[Reviewers] Prof. Guy Moore\\
    Institut f\"{u}r Kernphysik\\
    Technische Universit\"{a}t Darmstadt\\
    Germany

    Prof. Konrad Tywoniuk\\
    Department of Physics and Technology\\
    University of Bergen\\
    Norway

  \item[Opponent] Senior Researcher Dr. Fran\c{c}ois Gelis\\
    Institut de Physique Th\'{e}orique\\
    Universit\'{e} Paris-Saclay, CEA/Saclay\\
    France
}

\usepackage{pdfpages}

\makeatletter
  {\setbox\z@\vbox\bgroup}
  {\egroup
   \includepdf[pages=1-3]{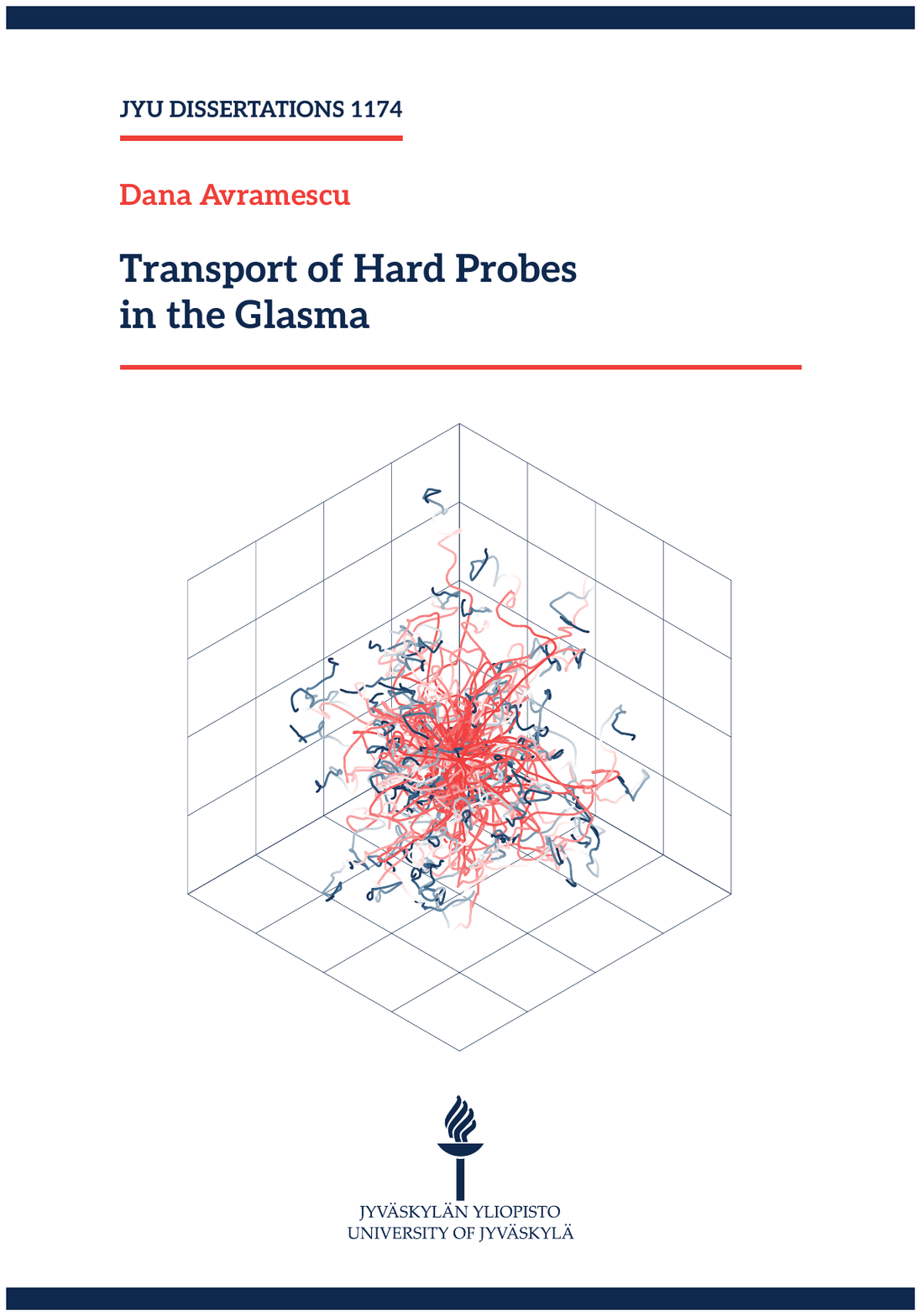}%
   \setcounter{page}{3}}
\makeatother
\makeindex
\begin{document}

\preface

\vspace{-0.5em}
\noindent\hfill\begin{minipage}{0.85\textwidth}
\raggedleft
\textit{``Science? Nonsense!\\
We don't need other worlds.\\
Man needs man.''}

\vspace{1.2em}
\textit{Solaris}, by Andrey Tarkovsky \textit{\&} Stanislaw Lem
\end{minipage}\par
\medskip

\noindent In \textit{Solaris}, the researchers reach for the unknown, but what they study so meticulously only reflects back their own limitations and need for otherness. This was my experience too, and this work would not have been possible without so many of you who offered guidance, support and care.

I would like to express my sincere gratitude to my supervisors. I thank Prof. Tuomas Lappi for his vast knowledge of physics, his passion for intuitively explaining any concept on the whiteboard, his utter geekiness which always made every discussion lively and non-Abelian, and whose amazing lecture notes I was studying long before I arrived in Jyväskylä. I thank Assoc. Prof. Heikki Mäntysaari for his impeccable approach to physics, his clear pragmatism in problem solving, his skill for having an answer ready before I even finished asking the question, and the supportive joviality present in all our meetings. It was a real privilege to learn the craft of research from both of you. Kiitos paljon!

I would like to thank the pre-examiners Prof. Guy Moore and Prof. Konrad Tywoniuk for their thorough reading and careful comments on the thesis. My gratitude goes to Dr. Fran\c{c}ois Gelis for agreeing to be my opponent during the public defence, and for unknowingly teaching me this topic through his masterfully crafted lecture notes. I look forward to our discussion during the defence!

I thank my collaborators Prof. Vincenzo Greco, Dr. Andreas Ipp, Carlos Lamas, Dr. Meijian Li, Dr. David Müller, Dr. Pooja, Prof. Marco Ruggieri, and Prof. Carlos Salgado for their dedication to our projects, for always surprising me with new ways to approach research, and for enriching my academic life. Through long meetings full of conceptual questions, numerical curiosities and project planning, they stayed curious and kept my passion alive. My deepest gratitude to Dr. David Müller and Dr. Kirill Boguslavski for putting me on this path and for supporting me to pursue this PhD in Jyväskylä. I thank David, my oldest collaborator, for welcoming me into his research, for being the one from whom I first learned this topic and its numerics, for letting me build on it into what has become this entire thesis, and for believing in me all this time. Enormous thanks to Prof. Carlos Salgado for his continuous support and for being a pillar of a community that always made me feel so welcome.

I would like to thank the Research Council of Finland, Centre of Excellence in Quark Matter (projects 346324 and 364191), and the Vilho, Yrjö and Kalle Väisälä Foundation of the Finnish Academy of Science and Letters for financial support while working on this thesis. I am also grateful for the lectures given by the CoE seniors, Prof. Kari Eskola, Dr. Hannu Paukkunen, Dr. Dong Jo Kim and Prof. Tuomas Lappi. Moreover, the lectures at the GGI school in Florence and the QCD Masterclass in Saint-Jacut-de-la-Mer were wonderful learning experiences.

I was privileged to visit many places during my studies. Thank you to Dr. Meijian Li for hosting me in Santiago de Compostela, and to Andrey, Carlos, Clara, Marcos, Sergio, Wenyang and Xoán for making my visit welcoming and for introducing me to Galician culture. Thank you to Prof. Sören Schlichting for hosting me at Universität Bielefeld, and to Adrien, Eamon, Ismail, Renata and Yannick for making my visit so enjoyable. Thank you to Dr. João Barata for hosting me at CERN, and to Govert, Petja and Urs for the nice lunches and extended coffee conversations. And lastly, thank you to Dr. Björn Schenke and the BNL group for hosting my visit at the EIC Theory Institute, which is still ongoing and where I am writing these words.

I would like to thank the many people who made conferences, workshops and schools so enjoyable, for taking a wallflower like me out to have fun, and for teaching me their research topics along the way. The list is so long, but here we go: Adam, André, Andreas, Arjun, Arthur, Bruno, Carlos, Chiara, Diego, Eamon, Farid, Federica, Florian, Gabriele, Hannah, Jean, João ($\times 3$, Obrigado!), Kayran, Luke, Marcos, Markus, Nuno, Orestis, Rachel, Sara, Sergio, Travis, Vincenzo, Xoán. 

To FL347, the messiest office, the tea station \#$2$, and home to what is currently the only living houseplant on our hallway. Thank you Consti, Mirja, Nyati, Olli, Pyry and Sami D. for being the most awesome office mates. I have so many fond memories of talking with you, doing mischief or sharing a warm cup of tea. Thank you to the CoE junior members from the late lunch group with whom I shared countless fun moments. Special thanks to Carlisle and Sami Y. for their companionship during our PhDs, you have been my team. Look at us defending almost together, and let us not forget, \textit{``the work is mysterious and important''}.

To Dr. Carlota Andrés, who has inspired me since her plenary talk at Hard Probes, my very first conference, and who I later met in person at my first poster at Initial Stages, where she engaged genuinely with the physics. Her support in preparing plenary talks in later editions of the same conferences, her very smart insight and careful dedication to physics and community, are shaping how I do research. I already feel so welcomed in your group and am looking forward to our next years together. Gracias and merci!

Thank you to my beautiful girlfriends, Andreea, Cătălina, Cornelia, Daria and Silvia, who have supported me throughout the years. Your fierceness, elegance, affinity to beauty and womanhood were an inspiration to me. To Hanna, whose inner strength and luminous presence guided me through dark moments. To Aura and Andrei, for supporting me tremendously in this journey and always believing in me. Thank you to our friends in Jyväskylä, Josh with Fabienne and dear Myra, who have lightened up my days and sparked so much joy every time we spent time together.

Lastly, to my parents, Ina and Dan, for their immense love and for nurturing my passion for knowledge. To Gabriel, my beloved fianc\'e, for being my anchor, lighthouse and kindred spirit. Vă \BeginAccSupp{method=hex,unicode,ActualText=006D0075006C021B0075006D006500730063}mulțumesc\EndAccSupp{} din tot sufletul!

\bigskip

\begin{flushright}
\textit{Brookhaven \& Helsinki}, August 2026\\
\textit{Dana Avramescu}
\end{flushright}

\clearpage
\thispagestyle{empty}
\vspace*{\fill}
\begin{flushright}
\begin{minipage}{0.85\textwidth}
\raggedleft
\itshape\large Dedicated to my parents and Gabriel.
\end{minipage}
\end{flushright}
\vspace*{\fill}
\clearpage

%

\unnumberedchapter{List of Publications}
\addcontentsline{toc}{chapter}{LIST OF PUBLICATIONS}
\booltrue{titlefirst}
\printbibliography[category=ownpapers,heading=none,prenote=lopdescr,postnote=lopcontr]
\boolfalse{titlefirst}

\noindent Other articles not used in this thesis, but to which the author has contributed:

\booltrue{titlefirst}
\printbibliography[category=authoredother,heading=none]
\boolfalse{titlefirst}

\nocite{ar1,ar2,ar3,ar4,ar5,Pooja:2026hjg}

\hypersetup{linkcolor=black}
\let\origcleardoublepage\cleardoublepage
\let\cleardoublepage\clearpage
\mainmatter
\let\cleardoublepage\origcleardoublepage
\hypersetup{linkcolor=myred}

\chapter{Introduction}

A well-established discovery in relativistic heavy-ion collisions is the fluid nature of the \emph{quark-gluon plasma} (QGP) produced at the \emph{Relativistic Heavy Ion Collider} (RHIC) and the \emph{Large Hadron Collider} (LHC). The QGP resembles an almost perfect liquid, with the ratio of shear viscosity to entropy density close to the quantum lower bound. However, there is still no consensus on the origin of this medium, and how exactly it is produced from the underlying theory of strong interactions, \emph{Quantum Chromodynamics} (QCD). In a weak coupling picture, the QGP emerges from a far-from-equilibrium state of strong classical color fields known as the \emph{glasma}. Additionally, hard probes such as heavy quarks and jets, as well as electromagnetic probes, are produced in the first stages of the collision and experience the entire evolution of the medium. Recently, there has been ongoing interest in understanding how these probes interact with the early-stage color fields. Historically, these were neglected, mostly due to the challenging theoretical and numerical complexity of the problem.
But \emph{how much are hard probes affected by the glasma fields?} And more importantly, \emph{what are the signatures of this interaction on experimental observables?} These questions are carving out a new subfield in heavy-ion physics, and are the main focus of this thesis.

Tackling these questions has both theoretical and phenomenological implications. On the theoretical side, answering them requires developing both the transport framework to capture the relevant physics of how hard probes interact with the glasma fields, and the numerical tools to implement this in realistic glasma simulations. 
The interaction between hard probes and the glasma fields modifies their momentum distribution before reaching the QGP stage, which has phenomenological consequences. Notably, it may cause large modifications of the transport coefficients of the QGP, which are typically obtained from data that completely neglect hard probes during pre-equilibrium stages. 
Despite growing interest and promising progress, existing calculations have not yet connected the transport of hard probes from the glasma to the QGP stage; there is still no clear estimate of the magnitude of energy loss in the glasma; and no observable exclusively sensitive to the glasma stage has been identified. This thesis addresses some of these gaps and lays the groundwork for future studies in these directions.

\section{Heavy-ion collisions}

\begin{figure}[t]
    \centering
    \includegraphics[width=0.9\textwidth]{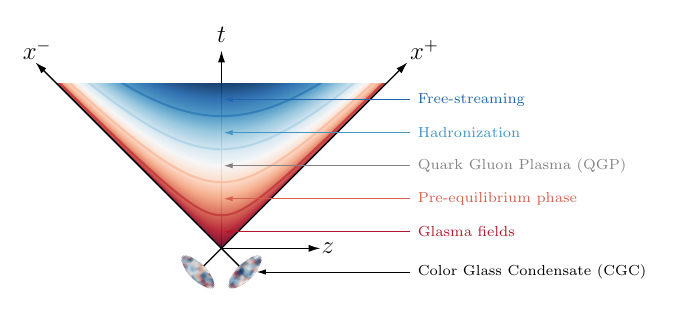}
    \caption{Stages of a relativistic heavy-ion collision. Before the collision, each nucleus is described by the Color Glass Condensate effective theory. In the forward light-cone after the collision, one obtains the glasma fields. The pre-equilibrium stage undergoes thermalization, captured by effective kinetic theory. Then, relativistic viscous hydrodynamics describes the QGP expansion until hadronization and freeze-out.}
    \label{fig:hic-stages}
\end{figure}

Understanding the properties of QCD matter under extreme conditions is an ongoing challenge in theoretical physics. QCD exhibits two fundamental properties, namely \emph{asymptotic freedom}~\cite{Gross:1973id,Politzer:1973fx}, where the coupling decreases at short distances, making high energy processes perturbative, and \emph{color confinement}~\cite{Wilson:1974sk}, which prevents the existence of isolated quarks or gluons. Such conditions may be reached in \emph{relativistic heavy-ion collisions}~\cite{Gelis:2021zmx}.  Lattice QCD, the numerical implementation of QCD, is an important tool~\cite{Aoki:2006we}, but due to its known sign problem, other effective theories are needed to gain theoretical insight into heavy-ion collisions at high energies. 

\noindent\textbf{\textit{Collision stages.}} The collision stages are depicted in~\cref{fig:hic-stages}. Before the collision, each nucleus propagates along the light-cone and is described by the \emph{Color Glass Condensate} (CGC), an effective theory for high-energy QCD~\cite{Iancu:2000hn,Iancu:2003xm,Gelis:2010nm,Gelis:2012ri}. Immediately after impact, the glasma, a state of strong classical color fields, is produced in the forward light-cone. The pre-equilibrium matter thermalizes rapidly, as described by effective kinetic theory of QCD~\cite{Baier:2000sb,Arnold:2002zm,Ghiglieri:2015ala,Kurkela:2015qoa,Kurkela:2018oqw,Du:2020dvp,Gelis:2021zmx}. The resulting QGP~\cite{Busza:2018rrf,Shuryak:2004cy} expands due to pressure gradients and is modeled by viscous relativistic hydrodynamics~\cite{Heinz:2013th,Heinz:2013wva,Gale:2013da,Schenke:2010nt,Schenke:2010rr}. The QGP was found to behave as an almost perfect liquid with very small shear viscosity~\cite{Romatschke:2007mq}. Then quarks and gluons recombine into hadrons, kinetic freeze-out is reached, and particles free-stream to detectors. This stage-by-stage description, together with Bayesian analysis~\cite{Auvinen:2017fjw,Nijs:2020roc,JETSCAPE:2020mzn,Bernhard:2016tnd}, has been successful at reproducing experimental data and constraining bulk properties of the QGP. 

\newpage
\noindent\textbf{\textit{Initial stage models.}} Modelling the early stages involves describing the parton structure of the colliding nuclei, the color fields produced after the collision, or both. Parton distribution functions (PDFs) encode the parton content across $x$ regimes and are constrained by global fits to \emph{deep inelastic scattering} (DIS) data, though LHC and RHIC data also contribute, especially to nuclear PDFs. At small-$x$, gluon PDFs grow large and the system enters the saturation regime of the CGC~\cite{Iancu:2000hn,Iancu:2003xm,Gelis:2010nm,Gelis:2012ri}. In the CGC framework, the large-$x$ constituents act as static color sources, and the small-$x$ gluon fields are derived from the Yang-Mills equations~\cite{Yang:1954ek}. The simplest choice is the \textit{McLerran-Venugopalan} (MV) model~\cite{McLerran:1993ni,McLerran:1993ka,McLerran:1994vd}, a Gaussian distribution of color charges with variance proportional to the nuclear thickness. As $x$ decreases, soft modes are integrated out through the JIMWLK equations~\cite{JalilianMarian:1996xn,Jalilian-Marian:1997ubg,Ferreiro:2001qy,Iancu:2000hn}, or the Balitsky-Kovchegov (BK) equation~\cite{Balitsky:1995ub,Kovchegov:1999yj}. This non-linear evolution generates a semi-hard saturation scale $Q_s$ that grows with nuclear size and decreasing $x$, below which gluon distributions saturate~\cite{Albacete:2014fwa,Blaizot:2016qgz}. The CGC organizes collision systems by a power counting hierarchy: in dilute-dilute collisions ($pp$), collinear factorization with PDFs is recovered; in the dilute-dense regime ($pA$), the dense target is resolved via Wilson lines capturing multiple scatterings; in the dense-dense ($AA$) regime, the two nuclei are strong sources and their fields are solved numerically.

The dense-dense collision produces the glasma~\cite{Kovner:1995ja,Kovner:1995ts,Lappi:2006fp,Lappi:2008eq,Fujii:2008km,Fukushima:2011nq}, a state of strong classical color fields described by real-time lattice simulations of Yang-Mills dynamics~\cite{Krasnitz:1998ns,Lappi:2003bi,Lappi:2006hq}. This is implemented in the state-of-the-art IP-Glasma framework~\cite{Schenke:2012wb,McDonald:2017}, which includes impact parameter dependence. The positions of the nucleons are sampled using the Monte Carlo Glauber method, color charge currents are generated for each nucleon using the MV model as initial condition, where the $x$ dependence is parametrized by IP-Sat, but with the option to incorporate JIMWLK evolution~\cite{McDonald:2018wql,Mantysaari:2025kls, Mantysaari:2025tcg}, and the Yang-Mills equations are solved numerically. The color charge distribution is constrained by HERA DIS and exclusive data, providing initial conditions with few free parameters~\cite{Garcia-Montero:2025hys,Garcia-Montero:2023gex}. The fluctuating proton substructure, constrained by HERA data, is incorporated in these frameworks~\cite{Mantysaari:2017cni,Schlichting:2014ipa}. 

Models such as IP-Glasma generate the energy-momentum tensor that initializes viscous hydrodynamics~\cite{Schenke:2012wb,Schenke:2012fw,Muller:2019bwd}, successfully reproducing azimuthal anisotropy coefficients and higher-order flow harmonics~\cite{Gale:2012rq,McDonald:2016vlt,Schenke:2020mbo}. Extensions beyond boost invariance enable 3+1D glasma simulations capturing rapidity-dependent dynamics~\cite{Schenke:2016ksl,Ipp:2017lho,Schlichting:2020wrv,McDonald:2023qwc,Schmidt:2026nin}. A fully consistent NLO treatment of the dense-dense initial state remains an open problem~\cite{Epelbaum:2013waa,Garcia-Montero:2025hys}.
The glasma description provides an explanation for the \emph{ridge}~\cite{Dumitru:2008wn,Dusling:2009ni}, the long-range rapidity correlations observed in heavy-ion and small-system collisions~\cite{CMS:2010ifv}. Glasma flux tubes extending between the colliding nuclei imprint azimuthal correlations that survive into the final state. Other frameworks include the EKRT saturation model~\cite{Niemi:2015qia,Eskola:1999fc} and the T\textsubscript{R}ENTo nuclear geometry parametrization~\cite{Moreland:2014oya}. Coupled to viscous hydrodynamics or kinetic theory, these reproduce the majority of observables and inform Bayesian analyses.

\section{Thermalization}

\begin{figure}[t]
    \centering
    \includegraphics[width=\textwidth]{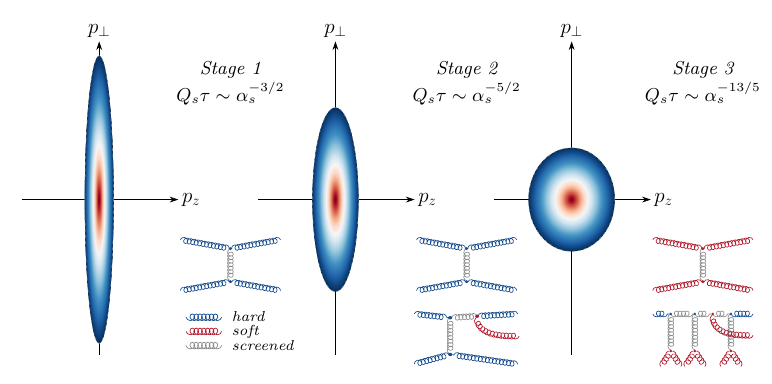}
    \caption{Stages of bottom-up thermalization, starting from the overoccupied glasma to the thermalized QGP, indicating the characteristic time scales and dominant processes at each stage. The distribution function $f(p_\perp,p_z)$ is schematically depicted at each stage, showing the evolution from a highly anisotropic distribution dominated by hard gluons to an isotropic thermal distribution. This uses the Romatschke-Strickland parametrization~\cite{Romatschke:2003ms} to illustrate the anisotropy. At the saturation scale $Q_s$ hard gluons initially isotropize through elastic scatterings \textit{(Stage 1)} until roughly $Q_s\tau \sim \alpha_s^{-3/2}$. A thermal bath of particles is generated through radiation \textit{(Stage 2)} until $Q_s\tau \sim \alpha_s^{-5/2}$. Finally, the particles transfer their energy to the medium \textit{(Stage 3)}, reaching thermalization at $Q_s\tau \sim \alpha_s^{-13/5}$.}
    \label{fig:therm-sequence}
\end{figure}

These initial-state frameworks provide the starting conditions for hydrodynamics, but do not address the essential question of how the glasma fields thermalize. The evolution of the far-from-equilibrium overoccupied glasma fields into the thermalized QGP may be described using the bottom-up thermalization scenario~\cite{Baier:2000sb}. At weak coupling, this proceeds through a sequence of physically different stages, driven by elastic and inelastic QCD scatterings~\cite{Arnold:2002zm}. The first stage of bottom-up may be described by classical statistical simulations~\cite{Schlichting:2013snv,Schlichting:2019abc,Berges:2020fwq}. 

\begin{emphbox}{myblue}{Bottom-up thermalization}
The glasma initial state is characterized by large gluon occupation numbers $f \sim 1/\alpha_s$ and starts with a strongly anisotropic momentum distribution. The key question is at what time $Q_s \tau \sim \alpha_s^{?}$ the longitudinal and transverse momenta equalize, $p_z \sim p_\perp$, and the occupation number reaches $f \sim 1$. Using Effective Kinetic Theory (EKT) of QCD, thermalization proceeds through three parametrically distinct stages, reaching completion at $Q_s \tau \sim \alpha_s^{-13/5}$.
\end{emphbox}
 
\noindent\textbf{\textit{Stages of thermalization.}} Initially, the gluon fields are described by the classical Yang-Mills equations, valid for large occupation numbers $f \sim 1/\alpha_s$. The highly anisotropic fields have $p_\perp \sim Q_s$ and large $p_z$. Below $f \sim 1/\alpha_s$, the Boltzmann equation for the gluon distribution $f(p_\perp, p_z, \tau)$ describes the system. The \emph{Debye mass} $m_D$ is a central quantity, as it sets the soft momentum scale and regularizes the infrared divergence of the elastic scattering cross section. When occupation numbers fall below $1/\alpha_s$, the kinetic description becomes valid, with a scale separation between hard modes at $p_\perp \sim Q_s$ and soft modes at $p_\perp \sim m_D$. This occurs at $Q_s\tau \sim 1$, marking the crossover from the classical glasma to the kinetic regime. The system evolves through three stages defined by which sector dominates the energy density $\varepsilon$, number density $n$, and Debye mass $m_D$, as the hard-soft separation shifts with growing soft particle production. The progress of thermalization is measured by the pressure anisotropy $p_z/p_\perp$, which reaches unity as the system thermalizes. The stages are depicted in~\cref{fig:therm-sequence} and described as follows:

\textit{Stage 1} ($1 \lesssim Q_s\tau \lesssim \alpha_s^{-3/2}$). All quantities are dominated by hard gluons. The longitudinal expansion redshifts $p_z$, driving the system toward greater anisotropy, while transverse momenta and the hard number density stay constant. Elastic collisions counteract the longitudinal redshift by isotropizing the distribution, driving the occupation number toward $f \sim 1$, which ends this stage at $Q_s\tau \sim \alpha_s^{-3/2}$.

\textit{Stage 2} ($\alpha_s^{-3/2} \lesssim Q_s\tau \lesssim \alpha_s^{-5/2}$). The occupation number is of order unity and hard modes control $\varepsilon$ and $n$. Bethe-Heitler inelastic scatterings produce an ensemble of soft gluons that ultimately dominates $m_D$. When the soft and hard number densities become comparable at $Q_s\tau \sim \alpha_s^{-5/2}$, the stage is finalized.

\textit{Stage 3} ($\alpha_s^{-5/2} \lesssim Q_s\tau \lesssim \alpha_s^{-13/5}$). The soft bath dominates both $m_D$ and $n$ and acts as an isotropic thermal medium, while hard gluons still carry most of the energy. Hard particles lose energy through medium-induced gluon radiation. This is suppressed by the \emph{Landau-Pomeranchuk-Migdal} (LPM) effect: when the formation time for a gluon exceeds the mean free path, successive emissions interfere destructively, and at most one scattering occurs within the formation time. The LPM rate is slower than the classical Bethe-Heitler rate. The stage ends when hard gluons deposit all their energy into the soft medium, at $Q_s\tau \sim \alpha_s^{-13/5}$.

\noindent\textbf{\textit{Instabilities \& attractors.}} The soft gluon exchange in the elastic scattering requires medium resummation. This may be handled by \emph{Hard Thermal Loop} (HTL) effective theory~\cite{Braaten:1989mz,Braaten:1990az,Frenkel:1989br}, which resums leading order medium corrections at soft momenta. A known limitation of current kinetic theory implementations~\cite{Kurkela:2015qoa,Kurkela:2018oqw,Du:2020dvp} is the isotropic Debye screening approximation. It neglects plasma instabilities that arise in the highly anisotropic pre-equilibrium plasma~\cite{Mrowczynski:1988dz,Mrowczynski:1993qm,Romatschke:2003ms,Romatschke:2004jh,Kurkela:2011ub,Mrowczynski:2016etf}. These \emph{Weibel instabilities} arise from the momentum anisotropy and generate exponentially growing transverse gauge field modes, but their effect on thermalization has been found to be subleading~\cite{Berges:2020fwq}. A feature of the pre-equilibrium evolution is that, for a wide range of initial conditions, the system rapidly approaches a common \emph{hydrodynamic attractor} before hydrodynamics becomes applicable~\cite{Heller:2015dha}. This attractor has been identified in weakly coupled kinetic theory~\cite{Kurkela:2015qoa,Kurkela:2018xxd,Du:2020zqg,Boguslavski:2023jvg,ScheihingHitschfeld:2024fmf} and in strongly coupled holographic models~\cite{Kurkela:2019set}, revealing universality across coupling regimes. This universal behavior is used to make predictions even with incomplete knowledge of the initial conditions. 

\section{Hard probes}

Hard probes, namely heavy quarks and jets, are produced in early hard scatterings. They traverse the medium from its earliest moments and are potentially sensitive to every stage of the evolution. Their kinematic regimes, the large mass of heavy quarks and high energy of jets, separate them from the bulk medium and enable them to probe its properties~\cite{vanHees:2005wb,Cao:2015hia,Cunqueiro:2021wls,Apolinario:2022vzg}. 

\noindent\textbf{\textit{Jets.}} Jets arise from hard parton scatterings and subsequently shower and fragment in the QGP, making them key probes of the medium. As a parton traverses the medium, it loses energy and acquires transverse momentum $\vec{p}_\perp$ through medium interactions, an effect known as \emph{jet quenching}~\cite{Gyulassy:1990ye,Casalderrey-Solana:2007knd,Majumder:2010qh,Mehtar-Tani:2013pia,Cao:2020wlm,Apolinario:2022vzg}, confirmed by measurements at RHIC and the LHC~\cite{CMS:2012aa}. In the QGP, medium-induced gluon radiation is described by the BDMPS-Z formalism~\cite{Casalderrey-Solana:2007knd,DEramo:2010wup,Hauksson:2021okc,Andres:2022ndd,Hauksson:2023tze,Kuzmin:2023hko} and elastic scatterings are described via Boltzmann transport~\cite{Baier:2008js}. Medium interactions are encoded in the collision kernel $C(\vec{p}_\perp)$, the probability per unit path length for a parton to receive a transverse kick $\vec{p}_\perp$. The jet quenching parameter $\hat{q} \equiv \int \diff^2\vec{p}_\perp/(2\pi)^2\, \vec{p}_\perp^{\,2}\, C(\vec{p}_\perp)$, the transverse momentum squared acquired per unit path length, varies with probe momentum and medium temperature, and has been extracted from data through global analyses~\cite{JET:2013cls,JETSCAPE:2021ehl,Apolinario:2022vzg}. It has been calculated in thermal equilibrium using perturbative QCD~\cite{Arnold:2002zm,Ghiglieri:2015ala}, in strongly coupled plasmas from AdS/CFT~\cite{Liu:2006ug,Liu:2006he}, non-perturbatively on the lattice~\cite{Panero:2013pla,Kumar:2020wvb,Moore:2021jwe} and in the pre-equilibrium, where the anisotropic medium requires $\hat{q}$ to generalize to a tensor $\hat{q}^{ij}$~\cite{Boguslavski:2023alu,Boguslavski:2023waw}. The effect of anisotropy on jet broadening and quenching has been studied~\cite{Hauksson:2021okc,Hauksson:2023tze,Kuzmin:2023hko,Barata:2023qds,Barata:2024bqp,MayoLopez:2025med}. Because energy loss is dependent on path length, in-plane partons traversing the medium lose less energy than their out-of-plane counterparts. This creates a finite azimuthal anisotropy $v_2$ even at high $p_T$~\cite{STAR:2003wqp}. Moreover, jet substructure observables probe how energy and color coherence are redistributed within the shower~\cite{Caucal:2017dpu,CMS:2018mqn,Apolinario:2022vzg}. 

\noindent\textbf{\textit{Heavy quarks.}} Charm and bottom quarks are produced in early hard scatterings and are not created thermally. Any modification of their momentum distribution comes from medium interactions. The \emph{dead cone effect} suppresses small angle radiation relative to massless partons, producing a mass hierarchy in energy loss, as observed experimentally~\cite{ALICE:2021aqk}. Heavy-flavor meson suppression and azimuthal anisotropy were measured and compared with theoretical predictions~\cite{vanHees:2005wb,Cunqueiro:2021wls,Apolinario:2022vzg}. Initial state nuclear effects, such as parton distribution modifications in nuclei~\cite{Eskola:2021nhw} and cold nuclear matter energy loss, must be disentangled from QGP effects to determine transport coefficients. In the QGP, heavy quark transport is described by a Langevin equation with $\eta_D$ the drag and $\kappa$ the momentum diffusion coefficients. The latter is related to the spatial diffusion coefficient using the Einstein relation $D_s = 2T^2/\kappa$. These are constrained from experimental data such as $R_{AA}$ and $v_2$~\cite{vanHees:2005wb,Romatschke:2006bb,Mrowczynski:2016etf}, or evaluated from lattice QCD~\cite{Banerjee:2011ra,Francis:2015daa}. Reproducing the $v_2$ and $R_{AA}$ of $D$ mesons within a single transport model across different collision systems remains an open problem~\cite{Das:2015ana,Noronha-Hostler:2016eow}. Moreover, the dimensionless combination $2\pi T D_s$ is found to be of order unity across QGP temperatures, much smaller than perturbative QCD predictions, suggesting heavy quarks are among the most strongly coupled constituents of the QGP~\cite{Moore:2004tg}. 

\begin{emphbox}{myblue}{Hard probes in the glasma}
The dynamics of hard probes through the glasma are studied using classical transport equations, either Wong's equations or the Fokker-Planck limit. The underlying glasma fields are evaluated analytically, in the proper time expansion or dilute approximation, or obtained from real-time lattice simulations. All transport methods, along with lattice field correlators, yield consistent results for $\hat{q}$ and $\kappa$ in the glasma, which are considerably larger than in the QGP. The effect of the pre-equilibrium stage on heavy quark $R_{AA}$, $v_2$ along with $q\overline{q}$ pair correlations and dissociation, has also been studied within these frameworks. Current limitations include the lack of back-reaction from the probe to the background fields, the absence of a complete quantum treatment of quark propagation and no estimate for the medium-induced radiation. None of the existing studies have connected the glasma transport to the QGP stage, and thus the relative contribution to observables is still undetermined. 
\end{emphbox}

Existing studies of hard probes in the glasma rely on \emph{Wong's equations}~\cite{Wong:1970fu} to describe the classical transport of particles in Yang-Mills fields. Jet momentum broadening and transport coefficient $\hat{q}$ in the glasma were extracted from field correlators on the lattice~\cite{Ipp:2020mjc,Ipp:2020nfu}, from Fokker-Planck equations with glasma fields evaluated analytically using the $\tau$ expansion~\cite{Carrington:2021dvw,Carrington:2022bnv,Carrington:2026qlg}, and from direct numerical simulations~\cite{Avramescu:2022vkd,Avramescu:2023qvv}. Heavy quark momentum broadening and transport coefficient $\kappa$ were studied using Wong's equations in glasma fields~\cite{Ruggieri:2018rzi,Liu:2019lac,Liu:2020cpj,Khowal:2021zoo,Ruggieri:2022kxv,Avramescu:2023qvv} or using field correlators on the lattice~\cite{Avramescu:2023qvv,Avramescu:2023vld}. In the static quark limit, $\kappa$ was evaluated from field correlators in an overoccupied Yang-Mills plasma~\cite{Boguslavski:2018beu,Boguslavski:2020tqz,Boguslavski:2020mzh,Backfried:2024rub} and shown to exhibit plasmon oscillations. All these studies find $\hat{q}$ and $\kappa$ in the glasma to be much larger than in the QGP. The large values in the glasma are consistent with estimates in the subsequent pre-equilibrium phase described by EKT~\cite{Boguslavski:2023fdm,Pandey:2023dzz,Du:2023izb,Boguslavski:2023alu,Boguslavski:2023waw,Boguslavski:2024ezg,Boguslavski:2024jwr,Boguslavski:2025ylx,Altenburger:2025iqa,Lindenbauer:2025ael,Barata:2025agq,Barata:2025zku,Danhoni:2026gve}. 

Most glasma studies extract $\hat{q}$ and $\kappa$, which are not directly measurable. For jets, the effect of various pre-equilibrium stages on $R_{AA}$ and $v_2$ was studied using the EKRT model~\cite{Andres:2019eus,Eskola:1999fc}, the BDMPS-Z framework adapted to pre-hydrodynamics~\cite{Andres:2022bql}, parametrizations of $\hat{q}$ that resemble pre-equilibrium stages~\cite{Adhya:2024nwx}, and different initial temperature profiles~\cite{Zigic:2019sth,Stojku:2020wkh,Ilic:2021ezl}. There is yet no consensus on the magnitude of the early stage effect on these observables~\cite{Pablos:2025cli,Singh:2025duj}. 
For heavy quarks, early calculations of heavy quark spectra in glasma fields found $R_{pA}$ consistent with $D$ meson data~\cite{Ruggieri:2018rzi,Liu:2019lac} and the glasma $R_{AA}$ slope opposite to standard QGP diffusion~\cite{Das:2017dsh,Sun:2019fud}, suggesting the glasma contributes to the $R_{AA}$ and $v_2$ puzzle. Recently, $R_{AA}$ in the glasma was shown to be a subleading effect compared to the contribution from nuclear PDFs~\cite{Avramescu:2024poa,Avramescu:2024xts}. More differential observables such as heavy-flavor angular correlations~\cite{Thomas:2024cso} may offer better sensitivity to the initial stage since they are sensitive to the heavy quark production mechanism~\cite{Altmann:2024icx}. Heavy quark azimuthal correlations in the QGP have been computed using Boltzmann transport~\cite{Nahrgang:2013saa,Scardina:2014lda} and Langevin dynamics~\cite{Cao:2014pka,Zhao:2024oma}, but no calculation other than~\cite{Avramescu:2024poa,Avramescu:2024xts} considers any pre-equilibrium stage. Nevertheless, these studies remain qualitative since the frameworks are not coupled to the QGP phase. 

\newpage
Many of these approaches treat the parton as a classical object, neglecting any quantum effects. While frameworks for both momentum broadening and medium-induced radiation are well-established in the QGP~\cite{Romatschke:2006bb,Baier:2008js,DEramo:2010wup,Hauksson:2021okc,Andres:2022ndd,Hauksson:2023tze,Kuzmin:2023hko}, glasma studies focus exclusively on momentum broadening. 
Collisions and energy loss were studied in non-Abelian plasmas and shown to drive plasma instabilities~\cite{Dumitru:2005gp,Dumitru:2006pz,Dumitru:2007rp,Bodeker:2007fw} but these results were never applied to the glasma. Medium-induced radiation was estimated for soft gluon emission by jets propagating through static electric field correlation domains that resemble the glasma~\cite{Barata:2024xwy} and found to resemble synchrotron radiation~\cite{Aurenche:2012qk}. The quantum Dirac equation for quarks in a glasma background was studied in the context of quark production~\cite{Gelis:2004jp,Gelis:2005pb,Gelis:2015eua,Tanji:2017xiw}, but not applied to estimate momentum broadening or radiation. Alternatively, the light-front quantization technique~\cite{Zhao:2013cma,Vary:2025yqo} was applied to the scattering of a quark off a classical CGC background field~\cite{Li:2020uhl} and to jet propagation in cold nuclear matter and the QGP~\cite{Li:2021zaw,Li:2023jeh,Li:2025wzq}, allowing the study of genuine quantum effects for probes in classical Yang-Mills background fields. Applications to quantum simulation have also been proposed~\cite{Barata:2022wim,Barata:2023clv,Qian:2024gph,Li:2024ufq}. All these frameworks are well suited to improve the treatment of probes in the glasma, but a fully consistent description of quarks in glasma fields is still missing. 

\section{This thesis}

The work presented in this thesis is based on Articles~\cite{ar1,ar2,ar3,ar4,ar5}. Article~\cite{ar1} introduces the \emph{colored particle-in-cell} (CPIC) method to simulate jets and heavy quarks in realistic SU(3) glasma fields and showcases results for momentum broadening and transport coefficients, which are found to be considerably larger than in the QGP. Articles~\cite{ar2,ar3} estimate the effect of the glasma on heavy-flavor spectra and angular correlations. Notably, back-to-back $q\bar{q}$ pairs are deflected by the glasma fields, significantly decorrelating their azimuthal angle. Article~\cite{ar4} investigates the distinction between in-medium kinetic momentum and the gauge-dependent canonical momentum, deriving both quantities from the Dirac Hamiltonian and quantifying their difference in the glasma. Article~\cite{ar5} provides the first quantum treatment of jet evolution in the glasma, using the light-front Hamiltonian approach, and obtains momentum broadening comparable to the classical results.

The rest of this thesis is structured as follows.~\Cref{chap:glasma} reviews the CGC formalism and explains how to obtain the glasma fields. In particular, it covers gluon saturation, classical pure gauge fields, Yang-Mills equations and their numerical solution.~\Cref{chap:hardprobes}, based on Articles~\cite{ar1}-\cite{ar3}, covers the theoretical framework for hard probes in glasma fields, presents the CPIC method along with predictions for the heavy-flavor correlations and spectra in the glasma.~\Cref{chap:quantum}, based on Articles~\cite{ar4} and~\cite{ar5}, addresses the kinetic and canonical momentum distinction and the light-front quantum treatment.~\Cref{chap:conclusion} summarizes the results and discusses remaining open questions. The accompanying~\cref{app:qcd,app:lightfront,app:wong} provide technical details on QCD, conventions, lattice gauge theory, light-front quantization, and the derivation of the classical transport equations.

\chapter{Glasma fields}
\label{chap:glasma}

\section{Color Glass Condensate}

\subsection{Background}

The \emph{Color Glass Condensate} (CGC) provides the theoretical framework to describe the gluon fields of high-energy nuclei. For detailed reviews, see~\cite{Kovchegov:2012mbw, Gelis:2010nm, Iancu:2003xm, Gelis:2015gza, Iancu:2002xk, Mueller:1999yb, Venugopalan:1998zd, Taels:2017shj, Salazar:2021pzs, Lappi:2023heqcd, Lappi:2010ykh, Weigert:2005us, Jalilian-Marian:2005ccm}. The following summarizes the key concepts of this effective theory and the phenomenon of gluon saturation.

\begin{figure}[tp]
    \centering
    \includegraphics[width=0.6\textwidth]{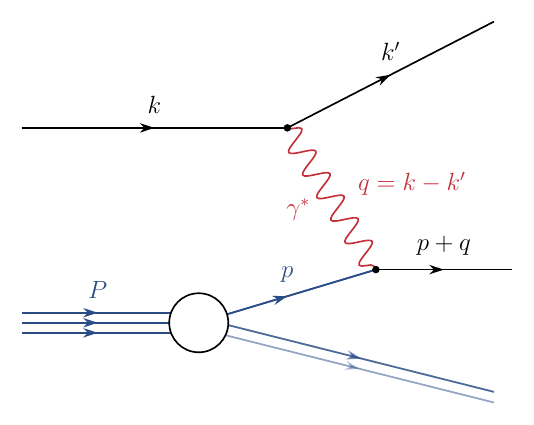}
    \caption{DIS diagram between an electron with momentum $k$ and a proton with momentum $P$. The process is mediated by a highly virtual photon $\gamma^*$, which then interacts with a parton $p$ within the proton. By measuring the final $k^\prime$ of the electron, one may infer the proton structure.}
    \label{fig:dis-diagram}
\end{figure}

\noindent\textbf{\textit{Deep inelastic scattering}}. The structure of matter may be studied in \emph{deep inelastic scattering} (DIS) experiments. In DIS, an electron with momentum $k$ scatters off a proton with momentum $P$ via the exchange of a highly virtual photon with momentum $q = k - k^\prime$, which interacts with a parton carrying momentum $p$ inside the proton. The final electron momentum is $k^\prime$, as depicted in Fig.~\ref{fig:dis-diagram}. The relevant kinematic invariants are the photon virtuality $Q^2 \equiv -q^2 > 0$, the center-of-mass energy squared $s=(k+P)^2$, and the invariant mass squared of the hadronic final state $W^2=(P+q)^2$. The Bjorken-$x$ variable is defined as $x_\mathrm{Bj}\equiv Q^2/(2P\cdot q)$, where $2P\cdot q=W^2+Q^2-M^2$ with $M$ the proton mass. The DIS regime is characterized by $Q^2\gg M^2$, such that the virtual photon resolves the internal structure of the proton rather than scattering off it as a whole.

A natural frame for the analysis is the \emph{infinite momentum frame} (IMF), obtained by boosting the proton to $P\rightarrow\infty$, such that $P^\mu=(P^+,\vec{P}_\perp,P^-)\xrightarrow{\mathrm{IMF}} (P^+, \vec{0}_\perp, 0)$ and all partons become collinear with the proton direction~\cite{Bjorken:1968dy,Susskind:1967rg}. In this frame each parton carries a longitudinal momentum fraction $x\equiv p/P \xrightarrow{\mathrm{IMF}} x_\mathrm{Bj}$. The high-energy limit $s\rightarrow\infty$ corresponds to the small-$x$ limit $x\rightarrow 0$, with $W^2\rightarrow\infty$, where the proton is probed at small longitudinal momentum fractions and is dominated by gluons~\cite{H1:2015ubc}.

\newpage
At \emph{leading order} (LO), the total $\gamma^*p$ cross section is expressed in terms of the structure function $F_2$ as
\begin{equation}
    \label{eq:dis-cross-section}
    \sigma_\mathrm{tot}^{\gamma^*p}=\frac{4\pi^2\alpha_\mathrm{em}}{Q^2}F_2(x,Q^2)\,,\quad\text{with}\,\,F_2(x,Q^2) \sim e_f^2 \left[x q_f(x,Q^2)+x \bar{q}_f(x,Q^2)\right]\,,
\end{equation}
where $\alpha_{\mathrm{em}}$ is the electromagnetic coupling constant, and $F_2$ is sensitive to the quark distribution functions, which depend on the quark charge $e_f$, the quark distribution function $q_f$, the anti-quark distribution function $\bar{q}_f$, with summation over quark flavors $f$ implied. The longitudinal cross section is given by
\begin{equation}
    \label{eq:dis-longitudinal-cross-section}
    \sigma_L^{\gamma^*p}=\frac{4\pi^2\alpha_\mathrm{em}}{Q^2}F_L(x,Q^2)\,,\quad\text{with}\,\,F_L(x,Q^2) \sim \alpha_s(Q^2) e_f^2\, x g(x,Q^2)\,,
\end{equation}
where $\alpha_s(Q^2)$ is the strong coupling constant, and $F_L$ is the longitudinal structure function $F_L(x,Q^2) \equiv F_2(x,Q^2) - 2xF_1(x,Q^2)$ sensitive to the gluon content of the proton. The Callan-Gross relation $F_2(x,Q^2) = 2xF_1(x,Q^2)$ holds for spin-$1/2$ quarks, implying $F_L=0$ at LO. A non-zero $F_L$ directly signals gluon contributions at higher orders. At moderate $x$ and $Q^2$, the structure function $F_2$ is approximately independent of $Q^2$, a phenomenon known as Bjorken scaling. Departures from this scaling are described by the DGLAP evolution~\cite{Gribov:1972ri, Altarelli:1977zs, Dokshitzer:1977sg} and confirmed experimentally at HERA~\cite{Abramowicz:2015mha}. However, DGLAP resums only logarithms of $Q^2$ and is unreliable at small $x$, where logarithms of $1/x$ are large and must be resummed. This is done through the BFKL equation introduced below.

At small-$x$, where the gluon density in the proton grows rapidly, the DIS cross section is more naturally described in the dipole picture. 

\begin{figure}[tp]
    \centering
    \includegraphics[width=0.65\textwidth]{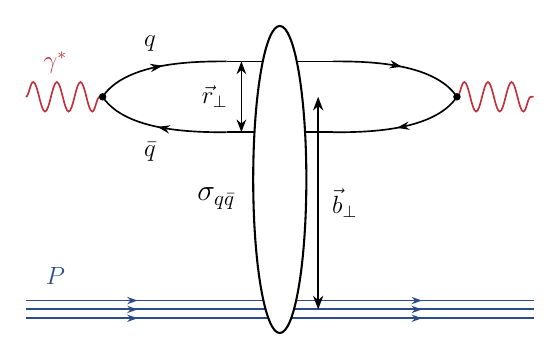}
    \caption{Diagram of the DIS dipole picture. A virtual photon $\gamma^*$ fluctuates into a quark-antiquark pair with transverse separation $\vec{r}_\perp$ and longitudinal momentum fraction $z$, which then interacts with the color field of the proton.}
    \label{fig:dipole-scattering}
\end{figure}

\noindent\textbf{\textit{Dipole picture}}. The cross section for virtual photon-proton scattering in the dipole picture, shown in Fig.~\ref{fig:dipole-scattering}, may be expressed as~\cite{Iancu:2003xm}

\newpage
\begin{equation}
    \label{eq:dipole-cross-section}
    \sigma^{\gamma^* p} = \int_0^1 \mathrm{d}z \int \mathrm{d}^2\vec{r}_\perp \, |\psi^{\gamma^\star\rightarrow q\bar{q}}(z, \vec{r}_\perp)|^2 \, \sigma_{q\bar{q}}(\vec{r}_\perp)\,,
\end{equation}
where $\psi^{\gamma^\star\rightarrow q\bar{q}}$ is the light-cone wavefunction of $\gamma^\star$ splitting into a $q\bar{q}$ with transverse separation $\vec{r}_\perp$ and longitudinal momentum fraction $z$, and $\sigma_{q\bar{q}}$ is the $q\bar{q}$ proton cross section, which encodes the interaction of the dipole with the gluon field of the proton. The cross section depends on $Q^2$ through the light-cone wavefunction.

The dipole cross section may be expressed in terms of the dipole scattering amplitude $N$ using the following expression
\begin{equation}
    \label{eq:dipole-cross-section-amplitude}
    \sigma_{q\bar{q}}(\vec{r}_\perp) = 2 \int \mathrm{d}^2\vec{b}_\perp\, N(\vec{r}_\perp, \vec{b}_\perp)\,,
\end{equation}
where $\vec{b}_\perp$ is the impact parameter of the dipole relative to the proton. This relation is based on the optical theorem, which relates the total cross section to the imaginary part of the scattering amplitude. The scattering amplitude is related to the $S$ matrix element for the dipole-proton scattering as $N(\vec{r}_\perp, \vec{b}_\perp) = 1 - S(\vec{r}_\perp, \vec{b}_\perp)$. The light-cone wavefunction for $\gamma^\star \to q\bar{q}$ splitting can be computed in perturbation theory in QED, with the leading order result given by the transverse $|\psi_T^{\gamma^* \to q\bar{q}}|^2 = \alpha_{\text{em}}/(2\pi^2)\, N_c e_f^2 \{\varepsilon^2[z^2 + (1-z)^2] K_1^2(\varepsilon r_\perp) + m_f^2 K_0^2(\varepsilon r_\perp)\}$ and longitudinal component $|\psi_L^{\gamma^* \to q\bar{q}}|^2 = 2\alpha_{\text{em}}/\pi^2 \, N_c e_f^2 \, Q^2 z^2 (1-z)^2 K_0^2(\varepsilon r_\perp)$, where $\varepsilon^2 \equiv Q^2 z(1-z) + m_f^2$, $e_f$ and $m_f$ are the electric charge and mass of quark flavor $f$, and $K_0$ with $K_1$ are modified Bessel functions of the second kind.

The dipole scattering matrix can be expressed in terms of the Wilson line correlator in the fundamental representation as $S_F(\vec{r}_\perp, \vec{b}_\perp) \equiv (1/N_c) \langle \mathrm{tr}[ U(\vec{b}_\perp + \vec{r}_\perp/2) U^\dagger(\vec{b}_\perp - \vec{r}_\perp/2)]\rangle$ or equivalently as $S_F(\vec{x}_\perp, \vec{y}_\perp) \equiv (1/N_c) \left\langle \mathrm{tr}\left[ U(\vec{x}_\perp) U^\dagger(\vec{y}_\perp) \right] \right\rangle$ in terms of the quark and antiquark transverse positions, with $\vec{r}_\perp = \vec{x}_\perp - \vec{y}_\perp$ and $\vec{b}_\perp = (\vec{x}_\perp + \vec{y}_\perp)/2$, where $U(\vec{x}_\perp)$ is the Wilson line in the fundamental representation along the light-cone direction. More generally, one may define the dipole correlator as
\begin{emphbox}{myblue}{Dipole correlator}
    The dipole correlator in representation $R$ is defined as
    \begin{equation}
        \label{eq:dipole-correlator-general}
        S_R(\vec{x}_\perp, \vec{y}_\perp) \equiv \frac{1}{D_R} \left\langle \mathrm{tr}\left[ U_R(\vec{x}_\perp) U_R^\dagger(\vec{y}_\perp) \right] \right\rangle\,,
    \end{equation}
    where $U_R$ is the Wilson line and $D_R$ is the dimension of representation $R$. The dipole scattering amplitude is related to the dipole correlator 
    \begin{equation}
        \label{eq:dipole-amplitude-correlator}
        N_R(\vec{x}_\perp, \vec{y}_\perp) = 1 - S_R(\vec{x}_\perp, \vec{y}_\perp)\,.
    \end{equation}
\end{emphbox}
The dipole correlator encodes the eikonal interaction of the quark or anti-quark with the gluon field of the proton, and the expectation value $\langle\,\cdots\,\rangle$ is taken over the color charge distribution of the proton, which requires external modeling.

In the eikonal approximation, a quark at transverse position $\vec{x}_\perp$ acquires a color rotation as it passes through the target's gluon field, with the accumulated phase given by
\begin{equation}
    \label{eq:wilson-line-eikonal}
    U(\vec{x}_\perp) = P\exp\!\left[-ig\int_{-\infty}^{+\infty} \diff x^+\, A^-(x^+,\vec{x}_\perp)\right]\,.
\end{equation}
Only the $A^-$ component of the gluon field contributes because the large $k^+$ momentum of the incoming particle selects this component in the high-energy limit. This may be seen by inserting the eikonal ansatz $\psi(x) = V(x)e^{-ip\cdot x}u(p)$ with $p^\mu=(p^+,\vec{0}_\perp,0)$ into the Dirac equation $(i\slashed{\partial}-g\slashed{A})\psi=0$. After using $\slashed{p}\,u(p)=0$ and multiplying by $\gamma^-$, only the component $\partial_+ V = -igA^- V$ survives, whose solution is the path-ordered exponential above.

The expectation value in Eq.~\eqref{eq:dipole-correlator-general} requires a model for the color charge distribution of the proton. A simple phenomenological parametrization of the resulting dipole cross section is the \emph{Golec-Biernat-Wüsthoff} (GBW) model~\cite{Golec-Biernat:1998zce, Golec-Biernat:1999qor}
\begin{equation}
    \label{eq:gbw-dipole-cross-section}
    \sigma_{q\bar{q}}(r_\perp) = \sigma_0 \left(1 - \exp\left[-\frac{r_\perp^2 Q_s^2(x)}{4}\right]\right)\,,
\end{equation}
where the saturation scale is given by $Q_s^2(x) = Q_0^2(x_0/x)^\lambda$ with $Q_0=1\,\mathrm{GeV}$ and parameters $(\sigma_0, x_0, \lambda)$ fitted to HERA data.

\begin{figure}[tp]
    \centering
    \includegraphics[width=\textwidth]{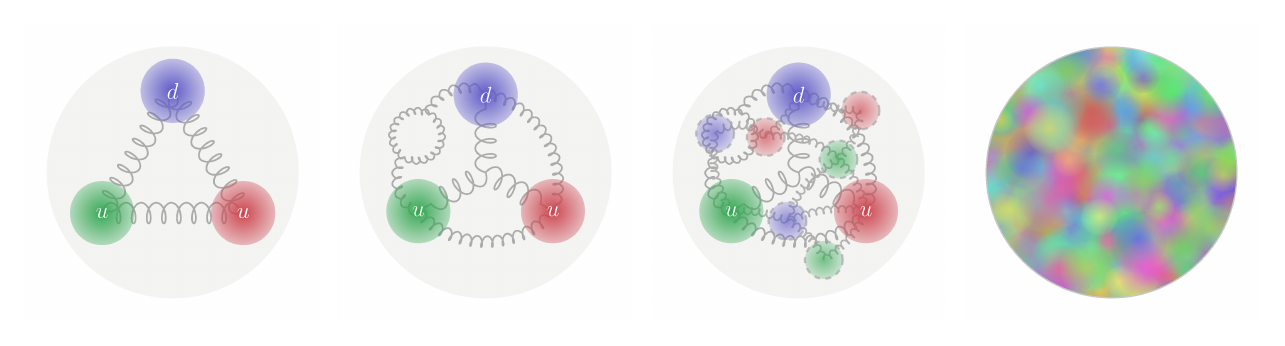}
    \caption{Illustration of parton content inside a proton evolving towards higher energies. A proton is composed of $3$ valence quarks, and as energy increases, it becomes populated with sea quarks and gluons. Going further, the gluon density increases rapidly, leading to a high occupation number of gluons, until saturation is reached.}
    \label{fig:small-x}
\end{figure}

\begin{figure}[tp]
    \centering
    \begin{subfigure}[b]{\textwidth}
        \centering
        \includegraphics[width=0.8\textwidth]{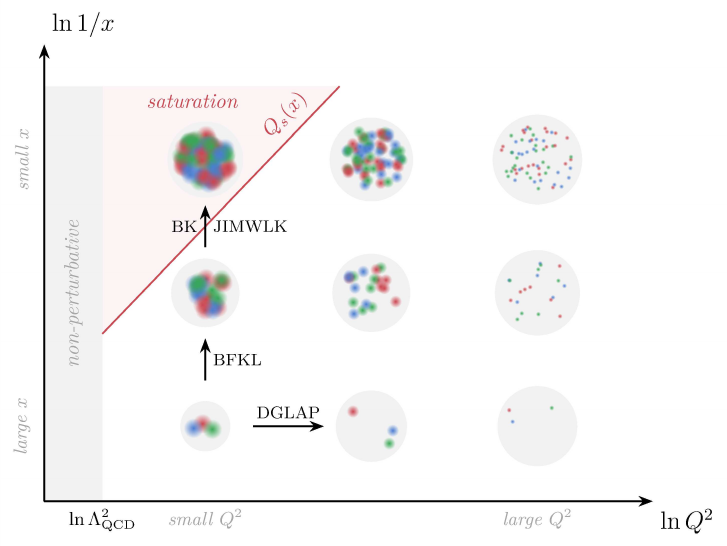}
        \caption{Phase space available through DIS, in terms of $\ln\,1/x$ and $\ln Q^2$. A proton at low energies, which consists of $3$ valence quarks, is shown in the lower left corner. As $Q$ increases, one may probe partons with smaller transverse resolutions. As $x$ decreases, more partons are produced, until the gluon saturation regime is reached at small $x$. Various regions are governed by distinct evolution equations, namely DGLAP towards large $Q^2$, BFKL towards small $x$, and the non-linear JIMWLK and BK equations towards the saturation regime.}
        \label{fig:parton-evolution}
    \end{subfigure}\\[1em]
    \begin{subfigure}[b]{\textwidth}
        \centering
        \includegraphics[width=0.9\textwidth]{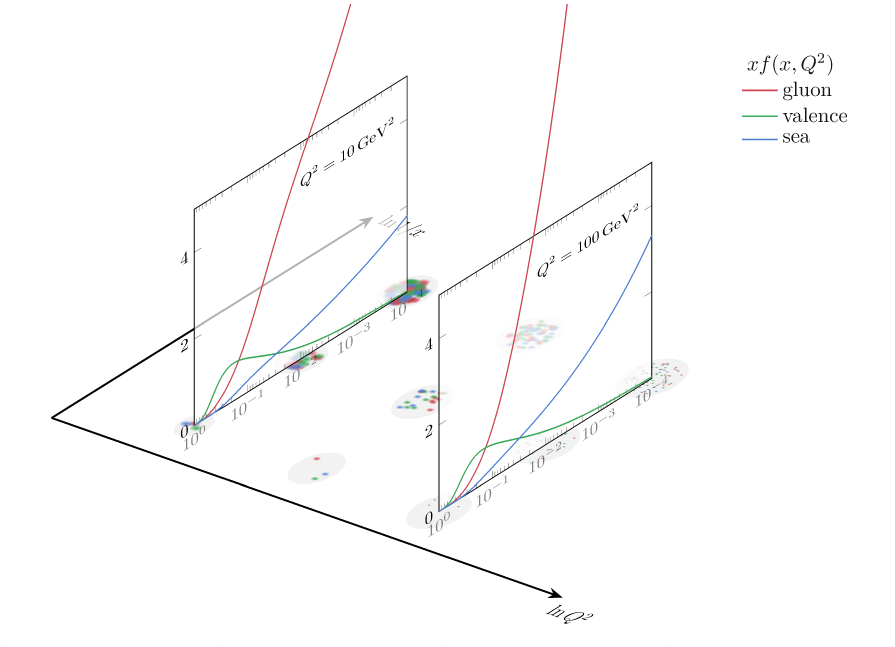}
        \caption{Parton distribution functions for gluons, along with valence and sea quarks \textit{(colored lines)} superimposed over the phase space diagram of DIS, for different values of $Q^2$. One may notice the dramatic growth of the gluon distribution function $xg(x,Q^2)$ when going to small-$x$ values.}
        \label{fig:parton-evolution-pdf}
    \end{subfigure}
    \caption{DIS phase space $(\ln\,1/x, \ln Q^2)$ and parton distribution functions.}
    \label{fig:dis-evolution}
\end{figure}

\noindent\textbf{\textit{Gluon saturation}}. HERA data~\cite{Abramowicz:2015mha} show that the gluon distribution $xg(x,Q^2)$ grows rapidly as $x$ decreases, as illustrated in Figs.~\ref{fig:small-x} and~\ref{fig:dis-evolution}. This small-$x$ evolution is governed by the BFKL equation~\cite{Kuraev:1977fs, Balitsky:1978ic, Forshaw:1997dc, Barone:2002cv}, which describes the rapidity evolution of the dipole amplitude by resumming logarithms of $\alpha_s \ln\,1/x$. Its solution grows exponentially with rapidity and eventually violates the unitarity of the $S$-matrix. This unbounded growth signals the breakdown of the linear approximation at high gluon densities, for which gluon recombination needs to be considered. Gluon saturation occurs when the gluon occupation number reaches $1/\alpha_s$. Non-linear effects tame the growth of the gluon distribution at small $x$. This occurs at the saturation scale $Q_s(x)$, which grows as $x$ decreases. The non-linear evolution equations that describe this saturation regime are the JIMWLK~\cite{JalilianMarian:1996xn, Jalilian-Marian:1997ubg, Ferreiro:2001qy, Iancu:2000hn} renormalization group equation for the rapidity evolution of the color charge distribution, and its large-$N_c$ mean-field limit, the BK equation.

The BK equation~\cite{Balitsky:1995ub, Kovchegov:1999yj} describes the rapidity $y \equiv \ln\,1/x$ evolution of the dipole scattering amplitude $N(\vec{r}_\perp)$, with $\vec{r}_\perp\equiv \vec{x}_\perp-\vec{y}_\perp$. In the large-$N_c$ limit, assuming translational invariance in the impact parameter, this evolution is given by
\begin{equation}
    \label{eq:bk}
    \begin{aligned}
        \partial_y N(\vec{r}_\perp) =& \frac{\alpha_s N_c}{2\pi^2}\int \mathrm{d}^2\vec{q}_\perp\;
        \mathcal{K}_{\rm BFKL}(\vec{r}_\perp,\vec{q}_\perp)\\
    &\times \left[N(\vec{q}_\perp) + N(\vec{r}_\perp - \vec{q}_\perp)
          - N(\vec{r}_\perp) - N(\vec{q}_\perp)\,N(\vec{r}_\perp - \vec{q}_\perp)\right]\,,
    \end{aligned}
\end{equation}
where the kernel
\begin{equation}
    \mathcal{K}_{\rm BFKL}(\vec{r}_\perp,\vec{q}_\perp) \equiv \left|\vec{K}(\vec{q}_\perp) + \vec{K}(\vec{r}_\perp{-}\vec{q}_\perp)\right|^2 = \frac{r_\perp^2}{q_\perp^2(\vec{r}_\perp{-}\vec{q}_\perp)^2}\,,
\end{equation}
is expressed in terms of the gluon emission amplitude $\vec{K}(\vec{r}_\perp) = \vec{r}_\perp/r_\perp^2$. The equation is derived by boosting the $q\bar{q}$ dipole to rapidity $y+\mathrm{d}y$, opening phase space for gluon emission at $\vec{z}_\perp$. In the large-$N_c$ limit, the $q\bar{q}g$ system decomposes into two dipoles with separations $\vec{q}_\perp \equiv \vec{x}_\perp-\vec{z}_\perp$ and $\vec{r}_\perp-\vec{q}_\perp = \vec{z}_\perp-\vec{y}_\perp$, and the mean-field factorisation $\langle S_{q\bar{q}g}\rangle \approx \langle S_{q\bar{q}}(\vec{q}_\perp)\rangle\langle S_{q\bar{q}}(\vec{r}_\perp-\vec{q}_\perp)\rangle$ closes the hierarchy. The non-linear term $N(\vec{q}_\perp)N(\vec{r}_\perp-\vec{q}_\perp)$ implements unitarisation. Dropping it in the dilute limit recovers the linear BFKL equation. 

\subsection{Small-\texorpdfstring{$x$}{x} physics}

\begin{emphbox}{myblue}{Color Glass Condensate}
An effective theory for QCD (the \textit{color}) at high-energy, used to describe the gluon fields of high-energy nuclei. It is based on the separation between large-$x$ partons, which are time dilated into frozen classical color sources (the \textit{glass}) and small-$x$ gluons, which generate a classical field with large occupation number (the \textit{condensate}). Varying the momentum separation cutoff gives renormalization group evolution which resums large logarithms in energy. This formalism predicts the phenomenon of gluon saturation.
\end{emphbox}

\begin{figure}[tp]
    \centering
    \includegraphics[width=0.75\textwidth]{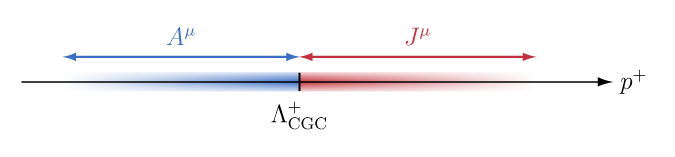}
    \caption{Separation of scales at an arbitrary cutoff $\Lambda_{\mathrm{CGC}}^+$ between large-$x$ partons (classical color sources) and small-$x$ partons (classical gluon fields).}
    \label{fig:momentum-separation}
\end{figure}

\newpage
\noindent\textbf{\textit{Separation of scales}}. In the IMF and light-cone coordinates, defined in~\cref{app:qcd}, there is a separation of scales between small-$x$ and large-$x$ partons inside a high-energy nucleus. The momentum fraction carried by a parton inside the proton in terms of light-cone momenta is $x=p^+/P^+$,
where $p^+$ is the parton light-cone momentum and $P^+$ is the proton light-cone momentum. One separates small-$x$ from large-$x$ degrees of freedom at an arbitrary momentum cutoff $\Lambda_{\mathrm{CGC}}^+$. The large-$x$ partons with $p^+>\Lambda_{\mathrm{CGC}}^+$ are integrated out and represented by a classical color current $J^\mu$, while small-$x$ partons with $p^+<\Lambda_{\mathrm{CGC}}^+$ give rise to classical gluon fields $A^\mu$, see Fig.~\ref{fig:momentum-separation}.

The Heisenberg uncertainty relation for the light-cone time gives
\begin{equation}
    \label{eq:light-cone-time-uncertainty}
    \delta x^+ \sim \frac{1}{p^-} = \frac{2 p^+}{p_\perp^2} \sim \frac{x P^+}{p_\perp^2}\Rightarrow \delta x^+_{\mathrm{small-}x}\ll \delta x^+_{\mathrm{large-}x}\,,
\end{equation}
therefore the large-$x$ partons are long-lived and appear frozen in light-cone time compared to the fast-moving small-$x$ partons. The small-$x$ partons therefore see the large-$x$ partons as static color sources. This may be modeled as $\rho^a = \rho^a(x^-,\vec{x}_\perp)$, with no $x^+$ dependence. This time dilation is the origin of the glassy behavior that gives the CGC its name, in analogy with spin glasses in condensed matter physics. Similarly, the Heisenberg uncertainty relation for the longitudinal coordinate gives
\begin{equation}
    \label{eq:light-cone-uncertainty}
    \delta x^- \sim \frac{1}{p^+} = \frac{1}{xP^+}\Rightarrow \delta x^-_{\mathrm{large-}x}\ll \delta x^-_{\mathrm{small-}x}\,,
\end{equation}
thus the large-$x$ partons are well localized in $x^-$, while the small-$x$ partons are spread over a much larger longitudinal extent. Consequently, the small-$x$ partons see the large-$x$ partons as an infinitely thin sheet of color charge. This justifies the approximation $\rho^a(x^-, \vec{x}_\perp) \sim \delta(x^-)\rho^a(\vec{x}_\perp)$. For a nucleus with $A$ nucleons, the nuclear radius scales as $R_A\sim A^{1/3}$ and the nuclear mass as $M_A= AM$. With Lorentz factor $\gamma=P_A^+/M_A=P^+/M$, the nucleus is Lorentz contracted in the longitudinal direction to a thickness
\begin{equation}
    \label{eq:nucleus-thickness}
    \delta x^-_A \sim \frac{R_A}{\gamma}\sim \frac{A^{1/3}}{P^+}\Rightarrow \delta x^-_{\mathrm{large-}x}\ll \delta x^-_A\ll \delta x^-_{\mathrm{small-}x}\,,
\end{equation}
and the condition for a parton to belong to the small-$x$ regime is $x \ll A^{-1/3}$.

\noindent\textbf{\textit{MV model}}. The \emph{McLerran-Venugopalan} (MV) model~\cite{McLerran:1993ni, McLerran:1993ka, McLerran:1994vd, Kovchegov:1996ty} provides an explicit ansatz for the statistical distribution of color charges in a large nucleus. The approximation for the MV model is well justified in the large-$A$ and moderate to small-$x$ regimes, before any quantum evolution is considered. Moreover, the MV model provides the initial condition $W_0[\rho]$ for the JIMWLK renormalization group flow. A nucleus moving along the $x^+$ direction generates a color current
\begin{equation}
    \label{eq:color-current}
    J^{\mu,a}(x) = \delta^{\mu+} \rho^a(x^-, \vec{x}_\perp)\,,
\end{equation}
where $\rho^a$ is the color charge density of the large-$x$ partons. Since a high-energy heavy nucleus contains many color charge constituents distributed across its transverse area, by the central limit theorem, the color charge density $\rho^a$ is treated as a Gaussian random variable with distribution $W[\rho]$.

\begin{emphbox}{myblue}{MV model}
    The generating functional for the color charge distribution is assumed to be Gaussian
    \begin{equation}
        \label{eq:mv-model}
        W[\rho] = \frac{1}{Z}\,\exp\left[-\frac{1}{2} \int \diff x^- \diff^2\vec{x}_\perp \frac{\rho^a(x^-, \vec{x}_\perp)\rho^a(x^-, \vec{x}_\perp)}{g^2\mu^2(x^-)}\right]\,,
    \end{equation}
    with $g^2\mu^2(x^-)$ the variance of the color charge density per unit transverse area and per unit $x^-$, and $Z$ the normalization factor. This leads to the one-point and two-point functions of the color charge density
    \begin{subequations}
        \label{eq:color-charge-correlators}
        \begin{align}
            &\langle \rho^a(x^-, \vec{x}_\perp) \rangle = 0 \\
            &\langle \rho^a(x^-, \vec{x}_\perp) \rho^b(y^-, \vec{y}_\perp) \rangle = g^2\mu^2(x^-)\delta^{ab} \delta(x^- - y^-) \delta^{(2)}(\vec{x}_\perp - \vec{y}_\perp)\,.
        \end{align}
    \end{subequations}
\end{emphbox}

The average of any observable $\mathcal{O}$ computed from the field $A_\mu$, which is generated with a color charge density $\rho$, is given by
\begin{equation}
    \label{eq:observable-average}
    \langle \mathcal{O}[A_\mu] \rangle = \frac{\int D[\rho]\, W[\rho]\, \mathcal{O}[A_\mu]}{\int D[\rho]\, W[\rho]}\,,
\end{equation}
where $D[\rho]$ is the functional integral over all possible color charge densities. Varying the cutoff $\Lambda_{\mathrm{CGC}}^+$ changes the separation between small-$x$ and large-$x$ degrees of freedom. Then, requiring that physical observables are independent of $\Lambda_{\mathrm{CGC}}^+$ generates the JIMWLK~\cite{JalilianMarian:1996xn, Jalilian-Marian:1997ubg, Ferreiro:2001qy, Iancu:2000hn} renormalization group evolution of the color charge distribution $W[\rho]$.

\subsection{Classical gluon fields}

The classical gluon fields generated by the color charge density of the large-$x$ partons are obtained by solving the \emph{classical Yang-Mills} (CYM) equations
\begin{equation}
    \label{eq:cyme}
    [D_\mu, F^{\mu\nu}] = J^\nu\,,
\end{equation}
with $[D_\mu,\chi] = \partial_\mu \chi - i g [A_\mu, \chi]$ the covariant derivative, $F^{\mu\nu} = \partial^\mu A^\nu - \partial^\nu A^\mu - i g [A^\mu, A^\nu]$ the field strength tensor, and $J^\nu$ the color current generated by the large-$x$ partons, which is given by Eq.~\eqref{eq:color-current}. More details about classical chromodynamics are given in~\cref{app:qcd}. Note that the classical treatment is justified because in the CGC framework gluon occupation numbers are very large.

\noindent\textbf{\textit{Pure gauge fields}}. The gauge fields generated by the static color charges are assumed to be independent of light-cone time $\partial_+ A^\mu = 0$. Imposing the covariant conservation of the light-cone current $[D_+, J^+] = 0$ and noting that $[D_+, J^+]=\partial_+ \rho -ig[A_+, \rho]$ with $\partial_+ \rho=0$, one gets $A^-=0$. These kinematic constraints imply $F^{+-}=F^{i-}=0$, therefore the CYM equation for $\nu=-$ is trivially satisfied.

One seeks a solution in the light-cone gauge $A^+_\mathrm{lc}=0$ and restricts to field configurations satisfying $F^{ij}=0$, that is transverse pure gauge fields. The remaining CYM equation for $\nu=i$ is then trivially satisfied, while the equation for $\nu=+$ reduces to $[D_i, F^{i+}_\mathrm{lc}] = \rho$ with $F^{i+}_\mathrm{lc}=-\partial^+ A^i_\mathrm{lc}$. The solution of the CYM equations is given by the transverse pure gauge fields
\begin{equation}
    \label{eq:pure-gauge-fields}
    A^i_\mathrm{lc}(x^-, \vec{x}_\perp) = \frac{i}{g} V(x^-, \vec{x}_\perp) \partial^i V^\dagger(x^-, \vec{x}_\perp)\,,
\end{equation}
with $V(x^-, \vec{x}_\perp)$ a Wilson line. As described in~\cref{app:qcd}, pure gauge fields may be gauge transformed to zero, with $V(x^-, \vec{x}_\perp)$ the gauge transformation that achieves this.

\begin{figure}[tp]
    \centering
    \includegraphics[width=0.6\textwidth]{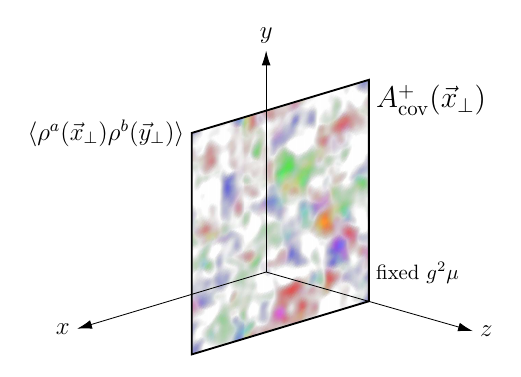}
    \caption{Transverse area of a large nucleus in the MV model, showing the field produced by $\rho^a(\vec{x}_\perp)$. The color charges are independently sampled from a Gaussian distribution with fixed $g^2\mu$ variance. The resulting density generates the covariant gauge field $A^+_\mathrm{cov}(\vec{x}_\perp)$ through the two-dimensional Poisson equation $\Delta_\perp A^+_\mathrm{cov} = -\rho_\mathrm{cov}$.}
    \label{fig:sheets1}
\end{figure}

\noindent\textbf{\textit{Covariant gauge fields}}. Transforming $A^i_\mathrm{lc}$ to the gauge where $\tilde{A}^i=0$, one also finds $\tilde{A}^-=0$ and $\partial_+ \tilde{A}^\mu=0$. The resulting fields satisfy the covariant gauge condition $\partial_\mu A^\mu_\mathrm{cov} = 0$. In the covariant gauge, the only non-zero component of the gauge field is $A^+_\mathrm{cov}(x^-, \vec{x}_\perp)$. The CYM equation $[D_i, F^{i+}]_\mathrm{cov}=0$ reduces to a two-dimensional Poisson equation in the transverse plane
\begin{equation}
    \label{eq:poisson-equation}
    \Delta_\perp A^+_\mathrm{cov}(x^-, \vec{x}_\perp) = -\rho_\mathrm{cov}(x^-, \vec{x}_\perp)\,,
\end{equation}
with $\Delta_\perp=\nabla_\perp^2 \equiv -\partial_i \partial^i$ the transverse Laplacian and $\rho_\mathrm{cov}$ the color charge density in the covariant gauge, related to the color charge density in the light-cone gauge by a gauge transformation $\rho_\mathrm{cov} = V \rho\, V^\dagger$. The gauge field generated by the color charge density $\rho^a(\vec{x}_\perp)$ sourcing Eq.~\eqref{eq:poisson-equation} is shown in Fig.~\ref{fig:sheets1}.

The solution of this equation is given by 
\begin{equation}
    \label{eq:covariant-gauge-fields}
    A^+_\mathrm{cov}(x^-, \vec{x}_\perp) = \int \diff^2\vec{y}_\perp\, G_\perp(\vec{x}_\perp - \vec{y}_\perp)\, \rho_\mathrm{cov}(x^-, \vec{y}_\perp)\,,
\end{equation}
namely the convolution of the color charge density with the two-dimensional Green's function $G_\perp(r_\perp)$. This function satisfies $\Delta_\perp G_\perp(\vec{r}_\perp) = -\delta^{(2)}(\vec{r}_\perp)$ with $\vec{r}_\perp \equiv \vec{x}_\perp - \vec{y}_\perp$, which is given by
\begin{equation}
    \label{eq:green-function}
    G_\perp(\vec{r}_\perp)\equiv \int \frac{\diff^2\vec{k}_\perp}{(2\pi)^2} \frac{1}{k_\perp^2}\exp[i\vec{k}_\perp \cdot \vec{r}_\perp] = \frac{1}{4\pi} \ln\left[\frac{1}{r_\perp^{2} m^2}\right]\,,
\end{equation}
where $m$ is an infrared cutoff.

\noindent\textbf{\textit{Gauge transformation}}. The gauge transformation between the light-cone gauge and the covariant gauge is given by the Wilson line 
\begin{equation}
    \label{eq:wilson-line}
    V(x^-, \vec{x}_\perp) = P \exp\left[-i g \int_{-\infty}^{x^-} \diff z^-\, A^+_\mathrm{cov}(z^-, \vec{x}_\perp)\right]\,,
\end{equation}
where $P$ denotes path ordering in the $x^-$ direction. This is obtained from the gauge transformation $A^+_\mathrm{lc} = V A^+_\mathrm{cov} V^\dagger + i/g\, V \partial^+ V^\dagger$ which gives $\partial^+ V^\dagger = i g\, A^+_\mathrm{cov} V^\dagger$ and whose solution is given by the aforementioned Wilson line.  

Combining the Poisson solution in Eq.~\eqref{eq:covariant-gauge-fields} with the Wilson line gauge transformation Eq.~\eqref{eq:wilson-line}, the classical gluon field of a single nucleus in the light-cone gauge reduces to a transverse pure gauge field, fully determined by the color charge distribution $\rho$. This result applies independently to each of the two colliding nuclei.

\begin{emphbox}{myblue}{Gluon fields before collision}
    For nucleus (1) propagating along $x^+$, the classical gluon field in the light-cone gauge $A^+_{lc}=0$ reduces to a transverse pure gauge field $\alpha^i\equiv A^i_{lc}$ given by
    \begin{equation}
        \label{eq:cgc-pure-gauge}
        \alpha^i_{(1)}(\vec{x}_\perp) = \frac{i}{g}\, V_{(1)}(\vec{x}_\perp)\, \partial^i V^\dagger_{(1)}(\vec{x}_\perp)\,,
    \end{equation}
    where $V_{(1)}(\vec{x}_\perp)$ is the Wilson line integrating the covariant gauge field $A^+_{\mathrm{cov},(1)}$ over the entire longitudinal extent of the nucleus
    \begin{equation}
        \label{eq:cgc-wilson-line}
        V_{(1)}(\vec{x}_\perp) = P\exp\!\left[-ig\int_{-\infty}^{+\infty} \diff x^-\, A^+_{\mathrm{cov},(1)}(x^-, \vec{x}_\perp)\right]\,,
    \end{equation}
    and with $A^+_{\mathrm{cov},(1)}$ sourced by the color current $J^+_{(1)}$ using the Poisson equation $\Delta_\perp A^+_{\mathrm{cov},(1)} = -\rho_{(1)}$. A similar solution holds for nucleus (2) traveling along $x^-$, with color current $J^-_{(2)}(\vec{x}_\perp)$, yielding the pure gauge field $\alpha^i_{(2)}(\vec{x}_\perp)$ from the corresponding Wilson line $V_{(2)}(\vec{x}_\perp)$.
\end{emphbox}

\begin{figure}[tp]
    \centering
    \includegraphics[width=0.7\textwidth]{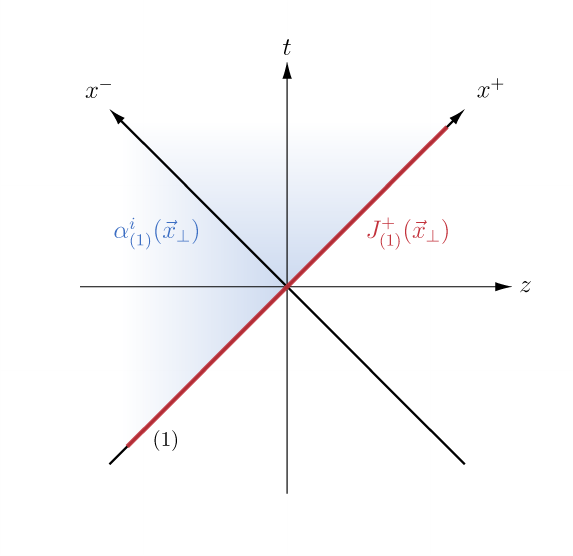}
    \caption{Light-cone diagram of a single nucleus $(1)$ moving along the $x^+$ direction with a classical color current $J^+$, generating a transverse pure gauge field $\alpha^i$.}
    \label{fig:lightcone1}
\end{figure}

These single nucleus pure gauge fields, shown as an example for nucleus $(1)$ in Fig.~\ref{fig:lightcone1}, fix the initial condition for the glasma evolution. The only physical quantity needed in this framework is the color charge density $g^2\mu$, which is related to the saturation momentum. Since the same Wilson line correlator $\langle VV^\dagger\rangle$ enters the DIS cross section in the dipole picture, $g^2\mu$ and equivalently $Q_s$ can be constrained from HERA data. To understand what $g^2\mu$ represents physically, we make a brief digression to extract the saturation momentum of a single nucleus from the MV model Wilson line correlator.

\subsection{Gluon saturation}

Using eikonal Wilson lines and the MV model, one may compute the dipole correlator and extract the saturation momentum $Q_s$, relevant for glasma physics.

\noindent\textbf{\textit{Dipole correlator}}. One starts from the adjoint dipole correlator
\begin{equation}
    \label{eq:dipole-correlator}
    S_A(x^-, \vec{r}_\perp) \equiv \frac{1}{N_c^2 - 1} \left\langle \mathrm{tr}\left[\, V_A(x^-, \vec{x}_\perp) V_A^\dagger(x^-, \vec{y}_\perp)\right] \right\rangle\,,
\end{equation}
with $V_{A}$ the Wilson line in the adjoint representation, which is related to the Wilson line in the fundamental representation by $V_{A}^{ab} = 2\, \mathrm{tr}\, [t^a V t^b V^\dagger]$ with $t^a$ the generators of the fundamental representation, and $D_A=N_c^2 - 1$ the dimension of the adjoint representation. The dipole correlator in the MV model is given by
\begin{equation}
    \label{eq:dipole-correlator-mv}
    S_{A}^{\mathrm{MV}}(r_\perp) = \exp\left[-\frac{r_\perp^2}{4}\mathcal{Q}_s^2(r_\perp)\right]\,,
\end{equation}
with the saturation scale $\mathcal{Q}_s$ expressed as
\begin{equation}
    \label{eq:saturation-momentum}
    \mathcal{Q}_s^2(r_\perp) \equiv N_c\,g^{4}\int_{-\infty}^{\infty}\mathrm{d}x^-\mu^2(x^-)\frac{1}{4\pi}\ln\left[\frac{1}{r_\perp^{2} m^2}\right]\,,
\end{equation}
where $r_\perp = |\vec{r}_\perp|$ is the transverse size of the dipole, and $m$ is an infrared cutoff, usually chosen as the confinement scale $\Lambda_\mathrm{QCD}$. The saturation momentum is usually treated as independent of $r_\perp$. It is conventionally defined by solving $S^{\mathrm{MV}}(r_\perp = c/Q_s) = S_c$ for $Q_s$, where $S_c$ is a constant. A common choice is $c = \sqrt{2}$ and $S_c = e^{-1/2}$. Alternatively, it may be defined at $c=2$ and $S_c = e^{-1}$, used below for the gluon occupation number.

The derivation of~\cref{eq:dipole-correlator-mv,eq:saturation-momentum} involves computing the $x^-$ evolution of the dipole correlator~\cite{Mantysaari:2015uca}, defined by the discretized partial derivative at fixed $\vec{r}_\perp$ as
\begin{equation}
    \label{eq:partial-derivative-dipole}
    \frac{\partial}{\partial x^-}S_{A}(x^-) = \lim_{\epsilon \to 0} \frac{1}{\epsilon}\left[S_{A}(x^- + \epsilon) - S_{A}(x^-)\right]\,.
\end{equation}
The Wilson line at $x^- + \epsilon$ is expanded as
\begin{equation}
    \label{eq:wilson-line-expansion}
    V_A(x^- + \epsilon) = V_A(x^-)\left\{\mathbbm{1} - ig\epsilon\,A^{+,a}_\mathrm{cov}(x^-)T^a - \frac{1}{2}g^2\epsilon^2\left[A^{+,a}_\mathrm{cov}(x^-)T^a\right]^2+\mathcal{O}(\epsilon^3)\right\}\,,
\end{equation}
where $T^a$ are the generators of the adjoint representation satisfying $T^aT^a=C_A$ with $C_A=N_c$ the quadratic Casimir. Assuming $\rho_\mathrm{cov}$ satisfies~\cref{eq:color-charge-correlators}, the covariant gauge fields sourced by~\cref{eq:poisson-equation} have $\langle A^{+,a}_\mathrm{cov}(x^-, \vec{x}_\perp) \rangle = 0$ and two-point correlator
\begin{equation}
    \label{eq:alpha-correlators}
    \langle A^{+,a}_\mathrm{cov}(x^-, \vec{x}_\perp)\, A^{+,b}_\mathrm{cov}(y^-, \vec{y}_\perp) \rangle = g^2 \mu^2(x^-)\,\delta^{ab}\, \delta(x^- - y^-)\, L(\vec{x}_\perp - \vec{y}_\perp)\,,
\end{equation}
where $L(\vec{x}_\perp - \vec{y}_\perp)$ is the convolution of the Green's function in Eq.~\eqref{eq:green-function}
\begin{equation}
    \label{eq:convolution-green-function}
    \begin{split}
        L(\vec{x}_\perp - \vec{y}_\perp) &\equiv \int \diff^2\vec{z}_\perp\, G_\perp(\vec{x}_\perp - \vec{z}_\perp) G_\perp(\vec{y}_\perp - \vec{z}_\perp)\\
        &= \int \frac{\diff^2\vec{k}_\perp}{(2\pi)^2} \frac{1}{k_\perp^4}\exp[i\vec{k}_\perp \cdot (\vec{x}_\perp - \vec{y}_\perp)]\,.
    \end{split}
\end{equation}
Discretising $x^-=n\epsilon$ and $y^-=m\epsilon$ with $n,m=0,1,2,\ldots$ and replacing $\delta(x^- - y^-)=\delta_{nm}/\epsilon$, one obtains
\begin{equation}
    \label{eq:dipole-correlator-epsilon}
    S_{A}^{\mathrm{MV}}(x^-+\epsilon, \vec{r}_\perp) = S_{A}^{\mathrm{MV}}(x^-, \vec{r}_\perp)\left\{1-\epsilon N_c g^4\mu^2(x^-)[L(\vec{0}_\perp)-L(\vec{r}_\perp)]\right\}\,.
\end{equation}
Solving the recursion in Eq.~\eqref{eq:partial-derivative-dipole} and taking the continuum limit gives
\begin{equation}
    \label{eq:dipole-correlator-solution}
    S_{A}^{\mathrm{MV}}(x^-, \vec{r}_\perp) = \exp\left[-N_c\,g^{4}\int_{-\infty}^{\infty}\mathrm{d}x^-\mu^2(x^-)\,[L(\vec{0}_\perp)-L(\vec{r}_\perp)]\right]\,.
\end{equation}
Restricting to the momentum range $m \leq k_\perp \leq 1/r_\perp$ relevant for small dipole sizes
\begin{equation}
    \label{eq:dipole-correlator-regularized}
    L(\vec{0}_\perp)-L(\vec{r}_\perp) \approx \frac{r_\perp^2}{16\pi} \ln\left[\frac{1}{r_\perp^2 m^2}\right]\,,
\end{equation}
yielding Eqs.~\eqref{eq:dipole-correlator-mv} and~\eqref{eq:saturation-momentum}. 

\noindent\textbf{\textit{Gluon occupation number}}. The light-cone gluon distribution function is defined as~\cite{Iancu:2003xm, Iancu:2005jft}
\begin{equation}
    \label{eq:gluon-distribution-function}
    xg(x, Q^2) \equiv \int \mathrm{d}^3\vec{k}\,\theta(Q^2 - k_\perp^2) \,x\delta\left(x - \frac{k^+}{P^+}\right) \frac{\mathrm{d}N}{\mathrm{d}^3\vec{k}}\,,
\end{equation}
where $\vec{k}\equiv (k_\perp, k^+)$ is the gluon momentum. The gluon occupation number in momentum space is given by the expectation value of the gluon number operator
\begin{equation}
    \label{eq:gluon-occupation-number}
    \frac{\mathrm{d}N}{\mathrm{d}^3\vec{k}} \equiv \langle a_a^{i\dagger}(x^+, \vec{k})\, a_a^i(x^+, \vec{k}) \rangle = \frac{2k^+}{(2\pi)^3} \langle A_{\mathrm{lc},a}^i(x^+, \vec{k})\, A_{\mathrm{lc},a}^i(x^+, -\vec{k}) \rangle\,,
\end{equation}
where the gauge fields quantized in light-cone coordinates are related to the gluon creation and annihilation operators, see~\cref{app:lightfront}. The field strength tensor in light-cone gauge and momentum space is given by $F^{i+} = i k^+ A^i$ and thus the gluon distribution function may be expressed in terms of the field strength tensor as
\begin{equation}
    \label{eq:gluon-distribution-function-field-strength}
    xg(x, Q^2) = \int \frac{\mathrm{d}^2\vec{k}_\perp}{4\pi^3}\, \theta(Q^2 - k_\perp^2) \langle F_{\mathrm{lc},a}^{i+}(x^+, \vec{k}_\perp) F_{\mathrm{lc},a}^{i+}(x^+, -\vec{k}_\perp) \rangle\,.
\end{equation}
Gauge invariance is enforced by inserting Wilson lines $U_\gamma(\vec{x},\vec{y})=P\exp\big[ig\int_\gamma \mathrm{d}\vec{z}\cdot \vec{A}_\mathrm{lc}(z^-,\vec{z}_\perp)\big]$, where $\vec{A}\equiv (A^+, \vec{A}_\perp)$. The gauge-invariant gluon distribution is computed from the operator $F_\mathrm{lc}^{i+}(\vec{x})\, U_\gamma(\vec{x},\vec{y})\, F_\mathrm{lc}^{i+}(\vec{y})$ with $\gamma$ a straight line connecting $\vec{x}$ and $\vec{y}$. In the light-cone gauge $A_\mathrm{lc}^+=0$, imposing the retarded boundary condition $A_\mathrm{lc}^i(x^-,\vec{x}_\perp)\rightarrow 0$ as $x^-\to -\infty$, the Wilson line reduces to identity $U_\gamma(\vec{x},\vec{y})\rightarrow\mathbbm{1}$, and the distribution reduces to Eq.~\eqref{eq:gluon-distribution-function-field-strength}.

For a nucleus that is homogeneous in the transverse plane, the gluon occupation number is
\begin{equation}
    \label{eq:gluon-occupation-number-homogeneous}
    f_g(k_\perp) = \frac{1}{(N_c^2-1) S_\perp}\big\langle F_{\mathrm{lc},a}^{i+}(\vec{k})\, F_{\mathrm{lc},a}^{i+}(-\vec{k})\big\rangle\,,
\end{equation}
where $S_\perp$ is the transverse area of the nucleus. This is obtained from the gluon number distribution for a transversely homogeneous nucleus $n_g(k_\perp)=xg(x,Q^2)/S_\perp$ by defining $f_g(k_\perp) = (2\pi)^3/[2(N_c^2-1)]\, n_g(k_\perp)$. The field strength tensor correlator in coordinate space can be factorised as
\begin{equation}
    \label{eq:field-strength-corr-fact}
    \big\langle F_{\mathrm{lc},a}^{i+}(\vec{x})\, F_{\mathrm{lc},a}^{i+}(\vec{y})\big\rangle_A = \langle V_A^{ab} (\vec{x}) \,V_A^{\dagger, ca}(\vec{y}) \rangle \langle\partial^i A^+_{\mathrm{cov}, b} (\vec{x})\,\partial^iA^+_{\mathrm{cov}, c} (\vec{y})\rangle\,.
\end{equation}
The relation $F^{i+,a}_\mathrm{lc} = V_A^{ab}\,\partial^i A^{+,b}_\mathrm{cov}$ is obtained from $gA^i_\mathrm{lc} = iV\partial^i V^\dagger$ by applying $\partial^-$ to the path-ordered $V^\dagger(\vec{x}_\perp)$ using the Leibniz rule, such that the derivative acts on each $x^-$ link in the path. Taking $F^{i+,a}_\mathrm{lc} = \partial^- A^{i,a}_\mathrm{lc}$ then isolates the contribution from the single sheet at $x^-$. In the MV model, color sources at different $x^-$ are uncorrelated. Therefore, the Wilson line $V_A^{ab}$, which integrates sources over all $z^- < x^-$, and $\partial^i A^{+,b}_\mathrm{cov}$, generated only by the sheet at $x^-$, are independent. The expectation value therefore factorises as in~\cref{eq:field-strength-corr-fact}. Using the convolution of the Green's function in Eq.~\eqref{eq:convolution-green-function} together with $-\nabla_\perp^2 L(\vec{r}_\perp)=G_\perp(\vec{r}_\perp)$, one derives
\begin{equation}
    \langle \partial^i A^+_{\mathrm{cov}, b} (x^-,\vec{x}_\perp)\,\partial^iA^+_{\mathrm{cov}, c} (y^-, \vec{y}_\perp) \rangle =\delta_{bc} \, \delta(x^- - y^-)g^2\mu^2(x^-)\frac{1}{4\pi} \ln \left[\frac{1}{r_\perp^2 m^2}\right]\,.
\end{equation}

\newpage
Substituting the color factors $\delta_{bc}$ and $\delta(x^- - y^-)$ into the field strength tensor correlator yields the dipole correlator $\langle \mathrm{tr}[V(x^-,\vec{x}_\perp) V^\dagger(x^-, \vec{y}_\perp)]\rangle$, computed in the adjoint representation in~\cref{eq:dipole-correlator,eq:dipole-correlator-mv}. Taking the Fourier transform then gives the adjoint field strength correlator in momentum space
\begin{align}
    \label{eq:field-strength-correlator-ft}
    \langle F_{\mathrm{lc},a}^{i+}(\vec{k}) F_{\mathrm{lc},a}^{i+}(-\vec{k}) \rangle_A
    &=\int \mathrm{d}^3\vec{x} \int \mathrm{d}^3\vec{y}\,\exp[i(\vec{x}-\vec{y})\cdot\vec{k}]\langle F_{\mathrm{lc},a}^{i+}(\vec{x}) F_{\mathrm{lc},a}^{i+}(\vec{y}) \rangle_A\\
    &=\frac{4S_\perp(N_c^2 - 1)}{g^2 N_c}\int \mathrm{d}^2\vec{r}_\perp \,\frac{\exp[i\vec{k}_\perp \cdot \vec{r}_\perp]}{r_\perp^2}\left(1 - \exp\!\left[-\frac{r_\perp^2}{4} \mathcal{Q}_s^2(r_\perp)\right]\right)\,.\notag
\end{align}
The saturation momentum $\mathcal{Q}_s$ is given by~\cref{eq:saturation-momentum}. The $1/r_\perp^2$ and $(1-S_A^{\mathrm{MV}})$ factors, with $S_A^{\mathrm{MV}}$ given by~\cref{eq:dipole-correlator-mv}, both arise from recognising the $x^-$ integrand as a total derivative. Denoting the partial integral as an $x^-$ dependent saturation momentum $\mathcal{Q}_s^2(x^-,r_\perp) \equiv N_c\,g^{4}/(4\pi)\int_{-\infty}^{x^-}\!\mathrm{d}z^-\,\mu^2(z^-)\ln[1/(r_\perp^2 m^2)]$, one notes that the integrand $g^2\mu^2(x^-)\ln[1/(r_\perp^2 m^2)]/(4\pi)\cdot\exp[-r_\perp^2\mathcal{Q}_s^2(x^-,r_\perp)/4]$ equals $-4/(r_\perp^2 g^2N_c)\,(\mathrm{d}/\mathrm{d}x^-)\exp[-r_\perp^2\mathcal{Q}_s^2(x^-,r_\perp)/4]$. Using $\mathcal{Q}_s^2(-\infty,r_\perp) = 0$ and $\mathcal{Q}_s^2(\infty,r_\perp) = \mathcal{Q}_s^2(r_\perp)$, the $x^-$ integral evaluates to $4/(r_\perp^2 g^2N_c)\,(1-S_A^{\mathrm{MV}}(r_\perp))$. An alternative derivation using the Weizsäcker-Williams gluon distribution is given in Ref.~\cite{Lappi:2017tfa}.

The gluon occupation number in the MV model becomes
\begin{equation}
    \label{eq:gluon-occupation-mv}
        f_g(k_\perp)=\frac{4}{g^2 N_c}\int \mathrm{d}^2\vec{r}_\perp \,\frac{\exp[i\vec{k}_\perp \cdot \vec{r}_\perp]}{r_\perp^2}\left(1 - \exp\!\left[-\dfrac{r_\perp^2}{4} \mathcal{Q}_s^2(r_\perp)\right]\right)\,.
\end{equation}

\begin{figure}[tp]
    \centering
    \includegraphics[width=0.55\textwidth]{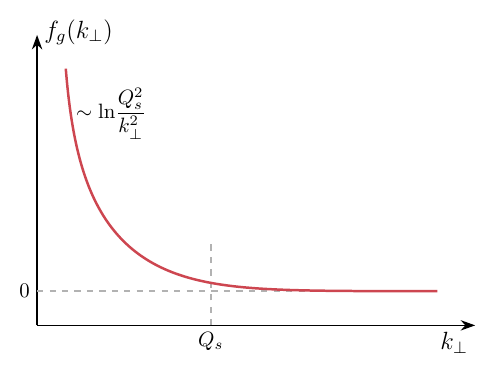}
    \caption{The gluon occupation number $f_g$ in the MV model as a function of transverse momentum $k_\perp$. Below the saturation scale $Q_s$, the occupation number grows only logarithmically, signalling the gluon saturation regime.}
    \label{fig:gbw_dist}
\end{figure}

The saturation scale $Q_s$ is defined by requiring the exponent to be unity at $r_\perp = 2/Q_s$, yielding $Q_s^2 = \mathcal{Q}_s^2(2/Q_s) \propto \ln\!\left[Q_s^2/m^2\right]$. The gluon distribution exhibits two distinct regimes, shown in Fig.~\ref{fig:gbw_dist}. For $k_\perp \gg Q_s$, the occupation number falls as $f_g(k_\perp) \sim Q_s^2/k_\perp^2$, while for $k_\perp \ll Q_s$, it grows slowly as $f_g(k_\perp) \sim \ln\!\left[Q_s^2/k_\perp^2\right]$. This signals the start of gluon saturation.

\begin{emphbox}{myblue}{Saturation momentum}
The saturation momentum scale gives the transition between the two regimes of the gluon distribution. It is defined in the MV model by requiring the exponent of the dipole correlator to be unity at a transverse separation of $r_\perp = 2/Q_s$. Above this scale, the occupation number falls as a power law, while below it grows logarithmically. 
\end{emphbox}

The saturation momentum $Q_s$ is the central physical scale of the glasma. It is connected to the color charge density $g^2\mu$, the only free parameter in the model. With the CGC framework established for each nucleus, we now turn to the collision of two such nuclei and explain how the glasma fields are generated.

\section{Boost invariant glasma}
\label{sec:glasma}

\begin{emphbox}{myblue}{Glasma fields}
The classical gluon fields generated in the forward light-cone after the collision of two high-energy nuclei in the CGC framework. Before the collision, each nucleus generates a pure gauge field in the transverse plane. At the collision, these pure gauge fields are patched together to give the initial condition, which is then evolved using the classical Yang-Mills equations. Each nucleus has a pure gauge field with vanishing field strength tensor and the collision generates the non-trivial color electric and magnetic field configuration of the glasma.
\end{emphbox}

\subsection{The glasma fields}
\label{sec:glasma-fields}

\noindent\textbf{\textit{Color sheet approximation}}. In the color sheet approximation, each nucleus is treated as an infinitely thin sheet of color charge, $\rho^a(x^-,\vec{x}_\perp)\sim \delta(x^-)\rho(\vec{x}_\perp)$. This gives $\langle \rho^a(\vec{x}_\perp) \rangle = 0$ together with 
\begin{equation}
    \label{eq:color-charge-correlators-sheet}
    \langle \rho^a(\vec{x}_\perp) \rho^b(\vec{y}_\perp) \rangle = g^2\mu^2 \delta^{ab} \delta^{(2)}(\vec{x}_\perp - \vec{y}_\perp)\,,
\end{equation}
where $\mu^2(x^-)=\mu^2 \delta(x^-)$. In this approximation, the transverse pure gauge field simplifies to $A^i_\mathrm{lc}(x^-, \vec{x}_\perp)=\theta(x^-)A^i_\mathrm{lc}(\vec{x}_\perp)$, where $\theta$ is the Heaviside step function. The field $A^i_\mathrm{lc}(\vec{x}_\perp)$ generated by the color sheet is given by
\begin{equation}
    \label{eq:sheet-pure-gauge-field}
    \alpha^i(\vec{x}_\perp) \equiv A^i_\mathrm{lc}(\vec{x}_\perp) = \frac{i}{g} V(\vec{x}_\perp) \partial^i V^\dagger(\vec{x}_\perp)\,,
\end{equation}
with the Wilson line having only transverse dependence
\begin{equation}
    \label{eq:sheet-wilson-line}
    V(\vec{x}_\perp) = P \exp\left[-i g \int_{-\infty}^{\infty} \diff x^-\, A^+_\mathrm{cov}(x^-, \vec{x}_\perp)\right]\,,
\end{equation}
where $A^+_\mathrm{cov}(x^-,\vec{x}_\perp)$ is the solution of the two-dimensional Poisson equation in Eq.~\eqref{eq:poisson-equation}. 

In numerical calculations, the longitudinal dependence in $\rho(x^-,\vec{x}_\perp)$ is not approximated by the strict delta function $\rho(x^-,\vec{x}_\perp)\rightarrow \delta(x^-)\rho(\vec{x}_\perp)$, but rather by a discretized version of the color charge density in $x^-$~\cite{Fukushima:2007ki}. The naive correlator in Eq.~\eqref{eq:color-charge-correlators-sheet} is replaced by
\begin{equation}
    \label{eq:color-charge-correlators-discretized}
    \langle \rho^a_n(\vec{x}_\perp) \rho^b_m(\vec{y}_\perp) \rangle = g^2\mu^2 \delta^{ab} \frac{\delta_{nm}}{N_s} \delta^{(2)}(\vec{x}_\perp - \vec{y}_\perp)\,,
\end{equation}
with $N_s$ the number of discretized sheets in $x^-$ and $\delta_{nm}$ the Kronecker delta for the discretized $x^-$ coordinate. Each individual color sheet obeys the Poisson equation $\Delta_\perp A^+_{\mathrm{cov},n}(\vec{x}_\perp) = -\rho_n(\vec{x}_\perp)$ with $n=1,\ldots,N_s$. The Wilson line is then given by the product of the Wilson lines for each sheet $V(\vec{x}_\perp) = \prod_{n=1}^{N_s} \exp\left[-i g A^+_{\mathrm{cov},n}(\vec{x}_\perp)\right]$. The pure gauge field solution for all color sheets is $\alpha^i(\vec{x}_\perp)=(i/g) V(\vec{x}_\perp) \partial^i V^\dagger(\vec{x}_\perp)$. Finally, each nucleus $(1,2)$ moving along the $x^\pm$ direction generates a color current $J^\mu_{(1,2)}$ as in Eq.~\eqref{eq:color-current} with color charge density $\rho^a_{(1,2)}(\vec{x}_\perp)$, which produces the corresponding pure gauge field $\alpha^i_{(1,2)}(\vec{x}_\perp)$. These serve as the input for the glasma initial condition.

\begin{figure}[tp]
    \centering
    \includegraphics[width=0.7\textwidth]{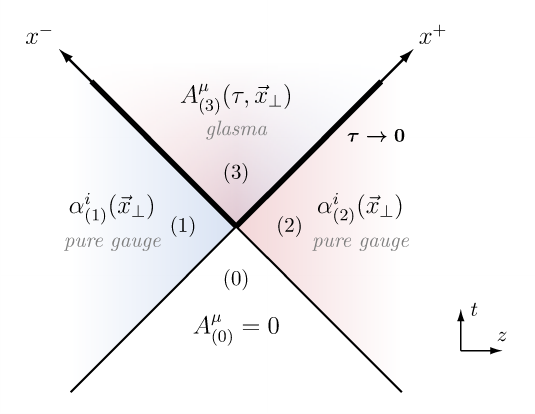}
    \caption{Light-cone diagram of the collision in the CGC framework. Region $(0)$ has the vacuum $A_{(0)}^\mu=0$. Regions $(1)$ and $(2)$ contain the two nuclei propagating along $x^+$ and $x^-$, each carrying a transverse pure gauge field $\alpha^i_{(1,2)}$. Region $(3)$ produces the glasma fields $A_{(3)}^\mu$ in the forward light-cone, with initial condition at $\tau\to 0$ given by patching the two pure gauge fields.}
    \label{fig:glasma-lightcone}
\end{figure}

\noindent\textbf{\textit{Initial condition}}. The collision of CGC fields is shown in Fig.~\ref{fig:glasma-lightcone}. Region $(0)$ is the vacuum, regions $(1)$ and $(2)$ contain the pure gauge fields of each nucleus before the collision, while region $(3)$, delimited by $x^+>0$ and $x^->0$, is where the glasma forms. The color currents of the two colliding nuclei $(1,2)$ moving along the $x^\pm$ directions combine into the total current
\begin{equation}
    \label{eq:total-color-current}
    J^\mu_{(3)} \equiv J^\mu_{(1)} + J^\mu_{(2)} = \delta^{\mu+} \delta(x^-) \rho_{(1)}(\vec{x}_\perp) + \delta^{\mu-} \delta(x^+) \rho_{(2)}(\vec{x}_\perp)\,,
\end{equation}
with $\rho_{(1,2)}(\vec{x}_\perp)$ the color charge densities of the two nuclei in the transverse plane. 

Each color current transforms under a longitudinal Lorentz boost as $J^\pm_{(1,2)}\to e^{\pm\beta} J^\pm_{(1,2)}$ with $\beta$ the rapidity of the boost, where $x^\pm \to e^{\pm\beta} x^\pm$ and $\vec{x}_\perp \to \vec{x}_\perp$. The total current is boost invariant $J^\mu_{(3)}\to J^\mu_{(3)}$, using $\delta(x^\mp)\to e^{\mp\beta} \delta(x^\mp)$. Thus, the initial condition and the resulting glasma fields are both boost invariant.

The solution in region ($3$) is formulated in the Fock-Schwinger gauge $x^+ A^-_{(3)}+x^- A^+_{(3)}=0$~\cite{Kovner:1995ja, Lappi:2003bi}, which enforces $A^\pm_{(3)}=0$ at $x^\mp=0$. In Milne coordinates $\tau = \sqrt{2x^+ x^-}$ and $\eta = \ln[x^+/x^-]/2$, see~\cref{app:milne} for conventions, this reduces to the temporal gauge $A^\tau_{(3)}=0$. The independent field components in region ($3$) are only $A^\eta_{(3)}$ and $A^i_{(3)}$. The glasma fields in region ($3$) are obtained by patching together the pure gauge fields of the two nuclei, characterized by $\theta_{(1)}\equiv \theta(-x^+)\theta(x^-)$, $\theta_{(2)}\equiv \theta(x^+)\theta(-x^-)$, and $\theta_{(3)}\equiv \theta(x^+)\theta(x^-)$, giving
\begin{equation}
    \label{eq:glasma-patched-fields}
    A^+ = \theta_{(3)}\, x^+ A_{(3)}^\eta, \quad A^- = -\theta_{(3)}\, x^- A_{(3)}^\eta, \quad A^i = \theta_{(1)} \alpha_{(1)}^i + \theta_{(2)} \alpha_{(2)}^i + \theta_{(3)} A_{(3)}^i\,.
\end{equation}
Note that in Milne coordinates $A^\eta=\theta_{(3)}A_{(3)}^\eta$. The fields are boost invariant, i.e., independent of $\eta$, and depend only on $\tau$ and $\vec{x}_\perp$ through $A^{\eta,i}_{(3)}(\tau,\vec{x}_\perp)$.

The glasma initial condition along the light-cone at $\tau=0$ is given by
\begin{subequations}
\label{eq:glasma-initial-condition}
\begin{align}
    \alpha^i(\tau, \vec{x}_\perp)\big|_{\tau=0} &= \alpha^i_1(\vec{x}_\perp) + \alpha^i_2(\vec{x}_\perp)\,, \\
    \alpha^\eta(\tau, \vec{x}_\perp)\big|_{\tau=0} &= \frac{ig}{2}[\alpha^i_1(\vec{x}_\perp), \alpha^i_2(\vec{x}_\perp)]\,,
\end{align}
\end{subequations}
where the fields are $\alpha^\eta\equiv A^\eta_{(3)}$ and $\alpha^i\equiv A^i_{(3)}$. The temporal derivatives of the fields vanish at the initial time
\begin{equation}
    \label{eq:glasma-initial-condition-derivative}
    \partial_\tau \alpha^i(\tau, \vec{x}_\perp)\big|_{\tau=0} = \partial_\tau \alpha^\eta(\tau, \vec{x}_\perp)\big|_{\tau=0} = 0\,.
\end{equation}
These correspond to purely longitudinal electric and magnetic fields at $\tau=0$, while the transverse components vanish at the initial time and build up dynamically as the glasma evolves. The initial conditions are obtained from the CYM equations $[D_\mu,F^{\mu\nu}]=J^\nu$ for $\nu=i,+$ along the light-cone, with the gauge fields from Eq.~\eqref{eq:glasma-patched-fields}, by imposing that the singular terms vanish~\cite{Kovner:1995ja,Kovner:1995ts}. For $\nu=i$, the terms in the field equation proportional to $\delta(x^-)\delta(x^+)$ coming from $\partial_-\partial_+ A^i$ are $\alpha^i(\tau, \vec{x}_\perp)-\alpha^i_1(\vec{x}_\perp)-\alpha^i_2(\vec{x}_\perp)$. Similarly, for $\nu=+$, the terms proportional to $\delta(x^-)\theta(x^+)$ arising from $\partial_-F^{-+}$ and $-[D_i,\partial_- A^-]$ give $2\alpha^\eta(\tau, \vec{x}_\perp)-ig[\alpha^i_1(\vec{x}_\perp), \alpha^i_2(\vec{x}_\perp)]$. These give the initial condition for the glasma fields in Eqs.~\eqref{eq:glasma-initial-condition}.

\noindent\textbf{\textit{Glasma field equations}}. The Yang-Mills action in Milne coordinates~\cite{Lappi:2003bi}, with the temporal gauge $A^\tau=0$ and boost invariance $\partial_\eta A_\mu=0$ imposed, takes the form
\begin{equation}
    \label{eq:ym-action-milne}
    S = \int \mathrm{d}\tau\, \mathrm{d}^2\vec{x}_\perp\, \mathrm{d}\eta\,\, \mathrm{tr}\left[\tau (\partial_\tau A_i)^2 + \frac{1}{\tau} (\partial_\tau A_\eta)^2 - \frac{\tau}{2} F_{ij}^2 - \frac{1}{\tau} [D_i,A_\eta]^2\right]\,.
\end{equation}
This expression is obtained from the Yang-Mills action in Minkowski spacetime by performing the coordinate transformation to Milne coordinates, see~\cref{app:milne}, and then imposing the temporal gauge and boost invariance. The conjugate field momenta are defined as
\begin{equation}
    \label{eq:conjugate-momenta}
    P^i \equiv \tau \partial_\tau A_i, \quad P^\eta \equiv \frac{1}{\tau} \partial_\tau A_\eta\,.
\end{equation}
These are related to the electric and magnetic fields in Milne coordinates at midrapidity $\eta=0$
\begin{equation}
    E_i\equiv \frac{1}{\tau} P^i,\quad E_z\equiv P^\eta, \quad B_i\equiv -\frac{1}{\tau}\epsilon_{ij} [D_j, A_\eta],\quad B_z\equiv -F_{xy}\,.
\end{equation}

The field equation for the temporal component gives the Gauss constraint
\begin{equation}
    \label{eq:gauss-law}
    [D_i, P^i] - ig[A_\eta, P^\eta] = 0\,.
\end{equation}
Varying the action with respect to $A_\eta$ and $A_i$ gives the field equations
\begin{equation}
    \partial_\tau P^i = \tau [D_j, F_{ji}] - \frac{ig}{\tau} [A_\eta, [D_i, A_\eta]]\,, \quad \partial_\tau P^\eta = \frac{1}{\tau} [D_i,[D_i, A_\eta]]\,.
\end{equation}
In region ($3$), the field equations become $\partial_\tau(\tau\,\partial_\tau \alpha_i) = \tau\, [D_j, F_{ji}] - ig[\alpha_\eta, [D_i, \alpha_\eta]]/\tau$ together with $\partial_\tau\!\left[\partial_\tau \alpha_\eta/\tau\right] = [D_i,[D_i,\alpha_\eta]]/\tau$. These two equations, supplemented by the Gauss constraint in Eq.~\eqref{eq:gauss-law} and the initial conditions from Eq.~\eqref{eq:glasma-initial-condition}, fully determine the evolution of the glasma fields.

\begin{emphbox}{myblue}{Lattice gauge theory}
The classical Yang-Mills theory may equivalently be formulated in terms of gauge potentials, components of the field strength tensor (electric and magnetic fields), or Wilson lines and loops. In the lattice discretization, the convenient degrees of freedom are the gauge links, namely Wilson lines connecting neighboring lattice sites, and the corresponding plaquettes, which are Wilson loops around elementary lattice squares. This formulation is gauge invariant upon discretization.
\end{emphbox}

The classical Yang-Mills equations for the glasma are coupled non-linear partial differential equations whose full solution is not known analytically. One may obtain approximate solutions either by performing a power series expansion in $\tau$, valid at early times, or in the dilute limit, which expands the fields in powers of the color charge density $\rho$~\cite{Leuthner:2025vsd}. To obtain the full boost-invariant solution, the field equations must be solved numerically. In practice, this is done using the real-time lattice formulation explained below.

\noindent\textbf{\textit{Lattice discretization}}. The lattice formulation of Yang-Mills theory in Minkowski spacetime is detailed in~\cref{app:qcd}. The discretization scheme for the glasma follows~\cite{Krasnitz:1998ns, Lappi:2004mbt, Muller:2019bwd}. In Milne coordinates, the lattice discretization is applied only to the transverse coordinates $\vec{x}_\perp$. Boost invariance means there are no rapidity gauge links, and $A^\eta$ transforms as a scalar under non-Abelian gauge transformations. In the temporal gauge $A^\tau=0$ there are no temporal gauge links. The transverse lattice spacing is $a_\perp \equiv a^x = a^y$, with $N_\perp$ sites per side and box size $L_\perp$, giving $a_\perp = L_\perp/N_\perp$. On the lattice, a field at transverse position $\vec{x}_\perp$ is written as $X^\mu(\tau,\vec{x}_\perp)\equiv X^\mu_{\boldsymbol{x}}(\tau)$, where $\boldsymbol{x}$ labels the corresponding lattice site. The independent degrees of freedom are the longitudinal gauge field $A^\eta_{\boldsymbol{x}}(\tau)$, the transverse gauge links $U_{\boldsymbol{x},i}(\tau)$ with associated plaquettes $U_{\boldsymbol{x}, ij}(\tau)$, and their conjugate momenta $P^\eta_{\boldsymbol{x}}(\tau)$ and $P^i_{\boldsymbol{x}}(\tau)$, which may be expressed as
\begin{equation}
    P_{\boldsymbol{x}}^\eta = \frac{1}{\tau}\partial_\tau A_{\boldsymbol{x},\eta}, \quad P_{\boldsymbol{x}}^i = -\frac{i\tau}{ga_\perp}\left(\partial_\tau U_{\boldsymbol{x},i}\right) U^\dagger_{\boldsymbol{x},i}\,.
\end{equation}

The forward and backward finite differences for the transverse covariant derivatives are defined as
\begin{equation}
    D^{F}_i X_{\boldsymbol{x},\mu} \equiv \frac{1}{a_\perp}\left(X^{+i}_{\boldsymbol{x},\mu} - X_{\boldsymbol{x},\mu}\right), \quad D^{B}_i X_{\boldsymbol{x},\mu} \equiv \frac{1}{a_\perp}\left(X_{\boldsymbol{x},\mu} - X^{-i}_{\boldsymbol{x},\mu}\right)\,,
\end{equation}
at the same $\tau$, where the parallel transported field along $i$ in the positive or negative direction is defined as $X^{\pm i}_{\boldsymbol{x},\mu}\equiv U_{\boldsymbol{x},\pm i}\, X_{\boldsymbol{x}\pm i,\mu}\, U_{\boldsymbol{x}\pm i,\mp i}$. This discretization is applied with $X_\mu=A_\eta$.

The Yang-Mills action in Milne coordinates, for boost-invariant fields in the temporal gauge, is given by
\begin{equation}
\begin{split}
    S = \int \mathrm{d}\tau\,\mathrm{d}\eta \sum_{\boldsymbol{x}} V_\perp \, \mathrm{tr} & \bigg[\frac{1}{\tau}\left(P_{\boldsymbol{x}}^i\right)^2 + \tau\left(P_{\boldsymbol{x}}^\eta\right)^2 \\
    & -  \frac{\tau}{2\left(g a_\perp^2\right)^2}\left(\mathbbm{2} - U_{\boldsymbol{x},ij} - U^\dagger_{\boldsymbol{x},ij}\right)- \frac{1}{\tau} \left(D_i^{F} A_{\boldsymbol{x},\eta}\right)^2\bigg]\,,
\end{split}
\end{equation}
where $V_\perp\equiv a_\perp^2$. The discretized Gauss constraint $D_i^{B} P_{\boldsymbol{x}}^i - ig\left[A_{{\boldsymbol{x}},\eta}, P_{\boldsymbol{x}}^\eta\right] = 0$ is
obtained from the variation of the action with respect to $A^\tau$ on the lattice, without imposing the temporal gauge condition $A^\tau=0$. The discretized field equations for the conjugate momenta are obtained by varying the lattice action. For $P^\eta$ one finds
\begin{equation}
    \partial_\tau P_{\boldsymbol{x}}^\eta = \frac{1}{\tau} D_i^2 A_{\boldsymbol{x},\eta}\,,\quad\text{with}\,\,D_i^2 \equiv D_i^F D_i^B\,,
\end{equation}
expressed in terms of the transverse lattice Laplacian. Using the variation $\delta U_{\boldsymbol{x},i} = ig a_\perp \delta A_{\boldsymbol{x},i}\, U_{\boldsymbol{x},i}$ yields the other field equation
\begin{equation}
    \partial_\tau P_{\boldsymbol{x}}^i = -\frac{\tau}{ga_\perp^3}\left[U_{\boldsymbol{x},ij} + U_{\boldsymbol{x},i-j}\right]_{\mathrm{ah}} - \frac{ig}{\tau}\left[A_{\boldsymbol{x},\eta}^{+i},\, D_i^{F} A_{\boldsymbol{x},\eta}\right]\,,
\end{equation}
where $[X]_{\mathrm{ah}} \equiv (1/(2i))(X - X^\dagger) - (1/N_c)\,\mathrm{tr}\!\left[(1/(2i))(X - X^\dagger)\right]\mathbbm{1}$ is the traceless anti-Hermitian part of a matrix $X$.

The time evolution is performed using a leapfrog algorithm. The gauge fields $A_{\boldsymbol{x},\mu}$ and gauge links $U_{\boldsymbol{x},i}$ are defined at integer time steps $\tau_n \equiv n \Delta \tau$ with $n=0,1,2,\ldots$, while the conjugate momenta $P_{\boldsymbol{x}}^\eta$ and $P_{\boldsymbol{x}}^i$ are defined at half-integer time steps $\tau_n+\Delta\tau/2$. The algorithm is second order accurate in $\Delta \tau$, is symmetric under time reversal, and preserves the Gauss constraint exactly at each time step. The discretized glasma equations of motion for the fields are
\begin{subequations}
    \label{eq:discretized-glasma-eom-1}
    \begin{align}
        A_{\boldsymbol{x},\eta}(\tau_{n+1}) &= A_{\boldsymbol{x},\eta}(\tau_n) + \Delta\tau\,\tau_{n+\frac{1}{2}}\, P_{\boldsymbol{x}}^\eta(\tau_{n+\frac{1}{2}})\,, \\
        U_{\boldsymbol{x},i}(\tau_{n+1}) &= \exp\!\left[\frac{iga_\perp\Delta\tau}{\tau_{n+\frac{1}{2}}}\, P_{\boldsymbol{x}}^i(\tau_{n+\frac{1}{2}})\right] U_{\boldsymbol{x},i}(\tau_n)\,,
    \end{align}
\end{subequations}
and for the conjugate momenta
\begin{subequations}
    \label{eq:discretized-glasma-eom-2}
    \begin{align}
        P_{\boldsymbol{x}}^\eta(\tau_{n+\frac{1}{2}}) =&\, P_{\boldsymbol{x}}^\eta(\tau_{n-\frac{1}{2}}) + \frac{\Delta\tau}{\tau_n}D_i^F D_i^B\, A_{\boldsymbol{x},\eta}(\tau_n)\,, \\
        P_{\boldsymbol{x}}^i(\tau_{n+\frac{1}{2}}) =& \,P_{\boldsymbol{x}}^i(\tau_{n-\frac{1}{2}}) - \frac{\Delta\tau\,\tau_n}{ga_\perp^3}\left[U_{\boldsymbol{x},ij}(\tau_n) + U_{\boldsymbol{x},i-j}(\tau_n)\right]_{\mathrm{ah}} \nonumber \\
        & - \frac{ig\Delta\tau}{\tau_n}\left[A_{\boldsymbol{x},\eta}^{+i}(\tau_n),\, D_i^{F} A_{\boldsymbol{x},\eta}(\tau_n)\right]\,.
    \end{align}
\end{subequations}

\noindent\textbf{\textit{Numerical implementation}}. The numerical procedure is explained in detail in Ref.~\cite{Muller:2019bwd} and the glasma solver is publicly available~\cite{curraun:glasma}. It proceeds as follows. The only physical scale is $g^2\mu$, the color charge density variance of the MV model, which is proportional to the saturation momentum $Q_s$, see Eq.~\eqref{eq:saturation-momentum} and the discussion before. All dimensionless observables in the simulation are expressed in units of $g^2\mu$. Random color charges are generated on the transverse lattice according to a Gaussian distribution $\langle \rho^a_{\boldsymbol{x},n} \rho^b_{\boldsymbol{y},m}\rangle = (g^2\mu^2/N_s) \delta_{nm}\delta^{ab}\delta^{2}_{\boldsymbol{x}\boldsymbol{y}}$ with color charge $\rho^a_{\boldsymbol{x},n}$ at each lattice site $\boldsymbol{x}$ and sheet $n$ having color index $a$. A \emph{Fast Fourier Transform} (FFT) of the color charge is then performed, where $\boldsymbol{k}\cdot\boldsymbol{x}=2\pi/L_\perp\,(k_x x + k_y y)$ and $\sum_{\boldsymbol{k}}\equiv \sum_{k_x, k_y}$, as
\begin{equation}
    \label{eq:lattice-fft}
    \tilde{\rho}^a_{\boldsymbol{k},n} = a_\perp^2 \sum_{\boldsymbol{x}}\rho^a_{\boldsymbol{x},n}\exp\!\left[-i\boldsymbol{k}\cdot\boldsymbol{x}\right]\,.
\end{equation}
The lattice momentum is extracted from the discretized transverse Laplacian
\begin{equation}
    \label{eq:lattice-momentum}
    k_\perp^2 = \sum_i^{x,y} \left[\frac{2}{a_\perp} \sin\left(\frac{k_i a_\perp}{2}\right)\right]^2\,.
\end{equation}

The Poisson equation for the covariant gauge field (dropping the gauge index in $A_\mathrm{cov}^+$) is solved in momentum space $k_\perp^2 \tilde{A}^{+a}_{\boldsymbol{k},n}=\tilde{\rho}^a_{\boldsymbol{k},n}$, with the solution
\begin{equation}
    \label{eq:lattice-poisson-solution}
    \tilde{A}^{+a}_{\boldsymbol{k},n} = \frac{\tilde{\rho}^a_{\boldsymbol{k},n}}{k_\perp^2+m^2}\,\theta(\Lambda^2-k_\perp^2)\,,
\end{equation}
where $m$ is the infrared (IR) and $\Lambda$ is the ultraviolet (UV) regulator. The lattice itself provides regulators $\lambda_\mathrm{UV} \equiv \pi/a_\perp$ and $\lambda_\mathrm{IR} \equiv 2\pi/L_\perp$, but additional regulators are required. The zero mode $k_\perp = 0$ causes a division by zero in the Poisson equation from~\cref{eq:lattice-poisson-solution} that the finite lattice volume does not regulate, but is fixed by introducing the infrared regulator $m$. Moreover, modes near the Brillouin zone boundary $k_\perp \sim \lambda_\mathrm{UV}$ have a dispersion relation in~\cref{eq:lattice-momentum} that deviates from $k_\perp^2$, thus one needs the ultraviolet regulator $\Lambda$. One requires $m > \lambda_\mathrm{IR}$ and $\Lambda < \lambda_\mathrm{UV}$ to ensure that $m$ and $\Lambda$ are the relevant IR and UV regulators. 

The inverse FFT $A^{+a}_{\boldsymbol{x},n} = (1/L_\perp^2)\sum_{\boldsymbol{k}} \tilde{A}^{+a}_{\boldsymbol{k},n} \exp\!\left[i\boldsymbol{k}\cdot\boldsymbol{x}\right]$ then gives the covariant gauge field for a single nucleus. One then performs a gauge transformation to the light-cone gauge to obtain the Wilson line entering the pure gauge field solution for each nucleus $(1,2)$ before the collision
\begin{equation}
    \label{eq:wilson-line-lattice}
    V_{\boldsymbol{x}}^{(1,2)} = \prod_n^{N_s}\exp\!\left[-igA^{+a\,(1,2)}_{\boldsymbol{x},n} t^a\right]\,,
\end{equation}
with $A^{+a}_{(1,2)}$ obtained for each nucleus $(1,2)$ according to the procedure described above. The gauge transformation is applied to the transverse gauge links, from the identity in the covariant gauge to the light-cone gauge as
\begin{equation}
    \label{eq:lattice-gauge-transformation}
    U_{\boldsymbol{x},i}^{(1,2)} = V_{\boldsymbol{x}+i}^{(1,2)} V_{\boldsymbol{x}}^{(1,2)\dagger}\,.
\end{equation}
This is done for each nucleus separately. The pure gauge fields for each nucleus are then used to construct the glasma initial condition according to Eq.~\eqref{eq:glasma-initial-condition}.

The lattice expression for the initial transverse gauge links in terms of the pure gauge fields
\begin{equation}
    \label{eq:lattice-initial-links}
    \left[(U_{\boldsymbol{x},i}^{(1)} + U_{\boldsymbol{x},i}^{(2)})\left(\mathbbm{1} + U_{\boldsymbol{x},i}^\dagger(\tau \to 0)\right)\right]_{\mathrm{ah}} = 0\,,
\end{equation}
has an exact solution for the SU($2$) gauge group but must be solved iteratively for SU($3$). The initial longitudinal conjugate momenta are given by
\begin{equation}
    \label{eq:lattice-initial-momenta}
    \begin{split}
        &P_{\boldsymbol{x}}^\eta(\tau \to 0) \\
        &= \frac{1}{2ga_\perp^2}\left[(U_{\boldsymbol{x},i} - \mathbbm{1})(U_{\boldsymbol{x},i}^{(2)\dagger} - U_{\boldsymbol{x},i}^{(1)\dagger}) + (U_{\boldsymbol{x},-i}^\dagger - \mathbbm{1})(U_{\boldsymbol{x}-i,i}^{(2)} - U_{\boldsymbol{x}-i,i}^{(1)})\right]_{\mathrm{ah}}\,,
    \end{split}
\end{equation}
with $U_{\boldsymbol{x},i}\equiv U_{\boldsymbol{x},i}(\tau \to 0)$. At the initial time, the longitudinal gauge fields and transverse conjugate momenta vanish: $A_{\boldsymbol{x}}^\eta(\tau \to 0) = 0$ and $P_{\boldsymbol{x}}^i(\tau \to 0) = 0$. These are the glasma initial conditions, expressed in a discretized form on the lattice, which are used to start the time evolution of the glasma fields according to the leapfrog algorithm in~\cref{eq:discretized-glasma-eom-1,eq:discretized-glasma-eom-2}.

\begin{figure}[tp]
    \centering
    \begin{subfigure}[b]{\textwidth}
        \centering
        \includegraphics[width=\textwidth]{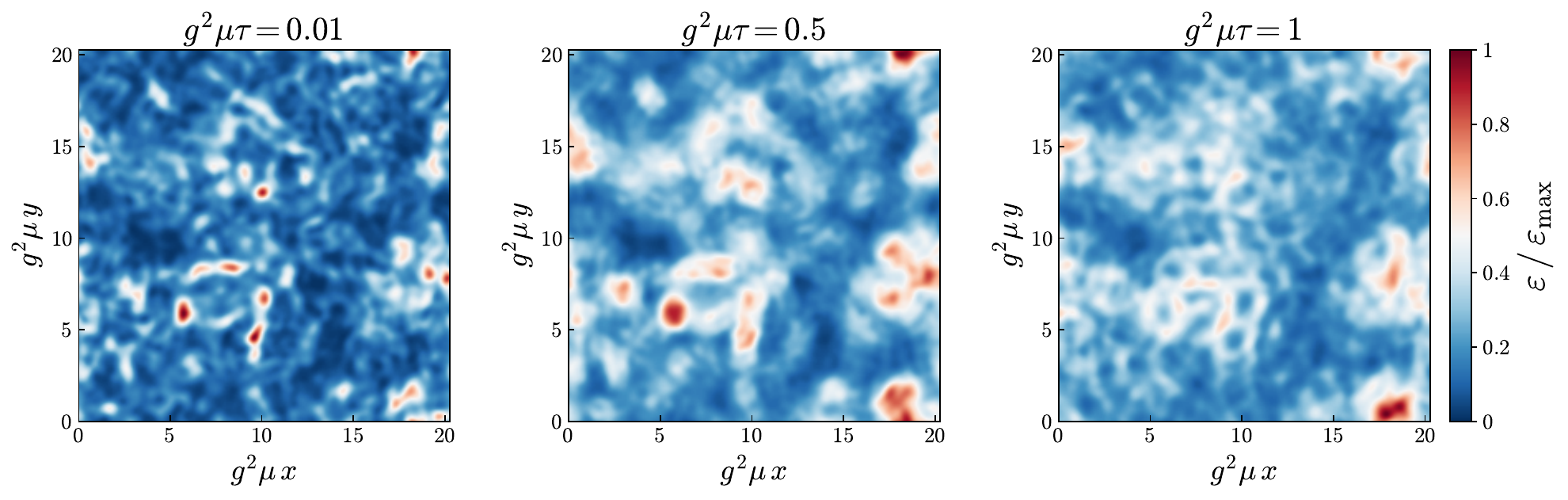}
        \caption{Normalized energy density $\varepsilon/\varepsilon_\mathrm{max}$ in the dimensionless transverse plane $(g^2\mu x, g^2\mu y)$ at various values of the proper time $g^2\mu\tau$ for fixed $g^2\mu=2\,\mathrm{GeV}$.}
        \label{fig:flux-tubes-auau}
    \end{subfigure}\\[1em]
    \begin{subfigure}[b]{\textwidth}
        \centering
        \includegraphics[width=\textwidth]{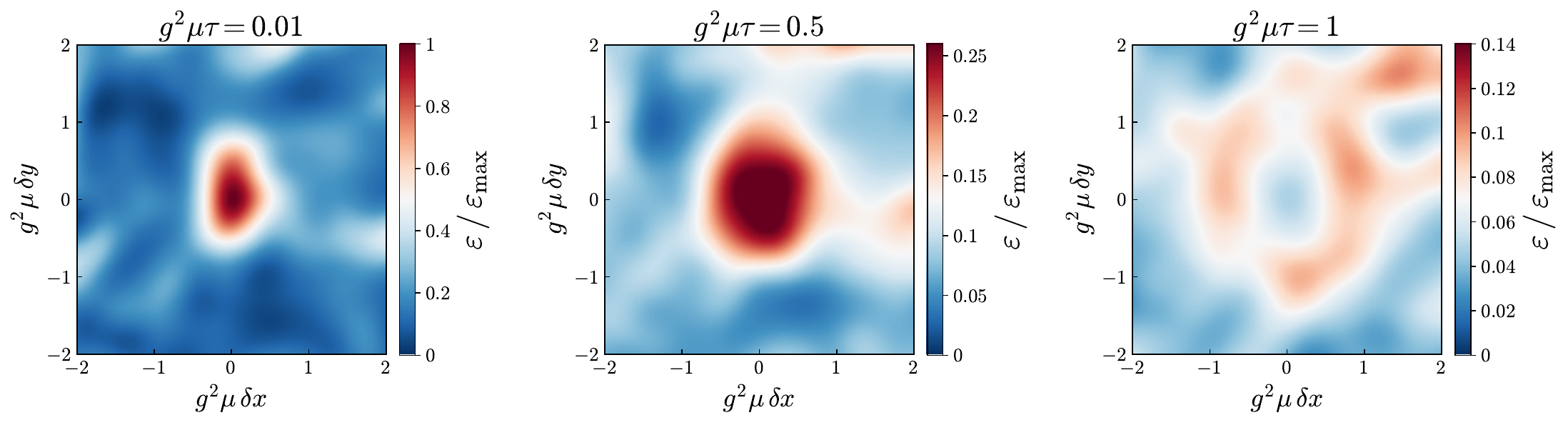}
        \caption{Zoom-in on a single correlation domain in the energy density in the transverse plane at different values of proper time. The energy density is normalized by the maximum value in each frame.}
        \label{fig:corr-domain}
    \end{subfigure}
    \caption{Glasma energy density in the transverse plane at different proper times.}
    \label{fig:energy-density-overview}
\end{figure}

The simulation is characterized by the physical scale $g^2\mu$, or equivalently $Q_s$ up to numerical proportionality factors~\cite{Lappi:2008eq}, along with lattice parameters: $N_\perp$ lattice sites, simulation length $L_\perp$ and the leapfrog time step $\Delta \tau=a_\perp/N_\tau$ with $N_\tau=1,2,\ldots$. The number of discretized color sheets $N_s$ is chosen to be large enough such that the continuum limit $N_s\rightarrow\infty$ is reached. To be in the dense regime one further requires $g^2\mu\, a_\perp \ll 1$, and leapfrog stability is ensured for large enough $N_\tau\gg 1$. One chooses $\lambda_\mathrm{IR} = 2\pi/L_\perp < m$ and $\Lambda < \lambda_\mathrm{UV} = \pi/a_\perp$ such that the IR and UV regulators are set by $m$ and $\Lambda$ and not determined by the lattice. The results depend on the dimensionless combinations $g^2\mu\, a_\perp$, $g^2\mu \,L_\perp$, and $m/(g^2\mu)$ at fixed $N_s$, $N_\tau$ and $\Lambda$. Typical values used in practice are $g=2$, $g^2\mu=2\,\mathrm{GeV}$, $m=0.2\,\mathrm{GeV}$, $\Lambda=10\,\mathrm{GeV}$, $L=10\,\mathrm{fm}$, $N_\perp=512$, $N_\tau=8$ and $N_s=50$.

\begin{figure}[tp]
    \centering
    \begin{subfigure}[b]{0.48\textwidth}
        \centering
        \includegraphics[width=0.9\textwidth]{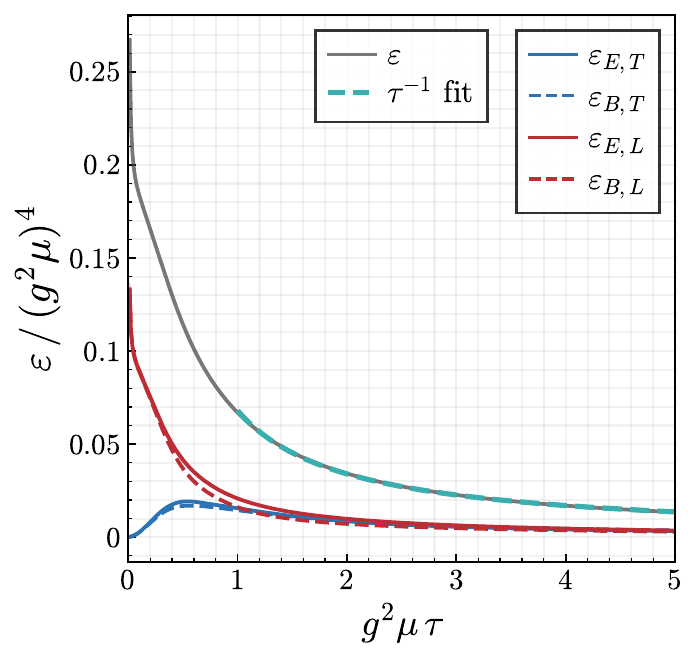}
        \caption{Longitudinal and transverse, electric and magnetic contributions to the energy density as a function of proper time.}
        \label{fig:components}
    \end{subfigure}
    \hfill
    \begin{subfigure}[b]{0.48\textwidth}
        \centering
        \includegraphics[width=0.9\textwidth]{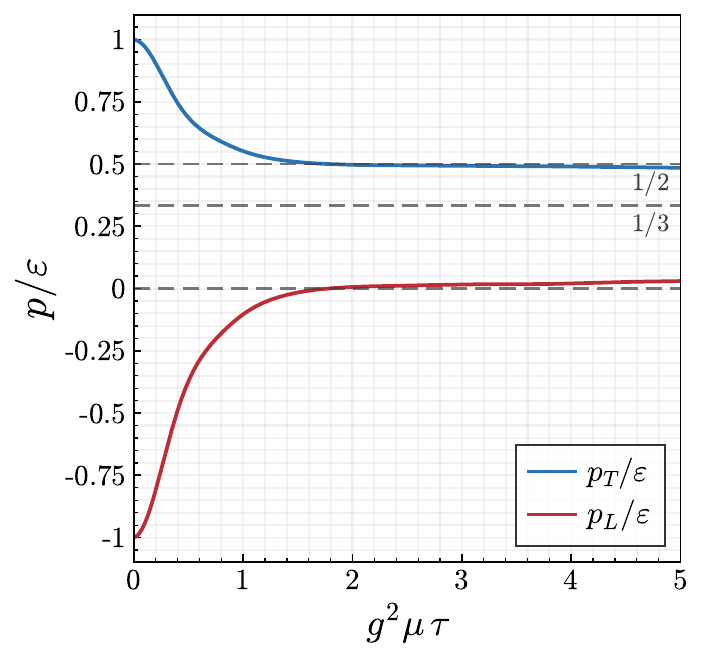}
        \caption{Longitudinal and transverse pressure to energy density ratios as a function of proper time.}
        \label{fig:ratio}
    \end{subfigure}
    \caption{Proper time evolution of the averaged energy density components and pressure ratios.}
    \label{fig:components-ratio}
\end{figure}

\subsection{Features of the glasma}

The energy-momentum tensor for the gluon fields is
\begin{equation}
    T^{\mu\nu} = \mathrm{tr}\left[2F^{\mu\rho}F_\rho^{\;\nu} + \frac{1}{2}g^{\mu\nu}F^{\rho\sigma}F_{\rho\sigma}\right]\,.
\end{equation}
It is derived from $L=-\mathrm{tr}\!\left[F_{\mu\nu}F^{\mu\nu}\right]/2$ by applying Noether's theorem for spacetime translation symmetry. This gives the conserved current $T^{\mu\nu}=[\delta L/\delta(\partial_\mu A_\rho)]\,\partial^\nu A_\rho -L g^{\mu\nu}$, which simplifies to the expression above. The diagonal temporal component of the energy-momentum tensor is
\begin{equation}
    \begin{split}
        T^{\tau\tau}&=\mathrm{tr}\left[\Big(\frac{1}{\tau}P^i\Big)^2+(P^\eta)^2+\frac{1}{2}F_{ij}^2+\Big(\frac{1}{\tau}[D_i,A_\eta]\Big)^2\right]\\
        &=\mathrm{tr}\left[E_T^2 + E_L^2 + B_T^2 + B_L^2\right]\,,
    \end{split}
\end{equation}
where $F_T^2\equiv F_x^2+F_y^2$ and $F_L\equiv F_z$ for $F=(E,B)$ fields. Similarly, the diagonal spatial components are given by
\begin{equation}
    T^{xx}=T^{yy}=\mathrm{tr}\left[E_L^2 + B_L^2\right], \quad T^{\eta\eta} = \frac{1}{\tau^2}\mathrm{tr}\left[E_T^2 + B_T^2 - E_L^2 - B_L^2\right]\,.
\end{equation}

Numerical results for the energy density in the transverse plane, for a single glasma event, are shown in Fig.~\ref{fig:flux-tubes-auau}. The fields are localized in the transverse plane in structures with a typical transverse size of $1/(g^2\mu)$. A zoom-in on a single correlation domain in the energy density is shown in Fig.~\ref{fig:corr-domain}, where one can see the expansion of a single glasma flux tube structure.

The averaged electric and magnetic energy densities are defined as $\varepsilon_{F,L/T} \equiv \langle\mathrm{tr}[F^2_{L/T}] \rangle$,
where $\langle\,\cdots\,\rangle$ denotes a combined average over the transverse lattice $\langle\,\cdots\,\rangle_{\boldsymbol{x}}$ and over the ensemble of MV model color charge configurations $\rho_{{(1,2)}}$ for each nucleus $(1,2)$, denoted $\langle\,\cdots\,\rangle_{\rho}$, namely $\langle X \rangle \equiv \big\langle\langle X_{\boldsymbol{x}}[\rho_{(1)}, \rho_{(2)}] \rangle_{\boldsymbol{x}}\big\rangle_{\rho}$.

The total energy density is $\varepsilon \equiv \langle T^{\tau\tau} \rangle = \varepsilon_{E,T} + \varepsilon_{E,L} + \varepsilon_{B,T} + \varepsilon_{B,L}$
and the transverse and longitudinal pressures are defined as $p_T\equiv \langle T^{xx} \rangle = \varepsilon_{E,L} + \varepsilon_{B,L}$ and $p_L \equiv \langle \tau^2 T^{\eta\eta} \rangle = \varepsilon_{E,T} + \varepsilon_{B,T} - \varepsilon_{E,L} - \varepsilon_{B,L}$ respectively.
Since the energy-momentum tensor is traceless, the energy density and pressures are related by $\varepsilon = 2p_T + p_L$.

\begin{figure}[tp]
    \centering
    \begin{subfigure}[b]{0.48\textwidth}
        \centering
        \includegraphics[width=\textwidth]{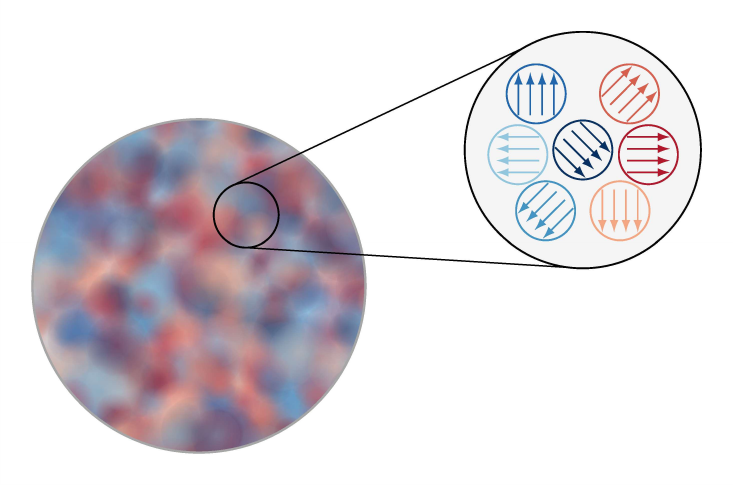}
        \caption{Glasma correlation domains in the transverse plane. The overoccupied fields arrange into domains of characteristic transverse size $\sim 1/(g^2\mu)$, with fields correlated inside each domain.}
        \label{fig:glasma-corr-dom}
    \end{subfigure}
    \hfill
    \begin{subfigure}[b]{0.48\textwidth}
        \centering
        \includegraphics[width=\textwidth]{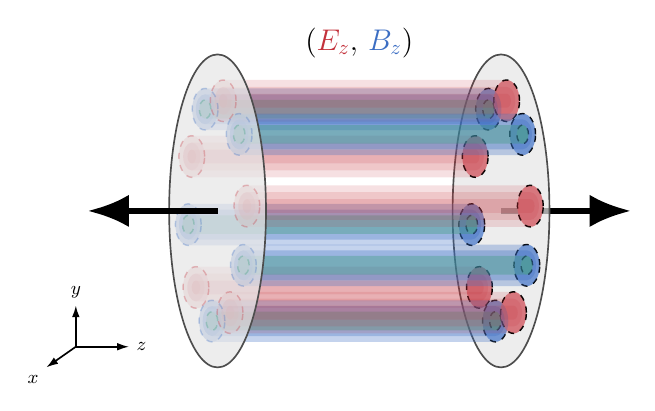}
        \caption{Electric $E_z$ and magnetic $B_z$ glasma fields forming flux tubes. Within each domain the $E_z$ fields are correlated \textit{(red)}, while the $B_z$ fields exhibit an anticorrelation \textit{(green)} surrounded by a correlated ring \textit{(blue)}.}
        \label{fig:glasma-fields-corr}
    \end{subfigure}
    \caption{Sketches of the transverse and longitudinal glasma correlation structure.}
    \label{fig:glasma-corr-structure}
\end{figure}

Fig.~\ref{fig:components-ratio} shows the energy density components and pressure ratios as a function of proper time, averaged over multiple field configurations. Initially, the glasma fields are purely longitudinal, but transverse components develop as the fields evolve in time. The strong fields in the glasma are diluted by the longitudinal expansion, which leads to a decrease in the energy density components at later times. The energy density free streams as $\varepsilon \sim 1/\tau$ after a time $\tau \sim 1/(g^2\mu)$, at which point all field components contribute equally. This $1/\tau$ behavior follows from the Bjorken expansion~\cite{Bjorken:1982qr}. Imposing that the total energy in a rapidity slice of fixed transverse area is conserved, that is $\varepsilon\cdot\tau\cdot S_\perp=\text{const}$, gives $\varepsilon\propto 1/\tau$. 

The longitudinal pressure $p_L$ is negative at early times, which is a feature of the glasma initial conditions from~\cref{eq:glasma-initial-condition,eq:glasma-initial-condition-derivative}, while the transverse pressure $p_T$ is positive, and initially there is an anisotropy. The glasma never reaches pressure isotropy during the classical evolution~\cite{Epelbaum:2013ekf,Epelbaum:2014xea}. Pressure isotropy $p_L\approx p_T$ is a physical prerequisite for the start of hydrodynamic evolution. For isotropization, additional mechanisms are required to drive the system toward the thermalized QGP, such as quantum fluctuations or plasma instabilities. In 3+1D glasma simulations, which relax the assumption of boost invariance, the longitudinal and transverse pressure ratios evolve differently and are sensitive to the finite longitudinal thickness of the colliding nuclei~\cite{Ipp:2020igo, Muller:2019bwd, Schlichting:2020wrv, McDonald:2023qwc, Schmidt:2026nin} and instabilities~\cite{Berges:2013eia,Berges:2013fga}.

\subsection{Field correlators}

\begin{figure}[tp]
    \centering
    \begin{subfigure}[b]{\textwidth}
        \centering
        \includegraphics[width=0.9\textwidth]{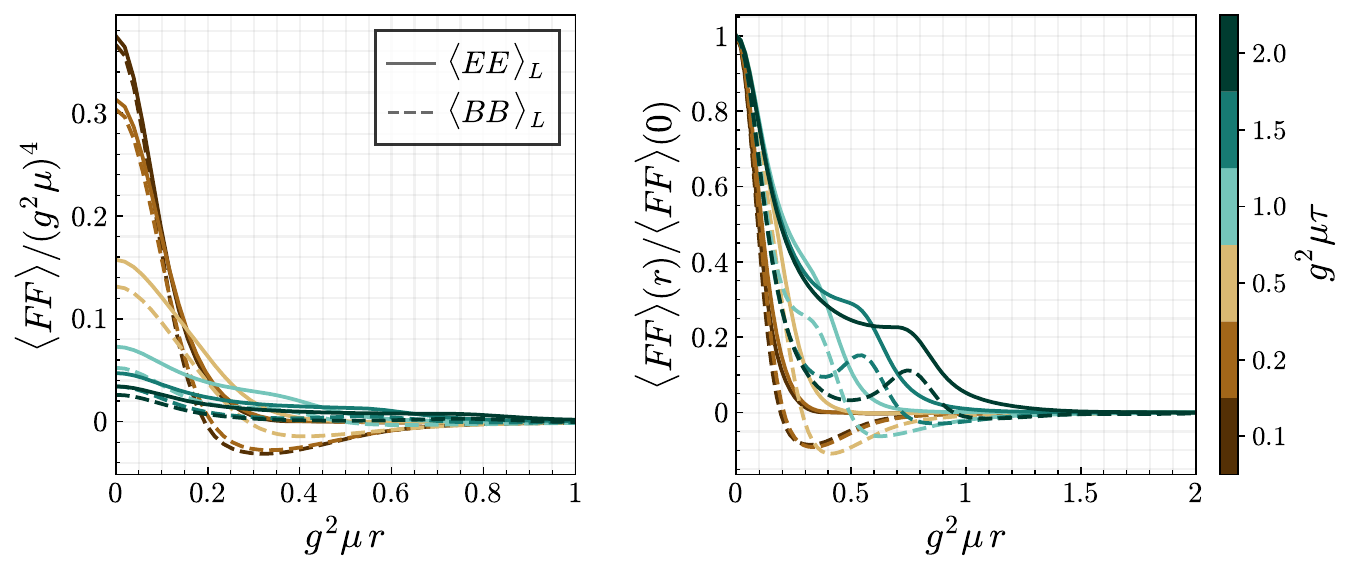}
        \caption{Longitudinal spatial correlation functions for electric and magnetic fields as a function of transverse separation $r$, for several values of proper time.}
        \label{fig:long-corr-spatial}
    \end{subfigure}\\[0.5em]
    \begin{subfigure}[b]{\textwidth}
        \centering
        \includegraphics[width=0.9\textwidth]{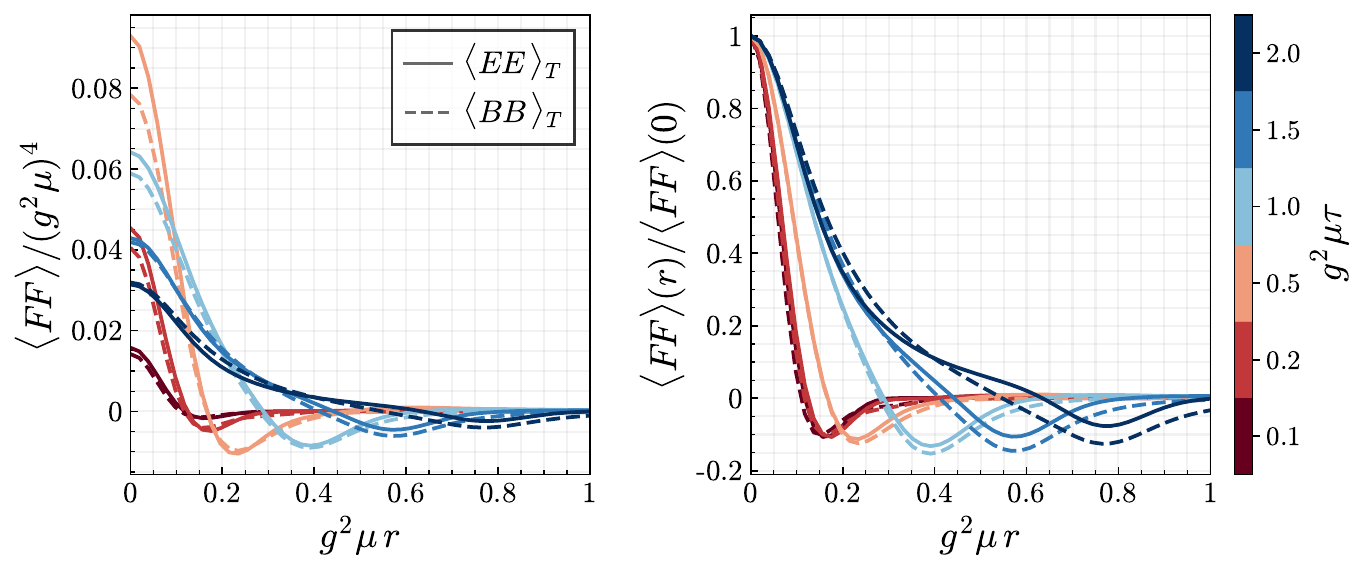}
        \caption{Transverse spatial correlation functions for electric and magnetic fields as a function of transverse separation $r$, for several values of proper time.}
        \label{fig:trans-corr-spatial}
    \end{subfigure}
    \caption{Longitudinal and transverse spatial correlation functions of the glasma fields, in dimensionless units and normalized by the value at the origin.}
    \label{fig:corr-spatial}
\end{figure}

The glasma fields organize into color domains of size $\sim 1/(g^2\mu)$, schematically illustrated in Fig.~\ref{fig:glasma-corr-dom}, with fields correlated inside a domain and uncorrelated between different domains. A sketch of the collision in Fig.~\ref{fig:glasma-fields-corr} shows the inner structure of the longitudinal flux tubes. At early times, the electric field $E_z$ is correlated inside the domain, while the magnetic field $B_z$ shows an anticorrelation ring followed by a correlation region. These correlation patterns were studied in Refs.~\cite{Dumitru:2014nka,Ruggieri:2017ioa,Ipp:2020mjc}. Beyond these sketches showing spatial structure, the domains evolve in time, and are expected to expand and decay together with the fields. To gain more insight, let us perform a more systematic study of the spatial and temporal correlation functions of the glasma fields. These correlators characterize the coherence length and lifetime of the glasma flux tubes. They are physically relevant for understanding how probes interact with the glasma, which is the topic of the next chapters.

\newpage

\begin{figure}[tp]
    \centering
    \begin{subfigure}[b]{\textwidth}
        \centering
        \includegraphics[width=0.9\textwidth]{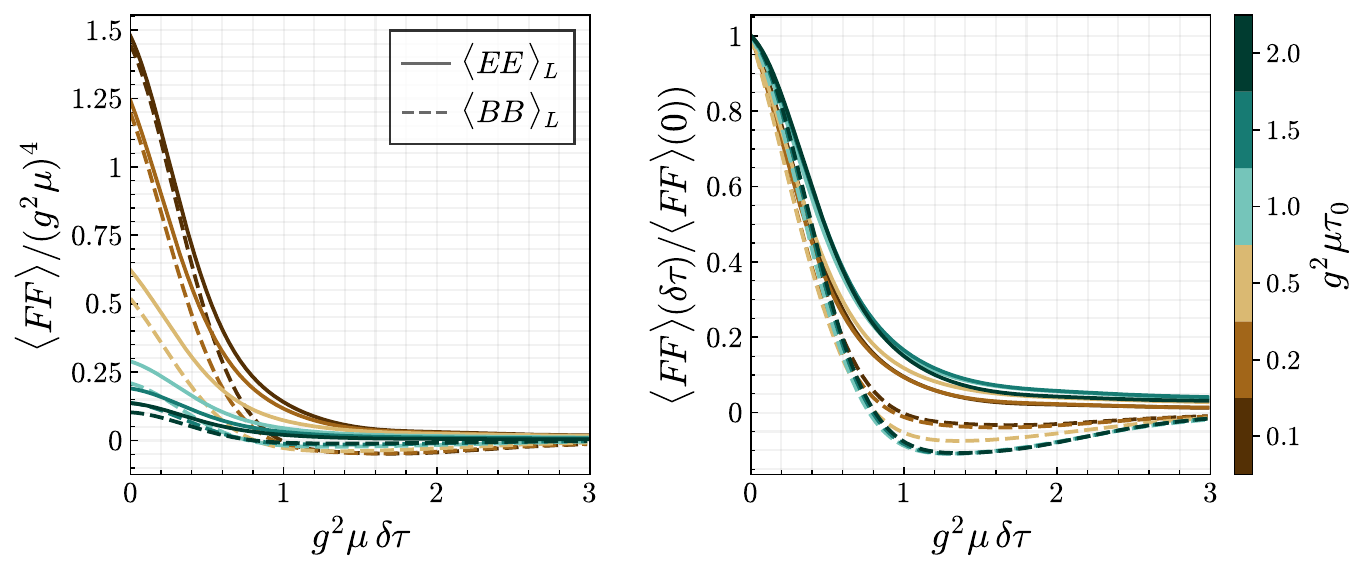}
        \caption{Longitudinal temporal correlation functions for electric and magnetic fields as a function of time separation $\delta\tau$, for various initial times $\tau_0$.}
        \label{fig:long-corr-time}
    \end{subfigure}\\[0.5em]
    \begin{subfigure}[b]{\textwidth}
        \centering
        \includegraphics[width=0.9\textwidth]{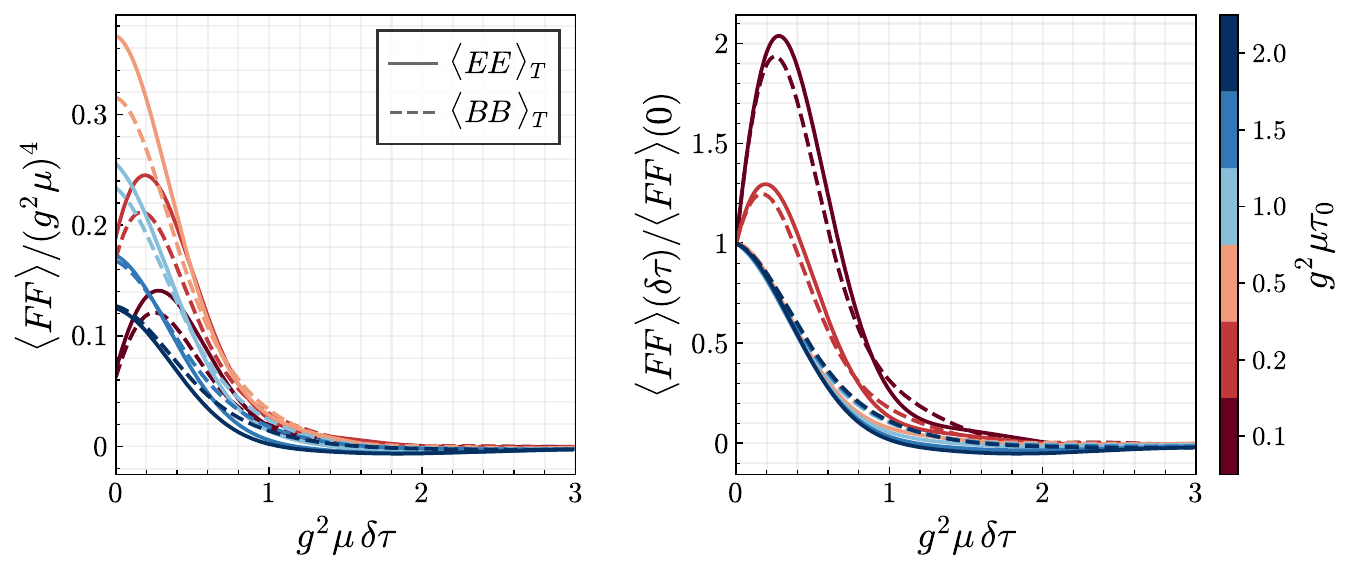}
        \caption{Transverse temporal correlation functions for electric and magnetic fields as a function of time separation $\delta\tau$, for various initial times $\tau_0$.}
        \label{fig:trans-corr-time}
    \end{subfigure}
    \caption{Longitudinal and transverse temporal correlation functions of the glasma fields, in dimensionless units and normalized by the value at the origin.}
    \label{fig:corr-time}
\end{figure}

\noindent\textbf{\textit{Spatial correlations}}. The glasma field setup from which spatial correlators are extracted is shown in Fig.~\ref{fig:glasma-fields-corr}. The gauge invariant longitudinal $L$ and transverse $T$ spatial correlation functions for the color electric and magnetic fields at a fixed proper time $\tau$ are defined as
\begin{equation}
    \langle FF\,\rangle_{L/T}(r) \equiv \left\langle \mathrm{tr} \left[ F_{L/T}(\vec{x}_\perp) U(\vec{x}_\perp ,\vec{y}_\perp) F_{L/T}(\vec{y}_\perp) U^\dagger(\vec{x}_\perp ,\vec{y}_\perp) \right] \right\rangle\,,
\end{equation}
where $r \equiv |\vec{x}_\perp - \vec{y}_\perp|$ for $F=(E,B)$. The longitudinal spatial correlation function is computed with $F_z$, while the transverse one is computed with $F_i$ with $i=(x,y)$ and summation over $i$ is implied. On the lattice, the spatial correlation is computed as $\langle \mathrm{tr} \left[ F_{\boldsymbol{x},T/L} U_{\boldsymbol{x} \rightarrow \boldsymbol{y}} F_{\boldsymbol{y},T/L} U^\dagger_{\boldsymbol{x} \rightarrow \boldsymbol{y}} \right] \rangle$. To ensure lattice gauge invariance, the field is parallel transported with
\begin{equation}
    U_{\boldsymbol{x} \rightarrow \boldsymbol{y}} = \prod_{\boldsymbol{z} \in \gamma(\boldsymbol{x},\boldsymbol{y})} U_{\boldsymbol{z},i}\,,
\end{equation}
where $\gamma(\boldsymbol{x},\boldsymbol{y})$ is the path connecting the lattice sites $\boldsymbol{x}$ and $\boldsymbol{y}$.

The spatial correlators are shown in Fig.~\ref{fig:corr-spatial}. Initially, the longitudinal components are much larger in magnitude than the transverse ones, and the longitudinal electric and magnetic correlators at any time exhibit different behaviors. Only the longitudinal magnetic correlator shows a clear anticorrelation inside a correlation domain $r< 1/(g^2\mu)$, while the electric fields are always correlated. The longitudinal magnetic anticorrelation region disappears with increasing time. The transverse electric and magnetic correlators behave similarly, and exhibit an anticorrelation region which gets pushed to larger separation. This is due to the longitudinal expansion which effectively increases the correlation length. The observed difference between the transverse and longitudinal correlators could be explained by the boost invariance, which breaks the symmetry, but the reason for these particular correlation patterns is not fully understood. 

\noindent\textbf{\textit{Temporal correlations}}. Similarly, the gauge invariant longitudinal and transverse temporal correlation functions for the color electric and magnetic fields, extracted as a function of the time separation $\delta\tau\equiv \tau-\tau_0$ from a fixed initial time $\tau_0$, are defined at fixed transverse position $\vec{x}_\perp$ as
\begin{equation}
    \langle FF\rangle_{L/T}(\delta\tau)\equiv \left\langle \mathrm{tr}\left[F_{L/T}(\tau_0,\vec{x}_\perp)F_{L/T}(\tau_0+\delta\tau,\vec{x}_\perp)\right]\right\rangle\,.
\end{equation}
On the lattice, the temporal correlation requires no gauge links since $A_\tau=0$, and is computed as $\langle \mathrm{tr}\left[F_{\boldsymbol{x},L/T}(\tau_0)F_{\boldsymbol{x},L/T}(\tau_0+\delta\tau)\right]\rangle$.

The temporal correlators are shown in Fig.~\ref{fig:corr-time}. The longitudinal components peak at the initial time and decay as the fields become more dilute. Similarly to the spatial case, the electric and magnetic correlators behave differently, with the magnetic correlator showing a negative dip after $\delta\tau>1/(g^2\mu)$. The transverse components show similar trends for both electric and magnetic contributions, with smaller values at early times, which then increase and overshoot their initial value before decaying. Unlike the longitudinal case, the transverse correlators remain positive throughout, and the electric and magnetic components track each other closely in shape.

\subsection{Coulomb gauge transformation}
\label{sec:coulomb-gauge}

\noindent\textbf{\textit{Coulomb gauge fixing}}. The Coulomb gauge is a convenient choice for defining the gluon occupation numbers and minimizing gauge artifacts. The glasma fields are evolved in the temporal gauge $A^\tau=0$, which leaves a residual freedom to perform time-independent gauge transformations. The transverse Coulomb gauge is imposed at a fixed $\tau$ and is given by
\begin{equation}\label{eq:CoulombGauge}
    \sum_{i\in(x,y)} \partial_i A_i(\tau,\vec{x}_\perp) = 0\,.
\end{equation}
Under boost invariance, this is equivalent to the generalized Coulomb condition $(1/\tau^2)\partial_\eta A_\eta + \sum_i \partial_i A_i = 0$. The Coulomb gauge minimizes the surface integral of the squared gauge potential~\cite{Gubarev:2000eu, Gubarev:2000nz}
\begin{equation}\label{eq:A2perp}
    \langle A_\perp^2(\tau)\rangle \equiv \int \diff^2\vec{x}_\perp\, \mathrm{tr}\!\left[\vec{A}_\perp^{\,2}(\tau,\vec{x}_\perp)\right]\,,
\end{equation}
with $\vec{A}_\perp = (A_x, A_y)$. The gauge condition is usually imposed only at $\tau=0$, but the fields slowly drift away from the Coulomb condition.

\begin{figure}[tp]
    \centering
    \begin{subfigure}[b]{0.48\textwidth}
        \centering
        \includegraphics[width=0.9\textwidth]{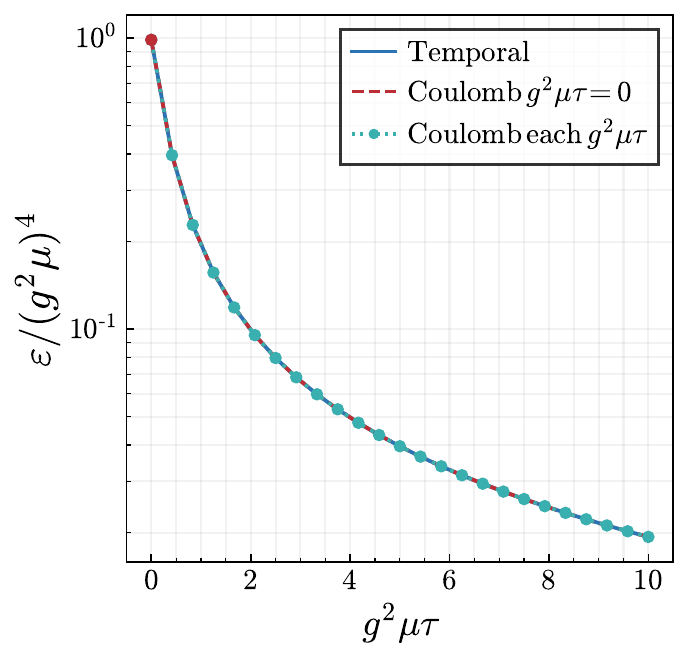}
        \captionsetup{justification=centering}
        \caption{Energy density.}
        \label{fig:energy-coulomb}
    \end{subfigure}
    \hfill
    \begin{subfigure}[b]{0.48\textwidth}
        \centering
        \includegraphics[width=0.9\textwidth]{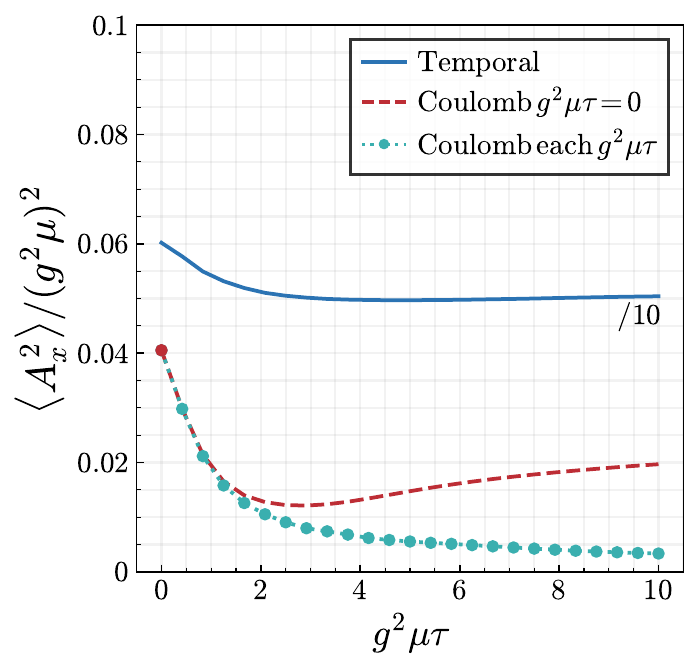}
        \captionsetup{justification=centering}
        \caption{Squared transverse gauge potential.}
        \label{fig:asq-x-coulomb}
    \end{subfigure}
    \caption{Energy density and squared transverse gauge potential as functions of proper time for different gauge choices. Quantities are represented in dimensionless units for temporal gauge \emph{(blue)}, Coulomb fixed at $\tau=0$ \emph{(red)} and Coulomb fixed at each $\tau$ \emph{(teal)}.}
    \label{fig:ed-asq-coulomb}
\end{figure}

\noindent\textbf{\textit{Lattice technique}}. The lattice Coulomb gauge fixing maximizes the functional
\begin{equation}\label{eq:FUG}
    \mathcal{F}_U[G] = \frac{1}{8N_c V_\perp}\sum_{\boldsymbol{x},\,i} \mathrm{tr}\!\left[U_{\boldsymbol{x},i}^{G} + U_{\boldsymbol{x},i}^{G\dagger}\right]\,,
\end{equation}
over SU($N_c$) gauge transformations $G_{\boldsymbol{x}} = \exp(ig_{\boldsymbol{x}})$ with $g_{\boldsymbol{x}} = g^a_{\boldsymbol{x}}\,t^a$, where the gauge-transformed links are $U_{\boldsymbol{x},i}^{G} = G_{\boldsymbol{x}}\,U_{\boldsymbol{x},i}\,G_{\boldsymbol{x}+i}^\dagger$.
Maximizing~\cref{eq:FUG} by varying $g^a$ at each $\boldsymbol{x}$ around $g^a \to 0$ yields the lattice Coulomb gauge condition
\begin{equation}
    \Delta_{\boldsymbol{x}} \equiv \sum_i \left[\Delta_{-i}U_{\boldsymbol{x},i}^{G} - \mathrm{h.c.} - \mathrm{trace}\right] = 0\,,
\end{equation}
with the backward difference $\Delta_{-i}U_{\boldsymbol{x},j} \equiv U_{\boldsymbol{x}-i,j} - U_{\boldsymbol{x},j}$. Taking the squared divergence averaged over the lattice, one may track the convergence through
\begin{equation}\label{eq:theta-coulomb}
    \theta = \frac{1}{V_\perp N_c}\sum_{\boldsymbol{x}}\mathrm{tr}\!\left[\Delta_{\boldsymbol{x}}^2\right]\,.
\end{equation}

The naive iteration of the gauge condition suffers from \emph{critical slowing down}, where the small $k_\perp$ modes converge much more slowly than large $k_\perp$ modes. This makes the algorithm slow on large lattices, which is a practical limitation. Fourier acceleration fixes this by rescaling the update step in momentum space such that all modes relax toward the gauge condition at roughly the same rate. This is achieved by applying the gauge transformation
\begin{equation}
    g_{\boldsymbol{x}}=\left[F^{-1}\left(\frac{\alpha}{2}\frac{k^2_{\perp,\mathrm{max}}}{k^2_\perp}\right)F\right]\Delta_{\boldsymbol{x}}\,,
\end{equation}
with $F$ being the Fourier transform, $F^{-1}$ the inverse Fourier transform, and where $k_\perp^2$ is the lattice momentum from Eq.~\eqref{eq:lattice-momentum}, giving $k_{\perp,\mathrm{max}}^2 = 8/a_\perp^2$. The gauge transformation is updated iteratively as $G_{\boldsymbol{x}}^\mathrm{new} = \exp(ig_{\boldsymbol{x}})\,G_{\boldsymbol{x}}^\mathrm{old}$ until $\theta$ reaches the desired numerical accuracy.

An adaptive convergence parameter is used, starting at $\alpha^{\mathrm{init}}=0.16$ near the Abelian estimate $\alpha=1/8$. It is halved when $\theta$ oscillates and increased by a factor of $1.1$ when convergence stalls. The resulting optimal gauge transformation at $\tau=0$, denoted $G_{\boldsymbol{x}}^\mathrm{opt}$, is applied to all gauge links and fields as $U_{\boldsymbol{x},i} \to G_{\boldsymbol{x}}^\mathrm{opt}\,U_{\boldsymbol{x},i}\,G_{\boldsymbol{x}+i}^{\mathrm{opt},\dagger}$ and $A_{\boldsymbol{x},\eta} \to G_{\boldsymbol{x}}^\mathrm{opt}\,A_{\boldsymbol{x},\eta}\,G_{\boldsymbol{x}}^{\mathrm{opt},\dagger}$, along with the
conjugate momenta $P_{\boldsymbol{x}}^\mu \to G_{\boldsymbol{x}}^\mathrm{opt}\,P_{\boldsymbol{x}}^\mu\,G_{\boldsymbol{x}}^{\mathrm{opt},\dagger}$ for $\mu \in (i,\eta)$. 

\noindent\textbf{\textit{Results in the glasma}}. After imposing the Coulomb gauge condition at $\tau=0$, one can monitor how well it is numerically maintained as the glasma fields evolve. For this purpose, Fig.~\ref{fig:ed-asq-coulomb} shows the energy density and the squared gauge potential $\langle A_x^2\rangle$ as a function of proper time for different gauge choices: temporal gauge, Coulomb fixed once at $\tau=0$, and the Coulomb condition imposed at each $\tau$. As expected, the energy density is gauge invariant, providing a numerical check of the Coulomb gauge fixing. On the other hand, the squared gauge potential, which is gauge dependent, gives in the temporal gauge a value one order of magnitude larger than either Coulomb choice. Fixing the Coulomb condition at $\tau=0$ reduces the initial $\langle A_x^2\rangle$, but the condition gradually drifts as the fields evolve, with $\langle A_x^2\rangle$ growing at late times. Imposing the Coulomb condition at every time step keeps $\langle A_x^2\rangle$ small, while simulations where the Coulomb gauge fixing is done only at $\tau=0$ start to deviate from this but
only from $g^2\mu\tau \gtrsim 2$, which is already beyond the typical lifetime of the glasma.

\begin{emphbox}{gray!50!black}{Summary}
\begin{itemize}[label=$\circ$, leftmargin=*]
    \item \textbf{Color Glass Condensate.} Contains a review of the CGC framework. It starts from DIS, discussing gluon saturation and small-$x$ evolution. Then, it describes the CGC separation of scales, obtains the pure gauge fields of a high-energy nucleus and introduces the MV model. Finally, the saturation momentum is derived from the dipole correlator.
    \item \textbf{Glasma fields \& features.} Derives the initial conditions from the pure gauge gluon fields and the patching prescription at initial time for boost-invariant fields. The classical Yang-Mills evolution is discretized using real-time lattice gauge theory. Constructs the energy-momentum tensor and shows numerical results for the energy density, along with longitudinal and transverse pressures in the glasma.
    \item \textbf{Correlators \& gauge choice.} Defines gauge invariant spatial and temporal two-point correlators for the electric and magnetic fields. Lattice results show these correlators for both longitudinal and transverse components, revealing different correlation patterns. Derives the transverse Coulomb condition, which minimizes the square of the transverse potential. The lattice implementation is described in detail.
\end{itemize}
\end{emphbox}

\chapter{Classical probes in the glasma}
\label{chap:hardprobes}

\section{Classical transport}
\label{sec:classical-transport}

\begin{emphbox}{myred}{Based on}
Article~\cite{ar1} presents the numerical framework used to simulate heavy quarks and jets propagating through the glasma. This is done by coupling test particles described by Wong's equations to classical real-time lattice gauge theory for the fields. A key technical contribution is the generation of random SU(3) classical color charges with group theory properties. The work also proposes a prescription to match classical to quantum quantities and it shows that momentum broadening must be divided by the dimension of the color representation to restore Casimir scaling. Results for heavy quarks show that the transport coefficient $\kappa$ is strongly peaked at early times, the momentum broadening is anisotropic, and damped oscillations manifest in the longitudinal direction. Analogous results for jets confirm a highly peaked $\hat{q}$ at early times and quantify finite mass and momentum effects beyond the eikonal limit.
\end{emphbox}

\subsection{Classical particles in Yang-Mills fields}

\noindent\textbf{\textit{Classical transport in the glasma}}. The gluon occupation numbers $f\sim 1/\alpha_s\gg 1$ in the glasma are large and thus justify the classical treatment of the fields. The produced hard partons are quantum, but since they are less abundant compared to the overoccupied fields, their backreaction is negligible. The classical trajectory approximation is justified by the \emph{Wentzel-Kramers-Brillouin} (WKB) condition~\cite{Litim:2001db} for the transverse wavelength $\lambda_T\sim 1/p_T\ll 1/Q_s$, where $Q_s$ sets the transverse correlation length of the boost-invariant glasma fields. This requires $p_T\gg Q_s$, but the classical framework is applied as a leading order estimate for any particle $p_T$. The hard partons are therefore described as classical colored test particles obeying transport equations in classical Yang-Mills fields~\cite{Wong:1970fu}.

\begin{emphbox}{myblue}{Wong's equations}
The equations of motion for a classical particle with mass $m$, color charge $Q^a$, and kinetic momentum $p^\mu$ in a background gauge field $A_\mu^a$ and corresponding field strength tensor $F_{\mu\nu}^a$ are given by
\begin{subequations} 
    \label{eq:Wong}
    \begin{align} 
        &\frac{\diff x^\mu}{\diff \lambda} = \frac{1}{m}p^\mu(\lambda)\,,\\
        &\frac{\diff p^\mu}{\diff \lambda}=\frac{g}{m} Q^a(\lambda) \,F^{\mu\nu,a}(\lambda)\,p_\nu(\lambda)\,,\\
        &\frac{\diff Q^a}{\diff \lambda}=\frac{g}{m} f^{abc}Q^c(\lambda)\,A_\mu^b(\lambda)\,p^\mu(\lambda)\,,
    \end{align}
\end{subequations}
where $\lambda$ is the affine parameter along the particle trajectory, which gives $(\diff\lambda)^2=g_{\mu\nu}\diff x^\mu\diff x^\nu$ as the infinitesimal worldline element, and $f^{abc}$ are the antisymmetric structure constants of the gauge group.
\end{emphbox}

Wong's equations are derived in~\cref{app:lightfront} in two ways: as the classical limit of the Heisenberg equations of motion starting from the Dirac Hamiltonian, and as equations of motion for test particles in a Boltzmann-Vlasov equation.

To express these equations in color matrix form, one defines $Q\equiv Q^a t^a$, $A_\mu=A_\mu^a t^a$, and $F_{\mu\nu}=F_{\mu\nu}^a t^a$, where $t^a$ are the generators of the gauge group in representation $R$. Using $\mathrm{tr}[t^at^b]=T_R\delta^{ab}$ with $T_R$ being the normalization constant for the generators in representation $R$, Wong's equations for the momentum and color charge may be written as
\begin{equation}
    \label{eq:WongAlg}
    \frac{\diff p^\mu}{\diff \lambda}=\frac{g}{T_R} \mathrm{tr}\left[Q(\lambda) \,F^{\mu\nu}(\lambda)\right] \frac{\diff x_\nu}{\diff \lambda}\,,\qquad \frac{\diff Q}{\diff \lambda}=ig [Q(\lambda),A_\mu(\lambda)]\frac{\diff x^\mu}{\diff \lambda}\,.
\end{equation}

The time dependence of the color charge may be expressed as a rotation in color space according to
\begin{align} \label{eq:WongColorRotation}
    Q(\lambda)=U(\lambda;\lambda_0)\,Q(\lambda_0)\,U^\dagger(\lambda;\lambda_0)\,,
\end{align}
with a Wilson line along the particle trajectory 
\begin{align} \label{eq:ParticleWilsonLine}
    U(\lambda;\lambda_0)=P\exp\left[-ig\int_{\lambda_0}^\lambda \diff\lambda^\prime \frac{\diff x^\mu}{\diff\lambda^\prime} A_\mu(\lambda^\prime)\right]\,.
\end{align}
This is obtained using $\diff U(\lambda;\lambda_0)/\diff\lambda=-ig\,\diff x^\mu/\diff\lambda\,A_\mu(\lambda)\,U(\lambda;\lambda_0)$, which is the parallel transport equation for a Wilson line. The classical color charge conserves the quadratic $q_2(R)$ and cubic $q_3(R)$ Casimir invariants along the trajectory in a given representation $R$ of the gauge group
\begin{equation}
    Q^aQ^a=q_2(R)\,, \qquad d^{abc}Q^aQ^bQ^c=q_3(R)\,,
\end{equation}
where the color charges are evaluated at the same $\lambda$ and $d^{abc}$ are the totally symmetric structure constants of the gauge group.

\newpage
It is convenient to express Wong's equations in Milne coordinates, see~\cref{app:milne}, namely $x^\mu=(\tau,x,y,\eta)$ and $p^\mu=(p^\tau,p^x,p^y,p^\eta)$. The glasma fields are boost invariant, but due to the presence of the longitudinal color electric and magnetic fields, the particle trajectories depend on the spacetime rapidity $\eta$. Wong's equations for coordinates read as 
\begin{equation}
    \frac{\mathrm{d} x^\mu}{\mathrm{d}\tau}=\frac{p^{\mu}}{p^\tau}\,,\qquad\text{with}\,\, (p^\tau)^2=(p^x)^2+(p^y)^2+\tau^2 (p^\eta)^2+m^2\,,
\end{equation}
where $x^\mu\in(x,y,\eta)$. The equations for momentum become
\begin{subequations}
    \label{eq:wonginmilne}
    \begin{align}
        &\tau\dfrac{\mathrm{d}p^\eta}{\mathrm{d}\tau}+2 p^\eta=\frac{g}{T_R}\left(\tr{Q E_\eta}\! - \! \tr{QB_x}\frac{p^y}{p^\tau} \! + \! \tr{QB_y} \frac{p^x}{p^\tau}\right)\,,\\
        &\frac{\mathrm{d}p^x}{\mathrm{d}\tau}=\frac{g}{T_R}\left(\tr{QE_x} \! + \! \tr{QB_\eta}\frac{p^y}{p^\tau}\! - \! \tr{QB_y} \frac{\tau p^\eta}{p^\tau}\right)\,,\\
        &\frac{\mathrm{d}p^y}{\mathrm{d}\tau}=\frac{g}{T_R}\left(\tr{QE_y}\! -\! \tr{QB_\eta} \frac{p^x}{p^\tau} \! + \! \tr{QB_x}\frac{\tau p^\eta}{p^\tau}\right)\,,
    \end{align}
\end{subequations}
with the longitudinal and transverse electric and magnetic fields defined as $E_i\equiv F_{\tau i}$,  $B_i\equiv \epsilon_{ij}F_{\eta j}/\tau$ for $i\in(x,y)$, $E_\eta\equiv F_{\tau\eta}/\tau$ and $B_\eta\equiv -F_{xy}$, which are the longitudinal fields $E_z$ and $B_z$ of~\cref{chap:glasma}. See~\Cref{app:milne} for a derivation of the field strength tensor in Milne coordinates. The color charge rotation takes the form
\begin{equation}
    Q(\tau)=U(\tau;\tau_0)\,Q(\tau_0)\,U^\dagger(\tau;\tau_0)\,,
\end{equation}
with the particle Wilson line given by
\begin{equation}
    U(\tau;\tau_0)=P\exp\Bigg[-ig\int_{\tau_0}^{\tau}\mathrm{d}\tau^\prime\,\frac{\mathrm{d}x^\mu}{\mathrm{d}\tau^\prime}\,A_\mu(x(\tau^\prime))\Bigg]\,.
\end{equation}
The evolution equation for the temporal component
\begin{align}
    \label{eq:ptaueq}
    \begin{split}
    \dfrac{\mathrm{d}p^\tau}{\mathrm{d}\tau}+\frac{\tau p^\eta}{p^\tau}p^\eta=\frac{g}{T_R}\left(\tr{Q E_\eta}\frac{\tau p^\eta}{p^\tau}+\tr{QE_x} \frac{p^x}{p^\tau}+\tr{QE_y} \frac{p^y}{p^\tau}\right)\,,
    \end{split}
\end{align}
is used in practice as a numerical check, since this component is extracted from the norm of the four-momentum.

\noindent\textbf{\textit{Classical color algebra}}. The color charges $Q^a$ which enter Wong's equations are classical random variables whose statistical properties must be fixed consistently with QCD and SU($N_c$) group theory. This is done by taking the classical limit of the SU($N_c$) Lie algebra, replacing the quantum commutators with Poisson brackets and imposing that the resulting $n$-point functions reproduce the quantum trace relations.

\begin{emphbox}{myblue}{Classical limit of a Lie algebra}
Working in the classical limit of the quantum Lie algebra of $SU(N_c)$, one derives the classical color charges. The quantum generators $\hat{T}^a$ obey the commutation relations
\begin{equation}
\big[\hat{T}^a,\hat{T}^b\big]=i f^{abc}\hat{T}^c\,.
\end{equation}
Taking the classical limit as $-i[\hat{X},\hat{Y}]\to\{X,Y\}_{PB}$, where $\{\cdot,\cdot\}_{PB}$ denotes the Poisson bracket, the classical color charges $Q^a$ satisfy
\begin{equation}
\left\{Q^a,Q^b\right\}_{PB}=f^{abc}Q^c\,.
\end{equation}
Treating $Q^a$ as functions on a manifold with phase space coordinates $(\phi_k,\xi_k)$, the Poisson bracket takes the explicit form
\begin{equation}
\label{eq:liepb-explicit}
\left\{Q^a,Q^b\right\}_{PB}=\frac{\partial Q^a}{\partial\phi_k}\frac{\partial Q^b}{\partial\xi_k}-\frac{\partial Q^a}{\partial\xi_k}\frac{\partial Q^b}{\partial\phi_k}\,,
\end{equation}
where summation over $k$ is implicit.
\end{emphbox}

This construction extends to any semisimple Lie algebra~\cite{Bulgac_1990}, but in particular for SU($2$) and SU($3$), it has an explicit solution in terms of Darboux variables~\cite{Johnson:1988qm,Litim:2001db}, see~\cref{app:su3} for more details.

In practice, we choose to initialize the color charges as classical random variables whose one-, two- and three-point functions mirror the quantum trace relations of SU($N_c$)~\cite{Heinz:1983nx,Kelly:1994dh,Litim:2001db}, namely $\mathrm{tr}[\hat{T}^a]=0$, $\mathrm{tr}[\hat{T}^a\hat{T}^b]=T_R\delta^{ab}$, and $\mathrm{tr}[\hat{T}^a\hat{T}^b\hat{T}^c]=A_R/4\,(d^{abc}+if^{abc})$, as
\begin{equation}
    \label{eq:qnpointfct}
    \langle Q^a \rangle_Q = 0\,,\quad \langle Q^a Q^b \rangle_Q = T_R\,\delta^{ab}\,,\quad \langle Q^a Q^b Q^c \rangle_Q = \frac{A_R}{4}\,d^{abc}\,.
\end{equation}
The subscript $Q$ denotes the average over the classical color charge configurations $\langle X \rangle_Q \equiv \int \diff Q\, X$, normalized as $\int \diff Q = 1$. Here, $T_R$ and $A_R$ are the Dynkin and anomaly coefficients, which for $R=F$ representation are given by $T_F=1/2$ and $A_F=1$, while for $R=A$ representation, are given by $T_A=N_c$ and $A_A=0$. These choices fix the quadratic and cubic Casimir invariants to
\begin{equation}
    q_{2,3}(R)=D_R\,C_{2,3}(R)\,,
\end{equation}
where $D_R$ is the dimension of the representation and $C_{2,3}(R)$ are the group theory quadratic and cubic Casimirs. Their explicit values are
\begin{subequations}
    \label{eq:classicalcasimirvals}
    \begin{alignat}{2}
        q_2(F) &= \frac{N_c^2-1}{2}\,, \qquad &
        q_3(F) &= \frac{(N_c^2-4)(N_c^2-1)}{4N_c}\,,\\
        q_2(A) &= N_c(N_c^2-1)\,, \qquad &
        q_3(A) &= 0\,.
    \end{alignat}
\end{subequations}
Notice that this is a choice that comes from the trace relations and a specific normalization, but is not unique. There is no preferred value for the classical Casimirs as they may be chosen only by matching to the quantum Casimirs. Curiously, for SU($3$), one may not generate classical charge components that simultaneously satisfy the group theory Casimirs for the fundamental representation, namely $q_{2,3}(F)=C_{2,3}(F)$. This is perhaps because the classical charges are taken in the limit of high occupation numbers, which may not be a good approximation for the fundamental representation. For this reason, we choose $q_{2,3}(F)=D_F\,C_{2,3}(F)$, then rescale the results such that in the end, the expected quantum color factor is recovered. 

Given a reference color vector $Q_0$, color charges are generated as $\mathrm{SU}(N_c)$ rotations $Q(U) = U Q_0 U^\dagger$,
with $U\in\mathrm{SU}(N_c)$ distributed according to the Haar measure, normalized as $\int\mathrm{d}U = 1$. The color components are extracted by projecting onto the generators $T^a$, namely $Q^a = \tr{QT^a}/T_R = Q_0^b\,U^{ab}$, where $U^{ab}$ is the adjoint representation matrix $U^{ab} \equiv \tr{T^a U T^b U^\dagger}/T_R$. By construction, the Casimir invariants $q_2$ and $q_3$ are preserved under these rotations. Thus, sampling random color charges $Q$ is equivalent to sampling random rotations $U$ of a reference color charge $Q_0$ that satisfies the Casimir constraints.

The resulting Haar measure averages of the color charges are
\begin{subequations}
    \label{eq:nptfcthaar}
    \begin{align}
        &\langle Q^a\rangle_U = Q_0^{a'}\int\diff U\, U^{aa'}\,,\\
        &\langle Q^aQ^b\rangle_U = Q_0^{a'}Q_0^{b'}\int\diff U\, U^{aa'}U^{bb'}\,,\\
        &\langle Q^aQ^bQ^c\rangle_U = Q_0^{a'}Q_0^{b'}Q_0^{c'}\int\diff U\, U^{aa'}U^{bb'}U^{cc'}\,.
    \end{align}
\end{subequations}
These integrals can be performed explicitly and they reproduce the one-, two- and three-point functions from Eqs.~\eqref{eq:qnpointfct}, with the result depending on $Q_0$ only through the Casimirs $q_2$ and $q_3$. Sampling over the Haar measure $\int\mathrm{d}U$ is therefore equivalent to direct sampling of classical color charges $\int\mathrm{d}Q$.

To generalize, one would define the color charge integration to match the quantum trace relations
\begin{equation}
    \int\diff Q\, Q^{a_1}\cdots Q^{a_n} = \tr{T^{(a_1}\cdots T^{a_n)}}\,,
\end{equation}
where $T^{(a_1}\cdots T^{a_n)}$ denotes the fully symmetrized product of generators. It is worth noting that this holds for $n=(1,2,3)$ with $Q^a$ satisfying the Casimir constraints, but fails for $n\geq 4$~\cite{Laine:2001my}. The classical framework is therefore consistent with pQCD only at the level of quadratic and cubic color moments and is not expected to capture higher order color correlations.

\noindent\textbf{\textit{Numerical implementation}}. Test particles are initialized randomly in the transverse simulation region with $x(\tau_0)\in[0,L_\perp)$ and $y(\tau_0)\in[0,L_\perp)$, where $\tau_0$ is the initial proper time, while $\eta(\tau_0)=0$. Momenta are initialized with fixed transverse values $p^x(\tau_0)$ and $p^y(\tau_0)$, while $p^\eta(\tau_0)=0$. The reference color charge $Q_0$ satisfies the Casimir constraints $Q_0^a Q_0^a=q_2(R)$ and $d^{abc}Q_0^a Q_0^b Q_0^c=q_3(R)$ and is used to generate random color charges $Q(\tau_0)=U Q_0 U^\dagger$ with $U$ a random SU($3$) matrix. An ensemble of $N_\mathrm{tp}$ test particles is initialized in a single glasma configuration and their trajectories are evolved using Wong's equations in Milne coordinates, while the glasma fields are obtained by solving the classical Yang-Mills equations. Performing ensemble averages over all test particles and glasma configurations gives the observables of interest. The particle solver is publicly available~\cite{curraun:wong}.

\begin{figure}[t]
    \centering
    \begin{subfigure}[b]{0.48\textwidth}
        \centering
        \includegraphics[width=\textwidth]{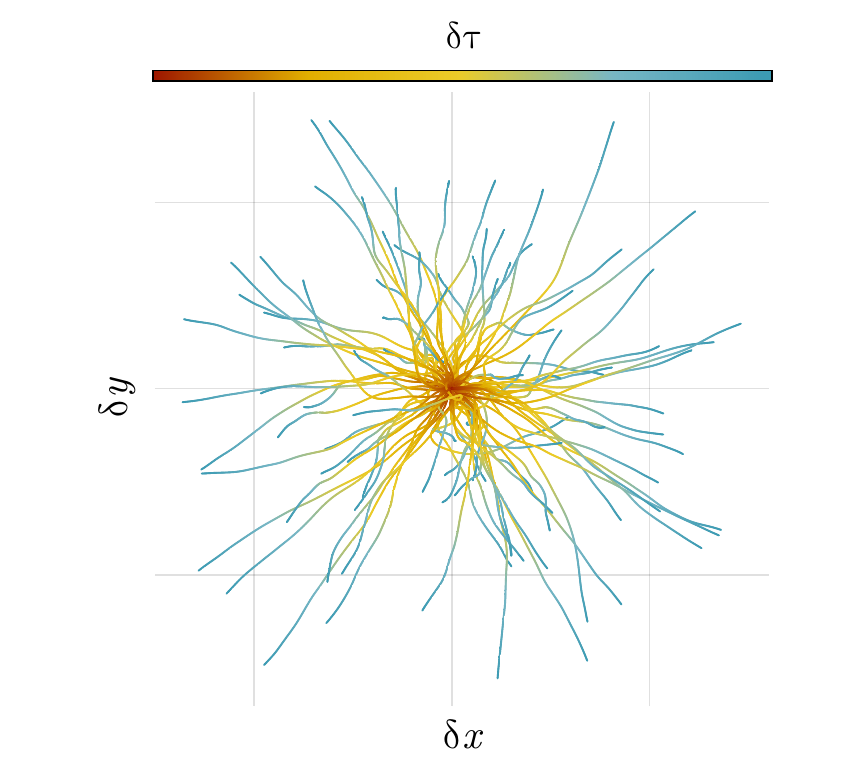}
        \caption{Trajectories in the transverse plane $(\delta x,\delta y)$ relative to the initial position.}
        \label{fig:glasma-2d}
    \end{subfigure}
    \hfill
    \begin{subfigure}[b]{0.48\textwidth}
        \centering
        \includegraphics[width=\textwidth]{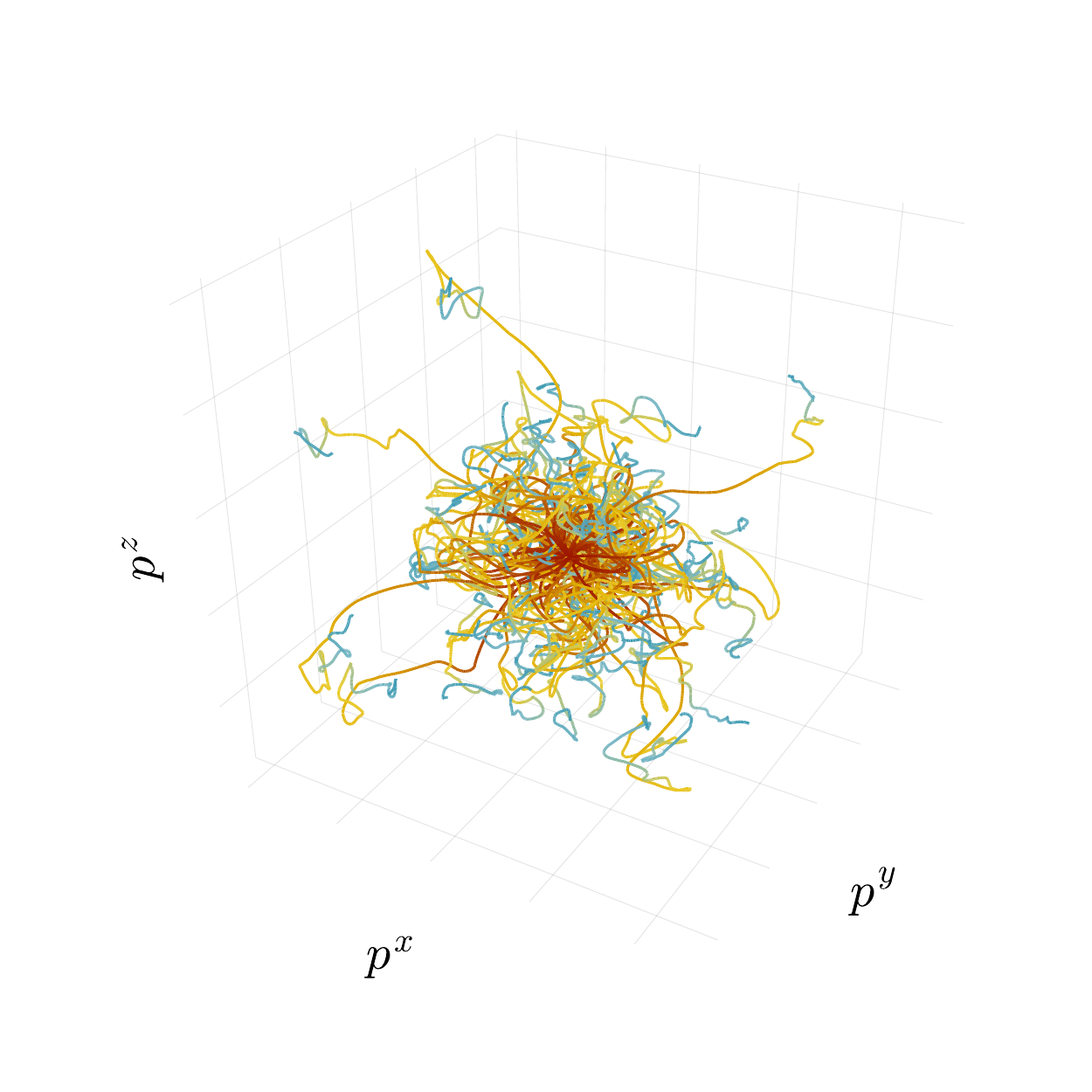}
        \caption{Trajectories in momentum space $(p^x, p^y, p^z)$ starting from null momenta.}
        \label{fig:glasma-3d}
    \end{subfigure}
    \caption{Trajectories for $N_\mathrm{tp}=100$ charm quark test particles propagating in glasma fields, solved using the particle solver. The color code indicates the proper time $\delta\tau\equiv\tau-\tau_0$ along the trajectory. Figure reproduced from Article~\cite{ar1} under license CC BY 4.0.}
    \label{fig:glasma-spaghetti}
\end{figure}

The location of each particle is approximated by the nearest grid point $\boldsymbol{x}_n(\tau_n)\equiv \mathrm{NGP}[\vec{x}_\perp(\tau_n)]$. The glasma electric and magnetic fields at the nearest lattice sites and integer time steps are then obtained through interpolation as
\begin{equation}
        \begin{aligned}
            &E_{i}\left(\tau_n,{\boldsymbol{x}_n}(\tau_n)\right)= \dfrac{1}{4 \tau_{n}}\left[P_{\boldsymbol{x}_n}^{i}(\tau_{n+\frac{1}{2}})+P_{\boldsymbol{x}_n}^{i}(\tau_{n-\frac{1}{2}})+ P_{\boldsymbol{x}_n}^{i, -i}(\tau_{n+\frac{1}{2}})+P_{\boldsymbol{x}_n}^{i, -i}(\tau_{n-\frac{1}{2}})\right]\,,\\
            &E_{\eta}\left(\tau_n,{\boldsymbol{x}_n}(\tau_n)\right) = \dfrac{1}{2}\left[P_{\boldsymbol{x}_n}^{\eta}(\tau_{n+\frac{1}{2}})+P_{\boldsymbol{x}_n}^{\eta}(\tau_{n-\frac{1}{2}})\right]\,,
        \end{aligned}
    \end{equation}
and similarly
\begin{align}
    &B_{i}\left(\tau_n,{\boldsymbol{x}_n}(\tau_n)\right)=-\dfrac{\epsilon_{ij}}{2 \tau_{n} a_{\perp}} \left[A_{\boldsymbol{x}_n, \eta}^{+j}\left(\tau_{n}\right) - A_{\boldsymbol{x}_n, \eta}^{-j}\left(\tau_{n}\right) \right]\,,\\
    &B_{\eta}\left(\tau_{n}, {\boldsymbol{x}_n}(\tau_n)\right)=- \dfrac{1}{4 g a_{\perp}^{2}}\Big[U_{\boldsymbol{x}_n, xy}\left(\tau_{n}\right)+U_{\boldsymbol{x}_n, y-x}\left(\tau_{n}\right)\notag\\
    &\hspace{10.5em}+U_{\boldsymbol{x}_n, -x-y}\left(\tau_{n}\right)+U_{\boldsymbol{x}_n, -yx}\left(\tau_{n}\right)\Big]_{\mathrm{ah}}\,.\notag
\end{align}
The lattice notation for the glasma fields and conjugate momenta ($U_{\boldsymbol{x},i}$, $A_{\boldsymbol{x},\eta}$, $P^i_{\boldsymbol{x}}$, $P^\eta_{\boldsymbol{x}}$) follows the conventions introduced in~\cref{sec:glasma-fields}. The particle Wilson line is approximated as a product of infinitesimal Wilson lines along the particle trajectory 
\begin{equation}
    \label{eq:colorrotnum}
    U(\tau_n;\tau_0) = \prod_{m=0}^{n-1} U(\tau_{m+1};\tau_{m})\,,\quad\text{with}\,\, U(\tau_{m+1};\tau_{m}) = U_{\boldsymbol{x}_{m},i}\, U_{\boldsymbol{x}_{m+1},\eta}\,,
\end{equation}
where $U_{\boldsymbol{x}_{m},i}(\tau_{m+1})$ is the transverse gauge link. One constructs a gauge link in rapidity as $U_{\boldsymbol{x}_{m+1},\eta}(\tau_{m+1})\equiv\exp\left[ig\,\delta\eta_{m+1}\,A_{\boldsymbol{x}_{m+1},\eta}(\tau_{m+1})\right]$, which is obtained from $A_\eta$ using matrix exponential. The factorization between transverse and rapidity gauge fields holds for small time steps $\delta\tau_{m+1}=\tau_{m+1}-\tau_m$ such that the relative rapidity $\delta\eta_{m+1}$ is also small. Consequently, one may neglect $\left[\int\diff x^i A_i,\,\delta\eta_{m+1} A_\eta\right]$ in the path-ordered exponential. In the \emph{Baker-Campbell-Hausdorff} (BCH) equation for the matrix exponential $\exp[A]\exp[B]\approx \exp[A+B+[A,B]/2+\ldots]$, higher order terms are therefore suppressed. The NGP is only used to evaluate the fields entering the Lorentz force, while the particle coordinates $(\vec{x}_\perp, \eta)$ remain continuous. The color charge is parallel transported using gauge links between lattice sites, and is thus also evaluated at the NGP. This prescription ensures numerical gauge invariance.

The resulting numerical trajectories obtained with the colored particle-in-cell solver are visualized in Fig.~\ref{fig:glasma-spaghetti}. The test particles are randomly distributed in the transverse simulation plane at formation time $\tau_0$ with vanishing initial momenta. The positions and momenta evolve according to Wong's equations, while the color charge gets rotated at each time step using transverse gauge links and a rapidity Wilson line. The spread of trajectories in both position and momentum space shows the chaotic behavior of the glasma fields. More precisely, particles initialized with the same momentum magnitude but different directions quickly diverge as they propagate through different correlation domains and accumulate momentum deflections from the glasma fields. 

\begin{figure}[t]
    \centering
    \includegraphics[width=0.5\textwidth]{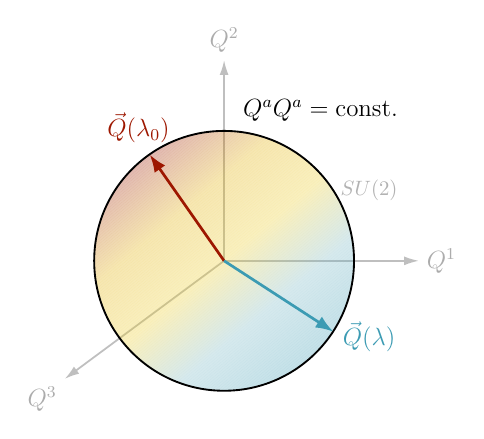}
    \caption{The color charge vector $\vec{Q}=(Q^1,Q^2,Q^3)$ for the classical algebra of SU(2) is constrained on a sphere of radius $\sqrt{Q^aQ^a}$ fixed by the quadratic Casimir $q_2$. The initial charge $\vec{Q}(\lambda_0)$ \textit{(red)} color rotates to $\vec{Q}(\lambda)$ \textit{(blue)}.}
    \label{fig:charge-sphere}
\end{figure}

Figure~\ref{fig:charge-sphere} depicts the classical color algebra for SU($2$) and also its SU($2$) subalgebra for SU($3$). In $\mathbbm{R}^{N_c^2-1}$ the classical color charges $Q^a$ are not free vectors, but occupy a fixed orbit determined by the Casimir invariants. The orbit is a $2$-sphere for SU($2$), whereas for SU($3$), it is a $6$-dimensional manifold parametrized by both $q_2$ and $q_3$. The classical particle color charges undergo SU($N_c$) rotations. The more convenient numerical approach is working with classical color charges stored as rotated color vectors, because this explicitly conserves the Casimir invariants.

\begin{figure}[t]
    \centering
    \includegraphics[width=\textwidth]{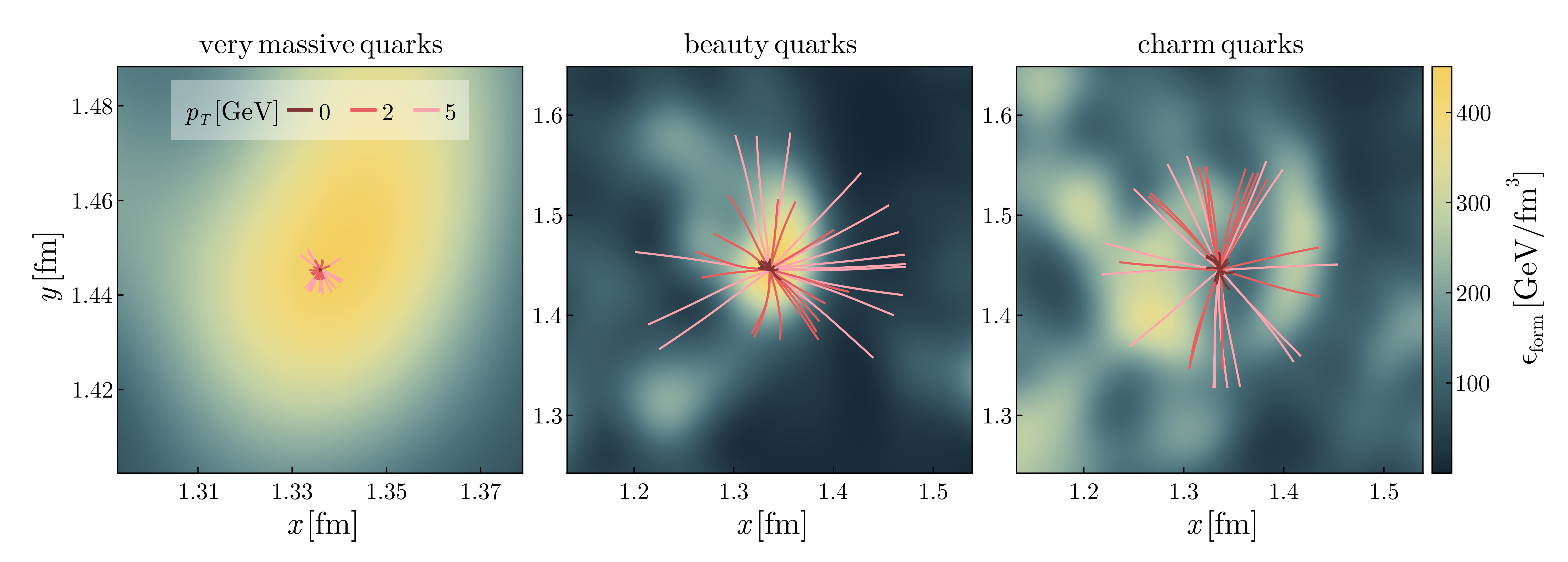}
    \caption{Trajectories of heavy quarks produced at the center of a glasma flux tube and their evolution up to $\tau=0.2\,\mathrm{fm}/c$, for \textit{(left)} very massive quarks close to the static limit, \textit{(middle)} beauty quarks, and \textit{(right)} charm quarks, with color coded different initial $p_T$ values. The background depicts the glasma energy density at the time of formation for each quark. Figure reproduced from Article~\cite{ar1} under license CC BY 4.0.}
    \label{fig:hqs-flux}
\end{figure}

\begin{figure}[t]
    \centering
    \includegraphics[width=0.5\textwidth]{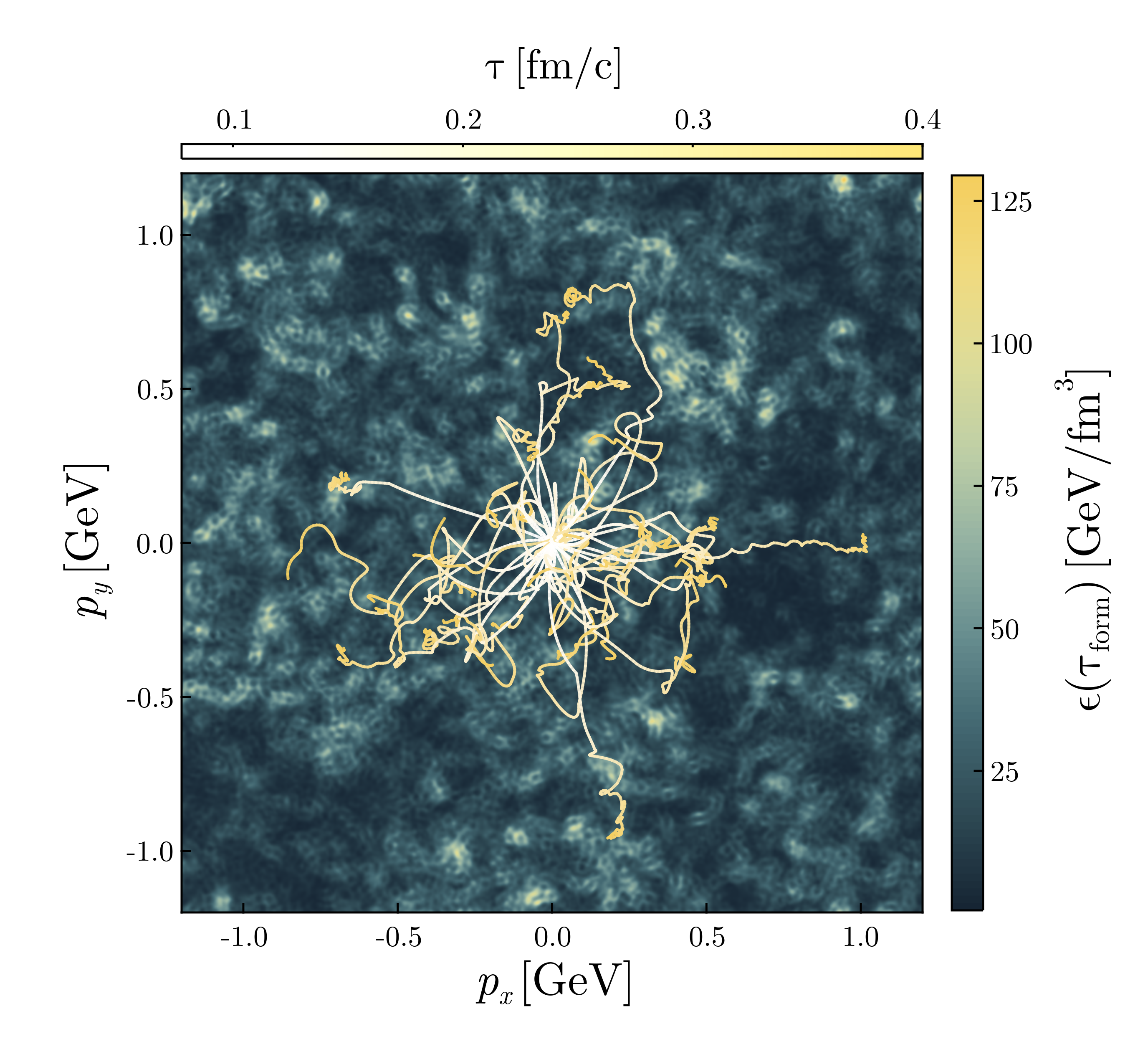}
    \caption{Trajectories of charm quark test particles in the transverse momentum plane $(p_x, p_y)$, initialized with vanishing momenta and evolved in a single glasma event. The tracks are color coded using the proper time $\tau$. The background displays the glasma energy density at the time of formation, $\varepsilon(\tau_0)$ in the transverse coordinate plane $(x, y)$, not shown in this visualization.}
    \label{fig:hqs-traj}
\end{figure}

A zoom-in on the formation region of heavy quarks is shown in Fig.~\ref{fig:hqs-flux}. This illustrates how the quark mass and initial transverse momentum determine the region of the glasma that a hard probe traverses. Almost static quarks barely move from their production point and remain inside the initial flux tube, which expands as the glasma evolves. Then, particles are formed at $\tau_0\approx 1/(2m)$, where $m$ is the quark mass. Dynamical beauty quarks, formed earlier at $\tau_{0,b}=1/(2m_b)\approx 0.02\,\mathrm{fm}/c$, still experience the highly correlated initial color fields. Although beauty quarks are heavier than charm quarks, which would naively suggest less momentum accumulation, they form earlier, when the glasma fields are stronger. Charm quarks are formed later at $\tau_{0,c}=1/(2m_c)\approx 0.06\,\mathrm{fm}/c$, close to the transition to the free streaming regime at $\tau\sim 1/Q_s \approx 0.1\,\mathrm{fm}/c$. Quarks with larger initial $p_T$ escape the correlation domains more rapidly, and therefore the initial quark $p_T$ is expected to affect the momentum broadening. The evolution in momentum space of individual charm quarks in a single glasma background event is plotted in Fig.~\ref{fig:hqs-traj}. This is complementary to the position space illustration in Fig.~\ref{fig:hqs-flux}. Each particle undergoes a series of correlated momentum kicks driven by the glasma fields and produces a spread in the $(p_x, p_y)$ plane that resembles a Brownian motion but whose momentum broadening will not follow a simple diffusion process. 

\subsection{Transport coefficients}

\noindent\textbf{\textit{Momentum broadening}}. 
The momentum broadening $\delta p_\mu^2(\tau) \equiv p_\mu^2(\tau) - p_\mu^2(\tau_0)$
measures the momentum accumulated since the initial time $\tau_0$ at which the particle enters the glasma fields with initial momentum $p_\mu(\tau_0)$. The averaged momentum broadening $\langle \delta p_\mu^2(\tau) \rangle$ is then given by $\langle X\rangle\equiv \langle\langle\langle X\rangle_{\mathrm{tp}}\rangle_Q\rangle_\rho$,
where $\langle\,\cdots\,\rangle_{\mathrm{tp}}$ is the average over the ensemble of test particles, $\langle\,\cdots\,\rangle_Q$ over the initial color charge in a given representation, and $\langle\,\cdots\,\rangle_\rho$ over the glasma field configurations. Using the equation of motion for momentum expressed as $\diff p_\mu/\diff\tau=g/T_R\,Q_0^a\,\mathrm{tr}[T^a\tilde{\mathcal{F}}_\mu]$,
where $Q_0$ is the initial color charge and $\tilde{\mathcal{F}}_\mu$ is the parallel transported color Lorentz force $\tilde{\mathcal{F}}_\mu(\tau)\equiv U(\tau;\tau_0)\,\mathcal{F}_\mu(\tau)\,U^\dagger(\tau;\tau_0)$ with the color Lorentz force defined as $\mathcal{F}_\mu\equiv F_{\mu\nu}\,p^\nu/p^\tau$, the momentum broadening is then

\begin{align}
    \label{eq:mombroadforcecorr}
    \big\langle\delta p_\mu^2(\tau)\big\rangle_Q=g^2\int_{\tau_0}^{\tau}\diff\tau'\int_{\tau_0}^{\tau}\diff\tau''\,\Big\langle\tr{\tilde{\mathcal{F}}_\mu(\tau')\tilde{\mathcal{F}}_\mu(\tau'')}\Big\rangle\,,
\end{align}
obtained using the average $\langle Q_0^aQ_0^b\rangle_Q=T_R\delta^{ab}$ and $\tr{T^a X}\tr{T^a Y}=T_R\,\tr{X Y}$ from the Fierz identity. Essentially, the momentum broadening is given by an integral over parallel transported color Lorentz force correlators, which varies for each trajectory.

\noindent\textbf{\textit{Field correlators}}. For a static heavy quark in temporal gauge, the Wilson line is trivial, $U(\tau;\tau_0)=\mathbbm{1}$ and the parallel transported force reduces to $\tilde{\mathcal{F}}_\mu=\mathcal{F}_\mu$. In the infinite mass limit $m\to\infty$ the color Lorentz force retains only electric field contributions $\mathcal{F}_i = F_{i\mu}\,p^\mu/p^\tau\rightarrow F_{i\tau}=-E_i$. For particles in the infinite mass limit, the momentum broadening simplifies to an integral over the electric field correlators~\cite{Banerjee:2011ra,Boguslavski:2020tqz}
\begin{align}
    \label{eq:mombroadhqs}
    \big\langle\delta p_i^2(\tau)\big\rangle_{m\to\infty}=g^2\int_{\tau_0}^{\tau}\diff\tau'\int_{\tau_0}^{\tau}\diff\tau''\,\Big\langle\tr{E_i(\tau')E_i(\tau'')}\Big\rangle\,.
\end{align}

For a highly energetic jet traveling along the $x$-axis, taking the limit $p^x\to\infty$ and setting $\tau_0=0$, the momentum broadening takes a similar expression~\cite{Ipp:2020mjc,Ipp:2020nfu}
\begin{align}
    \label{eq:mombroadjets}
    \big\langle\delta p_i^2(\tau)\big\rangle_{p^x\to\infty}=g^2\int_{0}^{\tau}\diff\tau'\int_{0}^{\tau}\diff\tau''\,\Big\langle\mathrm{tr}\big[\tilde{f}_i(\tau')\tilde{f}_i(\tau'')\big]\Big\rangle\,,
\end{align}
where the Lorentz force components experienced by the jet are $f_x\equiv E_x$, $f_y\equiv E_y-B_z$, $f_z\equiv E_z+B_y$
and the parallel transported force is $\tilde{f}_i(\tau)\equiv U_x(\tau;0)\,f_i(\tau)\,U_x^\dagger(\tau;0)$, with a Wilson line built along the jet trajectory
\begin{align}
    \label{eq:wilsonlinejet}
    U_x(\tau;0)=P\exp\left[-ig\int_{0}^{\tau}\diff\tau'\,A_x(\tau')\right]\,.
\end{align}

\begin{figure}[H]
    \centering
    \begin{subfigure}[b]{0.48\textwidth}
        \centering
        \includegraphics[width=\textwidth]{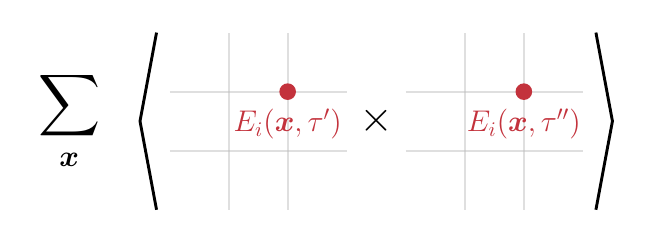}
        \caption{Static quark correlator, with the electric fields evaluated at the same lattice site $\boldsymbol{x}$.}
        \label{fig:corr-hqs}
    \end{subfigure}
    \hfill
    \begin{subfigure}[b]{0.48\textwidth}
        \centering
        \includegraphics[width=\textwidth]{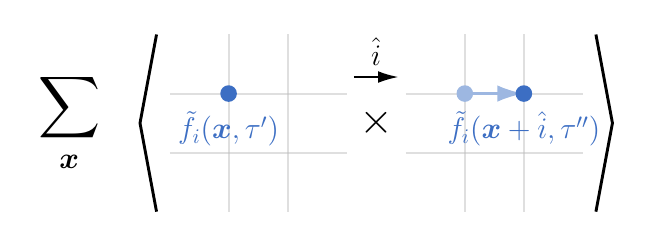}
        \caption{Eikonal quark correlator, with the force evaluated at neighboring sites $\boldsymbol{x}$ and $\boldsymbol{x}+\hat{i}$.}
        \label{fig:corr-jets}
    \end{subfigure}
    \caption{Lattice representation of the force correlators used to extract the momentum broadening for quarks in limiting kinematic regimes.}
    \label{fig:corr-hqs-jets}
\end{figure}

The relevant force correlators for the limiting cases are shown in Fig.~\ref{fig:corr-hqs-jets}. For static heavy quarks, the electric field correlator at the same lattice site shown in Fig.~\ref{fig:corr-hqs-jets}\subref{fig:corr-hqs} is summed over all lattice sites and averaged to give the momentum broadening from Eq.~\eqref{eq:mombroadhqs}. For eikonal jets, the parallel transported color Lorentz force $\tilde{f}_i$ in Fig.~\ref{fig:corr-hqs-jets}\subref{fig:corr-jets} is computed at sites shifted by one unit along the jet direction $\hat{i}$, including the color rotation, to give the momentum broadening from Eq.~\eqref{eq:mombroadjets}. These two limiting cases are benchmarks for validating the CPIC particle solver.

\begin{figure}[H]
    \centering
    \begin{subfigure}[b]{0.48\textwidth}
        \centering
        \includegraphics[width=\textwidth]{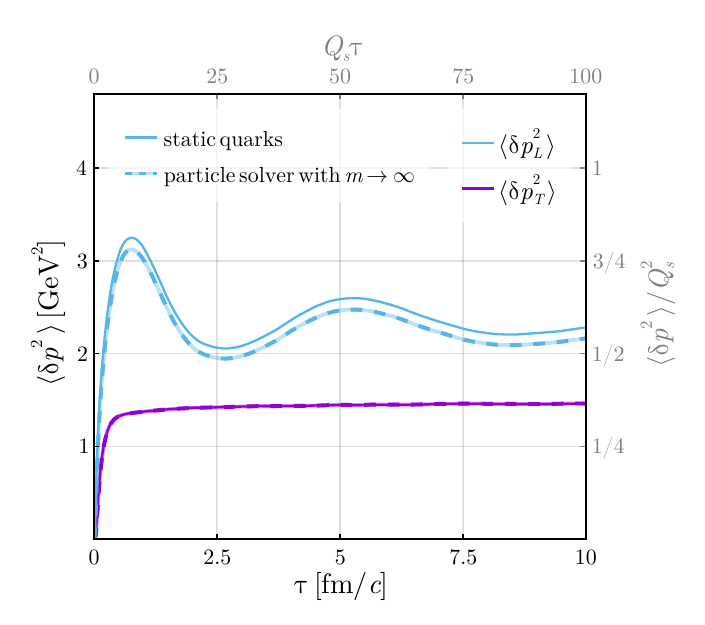}
        \captionsetup{justification=centering}
        \caption{Infinitely massive heavy quarks.}
        \label{fig:wong-kappa}
    \end{subfigure}
    \hfill
    \begin{subfigure}[b]{0.48\textwidth}
        \centering
        \includegraphics[width=\textwidth]{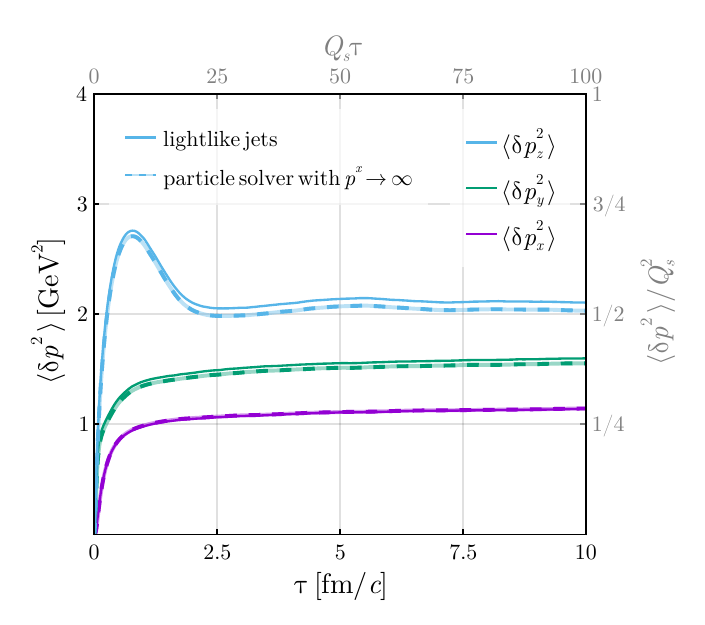}
        \captionsetup{justification=centering}
        \caption{Highly energetic lightlike jets.}
        \label{fig:wong-qhat}
    \end{subfigure}
    \caption{Comparison of momentum broadening components extracted from field correlators \textit{(full line)} or the particle solver in limiting cases \textit{(dashed line)}. The \textit{grey axes} indicate the corresponding dimensionless quantities $\langle\delta p^2\rangle/Q_s^2$ and $Q_s\tau$. Figure reproduced from Article~\cite{ar1} under license CC BY 4.0.}
    \label{fig:wong-kappa-qhat}
\end{figure}

Figure~\ref{fig:wong-kappa-qhat} contains a non-trivial validation of the CPIC particle solver in limiting cases against the field correlators results. For both static quarks and eikonal jets, the particle solver reproduces the results obtained purely from glasma field correlators with high accuracy. The small differences come from the different lattice spacing used in the correlator integrals, which is not as fine as the one used in the particle solver. Notably, the longitudinal momentum broadening for both static and eikonal quarks grows rapidly at early times because of the strong coherent longitudinal electric fields and reaches a maximum around $Q_s\tau\approx 1$. Then, for heavy quarks it undergoes damped oscillations, which are attributed to plasmon modes present in the glasma fields~\cite{Boguslavski:2020tqz,Backfried:2024rub}. For eikonal jets, the longitudinal broadening exhibits a single peak but then it saturates, without the oscillatory behavior. The transverse momentum broadening components grow more slowly and saturate at later times for both cases.

\noindent\textbf{\textit{Transport coefficients}}. The rate of momentum broadening accumulated by a classical particle in the glasma fields defines instantaneous transport coefficients. The underlying physical origin of these transport coefficients in the glasma differs from the diffusive dynamics of heavy quarks or independent multiple scatterings of jets in the QGP.

\begin{emphbox}{myblue}{Transport coefficients in glasma fields}
The momentum broadening rate defines the transport coefficients
\begin{align}
    \label{eq:transportcoeffs}
    \frac{\diff}{\diff\tau}\langle\delta p_i^2(\tau)\rangle \equiv
    \begin{cases}
        \kappa_i(\tau), & \text{heavy quarks}\,,\\[4pt]
        \hat{q}_i(\tau), & \text{jets}\,.
    \end{cases}
\end{align}
These are separately extracted along each direction $i=(x,y,z)$. The transport coefficients in the glasma are calculated from the force correlators or classical transport equations, and are inherently non-perturbative and out-of-equilibrium quantities.
\end{emphbox}

\begin{figure}[t]
    \centering
    \includegraphics[width=\textwidth]{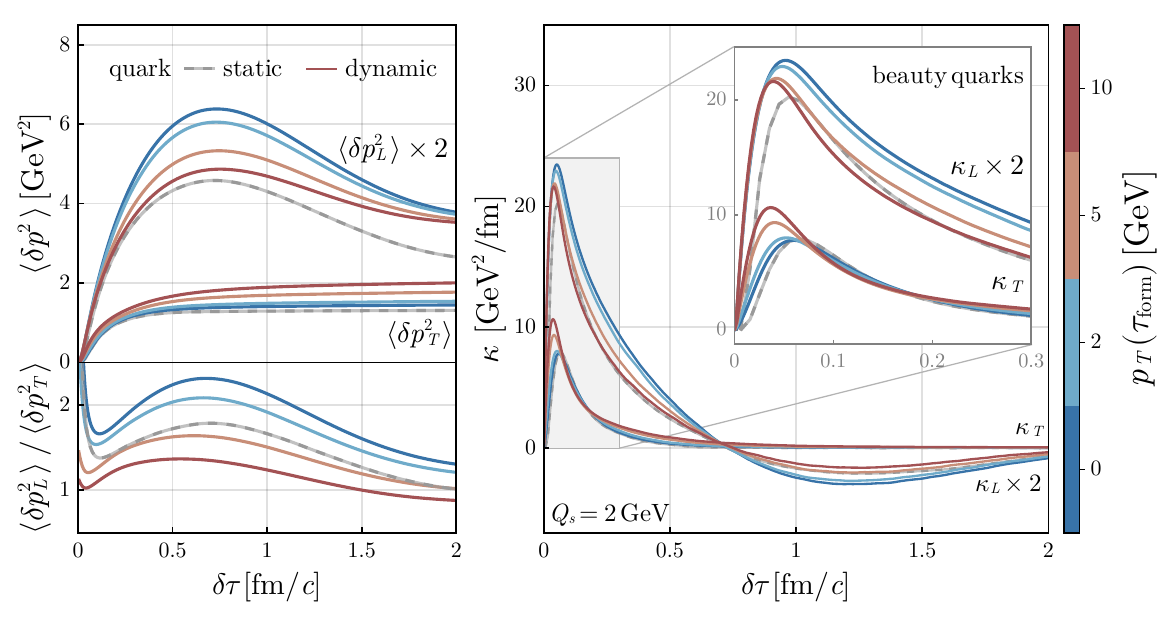}
    \caption{Proper time evolution of \textit{(top left)} the longitudinal and transverse momentum broadening components, \textit{(bottom left)} their ratio which quantifies the momentum anisotropy, and \textit{(right)} the instantaneous transport coefficients with \textit{(inset)} a zoom-in on the very early stage, as a function of the relative $\delta\tau\equiv\tau-\tau_0$. All simulations are for beauty quarks in glasma fields with $Q_s=2\,\mathrm{GeV}$. The \textit{colored lines} show dynamical quarks evolved with the particle solver at fixed $p_T(\tau_0)$, while the \textit{grey dashed lines} show the static quark limit from the lattice field correlators. Figure reproduced from Ref.~\cite{Avramescu:2023vld} under license CC BY-NC-ND 4.0.}
    \label{fig:hqs-kappa-anis}
\end{figure}

\begin{figure}[t]
    \centering
    \includegraphics[width=\textwidth]{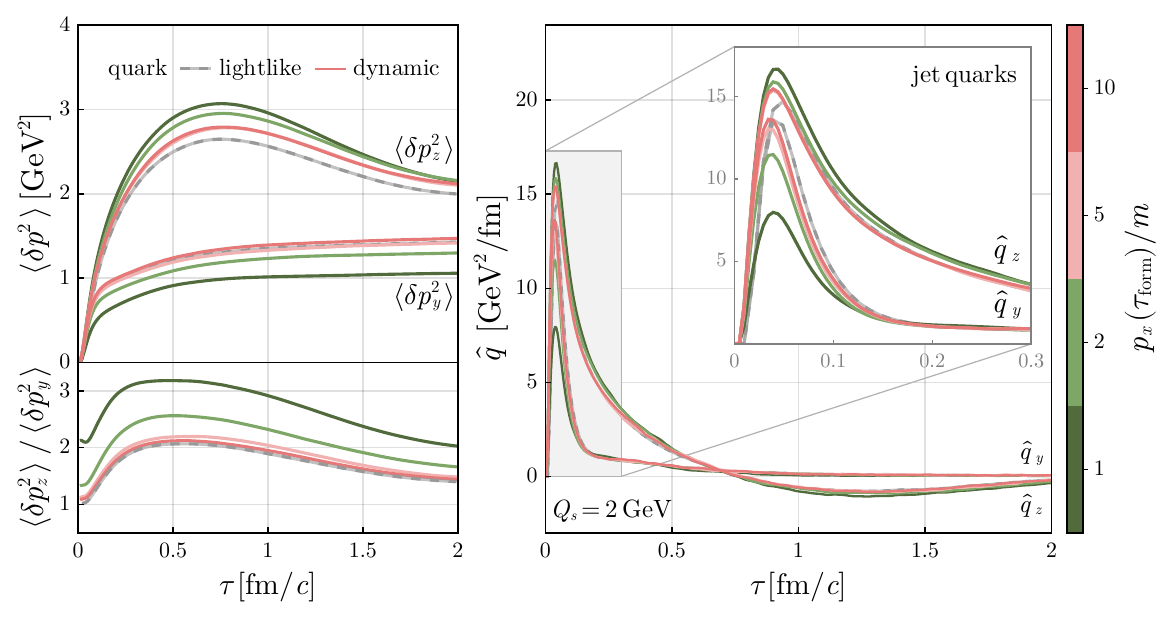}
    \caption{Proper time evolution of \textit{(top left)} longitudinal and transverse momentum broadening components, \textit{(bottom left)} jet momentum anisotropy and \textit{(right)} transport coefficients with \textit{(inset)} a zoom-in on early times, as a function of proper time. The \textit{colored lines} correspond to dynamical jet quarks with various $p_x(\tau_0)/m$ ratios, while the \textit{grey dashed lines} show the eikonal limit from lattice glasma field correlators. Figure reproduced from Ref.~\cite{Avramescu:2023vld} under license CC BY-NC-ND 4.0.}
    \label{fig:jets-qhat-anis}
\end{figure}

The momentum broadening and transport coefficient for beauty quarks in the glasma, extracted using the dynamical evolution with several initial momenta and the static quark limit from lattice correlators, are shown in Fig.~\ref{fig:hqs-kappa-anis}. Both longitudinal and transverse components increase substantially in the early stage as quarks are accelerated by the coherent glasma fields, with the corresponding $\kappa$ exhibiting a sharp peak of order $\kappa_\mathrm{peak}\approx 20\,\mathrm{GeV}^2/\mathrm{fm}$ at $\delta\tau_\mathrm{peak}\approx 0.04\,\mathrm{fm}/c$. This is consistent with earlier numerical studies of heavy quark transport in an SU($2$) glasma~\cite{Ruggieri:2018rzi,Sun:2019fud}. The accumulated momentum is anisotropic, namely the longitudinal broadening exceeds the transverse one, with the anisotropy inversely correlated with $p_T(\tau_0)$. At late times, the longitudinal momentum broadening decreases, yielding a negative $\kappa$. This is attributed to plasmon oscillations of the glasma fields~\cite{Backfried:2024rub}.

Similarly, Fig.~\ref{fig:jets-qhat-anis} shows the jet momentum broadening and the corresponding transport coefficient in the glasma, comparing dynamical simulations to the eikonal limit. All resulting time dependences resemble the heavy quark results from Fig.~\ref{fig:hqs-kappa-anis} and are consistent with previous estimates in the glasma~\cite{Ipp:2020nfu,Ipp:2021lwz,Carrington:2020sww,Carrington:2022bnv}. There is a rapid growth at early times, which gives a sharp $\hat{q}$ peak of $\hat{q}_\mathrm{peak}\approx 25\,\mathrm{GeV}^2/\mathrm{fm}$ at $\tau_\mathrm{peak}\approx 0.04\,\mathrm{fm}/c$, after which the coefficients decay as the fields become dilute. The anisotropy is consistently larger for dynamical jets than for the lightlike limit. Unlike heavy quarks, the oscillations in the longitudinal momentum broadening do not persist at later times. Notably, the jet $\hat{q}$ results obtained in the glasma for eikonal jets are compatible with values extracted during the EKT stage, while the heavy quark $\kappa$ values in the glasma from field correlators show deviations from the EKT results~\cite{Boguslavski:2023alu,Boguslavski:2023fdm,Boguslavski:2024ezg}.

\subsection{Properties of transport coefficients}

\begin{figure}[t]
    \centering
    \includegraphics[width=0.6\textwidth]{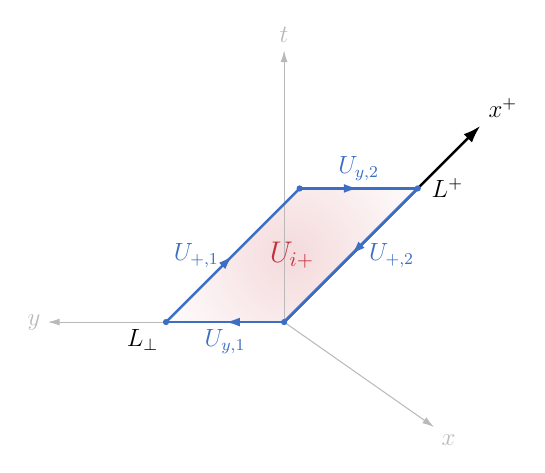}
    \caption{Rectangular Wilson loop $U_{i+}$ in the $(x^+, y)$ plane, consisting of two lightlike segments $U_{+,1}$, $U_{+,2}$ of extent $L^+$ along $x^+$ and two transverse segments $U_{y,1}$, $U_{y,2}$ of separation $L_\perp$.}
    \label{fig:wilson-loop}
\end{figure}

\begingroup\relpenalty=0
\noindent\textbf{\textit{Wilson loop formulation}}. 
The momentum broadening of a lightlike parton in the eikonal approximation may be expressed from a rectangular Wilson loop $U_{i+}$ with one lightlike side of extent $L^+$ and one transverse side of extent $L_\perp$~\cite{Liu:2006ug,Casalderrey-Solana:2007ahi,DEramo:2010wup}. The Wilson loop is $U_{i+}=U_{+,2}\,U_{i,2}\,U_{+,1}\,U_{i,1}$, with $q$ the Abelian charge, where the lightlike segments are $U_{+,1}=\exp\big[-iq\int_0^{L^+}\!\diff x^+\,A_+(x^+,0)\big]$ and $U_{+,2}=\exp\big[iq\int_0^{L^+}\!\diff x^+\,A_+(x^+,L_\perp)\big]$ and the transverse segments are $U_{i,1}=\exp\big[-iq\int_0^{L_\perp}\!\diff x^i\,A_i(0,x^i)\big]$ and $U_{i,2}=\exp\big[iq\int_0^{L_\perp}\!\diff x^i\,A_i(L^+,x^i)\big]$. The rectangular Wilson loop $U_{i+}$ is sketched in Fig.~\ref{fig:wilson-loop}.
\par\endgroup

In the Abelian case, Stokes' theorem reduces this product to an area integral over the field strength tensor
\begin{align}
    \label{eq:stokes-wilson}
    U_{i+} = \exp\left[-iq\int_0^{L^+}\!\diff x^+\int_0^{L_\perp}\!\diff x^i\,F_{i+}(x^+,x^i)\right]\,.
\end{align}
For small $L_\perp$, one expands $U_{i+}=\mathbbm{1}+L_\perp U_{i+}^{(1)}+(L_\perp^2/2)U_{i+}^{(2)}+\mathcal{O}(L_\perp^3)$ with coefficients $U_{i+}^{(1)}=-iq\int_0^{L^+}\!\diff x^+\,F_{i+}$ and $U_{i+}^{(2)}=-q^2(\int_0^{L^+}\diff x^+\,F_{i+})^{2}$. The relation between the Wilson loop and the momentum broadening is $\big\langle\mathrm{Re}\,U_{i+}\big\rangle = \exp[-L_\perp^2\langle p_i^2\rangle/2]$.
Using the dipole approximation for $L_\perp\ll L^+$, one matches the $L_\perp^2$ coefficients on both sides to find the momentum broadening in terms of field strength correlators
\begin{align}
    \langle p_i^2(x^+)\rangle = -\big\langle\mathrm{Re}\,U_{i+}^{(2)}\big\rangle \approx q^2\int_0^{x^+}\diff \xi^\prime\int_0^{x^+}\diff \xi^{\prime\prime}\,\big\langle F_{i+}(\xi^{\prime})F_{i+}(\xi^{\prime\prime})\big\rangle\,,
\end{align}
with $\xi$ corresponding to the light-cone time $\xi = t/\sqrt{2}$ and $t=\tau$ for mid-rapidity particles $\eta=0$.

This result can be generalized to the non-Abelian case, where the Wilson loop carries a color trace and the $D_R$ prefactor ensures identity when $U_{i+}\to\mathbbm{1}_{D_R}$, namely
\begin{equation}
    \begin{aligned}
        \label{eq:mombroadwilson}
        \langle p_i^2(x^+)\rangle &= -\frac{1}{D_R}\Big\langle\mathrm{Re}\left[\mathrm{tr}\big[U_{i+}^{(2)}\big]\right]\Big\rangle\\
        &\approx\frac{g^2}{D_R}\int_{0}^{x^+}\diff\xi^\prime\int_{0}^{x^+}\diff\xi^{\prime\prime}\,\Big\langle\mathrm{tr}\big[\tilde{F}_{i+}(\xi')\tilde{F}_{i+}(\xi'')\big]\Big\rangle\,,
    \end{aligned}
\end{equation}
where $\tilde{F}_{i+}(x^+)=U_+(x^+;0)F_{i+}(x^+)U_+^\dagger(x^+;0)$ with the lightlike Wilson line expressed as $U_+(\xi_2,\xi_1)=P \exp\big[-ig\int_{\xi_1}^{\xi_2}\diff \xi^+A_+(\xi^+)\big]$. This matches the classical result up to the factor $1/D_R$, which comes from the color charge averaging $\langle Q_0^aQ_0^b\rangle_Q=T_R\delta^{ab}$ used in the classical calculation~\cite{Majumder:2009cf,Carrington:2016mhd}.

\noindent\textbf{\textit{Casimir dependence}}. The momentum broadening depends on the quark color representation $R$ through the Casimir invariants of the gauge group. In perturbative QCD, this dependence takes the particularly simple form of Casimir scaling. In the classical transport framework, the Casimir dependence is more intricate and does not generally produce the expected Casimir scaling.

\begin{emphbox}{myblue}{Casimir scaling}
Casimir scaling in pQCD calculations comes from group theory properties of color factors in Feynman diagrams. It implies that the ratio of momentum broadening between the adjoint and fundamental representations equals the ratio of the group theory quadratic Casimir invariants
\begin{align}
    \label{eq:casimirscaling}
    \langle p^2\rangle_A^{\mathrm{pQCD}} \big/ \langle p^2\rangle_F^{\mathrm{pQCD}} = C_2(A) \big/ C_2(F)\,.
\end{align}
As a result, the gluon momentum broadening is larger than the quark momentum broadening by a factor of $C_2(A)/C_2(F)=9/4$ for SU(3).
\end{emphbox}

Casimir scaling also holds classically, but only in the limiting cases. For these, the parallel transport is done along a fixed direction and the color charge average $\langle Q_0^aQ_0^b\rangle_Q=T_R\delta^{ab}$ factorizes cleanly from the field correlator. For general trajectories, the color charge evolves along the particle path and stays correlated with the background fields, preventing this factorization.

In the limiting cases, this property is already implied from Eq.~\eqref{eq:mombroadforcecorr}. Using the relation $\mathrm{tr}\big[\tilde{\mathcal{F}}_\mu\tilde{\mathcal{F}}_\mu\big]=T_R\,\tilde{\mathcal{F}}_\mu^a\tilde{\mathcal{F}}_\mu^a$, the color-averaged momentum broadening becomes 
\begin{align}
    \label{eq:mombroadTR}
    \langle\delta p_\mu^2(\tau)\rangle_Q = T_R\,g^2\int_{\tau_0}^{\tau}\diff\tau'\int_{\tau_0}^{\tau}\diff\tau''\,\Big\langle\tilde{\mathcal{F}}_\mu^a(\tau')\tilde{\mathcal{F}}_\mu^a(\tau'')\Big\rangle\,,
\end{align}
where the integral is representation independent and $T_R$ carries all the $R$ dependence. 

\newpage
\begin{figure}[H]
    \centering
    \begin{subfigure}[b]{\textwidth}
        \centering
        \includegraphics[width=\textwidth]{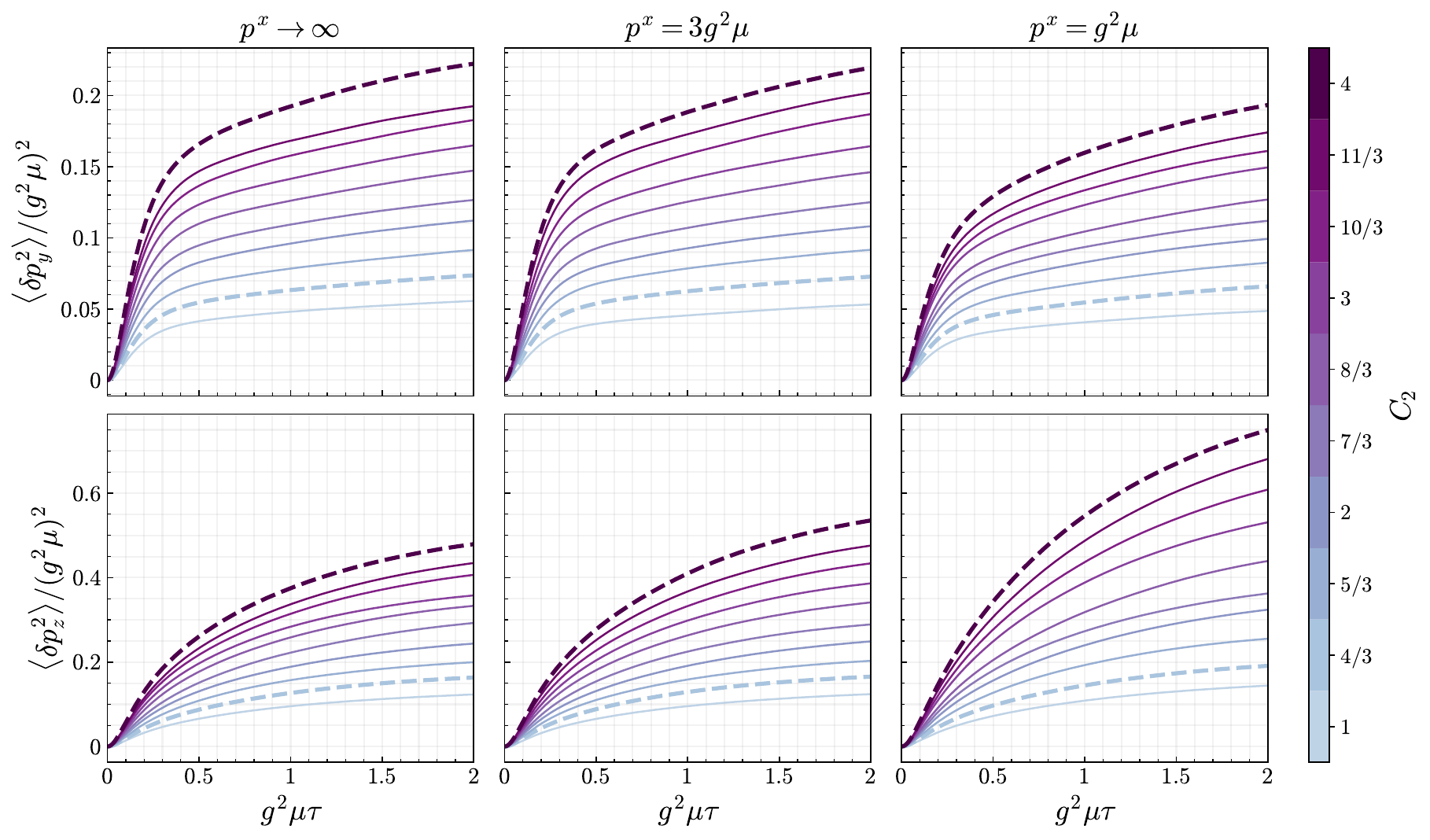}
        \caption{Transverse \textit{(top)} and longitudinal \textit{(bottom)} jet momentum broadening as a function of proper time, for $p^x\to\infty$ \textit{(left)}, $p^x=3g^2\mu$ \textit{(middle)}, and $p^x=g^2\mu$ \textit{(right)}. The curves are color coded by the quadratic Casimir $C_2\in[1,4]$, with the \textit{dashed lines} corresponding to the fundamental and adjoint representations. The momentum broadening is scaled by $g^2\mu$ and the time as $g^2\mu\tau$ for fixed $g^2\mu=2\,\mathrm{GeV}$.}
        \label{fig:mom-broad-q2-dep}
    \end{subfigure}\\[2em]
    \begin{subfigure}[b]{\textwidth}
        \centering
        \includegraphics[width=\textwidth]{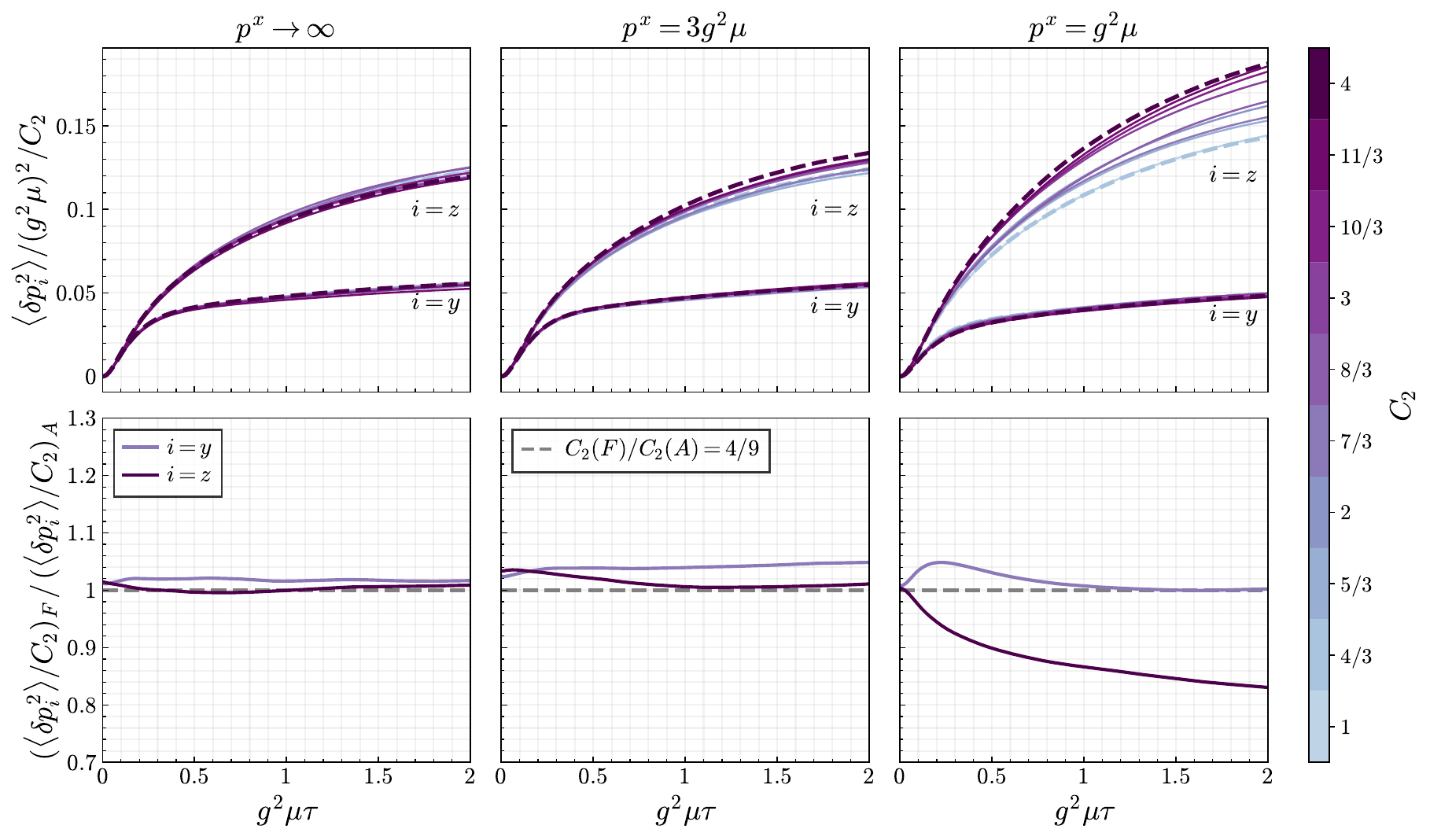}
        \caption{\textit{(Top)} momentum broadening rescaled by $C_2$ and \textit{(bottom)} ratio of the Casimir scaled broadening in the fundamental to adjoint representation, with the \textit{grey dashed line} marking the ratio expected from the exact Casimir scaling.}
        \label{fig:mom-broad-q2-dep-scaled}
    \end{subfigure}
    \caption{Jet momentum broadening for different color representations.}
    \label{fig:mom-broad-q2-dep-all}
\end{figure}

Since $T_A/T_F=q_2(A)/q_2(F)$, the classical ratio reads 
\begin{equation}
    \langle\delta p_\mu^2\rangle_A^\mathrm{classic} \big/ \langle\delta p_\mu^2\rangle_F^\mathrm{classic} = q_2(A) \big/ q_2(F)\,.
\end{equation}
Dividing by $D_R$ and using $q_2(R)/D_R=C_2(R)$ recovers the quantum Casimir scaling in Eq.~\eqref{eq:casimirscaling}.

Figures~\ref{fig:mom-broad-q2-dep} and \ref{fig:mom-broad-q2-dep-scaled} examine how the jet momentum broadening scales with the representation. In Fig.~\ref{fig:mom-broad-q2-dep}, the broadening grows with $C_2$ and Fig.~\ref{fig:mom-broad-q2-dep-scaled} makes the Casimir scaling explicit. In the lightlike limit $p^x\to\infty$, dividing by $C_2$ collapses all curves to a single line and the fundamental to adjoint ratio is unity, which confirms the exact classical Casimir scaling. For dynamical quarks at finite $p^x$, the transverse component retains approximate Casimir scaling, while the longitudinal component develops increasing deviations at late times. Notably, violations of the Casimir scaling were also reported for jets in media simulated using JEWEL~\cite{Apolinario:2020nyw}.

\section{Heavy quark phenomenology}

After studying the properties of the pre-equilibrium transport coefficients $\kappa$ and $\hat{q}$, this section turns to their phenomenological consequences. More specifically, it studies how the momentum broadening in the glasma stage changes the heavy flavor nuclear modification factor $R_{AA}$ and the angular decorrelation of $Q\bar{Q}$ pairs. In heavy-ion collision experiments, both quantities can be measured. Unlike~\cref{sec:classical-transport}, where test particles were initialized with fixed $p_T$ values, here the initial heavy quark $p_T$ spectra are sampled from pQCD calculations.

\begin{emphbox}{myred}{Based on}
Articles~\cite{ar2,ar3} study the phenomenological impact of the glasma pre-equilibrium stage on heavy quark observables, using the same CPIC numerical framework as~\cite{Avramescu:2023qvv}. Heavy quark pairs are initialized back-to-back with fixed initial $p_T$ and evolved in the SU(3) glasma background. The main result is that the glasma induces a large and rapid decorrelation of $c\bar{c}$ and $b\bar{b}$ pairs in azimuthal angle, with the decorrelation width $\sigma_{\Delta\phi}$ growing with $Q_s$ and decreasing with the initial quark $p_T$. By contrast, the effect on the single inclusive spectrum is moderate, where the $p_T$ spectra are initialized with the FONLL pQCD calculation. The glasma causes a $p_T$ migration that suppresses $R_{AA}$ at low $p_T$ and enhances it at high $p_T$, but the main contribution to $R_{AA}$ comes from the nuclear PDF.
\end{emphbox}

\subsection{Nuclear modification factor}

The first single inclusive observable sensitive to glasma momentum broadening is the nuclear modification factor $R_{AA}$, which compares the $p_T$ spectrum of heavy quarks in nuclear collisions to the proton baseline. The transverse momentum change induced by the glasma shifts quarks between $p_T$ bins, thereby modifying $R_{AA}$.

\begin{figure}[t]
    \centering
    \begin{subfigure}[b]{0.48\textwidth}
        \centering
        \includegraphics[viewport={222 12 432 191},clip,width=0.9\textwidth]{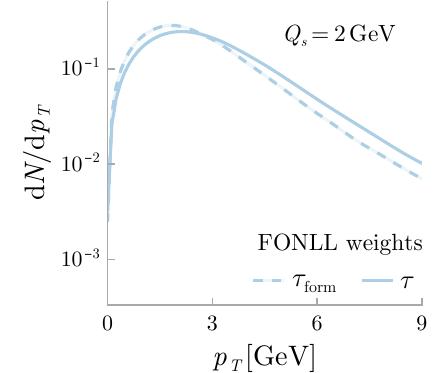}
        \caption{$p_T$ spectrum weighted with FONLL and evolved in the glasma.}
        \label{fig:raa-sketch-mid}
    \end{subfigure}
    \hfill
    \begin{subfigure}[b]{0.48\textwidth}
        \centering
        \includegraphics[viewport={432 12 643 191},clip,width=0.9\textwidth]{images/sketch_raa_gl_fonll.pdf}
        \caption{$R_{AA}(p_T)$ extracted using flat or FONLL weighted spectra.}
        \label{fig:raa-sketch-right}
    \end{subfigure}
    \caption{Charm quark $p_T$ spectrum at different times $\tau$ in the glasma and the corresponding $R_{AA}(p_T)$. Figure reproduced from Article~\cite{ar2} under license CC BY 4.0.}
    \label{fig:raa-sketch}
\end{figure}

\begin{emphbox}{myblue}{Nuclear modification factor}
    The color Lorentz force deflects classical heavy quarks in glasma and changes their transverse momentum, consequently broadening their $p_T$ distribution and modifying $R_{AA}$ defined as
    \begin{equation}
        \label{eq:raa}
        R_{AA}(\tau) \equiv \frac{1}{A^2}\frac{\sigma_\mathrm{tot}^{AA}}{\sigma_\mathrm{tot}^{pp}} \left.\dfrac{\diff N}{\diff p_T}\right|^{pp/AA}_{\mathrm{glasma}}(\tau) \Bigg/ \left.\dfrac{\diff N}{\diff p_T}\right|^{pp}(\tau_\mathrm{form})\,.
    \end{equation}
    This ratio compares the heavy quark $p_T$ spectrum in $pp$ or $AA$ collisions evolved in the glasma, to the $pp$ baseline, normalized to the number of binary nucleon-nucleon collisions $A^2$. 
\end{emphbox}

 The superscript $pp/AA$ labels the choice of parton distribution functions used to initialize the heavy quark spectrum, either a standard PDF for $pp$ collisions, or a nuclear PDF (nPDF) for $AA$ collisions. CT14NLO~\cite{Dulat:2015mca} is used as the PDF for $pp$ and EPPS16~\cite{Eskola:2016oht} as the nPDF in PbPb collisions. When the $pp$ FONLL initialization is used, $\sigma_\mathrm{tot}^{AA} = A^2\sigma_\mathrm{tot}^{pp}$ and $R_{AA}$ is simply the ratio of the evolved to the initial spectrum, obeying a sum rule, namely that suppression at low $p_T$ is compensated by enhancement at high $p_T$. For the $AA$ initialization, cold nuclear matter effects in the nPDF give $\sigma_\mathrm{tot}^{AA} < A^2\sigma_\mathrm{tot}^{pp}$ and $R_{AA} < 1$ even without glasma evolution.

\noindent\textbf{\textit{Initial production}}. 
The initial heavy quark spectrum is taken from pQCD \textit{Fixed Order + Next-to-Leading Logarithm} (FONLL) calculations~\cite{Cacciari:1998it,Cacciari:2001td} of the $Q\overline{Q}$ differential cross section. This is extracted for bare quarks, without fragmentation into subsequent mesons. The FONLL cross section is computed at different center-of-mass energies $\sqrt{s}$ and with different PDF/nPDF choices. In practice, the FONLL result is approximated by a parametric fit, chosen to lie within the FONLL scale and mass uncertainties, as
\begin{equation}
    \label{eq:fonllsigma}
    \left.\frac{\diff\sigma_{Q\bar{Q}}}{\diff p_T}\right|^{pp/AA} \equiv \frac{x_0\, p_T}{\left(1 + x_3\, p_T^{\,x_1}\right)^{x_2}}\,,
\end{equation}
where the fit parameters $(x_0,x_1,x_2,x_3)$ depend on $\sqrt{s}$, quark flavor, and the PDF/nPDF choice. The initial probability distribution of heavy quarks at the formation time is given by
\begin{equation}
    \label{eq:fonlldndpt}
    \left.\frac{\diff N}{\diff p_T}\right|^{pp/AA}(\tau_\mathrm{form}) = \frac{1}{\sigma_\mathrm{tot}^{pp/AA}}\left.\frac{\diff\sigma_{Q\bar{Q}}}{\diff p_T}\right|^{pp/AA}\,,
\end{equation}
and normalized such that $\int \diff p_T\, \diff N/\diff p_T = 1$, with $\sigma_\mathrm{tot}^{pp/AA} = \int \diff p_T\, \diff\sigma_{Q\bar{Q}}^{pp/AA}/\diff p_T$.

\noindent\textbf{\textit{Effect of glasma}}. 
Heavy quark test particles are initialized with transverse momenta sampled from Eq.~\eqref{eq:fonlldndpt} and evolved in the glasma fields using Wong's equations. At a given proper time $\tau$, the distribution $\diff N/\diff p_T(\tau)$ is extracted numerically using a \textit{kernel density estimator} (KDE) and averaged over glasma events. Then, $R_{AA}$ is computed from Eq.~\eqref{eq:raa}. To obtain sufficient statistics at high $p_T$, the sampling is implemented using a reweighting procedure. The test particles are initialized with a flat $p_T$ distribution and the KDE is weighted by the initial FONLL spectrum evaluated at each $p_T$ value. This procedure is shown in Fig.~\ref{fig:raa-sketch}. The glasma color fields broaden the $p_T$ distribution of the heavy quarks. When convolved with the steeply falling FONLL spectrum, the quarks migrate from low to high $p_T$, causing a depletion $R_{AA}<1$ below $p_T\approx 3\,\mathrm{GeV}$ and an enhancement $R_{AA}>1$ at larger values. This effect grows more pronounced with proper time.

\begin{figure}[t]
    \centering
    \includegraphics[width=0.5\textwidth]{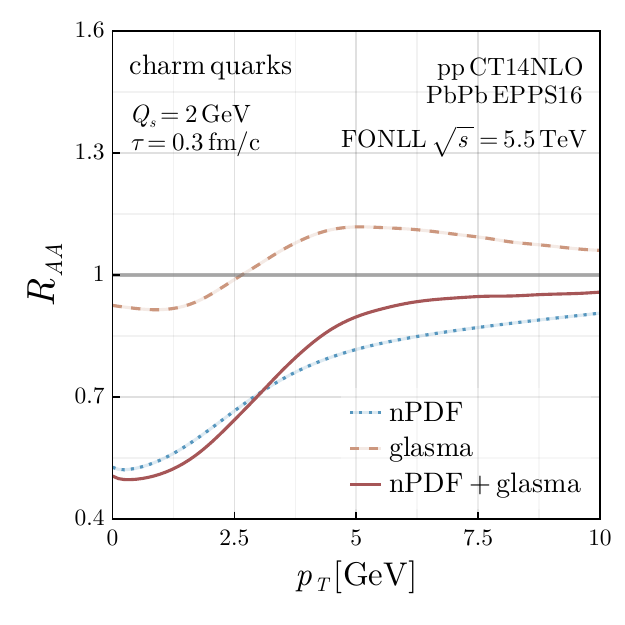}
    \caption{$R_{AA}(p_T)$ for charm quarks at $\tau=0.3\,\mathrm{fm}/c$ comparing the glasma result \textit{(pink dashed)} with the nPDF one \textit{(blue dashed)} and the combined nPDF with glasma effects \textit{(red solid)}. Figure reproduced from Article~\cite{ar2} under license CC BY 4.0.}
    \label{fig:raa-tau-npdf}
\end{figure}

Figure~\ref{fig:raa-tau-npdf} compares the two distinct initial state effects that influence the $p_T$ spectrum in nuclear collisions. One comes from gluon shadowing present in the nPDFs, while the other is due to gluon saturation inherent to the glasma. These effects are physically related, since gluon saturation also explains the suppression of the nuclear gluon distribution at small-$x$. However, for simplicity, here they are assumed to factorize. The glasma induces $p_T$ migration, while nuclear shadowing from nPDFs suppresses the production of small-$x$ gluons, giving $R_{AA}<1$ even without the glasma. The combined nPDF and glasma effect yields a steeper $p_T$ slope than either alone, and even though the glasma effect is moderate, it is non-negligible and should be included for a complete description of the initial state.

\noindent\textbf{\textit{Gaussian toy model}}. 
In addition to the numerical simulation, the transverse momentum broadening in the glasma may be described analytically using the Gaussian broadening approximation.

\begin{emphbox}{myblue}{Gaussian broadening model}
    Treating each $p_T$ bin as being independent, the evolved spectrum $\mathcal{N}_\tau(p_T) \equiv \diff N/\diff p_T(\tau)$ is the convolution of the initial distribution $\mathcal{N}_0 \equiv \mathcal{N}_{\tau_\mathrm{form}}$ with a broadening kernel
    \begin{equation}
        \label{eq:dndptconv}
        \mathcal{N}_\tau(p_T) = \int \diff^2\vec{q}_T\, \mathcal{K}_\tau(\vec{q}_T,\vec{p}_T)\, \mathcal{N}_0(q_T)\,.
    \end{equation}
    For isotropic $2$-dimensional Gaussian momentum kicks, a quark initialized at $\vec{q}_T$ acquires the final momentum $\vec{p}_T$ with probability
    \begin{equation}
        \label{eq:gauss2dxy}
        \mathcal{K}_\tau(\vec{q}_T,\vec{p}_T) = \frac{1}{2\pi\sigma_\tau^2}\exp\!\left\{-\frac{(\vec{p}_T-\vec{q}_T)^2}{2\sigma_\tau^2}\right\}\,,
    \end{equation}
    where $\sigma_\tau$ is the width of the Gaussian at a time $\tau$, and is extracted from numerical simulations as the Gaussian variance of the $p_T$ broadening.
\end{emphbox}

\begin{figure}[t]
    \centering
    \begin{subfigure}[b]{0.5\textwidth}
        \centering
        \includegraphics[width=\textwidth]{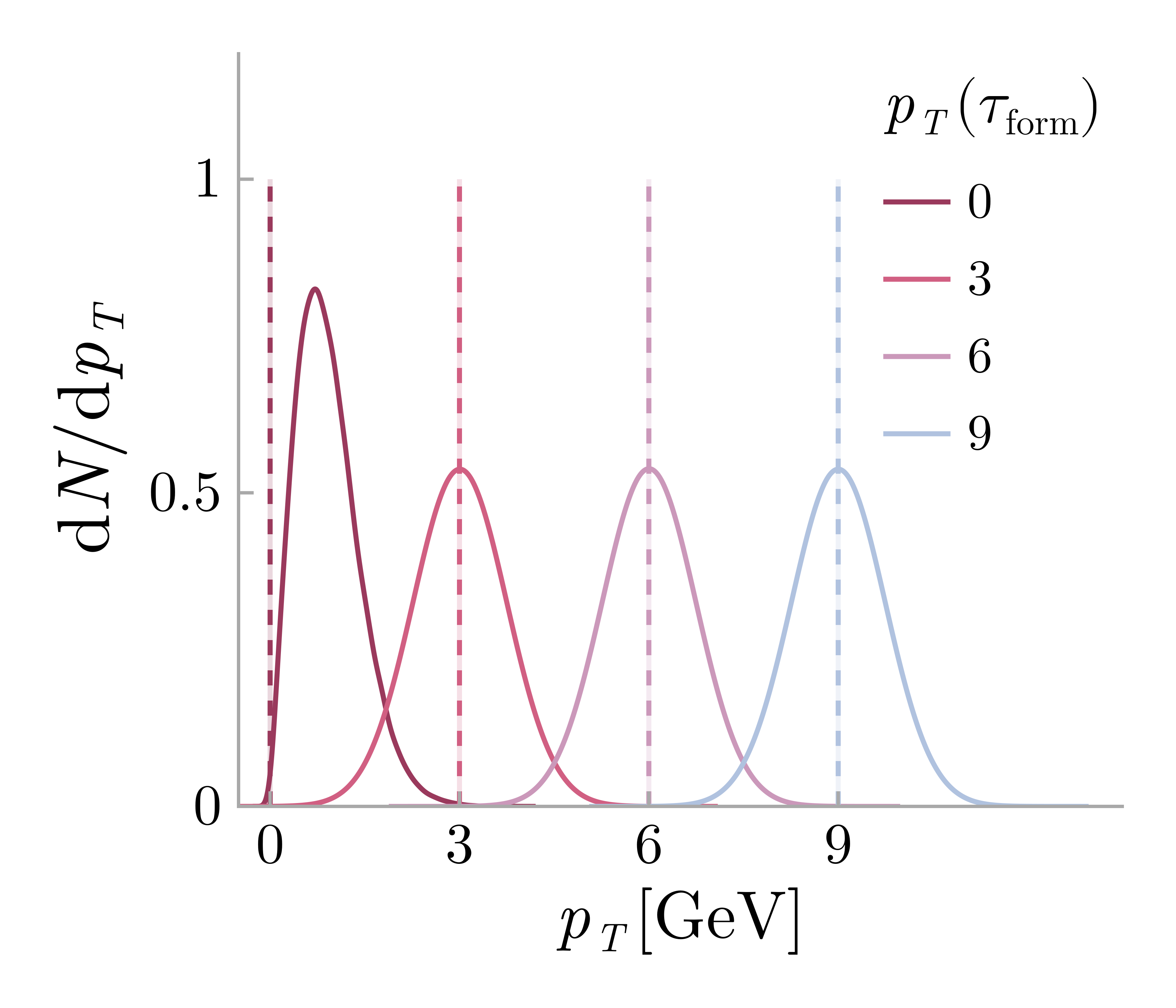}
        \caption{Time evolution of the $p_T$ spectrum from numerical simulations for distributions with fixed initial $p_T$ values.}
        \label{fig:raa-gauss}
    \end{subfigure}
    \hfill
    \begin{subfigure}[b]{0.46\textwidth}
        \centering
        \includegraphics[width=\textwidth]{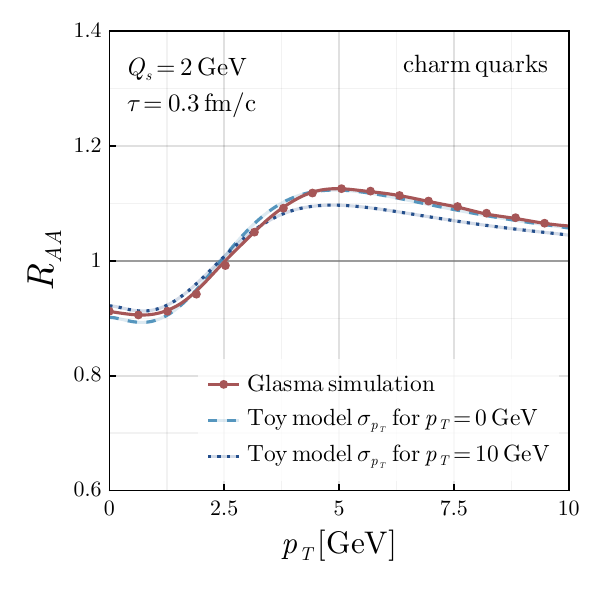}
        \caption{$R_{AA}(p_T)$ from the glasma simulation and the Gaussian broadening toy model with different $\sigma_{p_T}$.}
        \label{fig:raa-analytical}
    \end{subfigure}
    \caption{Charm quark $p_T$ spectrum and $R_{AA}$ from both the particle solver and extracted using the analytical Gaussian broadening toy model. Figure reproduced from Article~\cite{ar2} under license CC BY 4.0.}
    \label{fig:raa-comparison}
\end{figure}

After expressing Eq.~\eqref{eq:dndptconv} in polar coordinates and using the Bessel function identity $\int_0^{2\pi}\diff\varphi\, e^{a\cos\varphi} = 2\pi I_0(a)$~\cite{AbramowitzStegun} to perform the azimuthal angle integral analytically, the evolved spectrum becomes
\begin{align}
    \label{eq:raatoyanaly}
    \mathcal{N}_\tau(p_T) = \int_0^\infty \diff q_T\, \frac{\pi q_T}{2\sigma_\tau^2}\exp\!\left\{-\frac{p_T^2+q_T^2}{2\sigma_\tau^2}\right\} I_0\!\left(\frac{p_T q_T}{\sigma_\tau^2}\right)\mathcal{N}_0(q_T)\,,
\end{align}
where $I_0$ is the modified Bessel function of the first kind. Once $\sigma_\tau$ is fixed and $\mathcal{N}_0$ is known from Eq.~\eqref{eq:fonllsigma}, the integral is evaluated numerically and $R_{AA}$ is extracted from Eq.~\eqref{eq:raa}. The Gaussian approximation is motivated by Fig.~\ref{fig:raa-comparison}\subref{fig:raa-gauss} which shows the time evolved $p_T$ spectrum for charm quarks all initialized with a given $p_T$, with Gaussian distributions obtained for most $p_T$ values. Figure~\ref{fig:raa-comparison}\subref{fig:raa-analytical} compares the results obtained from the glasma particle solver to the proposed Gaussian toy model, which approximates broadening as isotropic kicks of width $\sigma_{p_T}$ extracted from the numerical transverse broadening $\langle\delta p_T^2\rangle$. The toy model captures both the shape and magnitude of $R_{AA}$ but is expected to break down once any energy loss mechanism is included and the $p_T$ bins become coupled. This requires a large number of heavy quarks and their backreaction in the CYM field equations, which is currently not considered.

\subsection{Two-particle correlations} 

The simultaneous description of $R_{AA}$ and $v_2$ remains a challenge for existing heavy quark transport models, and more differential observables such as heavy flavor angular correlations are expected to offer better discrimination. Existing calculations of $Q\bar{Q}$ azimuthal correlations in the QGP, using Langevin dynamics or the Boltzmann equation in a hydrodynamic medium, all lack a treatment of the pre-equilibrium stage. 

\noindent\textbf{\textit{Initial production}}.
Heavy quark and antiquark $Q\bar{Q}$ pairs are initialized back-to-back in azimuthal angle, as predicted by leading order pQCD production~\cite{Andronic:2015wma}. This assumes that the perturbative production of the pair can be factorized from the subsequent glasma evolution. Each quark of mass $m$ is assigned a fixed initial transverse momentum $p_T$ and is formed at proper time $\tau_\mathrm{form}=1/(2m_T)$, where $m_T^2=m^2+p_T^2$~\cite{Avramescu:2024poa}. Note that in~\cref{sec:classical-transport} the simpler estimate $\tau_\mathrm{form}=1/(2m)$ was used. The quark and antiquark have opposite transverse momenta, $\vec{p}_T^{\,Q}(\tau_\mathrm{form})=-\vec{p}_T^{\,\bar{Q}}(\tau_\mathrm{form})$, with $p^\eta(\tau_\mathrm{form})=0$. Both are produced at the same transverse position $x_T=(x,y)$, randomly distributed in the simulation region, with $\eta(\tau_\mathrm{form})=0$. The initial color charges of $Q$ and $\bar{Q}$ are random, since the color degrees of freedom decorrelate rapidly in the glasma regardless of their initial values.

\begin{figure}[H]
    \centering
    \includegraphics[width=\textwidth]{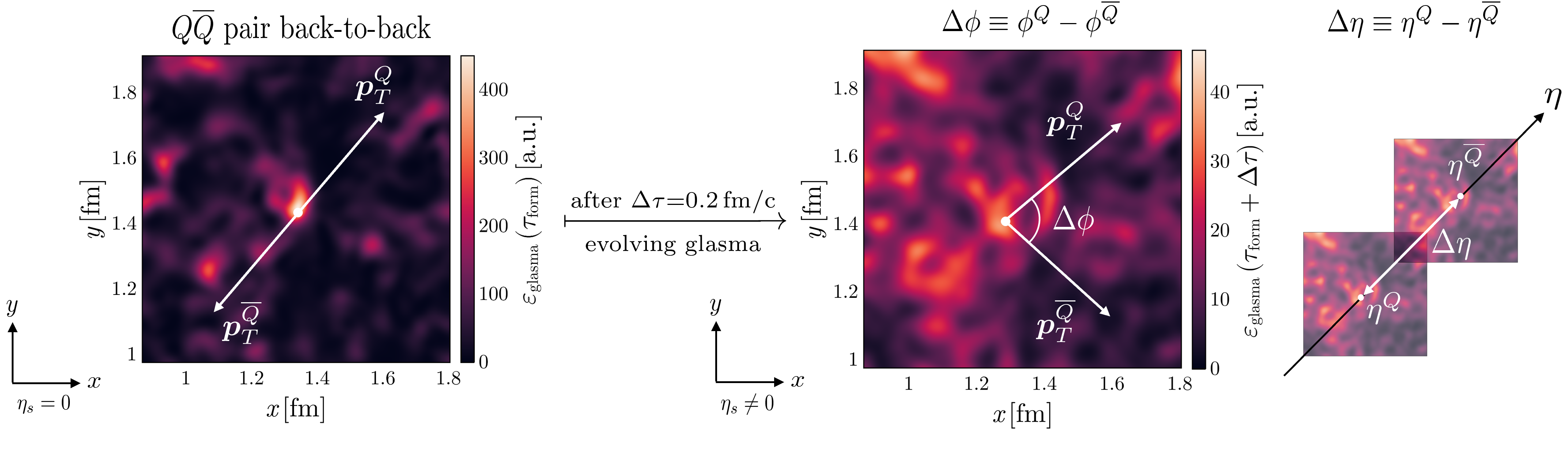}
    \caption{Schematics of a $Q\bar{Q}$ pair initialized back-to-back with $\vec{p}_T^{\,Q}=-\vec{p}_T^{\,\bar{Q}}$ at $\tau_\mathrm{form}$ and evolved in the glasma. Figure reproduced from Article~\cite{ar2} under license CC BY 4.0.}
    \label{fig:quark-pair-evo}
\end{figure}

\noindent\textbf{\textit{Effect of glasma}}. 
The pair is evolved in the glasma background using the classical particle solver. The laboratory frame longitudinal momentum is $p^z=\sinh\eta\, p^\tau+\cosh\eta\,\tau p^\eta$. The pseudorapidity of each quark is $\eta^q=\ln[(p^q+p^q_z)/(p^q-p^q_z)]/2$,
with $(p^q)^2=(p^q_T)^2+(p^q_z)^2$ for $q\in\{Q,\bar{Q}\}$. The relative pseudorapidity and azimuthal angle between the pair are $\Delta\eta \equiv \eta^Q-\eta^{\bar{Q}}$ and $\Delta\phi\equiv \angle(\vec{p}_T^{\,\,Q},\vec{p}_T^{\,\,\bar{Q}})$. This setup is sketched in Fig.~\ref{fig:quark-pair-evo}.

\begin{figure}[H]
    \centering
    \begin{subfigure}[b]{0.48\textwidth}
        \centering
        \includegraphics[width=0.8\textwidth]{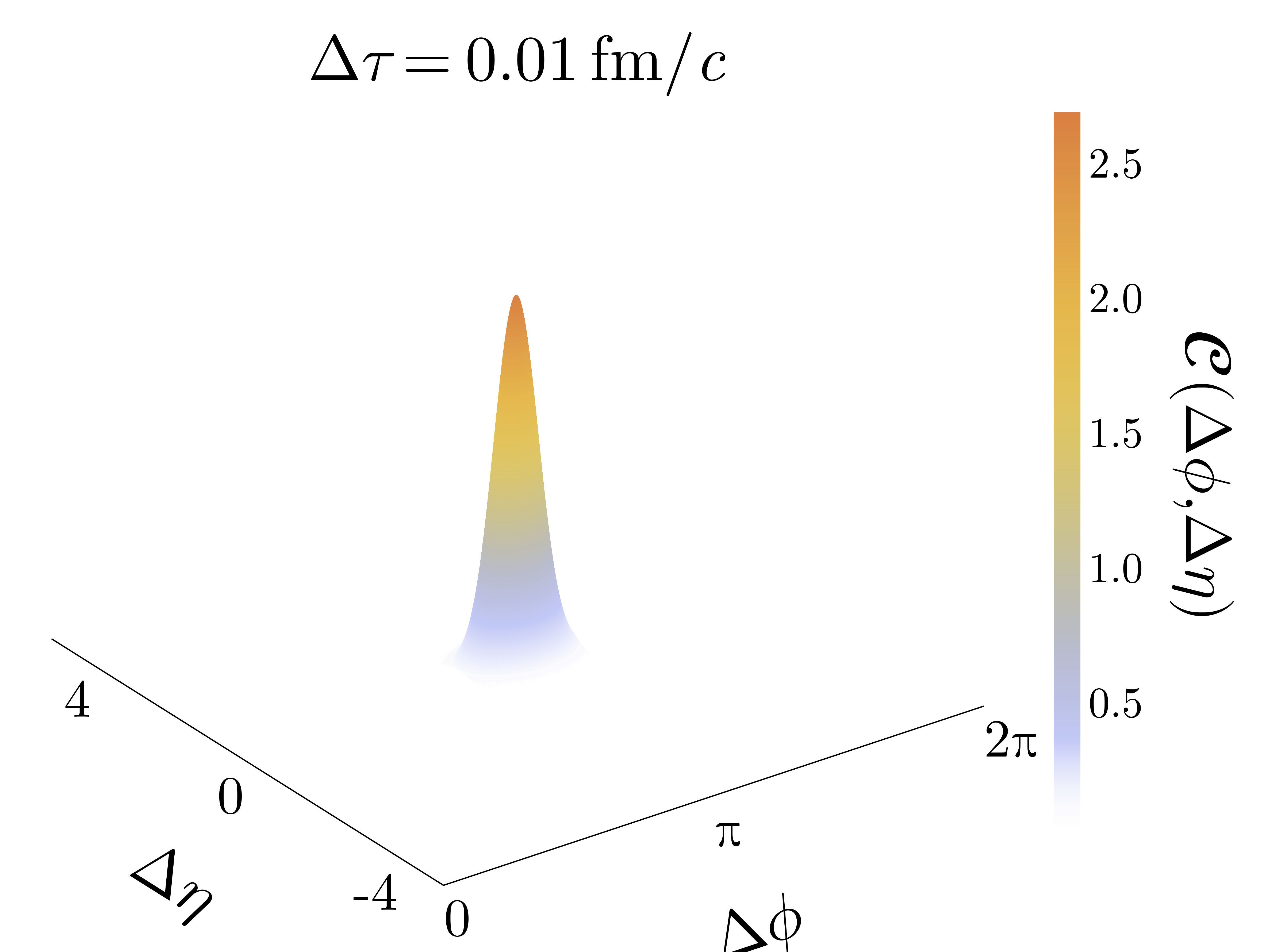}
        \captionsetup{justification=centering}
        \caption{Early time decorrelation.}
        \label{fig:corr-3d-tau01}
    \end{subfigure}
    \hfill
    \begin{subfigure}[b]{0.48\textwidth}
        \centering
        \includegraphics[width=0.8\textwidth]{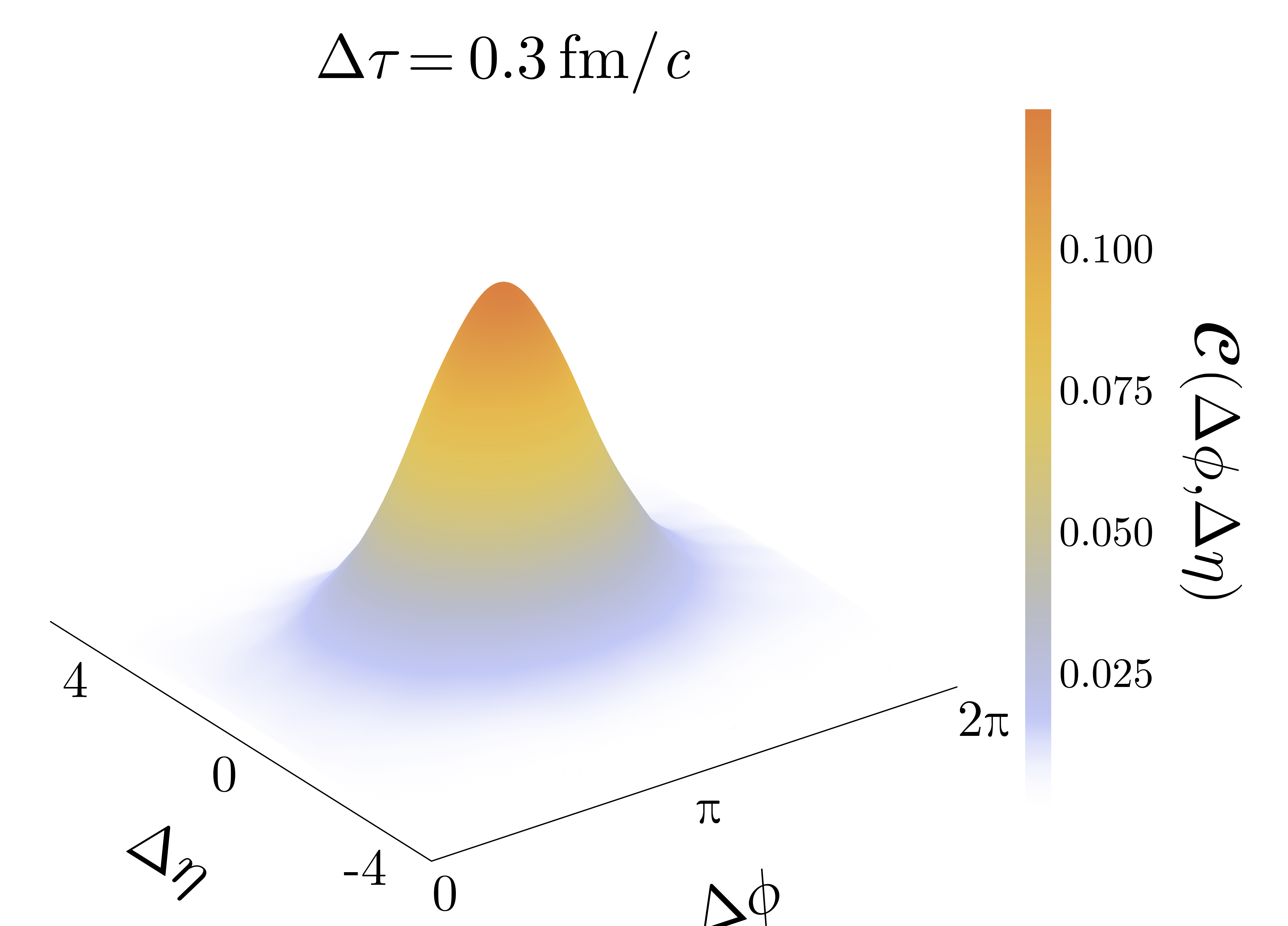}
        \captionsetup{justification=centering}
        \caption{Late time decorrelation.}
        \label{fig:corr-3d-tau3}
    \end{subfigure}
    \caption{Two-particle correlations $\mathcal{C}(\Delta\eta,\Delta\phi)$ for charm quark pairs initialized with $p_T(\tau_\mathrm{form})=1\,\mathrm{GeV}$ in a glasma with $Q_s=2\,\mathrm{GeV}$, at early and late values of proper time $\Delta\tau$. Figure reproduced from Article~\cite{ar3} under license CC BY 4.0.}
    \label{fig:corr-3d-tau}
\end{figure}

After collecting $(\Delta\eta_i,\Delta\phi_i)$ over $N_\mathrm{tp}$ test particles, the two-particle correlation function
\begin{align}
    \label{eq:twopartcorr}
    \mathcal{C}(\Delta\eta,\Delta\phi)\equiv\frac{1}{N_\mathrm{pairs}}\frac{\diff^2 N}{\diff\Delta\eta\,\diff\Delta\phi}\,,
\end{align}
is computed as a function of relative proper time $\Delta\tau\equiv\tau-\tau_\mathrm{form}$. Integrating over one variable gives the distributions $\mathcal{C}(\Delta\phi) = 1/N_\mathrm{pairs}\,\diff N/\diff \Delta\phi$ and similarly for $\mathcal{C}(\Delta\eta)$, where $N_\mathrm{pairs}$ is the total number of pairs. 
The momentum kicks from the glasma smear out the initial back-to-back peak into a roughly Gaussian distribution in both $\Delta\phi$ and $\Delta\eta$. The degree of decorrelation is quantified by the widths $\sigma_{\Delta\phi}$ and $\sigma_{\Delta\eta}$, extracted as standard deviations of $\mathcal{C}(\Delta\phi)$ and $\mathcal{C}(\Delta\eta)$.

The resulting two-particle correlations are depicted in Fig.~\ref{fig:corr-3d-tau}. The initial back-to-back peak gets washed out rapidly in both $\Delta\eta$ and $\Delta\phi$, and by $\Delta\tau=0.3\,\mathrm{fm}/c$, the typical end of the glasma stage at the LHC~\cite{Kurkela:2018vqr}, it is heavily suppressed.

\begin{figure}[H]
    \centering
    \hspace{0.05\textwidth}%
    \begin{subfigure}[b]{0.4\textwidth}
        \centering
        \includegraphics[trim={0 0 255 0}, clip, width=\linewidth]{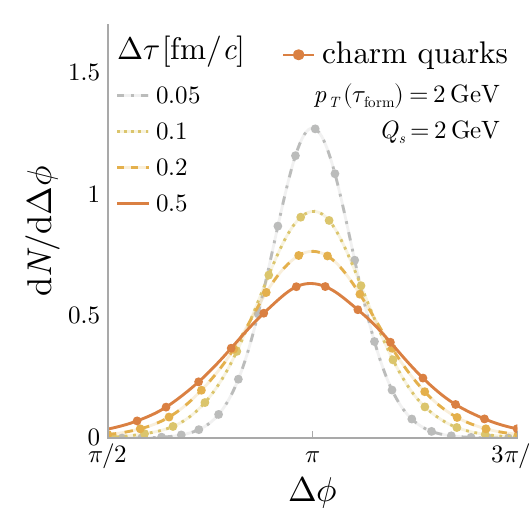}
        \captionsetup{justification=centering}
        \caption{Charm quark pairs.}
        \label{fig:dndphi-tau-charm}
    \end{subfigure}%
    \hfill
    \begin{subfigure}[b]{0.4\textwidth}
        \centering
        \includegraphics[trim={255 0 0 0}, clip, width=\linewidth]{images/dNdphi_tau_dep_charm_beauty.pdf}
        \captionsetup{justification=centering}
        \caption{Beauty quark pairs.}
        \label{fig:dndphi-tau-beauty}
    \end{subfigure}%
    \hspace{0.05\textwidth}
    \caption{Azimuthal angle correlation $\mathcal{C}(\Delta\phi)$ for charm and beauty quark pairs initialized with $p_T(\tau_\mathrm{form})=2\,\mathrm{GeV}$ and $Q_s=2\,\mathrm{GeV}$, extracted at different values of $\Delta\tau$. Figure reproduced from Article~\cite{ar2} under license CC BY 4.0.}
    \label{fig:dndphi-tau}
\end{figure}

\begin{figure}[H]
    \centering
    \begin{subfigure}[b]{0.49\textwidth}
        \centering
        \includegraphics[trim={0 0 325 0}, clip, width=0.831\linewidth]{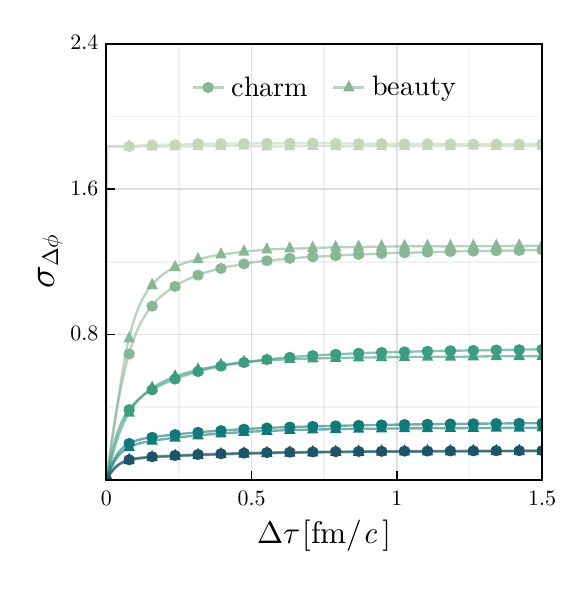}%
        \includegraphics[trim={544 0 0 0}, clip, width=0.169\linewidth]{images/sigma_dphideta_tau_charm_beauty_pT_dep.pdf}
        \caption{$p_T(\tau_\mathrm{form})$ dependence at $Q_s=2\,\mathrm{GeV}$.}
        \label{fig:sigma-phi-pt-dep}
    \end{subfigure}%
    \hfill
    \begin{subfigure}[b]{0.49\textwidth}
        \centering
        \includegraphics[trim={0 0 328 0}, clip, width=0.822\linewidth]{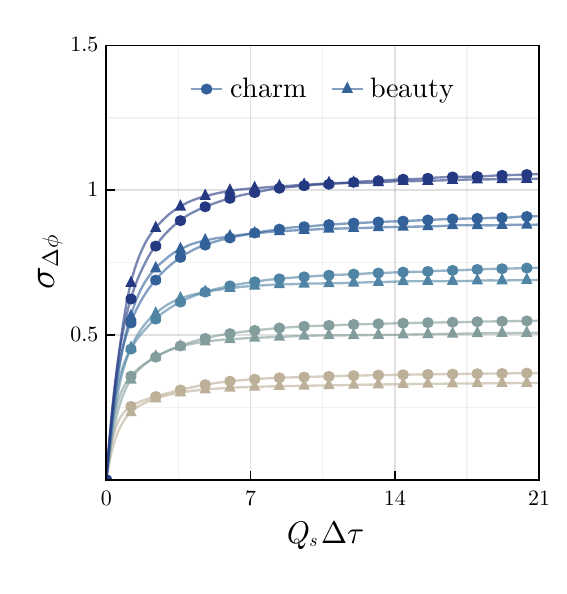}%
        \includegraphics[trim={541 0 0 0}, clip, width=0.178\linewidth]{images/sigma_dphideta_tau_charm_beauty_Qs_dep_scaled.pdf}
        \caption{$Q_s$ dependence at $p_T(\tau_\mathrm{form})=2\,\mathrm{GeV}$.}
        \label{fig:sigma-phi-qs-dep}
    \end{subfigure}
    \caption{Azimuthal decorrelation width $\sigma_{\Delta\phi}$ for charm \textit{(circles)} and beauty \textit{(triangles)} quark pairs as a function of glasma evolution time, for varying $p_T(\tau_\mathrm{form})$ at fixed $Q_s=2\,\mathrm{GeV}$, and varying $Q_s$ plotted against the dimensionless time $Q_s\Delta\tau$ at fixed $p_T(\tau_\mathrm{form})=2\,\mathrm{GeV}$. Figure reproduced from Article~\cite{ar2} under license CC BY 4.0.}
    \label{fig:sigma-phi-dep}
\end{figure}

Figure~\ref{fig:dndphi-tau} focuses on the azimuthal decorrelation, showing the rapid suppression of the initial $\delta(\Delta\phi-\pi)$ peak during the first $\Delta\tau\lesssim 0.1\,\mathrm{fm}/c$ when the glasma fields are strongest, followed by a much slower evolution at later times as the longitudinal expansion dilutes the fields. The decorrelation is comparable for charm and beauty quarks at this $p_T$, despite their different masses, because beauty quarks form earlier and experience the stronger early time fields. This shows that the decorrelation is not simply a monotonic function of mass, but depends on the interplay between formation time and field strength. Lastly, Fig.~\ref{fig:sigma-phi-dep} shows the azimuthal decorrelation width $\sigma_{\Delta\phi}$ for the two main dependencies of the glasma stage. The resulting $\sigma_{\Delta\phi}$ decreases monotonically with increasing $p_T(\tau_\mathrm{form})$. The low $p_T$ pairs are easily deflected by the glasma kicks, while pairs with larger $p_T$ retain more of the initial back-to-back structure. Despite being heavier, beauty quarks decorrelate more than charm quarks at the same $p_T$. Plotting against the dimensionless time $Q_s\Delta\tau$ makes the curves for different $Q_s$ collapse and confirms that $1/Q_s$ sets the relevant decorrelation timescale.

Nevertheless, the $Q\bar{Q}$ pair at NLO is not produced back-to-back, due to gluon splitting and flavor excitation processes. Such NLO effects are included in Monte Carlo event generators such as PYTHIA~\cite{Sjostrand:2014zea}, and the resulting initial $\mathcal{C}(\Delta\phi)$ is a broader distribution peaked at $\Delta\phi=\pi$ rather than $\delta(\Delta\phi-\pi)$. The glasma kicks will further broaden this distribution, but the relative contribution of the glasma to the total decorrelation is expected to be smaller than in the LO case. Moreover, the transport of the $Q\bar{Q}$ pair through the QGP further decorrelates the pair, and the relative contribution of the glasma will depend on the relative timescales of the two stages. A full quantitative prediction for heavy flavor correlations in nuclear collisions requires coupling the glasma initial conditions to subsequent stages.

\begin{emphbox}{gray!50!black}{Future directions}
Several limitations of the classical probe framework presented in this chapter point to future extensions~\cite{Avramescu:2023qvv,Avramescu:2024poa,Avramescu:2024xts}:
\begin{itemize}[label=$\circ$, leftmargin=*]
    \item \textbf{Backreaction.} Hard probes are treated as test particles, but the color current they generate in the glasma is neglected. Including backreaction would couple the particle solver to the field equations. 
    However, fast moving charged particles on the lattice are subject to the numerical Cherenkov instability.
    \item \textbf{Collision kernel.} The present approach studies only the second moment $\langle\delta p^2_\perp\rangle$ of the momentum broadening distribution $P(\vec{p}_\perp)$. This could be extracted from the Wilson loops in the glasma and would provide access to the collision kernel in the glasma.
    \item \textbf{Beyond boost invariance.} The boost-invariant glasma captures the rapidity dependence only through the longitudinal Lorentz force, whereas a 3+1D glasma would give more insight into the longitudinal structure of angular decorrelations $\mathcal{C}(\Delta\eta)$.
    \item \textbf{Realistic nuclear geometry.} Finite size effects, impact parameter dependent initial conditions and subnucleonic fluctuations or hot spots are currently absent. Incorporating these is required for making more quantitative predictions in $AA$, $pA$, and $pp$ collisions.
    \item \textbf{Coupling to subsequent stages.} The glasma contributions to $R_{AA}$ and $Q\bar{Q}$ angular correlations need to be matched to hydrodynamic evolution for a direct comparison with experimental data. A systematic multi-stage framework coupling all phases of the collision is not yet available.
\end{itemize}
\end{emphbox}

\chapter{Quantum probes in the glasma}
\label{chap:quantum}

In this chapter, three notations for the gauge field are employed, distinguished by color structure. The gauge field $\mathcal{A}_\mu$ is a color matrix, $\mathcal{A}_\mu \equiv \mathcal{A}_\mu^a t^a$, with $t^a$ the $\mathrm{SU}(3)$ generators. For a classical probe with color charge $Q^a$, the gauge field is written as $A_\mu \equiv \mathcal{A}_\mu^a Q^a$, which is not the full gauge field matrix but the color field component $\mathcal{A}_\mu^a$ projected onto $Q^a$. Within the quantum framework, the gauge field is $\hat{A}_\mu \equiv \mathcal{A}_\mu^a \hat{T}^a$, with $\hat{T}^a$ the color charge operators. In all three cases, $\mathcal{A}_\mu^a$ refers to the same classical color field components. Similarly, the field strength tensor is denoted $\mathcal{F}_{\mu\nu}^a$ for classical components, $F_{\mu\nu}$ for the color matrix, and $\hat{F}_{\mu\nu}$ for the quantum operator, with the same decomposition. The kinetic momentum from~\cref{chap:hardprobes} is denoted $p^\mu_{kin}$ and the canonical momentum $p^\mu$.
The electric and magnetic fields also acquire the calligraphic notation, $\mathcal{E}_i$ and $\mathcal{B}_i$, and similarly for other quantities, to be consistent with the notations used in Articles~\cite{ar4} and~\cite{ar5}.

\section{In-medium momentum broadening}

\begin{emphbox}{myred}{Based on}
Article~\cite{ar4} studies classical eikonal and static quarks propagating through the glasma fields. Notably, it clarifies the distinction between the kinetic and canonical momentum broadening. The former is the physical gauge invariant quantity, while the latter is gauge dependent. In the glasma, the longitudinal kinetic momentum is larger than the transverse one, while the canonical momentum broadening exhibits a much larger transverse component. Additionally, this work investigates the numerical effect of the gauge choice and shows that imposing the Coulomb gauge at the initial time reduces significantly the gauge artifacts and numerical errors for eikonal quarks. This further motivates the use of the Coulomb gauge in the subsequent quantum calculation for eikonal quarks. 
\end{emphbox}

\subsection{Quantum to classical correspondence}

\noindent\textbf{\textit{Quantum equations of motion}.} The derivation is done in a covariant manner and starts from the quadratic form of the Dirac Hamiltonian~\cite{Heinz:1984yq, Elze:1989un}
\begin{align} \label{eq:QuadDirac}
    \hat{H} = \frac{1}{2m}\left(m^2 - (\hat{p}^\mu - g \hat{A}^\mu)^2\right) + \frac{g}{2m} \hat{S}^{\mu\nu} \hat{F}_{\mu\nu}\,,
\end{align}
where the minimal coupling $\partial_\mu \to \partial_\mu + ig\hat{A}_\mu$ contains the interaction of the quantum particle with the classical background field. The quantum operator for coupling the particle to the gauge field $\hat{A}^\mu = \mathcal{A}^{\mu,a}\hat{T}^a$ is given in terms of the classical color components of the field $\mathcal{A}_\mu^a$ and the quantum color charge operators $\hat{T}^a$, and similarly for the field strength tensor $\hat{F}_{\mu\nu} = \mathcal{F}_{\mu\nu}^a\hat{T}^a$. The spin structure of the particle is contained in $\hat{S}^{\mu\nu} = \hat{\sigma}^{\mu\nu}/2$, where $\sigma^{\mu\nu} = i[\gamma^\mu, \gamma^\nu]/2$ is the antisymmetric tensor constructed from the Dirac gamma matrices.

The classical limit may be conveniently obtained by working in the Heisenberg picture. Here, the time dependence is contained in the operators $\hat{\mathcal{O}}_H(\lambda)$, with $\lambda$ being the worldline parameter. They satisfy the evolution equation
\begin{align}\label{eq:Ehrenfest}
    \frac{\diff \hat{\mathcal{O}}_H}{\diff \lambda} = i [\hat{H}, \hat{\mathcal{O}}_H] + \frac{\partial \hat{\mathcal{O}}_H}{\partial \lambda}\,.
\end{align}
The Heisenberg picture quantum state $|\psi\rangle_H \equiv |\psi;0\rangle$ is then used to evaluate the quantum expectation value $\langle\hat{X}\rangle_\psi \equiv {}_H\langle\psi|\hat{X}_H|\psi\rangle_H$, from which the classical limit is inferred. From now on, the $H$ subscript is omitted for notation brevity. To derive Wong's equations, one applies~\cref{eq:Ehrenfest} to the coordinate, momentum, color charge and spin operators $\hat{\mathcal{O}}\in(\hat{x}^\mu,\hat{p}^\mu,\hat{Q}^a,\hat{S}^{\mu\nu})$.

The velocity operator is defined as the worldline proper time derivative of the coordinate operator and is obtained as
\begin{align}\label{eq:VelocityOperator}
    \hat{v}^\mu \equiv \frac{d\hat{x}^\mu}{d\lambda} = i[\hat{H}, \hat{x}^\mu] = \frac{\hat{p}^\mu - g\hat{A}^\mu}{m}\,.
\end{align}

\begin{emphbox}{myblue}{Kinetic and canonical momentum}
The kinetic momentum operator of a particle subject to a background gauge field is given by
\begin{equation}\label{eq:KinCanMom}
    \hat{p}_{kin}^\mu \equiv m\hat{v}^\mu = \hat{p}^\mu - g\hat{A}^\mu\,.
\end{equation}
This differs from the canonical momentum $\hat{p}^\mu$, which is conjugate to the coordinates and satisfies the canonical commutation relation $[\hat{x}^\mu, \hat{p}^\nu] = -ig^{\mu\nu}$. The kinetic momentum is the gauge invariant physical quantity that corresponds to the classical momentum of the particle.
\end{emphbox}

The equation of motion for the kinetic momentum is obtained in the same way 
\begin{equation} \label{eq:HeisenbergKineticMomentum}
    \frac{\diff \hat{p}_{kin}^\mu}{\diff \lambda} = i[\hat{H}, \hat{p}^\mu_{kin}] = \frac{g}{2m}\left( \hat{p}_\nu^{kin} \hat{F}^{\mu\nu} + \hat{F}^{\mu\nu} \hat{p}_\nu^{kin} + \hat{S}^{\nu\rho} \hat{D}^\mu \hat{F}_{\nu\rho}\right)\,,
\end{equation}
where $\hat{D}_\mu = \partial_\mu + ig\hat{A}_\mu$ and the last term contains the coupling between the spin and the background field. Note that the sign convention for the minimal coupling used in this chapter is opposite to the one of~\cref{chap:glasma} and~\cref{app:qcd}, where $D_\mu=\partial_\mu-igA_\mu$. 
 The equation for the time evolution of the color charge is 
\begin{equation}\label{eq:HeisenbergColor}
    \frac{\diff \hat{T}^a}{\diff \lambda} = i[\hat{H}, \hat{T}^a] = -\frac{g}{2m} f^{abc}\left(- \mathcal{A}^b_\mu (\hat{p}^\mu_{kin} \hat{T}^c + \hat{T}^c \hat{p}^\mu_{kin}) - \mathcal{F}^b_{\mu\nu} \hat{S}^{\mu\nu} \hat{T}^c\right)\,.
\end{equation}
Lastly, the time evolution of the particle spin operator takes the form
\begin{equation}\label{eq:HeisenbergSpin}
    \frac{\diff \hat{S}^{\mu\nu}}{\diff \lambda} = i[\hat{H}, \hat{S}^{\mu\nu}] = \frac{2g}{m} \hat{S}^{\rho[\mu} \hat{F}^{\nu]}_{\ \ \rho}\,,\quad \text{with}\; X^{[\mu\nu]} \equiv \frac{1}{2}(X^{\mu\nu}-X^{\nu\mu})\,,
\end{equation}
where the anti-symmetrization is denoted as $X^{[\mu\nu]}$. In this derivation, note that $\hat{A}^\mu$ and $\hat{F}^{\mu\nu}$ do not explicitly depend on $\lambda$ but only through $x^\mu(\lambda)$, therefore $\partial \hat{\mathcal{O}}_H/\partial \lambda = 0$ for these operators and the evolution is given entirely by the commutator with the Hamiltonian. 

The evolution equation for the canonical momentum is
\begin{equation} \label{eq:CanonicEqMotion}
    \frac{\diff \hat{p}^\mu}{\diff \lambda} = i[\hat{H}, \hat{p}^\mu] = \frac{g}{2m} \left(\hat{p}_\nu^{kin} (\partial^\mu \hat{A}^\nu) + (\partial^\mu \hat{A}^\nu) \hat{p}_\nu^{kin} + \hat{S}^{\nu\rho} \partial^\mu \hat{F}_{\nu\rho}\right)\,,
\end{equation}
while the time evolution equation of the gauge field operator along the particle trajectory is given by
\begin{align} \label{eq:FieldEqMotion}
    \frac{\diff \hat{A}^\mu}{\diff \lambda} = i[\hat{H}, \hat{A}^\mu] = \frac{1}{2m}\left( \hat{p}_\nu^{kin} (\hat{D}^\nu\hat{A}^\mu) + (\hat{D}^\nu\hat{A}^\mu)\hat{p}_\nu^{kin} - ig\,[\hat{A}^\mu,\hat{F}_{\nu\rho}]\,\hat{S}^{\nu\rho} \right)\,.
\end{align}
The evolution of the canonical momentum and gauge field is consistent with the evolution of the kinetic momentum through $\diff\hat{p}^\mu/\diff\lambda - g\,\diff\hat{A}^\mu/\diff\lambda = \diff\hat{p}^\mu_{kin}/\diff\lambda$ according to~\cref{eq:KinCanMom}. 

\noindent\textbf{\textit{Classical limit}.} The classical equations of motion are obtained by taking the classical limit in the Heisenberg equations of motion. More precisely, this is done by replacing the expectation values of the quantum operators by their classical values $\langle\hat{\mathcal{O}}\rangle_\psi \to \mathcal{O}$ with $\mathcal{O} \in (x^\mu, p_{kin}^\mu, Q^a, S^{\mu\nu})$. In the intermediate steps, one assumes the factorization of the expectation values $\langle \hat{\mathcal{O}}\,\hat{T}^a\rangle_\psi \to \mathcal{O} Q^a$. The resulting classical equations of motion are obtained as
\begin{subequations} \label{eq:ClassicalLimHeisenbergEq}
    \begin{align}
        &\frac{\diff x^\mu}{\diff \lambda} = \frac{p_{kin}^\mu}{m}\,, && \frac{\diff p_{kin}^\mu}{\diff \lambda} = \frac{g}{m} Q^a \left(\mathcal{F}^{\mu\nu}_a p_\nu^{kin} + \frac{1}{2} S^{\nu\rho} \,\mathcal{D}^\mu \mathcal{F}_{\nu\rho}^a\right)\,, \\
        &\frac{\diff S^{\mu\nu}}{\diff \lambda} = \frac{2g}{m}Q^a S^{\rho[\mu}F^{\nu],a}_{\ \ \rho}\,, && \frac{\diff Q^a}{\diff \lambda} = \frac{g}{m} f^{abc} Q^c\left( \mathcal{A}^b_\mu\,  p^\mu_{kin} + \frac{1}{2} S^{\mu\nu} \mathcal{F}^b_{\mu\nu}\right)\,.
    \end{align}
\end{subequations}
These are Wong's equations for the coordinate, momentum and color charge from \cref{eq:Wong}, along with extra terms and an additional equation of motion involving the classical spin. One may adapt Wong's equations to the massless case by using $m\diff\mathcal{O}/\diff\lambda = E\diff\mathcal{O}/\diff t$, with $t$ the coordinate time.

Taking the classical limit $\langle \hat{A}^\mu\rangle_\psi = \mathcal{A}^{\mu,a}\langle\hat{T}^a\rangle_\psi \to \mathcal{A}^{\mu,a}Q^a = A^\mu$ gives the classical canonical momentum equation of motion
\begin{subequations} \label{eq:ClassCanonicEq}
    \begin{align}
        \frac{\diff p^\mu}{\diff \lambda} = \frac{g}{m} Q^a\left( (\partial^\mu \mathcal{A}^{\nu,a})\,p_\nu^{kin} + \frac{1}{2} S^{\nu\rho}\,( \partial^\mu \mathcal{F}^a_{\nu\rho})\right)\,.
    \end{align}
\end{subequations}
An alternative derivation in Minkowski coordinates, starting from the linear color Dirac Hamiltonian, but neglecting spin, is provided in~\cref{app:lightfront}. It also contains explicit expressions for the kinetic and canonical momentum equations of motion in the eikonal limit, and applies the expressions to the case of glasma, which are boost-invariant gauge fields in the temporal gauge.

\subsection{Static and eikonal quarks}

For boost-invariant glasma fields $\mathcal{A}^\mu$ and the worldline parameter $\lambda$ being the Milne proper time $\tau$, the classical kinetic momentum broadening averaged over color charge configurations $\langle\,\cdots\,\rangle_Q$ is given by
\begin{align} \label{eq:KinMomLorentzForce}
    \langle (\delta p_i^{kin}(\tau))^2\rangle_Q=g^2\int_0^\tau \diff \tau^\prime \int_0^\tau \diff\tau^{\prime \prime}\,\tr{ \tilde{f}^i(\tau^\prime)\tilde{f}^i(\tau^{\prime \prime})}\,,
\end{align}
where $\tilde{f}^i$ is the color Lorentz force color rotated with the particle Wilson line $\tilde{f}^i=\mathcal{U}(\tau;0)f_i\,\mathcal{U}^\dagger(\tau;0)$, with $\tilde{f}^i$ depending on the quark trajectory. For static quarks with $m\rightarrow\infty$ it is denoted as $f^i_{stat}$ and contains only glasma electric fields, while for eikonal quarks with $p^x\rightarrow\infty$ it is denoted as $f^i_{eik}$ and contains both electric and magnetic fields
\begin{equation}
    f^i_{stat} = \mathcal{E}_i\,,\qquad f^i_{eik} = \mathcal{E}_i - \epsilon_{ij} \mathcal{B}_j\,,
\end{equation}
where $i\in(y,z)$, different from the lattice transverse index $\underline{i}\in(x,y)$, is taken over the directions transverse to the eikonal quark which propagates along $x$, and $\epsilon_{ij}$ is the antisymmetric tensor in the transverse plane $(y,z)$, giving $f^y_{eik} = \mathcal{E}_y - \mathcal{B}_z$ and $f^z_{eik} = \mathcal{E}_z + \mathcal{B}_y$. Since the eikonal quark propagates along $x$, only the transverse components in $i\in(y,z)$ experience momentum broadening. The color rotation of the Lorentz force is trivial for static quarks in the temporal gauge $\mathcal{A}_\tau=0$, while for eikonal quarks it picks up the longitudinal component of the gauge field along the trajectory
\begin{equation}
    \mathcal{U}_{stat} = \mathbbm{1}\,,\qquad \mathcal{U}_{eik}(\tau;0) = \mathcal{P}\exp\left[-ig\int_0^\tau \diff \tau^\prime \,\mathcal{A}_x(\tau^\prime)\right]\,.
\end{equation}

Similarly, the canonical momentum broadening is expressed as the squared integrated canonical Lorentz force
\begin{equation} \label{eq:CanMomLorentzForce}
    \langle(\delta p_i(\tau))^2\rangle_Q = g^2\int_0^\tau\diff\tau^\prime\int_0^\tau\diff\tau^{\prime\prime}\,\tr{\tilde{f}^i_p(\tau^\prime)\tilde{f}^i_p(\tau^{\prime\prime})}\,,
\end{equation}
where $\tilde{f}^i_p=\mathcal{U}f^i_p\,\mathcal{U}^\dagger$ is the parallel transported canonical Lorentz force.  As for the kinetic case, the canonical Lorentz force reduces to the electric field components and for eikonal jets it contains both electric and magnetic fields
\begin{equation}\label{eq:StatLorFor}
    f^{p,i}_{stat}=\mathcal{E}_i^p\,,\qquad f^{p,i}_{eik} = \mathcal{E}_i^p - \epsilon_{ij} \mathcal{B}_j^p\,,
\end{equation}
with the electric and magnetic field components contributing to the canonical momentum given by
\begin{equation}
    \mathcal{E}^p_y=0\,,\quad \mathcal{E}^p_z=\frac{1}{\tau^2}\mathcal{A}_\eta\,,\quad\mathcal{B}^p_y=0\,,\quad \mathcal{B}^p_z=\partial_y\mathcal{A}_x\,.
\end{equation}
These expressions are also derived in~\cref{app:lightfront}.

\subsection{Gauge dependence}

\noindent\textbf{\textit{Kinetic and canonical momentum}.} The non-Abelian gauge transformation of the gauge field is given by $\mathcal{A}_\mu\rightarrow \mathcal{G}\,\mathcal{A}_\mu \,\mathcal{G}^{-1}+(i/g)\,\mathcal{G}\,(\partial_\mu \mathcal{G}^{-1})$ with $\mathcal{G}\in\mathrm{SU}(N_c)$. The field strength tensor transforms covariantly as $\mathcal{F}_{\mu\nu}\rightarrow \mathcal{G}\,\mathcal{F}_{\mu\nu}\,\mathcal{G}^{-1}$ and similarly for the color charge $Q\rightarrow \mathcal{G}\,Q\,\mathcal{G}^{-1}$. The kinetic momentum broadening is thus gauge invariant. The canonical Lorentz force, however, is not gauge invariant, making the canonical momentum broadening gauge dependent. This makes it a useful diagnostic tool for quantifying the effects of gauge transformations on the glasma fields. 

For boost-invariant gauge transformations $\partial_\eta\mathcal{G}=0$, the rapidity component transforms as $\mathcal{A}_\eta\rightarrow \mathcal{G}\,\mathcal{A}_\eta\,\mathcal{G}^{-1}$. The longitudinal canonical Lorentz force, which is determined only by $\mathcal{A}_\eta$, transforms as $f_p^z = \mathcal{A}_\eta/\tau^2 \rightarrow \mathcal{G}\,\mathcal{A}_\eta\,\mathcal{G}^\dagger/\tau^2$. Therefore, the trace in the canonical broadening $\langle(\delta p_z)^2\rangle_Q$ is gauge invariant. For the transverse component, the term $\mathcal{G}\,(\partial_y\mathcal{G}^\dagger)$ generates a genuine gauge dependence and only $\langle(\delta p_y)^2\rangle_Q$ is affected by gauge transformations. To show this effect, the kinetic and canonical momentum broadening are evaluated in the temporal gauge $\mathcal{A}_\tau=0$ and transverse Coulomb gauge $\partial_{\underline{i}} \mathcal{A}_{\underline{i}}=0$, with ${\underline{i}}\in(x,y)$.

The glasma fields are obtained using a SU($3$) numerical solver based on real-time lattice gauge theory, as described in~\cref{sec:glasma}. For the results shown here, the numerical parameters are $g^2\mu = 1.5\,\mathrm{GeV}$, roughly corresponding to LHC energies, number of transverse lattice points $N_\perp = 1024$, transverse length $L_\perp = 5\,\text{fm}$, which correspond to $g^2\mu L_\perp \approx 38$ and $g^2\mu a_\perp \ll 1$, along with the IR and UV regulators $m = 0.3\,\mathrm{GeV}$ and $\Lambda = 10\,\mathrm{GeV}$. Unless otherwise stated, observables are averaged over the quark color charge $\langle\,\cdots\,\rangle_Q$, glasma field configurations $\langle\,\cdots\,\rangle_{\rho}$, and transverse lattice sites $\langle\,\cdots\,\rangle_{\vec{x}_\perp}$.

The glasma fields are solved in temporal gauge $\mathcal{A}_\tau=0$, which leaves a residual freedom to perform time-independent gauge transformations. One can exploit this freedom to impose the transverse Coulomb gauge condition $\sum_{\underline{i}}\partial_{\underline{i}}\mathcal{A}_{\underline{i}}=0$ at $\tau=0$ with $\underline{i}\in(x,y)$. This gauge minimizes the squared transverse gauge potential $\sum_{\underline{i}}\langle\mathcal{A}_{\underline{i}}^2\rangle$~\cite{Gubarev:2000eu, Gubarev:2000nz} and reduces the overall magnitude of the transverse fields. The fields are then free to evolve without any further gauge fixing. The Coulomb gauge fixing is explained in more detail in~\cref{sec:coulomb-gauge}.

\begin{figure}[t]
    \centering
    \begin{subfigure}[b]{0.48\textwidth}
        \centering
        \includegraphics[width=\textwidth]{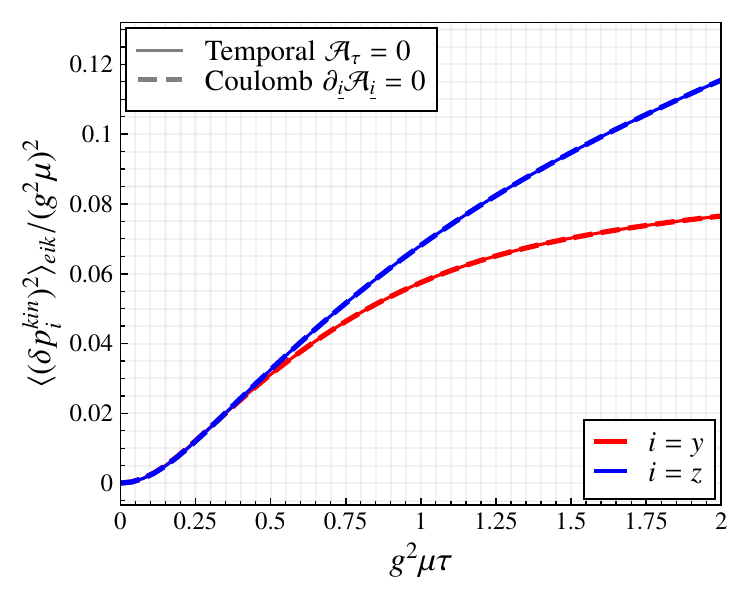}
        \caption{Kinetic momentum broadening.}
        \label{fig:pkin-coulomb}
    \end{subfigure}
    \hfill
    \begin{subfigure}[b]{0.48\textwidth}
        \centering
        \includegraphics[width=\textwidth]{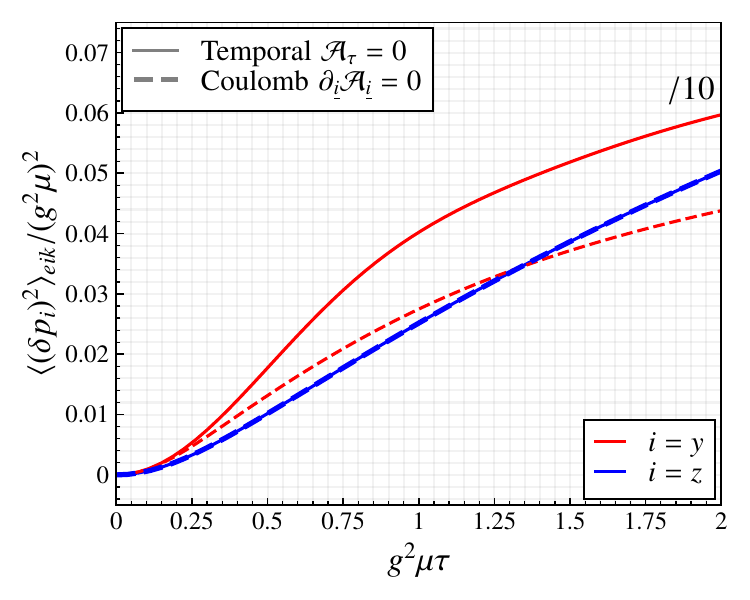}
        \caption{Canonical momentum broadening.}
        \label{fig:pcan-coulomb}
    \end{subfigure}
    \caption{Kinetic \textit{(left)} and canonical \textit{(right)} momentum broadening for eikonal quarks, rescaled by $(g^2\mu)^2$, as a function of  $g^2\mu\tau$. The $y$ component is colored in \textit{red} and the $z$ component in \textit{blue}. The \textit{solid line} corresponds to the temporal gauge, while the \textit{dashed line} to the Coulomb gauge. Note that the $y$ component of the canonical momentum broadening in temporal gauge \textit{(solid red)} is divided by $10$. Figure reproduced from Ref.~\cite{Avramescu:2026fgv} under license CC BY 4.0.}
    \label{fig:pkin-pcan-coulomb}
\end{figure}

\cref{fig:pkin-pcan-coulomb} shows the numerical gauge invariance of the kinetic momentum broadening and the gauge dependence of the canonical momentum broadening. The gauge-dependent component $\langle(\delta p_y)^2\rangle_Q$ of the canonical momentum broadening is considerably larger in the temporal gauge than in the Coulomb gauge. In the temporal gauge, the kinetic broadening is anisotropic with more longitudinal than transverse momentum broadening, while the canonical broadening is anisotropic in the opposite way, with a much larger transverse component. 

\noindent\textbf{\textit{Gauge potential along trajectory}.} The time evolution of the gauge field along the particle trajectory is either given by the change in the gauge potential $\delta A_i(\tau)\equiv A_i(\tau)-A_i(0)$, or related to the difference between the canonical and kinetic Lorentz forces involved in the momentum decomposition $g\,\delta A_i^f(\tau)\equiv\delta p_i(\tau)-\delta p^{kin}_i(\tau)$. These expressions are equivalent and must satisfy $\delta A_i(\tau)=\delta A_i^f(\tau)$. The change in the gauge potential can be expressed in terms of color components $\delta A_i(\tau)= \mathcal{A}_i^a(\tau)\,Q^a(\tau)-\mathcal{A}_i^a(0)\,Q^a(0)$. Using $Q^a(\tau)=\tr{Q(\tau)\,t^a}/T_R$ and the rotation of the color charge along the trajectory, the variation of the field becomes $g\,\delta A_i(\tau)=g\,Q^a(0)\,\tr{\delta \mathcal{A}_i (\tau)\, t^a}/T_R$. Note that the gauge potential experienced by the particle projects onto only a single color direction $Q^a(0)$. The change of the gauge field
\begin{equation}\label{eq:deltaArot}
    \delta \mathcal{A}_i(\tau) \equiv \tilde{\mathcal{A}}_i(\tau) - \mathcal{A}_i(0)\,,
\end{equation}
is evaluated using the parallel transported gauge field $\tilde{\mathcal{A}}_i(\tau)\equiv \mathcal{U}(\tau;0)\,\mathcal{A}_i(\tau)\,\mathcal{U}^\dagger(\tau;0)$. The same quantity expressed through the Lorentz force has a similar form $g\,\delta A_i^f(\tau) = g\,Q^a(0)\,\mathrm{tr}[\delta \mathcal{A}^f_i (\tau)\, t^a]/T_R$, where the variation of the gauge field is
\begin{equation}\label{eq:deltaAforce}
    \delta \mathcal{A}_i^f(\tau) \equiv \int_0^\tau \diff\lambda \,\tilde{f}^A_i(\lambda)\,.
\end{equation}
Here $f^A_i = f_i - f_i^p$ are the Lorentz force components contributing to the kinetic and canonical momentum, and $\tilde{f}$ indicates the color rotation along the quark trajectory. Combining the two expressions yields the consistency relation
\begin{equation} \label{eq:deltaAdeltaAtilde}
    \delta \mathcal{A}_i(\tau)= \delta\mathcal{A}^f_i(\tau)\,.
\end{equation}
This gives two equivalent ways to evaluate the field evolution, either by integrating the color-rotated Lorentz force $\tilde{f}^A_i$ to obtain $\delta\mathcal{A}^f_i$, or using the color-rotated potential $\delta\mathcal{A}_i$ directly. The former approach corresponds to the classical calculation, while the latter mirrors the quantum calculation, where one evaluates the operator $\hat{A}_i$ directly. The two approaches are explicitly equivalent, but a reliable comparison of classical and quantum calculations requires them to agree also numerically. In practice, for the numerical glasma fields, the transverse gauge components $\mathcal{A}_{\underline{i}}$ are extracted from the gauge link $U_{\underline{i}}$ with $\underline{i}\in(x,y)$ and the longitudinal component is $\mathcal{A}_z=\mathcal{A}_\eta/\tau$ at midrapidity $\eta=0$.

\begin{figure}[H]
    \centering
    \begin{subfigure}[b]{0.48\textwidth}
        \centering
        \includegraphics[width=\textwidth]{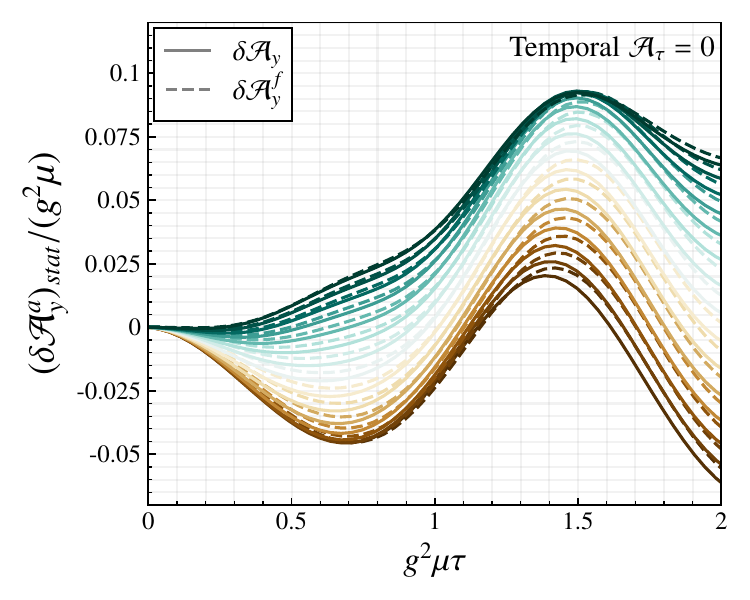}
        \caption{$y$ component.}
        \label{fig:deltaa-y-static}
    \end{subfigure}
    \hfill
    \begin{subfigure}[b]{0.48\textwidth}
        \centering
        \includegraphics[width=\textwidth]{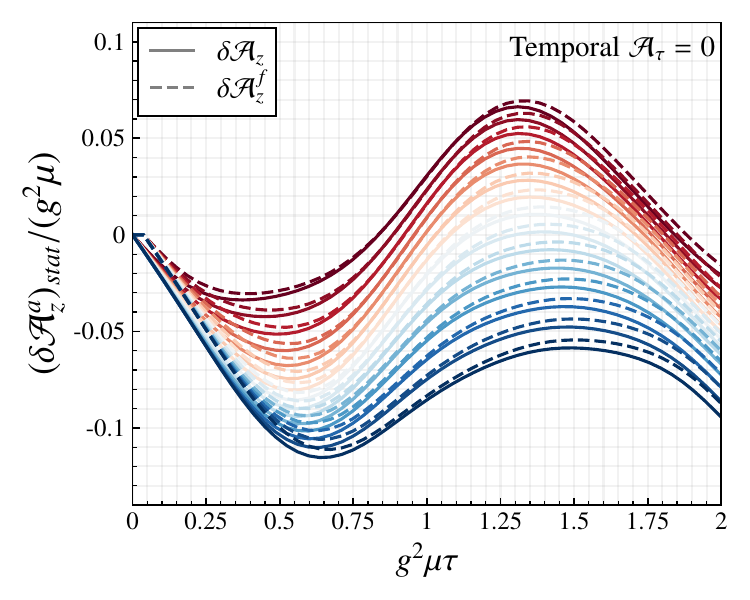}
        \caption{$z$ component.}
        \label{fig:deltaa-z-static}
    \end{subfigure}
    \caption{Single color component $a$ of $\delta\mathcal{A}_{i,stat}^a$ \textit{(solid)} and $\delta\mathcal{A}_{i,stat}^{f,a}$ from Lorentz force integration \textit{(dashed)}. Both are rescaled by $g^2\mu$ and studied as a function of $g^2\mu\tau$, for $n_{traj}=15$ static quark trajectories \textit{(different colors)} in temporal gauge. Figure reproduced from Ref.~\cite{Avramescu:2026fgv} under license CC BY 4.0.}
    \label{fig:deltaa-static}
\end{figure}

\begin{figure}[H]
    \centering
    \includegraphics[width=0.5\textwidth]{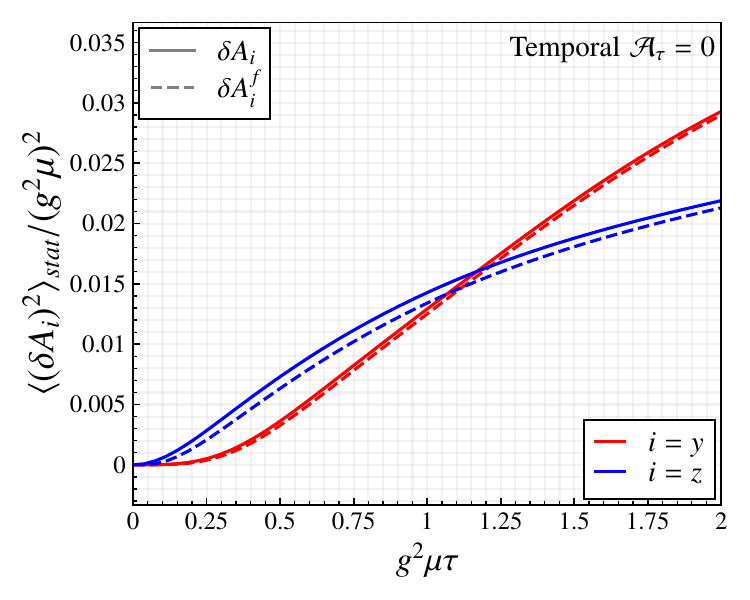}
    \caption{Averaged $\langle(\delta A_i)^2\rangle_{stat}$ from gauge field variation \textit{(solid)} and $\langle(\delta A_i^f)^2\rangle_{stat}$ \textit{(dashed)} from Lorentz force calculation, rescaled by $(g^2\mu)^2$, for the $y$ \textit{(red)} and $z$ components \textit{(blue)} of static quarks in temporal gauge. Figure reproduced from Ref.~\cite{Avramescu:2026fgv} under license CC BY 4.0.}
    \label{fig:a-static}
\end{figure}

\begin{figure}[H]
    \centering
    \begin{subfigure}[b]{0.48\textwidth}
        \centering
        \includegraphics[width=\textwidth]{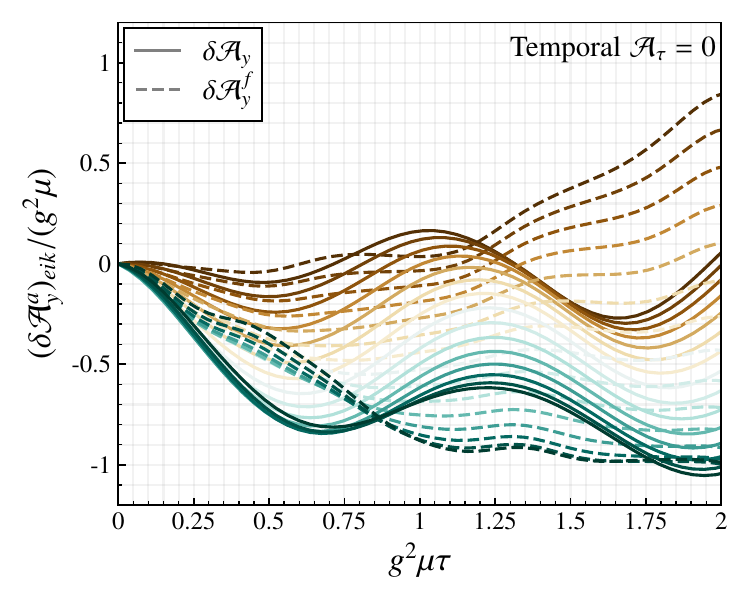}
        \caption{Temporal gauge.}
        \label{fig:deltaa-y-nc-jet}
    \end{subfigure}
    \hfill
    \begin{subfigure}[b]{0.48\textwidth}
        \centering
        \includegraphics[width=\textwidth]{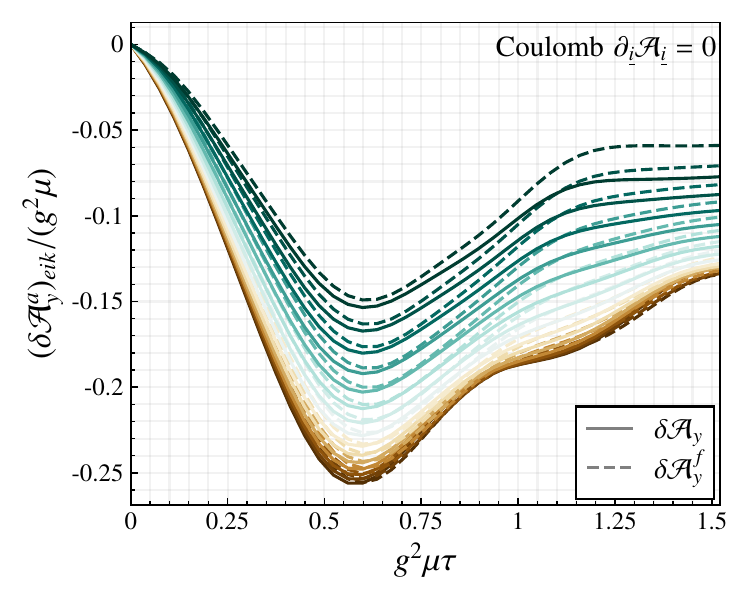}
        \caption{Coulomb gauge.}
        \label{fig:deltaa-y-c-jet}
    \end{subfigure}
    \caption{Single color component $a$ of $\delta\mathcal{A}_{y,eik}^a$ \textit{(solid)} and $\delta\mathcal{A}_{y,eik}^{f,a}$ \textit{(dashed)}, rescaled by $g^2\mu$ as a function of $g^2\mu\tau$, for $n_{traj}=15$ eikonal quark trajectories \textit{(different colors)} in temporal \textit{(left)} and Coulomb \textit{(right)} gauges. Figure reproduced from Ref.~\cite{Avramescu:2026fgv} under license CC BY 4.0.}
    \label{fig:deltaa-jet}
\end{figure}

\begin{figure}[H]
    \centering
    \begin{subfigure}[b]{0.48\textwidth}
        \centering
        \includegraphics[width=\textwidth]{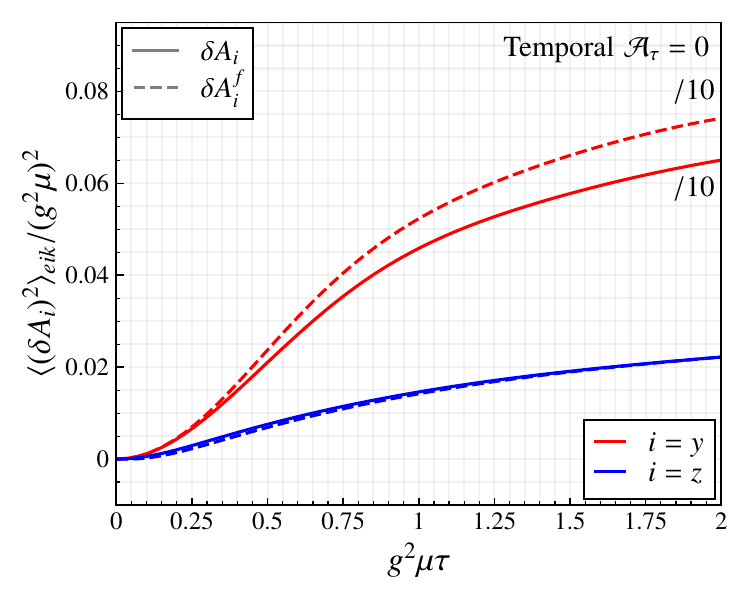}
        \caption{Temporal gauge.}
        \label{fig:a-nc-jet}
    \end{subfigure}
    \hfill
    \begin{subfigure}[b]{0.48\textwidth}
        \centering
        \includegraphics[width=\textwidth]{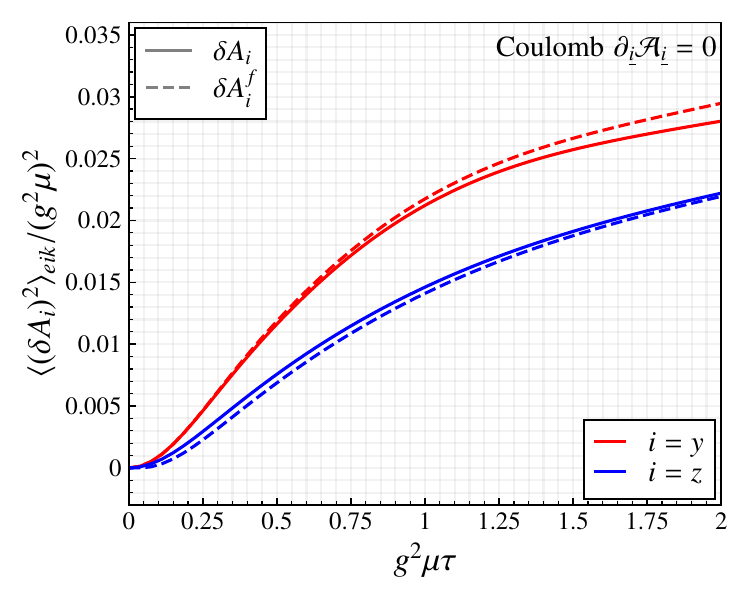}
        \caption{Coulomb gauge.}
        \label{fig:a-c-jet}
    \end{subfigure}
    \caption{Averaged $\langle(\delta A_i)^2\rangle_{eik}$ \textit{(solid)} and $\langle(\delta A_i^f)^2\rangle_{eik}$ \textit{(dashed)}, rescaled by $(g^2\mu)^2$, for the $y$ \textit{(red)} and $z$ \textit{(blue)} components of eikonal quarks in temporal \textit{(left)} and Coulomb \textit{(right)} gauge. Note that the $y$ component of the canonical momentum broadening in the temporal gauge \textit{(solid red)} is divided by $10$. Figure reproduced from Ref.~\cite{Avramescu:2026fgv} under license CC BY 4.0.}
    \label{fig:a-jet}
\end{figure}

For static quarks in the temporal gauge, the consistency relation for a single color component and distinct trajectories is shown in~\cref{fig:deltaa-static}, while the averaged quantity is shown in~\cref{fig:a-static}, for both the $y$ and $z$ components. The two methods for evaluating the gauge field evolution agree well for static quarks. The trivial color rotation $\mathcal{U}_{stat}=\mathbbm{1}$ in this case means that no path-ordered exponential is accumulated along the trajectory, and therefore numerical errors do not build up. This case confirms that the consistency check is reliable, before moving to the eikonal case. The same consistency checks yield a different outcome for eikonal quarks. The results for the single color component are shown in~\cref{fig:deltaa-jet}, while the averaged quantity is shown in~\cref{fig:a-jet}. 

For simplicity, this comparison is shown only for the $y$ component, but the qualitative behavior for the $z$ component is similar. In the temporal gauge, the two methods diverge at later times at both the single trajectory and averaged level. The $y$-component reaches values that are an order of magnitude larger than in Coulomb gauge. Imposing the Coulomb gauge condition minimizes the squared transverse gauge potential and substantially reduces $\mathcal{A}_x$ along the trajectory. This suppresses the color rotation in $\mathcal{U}_{eik}$ and thus limits the accumulation of errors. Compared to the static quark case, the Coulomb gauge fixing is specifically needed for the eikonal trajectory and obtaining accurate numerical results.


\begin{figure}[t]
    \centering
    \begin{subfigure}[b]{0.48\textwidth}
        \centering
        \includegraphics[width=\textwidth]{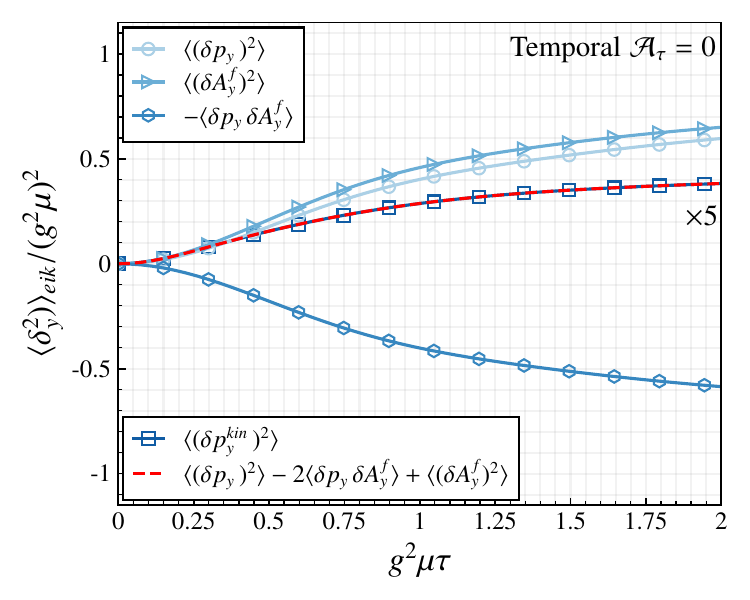}
        \caption{$y$ component.}
        \label{fig:decomp-y-nc}
    \end{subfigure}
    \hfill
    \begin{subfigure}[b]{0.48\textwidth}
        \centering
        \includegraphics[width=\textwidth]{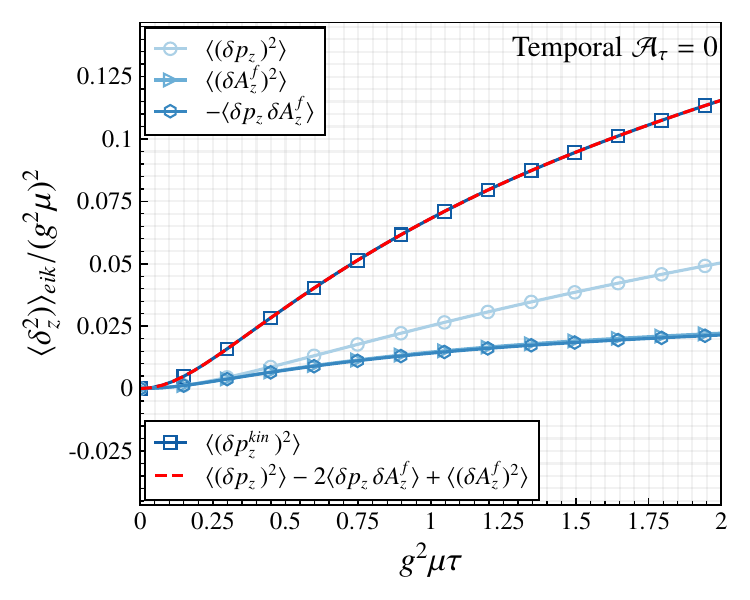}
        \caption{$z$ component.}
        \label{fig:decomp-z-nc}
    \end{subfigure}
    \caption{Kinetic momentum broadening $\langle(\delta p_i^{kin})^2\rangle$ \textit{(squares)} and its decomposition into canonical broadening $\langle(\delta p_i)^2\rangle$ \textit{(circles)}, squared gauge potential $g^2\langle(\delta A_i^f)^2\rangle$ \textit{(triangles)}, cross term $-2g\langle\delta p_i\,\delta A_i^f\rangle$ \textit{(diamonds)}, and their sum \textit{(red dashed)}. All terms are numerically evaluated using the corresponding Lorentz force. Each contribution is rescaled by $(g^2\mu)^2$, represented in terms of $g^2\mu\tau$, in the temporal gauge. Figure reproduced from Ref.~\cite{Avramescu:2026fgv} under license CC BY 4.0.}
    \label{fig:decomp-nc}
\end{figure}

\noindent\textbf{\textit{Momentum broadening decomposition}.} Squaring $g\,\delta A_i^f = \delta p_i - \delta p_i^{kin}$ and averaging over color charge configurations gives the decomposition of the momentum broadening
\begin{equation}\label{eq:KinCanMomBroadDecomp}
    \langle (\delta p_i^{kin})^2\rangle_Q=\langle (\delta p_i)^2\rangle_Q-2g\langle \delta p_i\,\delta A_i^f\rangle_Q+g^2 \langle (\delta A_i^f)^2\rangle_Q\,,
\end{equation}
where the cross terms can be evaluated in two equivalent ways. The first approach integrates the corresponding Lorentz force components
\begin{subequations}\label{eq:deltaAforceSquared}
    \begin{align}
        &\langle \delta p_i \,\delta A_i^f \rangle_Q=g\int_0^\tau \diff \tau^\prime \int_0^\tau \diff\tau^{\prime \prime}\,\tr{ \tilde{f}_p^i(\tau^\prime)\tilde{f}_A^i(\tau^{\prime\prime})}\,,\\
        &\langle (\delta A_i^f)^2\rangle_Q=\int_0^\tau \diff \tau^\prime \int_0^\tau \diff\tau^{\prime \prime}\,\tr{ \tilde{f}_A^i(\tau^\prime)\tilde{f}_A^i(\tau^{\prime\prime})}\,.
    \end{align}
\end{subequations}
The second uses the consistency relation $\delta A_i^f = \delta A_i$ to express the same terms directly through the parallel transported gauge field
\begin{subequations} \label{eq:deltaAfieldSquared}
    \begin{align}
        &\langle \delta p_i \,\delta A_i\rangle_Q=g\int_0^\tau \diff \tau^\prime \,\tr{ \tilde{f}_p^i(\tau^\prime)\,\delta \mathcal{A}_i(\tau)}\,,\\
        &\langle (\delta A_i)^2\rangle_Q=\tr{(\delta \mathcal{A}_i)^2}\,.
    \end{align}
\end{subequations}

In practice, each of the individual terms $(\delta p)^2$, $(\delta A)^2$ and $-2\delta p\,\delta A$ can be large, while their combination into $(\delta p_{kin})^2$ is small. This is achieved by careful numerical cancellations. Working in a gauge that minimizes these cancellations substantially improves the accuracy of the numerical results. This is crucial for developing a reliable extraction of the classical momentum broadening and a method which is then used for the quantum calculation.

\begin{figure}[t]
    \centering
    \begin{subfigure}[b]{0.48\textwidth}
        \centering
        \includegraphics[width=\textwidth]{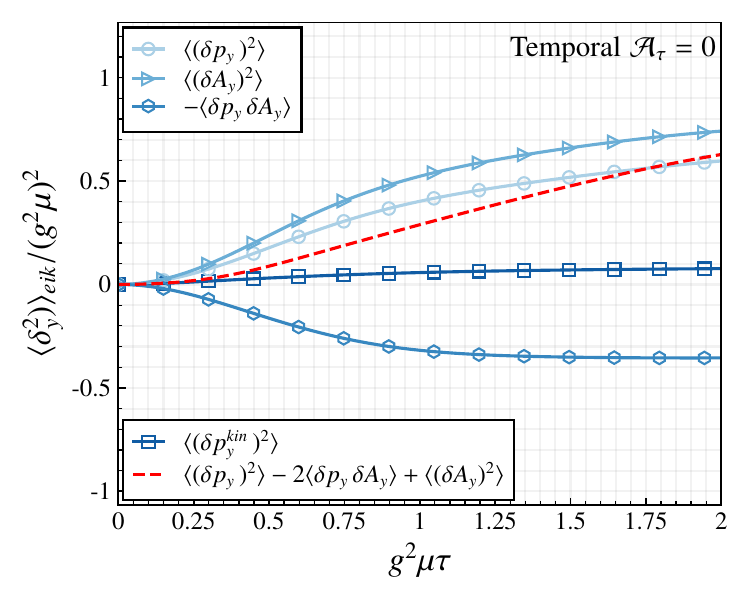}
        \caption{Temporal gauge.}
        \label{fig:decomp-atilde-nc}
    \end{subfigure}
    \hfill
    \begin{subfigure}[b]{0.48\textwidth}
        \centering
        \includegraphics[width=\textwidth]{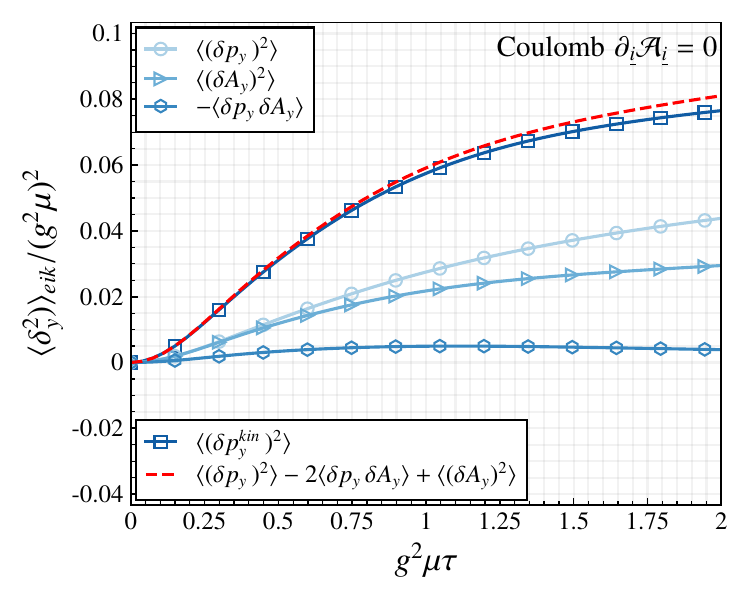}
        \caption{Coulomb gauge.}
        \label{fig:decomp-atilde-c}
    \end{subfigure}
    \caption{Same representation as in~\cref{fig:decomp-nc} for the $y$ component, but using $\delta A_y$ directly extracted from the gauge field instead of $\delta A_y^f$ evaluated from the Lorentz force, in temporal \textit{(left)} and Coulomb \textit{(right)} gauge. Figure reproduced from Ref.~\cite{Avramescu:2026fgv} under license CC BY 4.0.}
    \label{fig:decomp-atilde}
\end{figure}

Figure~\ref{fig:decomp-nc} shows the decomposition of the kinetic momentum broadening from~\cref{eq:KinCanMomBroadDecomp} extracted using the Lorentz force method in temporal gauge. The sum of the gauge-dependent terms, namely the canonical broadening, squared gauge potential and cross term, agrees well with the direct kinetic broadening in both the $y$ and $z$ directions. This provides a numerical validation of the decomposition. Notably, the individual terms in the $y$ direction are each roughly five times larger than the physical kinetic broadening, which is recovered only through large numerical cancellations. These significant contributions arise from the large $\mathcal{A}_y$ in temporal gauge, which is then reduced in magnitude in the Coulomb gauge.

The same decomposition but using the directly extracted gauge potential $\delta A_y$ for the gauge field terms, instead of $\delta A_y^f$ from Lorentz force integration, is shown in~\cref{fig:decomp-atilde}. The two methods are analytically equivalent, but numerically they differ due to accumulation of errors from the color rotation. In the temporal gauge, the decomposition fails to reproduce the kinetic broadening but the Coulomb gauge fixing restores the agreement. Notice that the magnitude of the individual terms is reduced in the Coulomb gauge. 

This result is directly relevant to the corresponding quantum calculation. In that case, the quantum eikonal probe propagating along the light-cone time direction couples to the light-front component of the background gauge field, therefore the quantum calculation accesses the gauge potential directly. This is different from the classical approach, where the kinetic broadening is evaluated from the field strength. Coulomb gauge fixing is therefore an important numerical tool which assures accuracy when bridging the classical and quantum frameworks.

\newpage
\section{Quantum transport}

\begin{emphbox}{myred}{Based on}
The main result in Article~\cite{ar5} is the application of the light-front Hamiltonian formalism to compute the real-time quantum evolution of a high energy quark in the classical glasma fields. The evolution of the state is performed using the time-dependent Basis Light-Front Quantization (tBLFQ). In this work, the Fock space expansion of the state is truncated to the single quark sector $|q\rangle$. The main finding is that the quantum momentum broadening and associated transport coefficient $\hat{q}$ are consistent with the classical results. This calculation provides the foundation to include gluon contributions from the $|qg\rangle$ sector, also incorporating subeikonal corrections.
\end{emphbox}

\subsection{Light-front Hamiltonian approach}

\begin{emphbox}{myblue}{Light-front evolution}
In light-front quantization, the quantum state $|\psi; x^+\rangle$ is described on the lightlike hypersurface at fixed $x^+$. The quark state evolution is described by the Schr\"{o}dinger equation
\begin{equation}\label{eq:LFSchrodinger}
    i\,\frac{\partial}{\partial x^+}|\psi;\, x^+\rangle = \frac{1}{2}P^-(x^+)\,|\psi;\, x^+\rangle\,,
\end{equation}
where the generator $P^-$ represents the QCD light-front Hamiltonian and $x^+$ the light-cone time. The time-dependent Hamiltonian $P^-(x^+)=P^-+V(x^+)$ includes the kinetic term of the quark $P^-$ and its interaction with the external time-dependent background field $V(x^+)$.
In practice, the state is expanded in a complete discrete basis of free eigenstates for $P^-$ and evolved in $x^+$. 
\end{emphbox}

More details about the light-cone quantization and the light-front Hamiltonian formalism are provided in~\cref{app:lightfront}. Note that here the light-cone coordinates are defined as $x^\pm = x^0\pm x^3$ \emph{(Brodsky-Lepage)}, which gives the Schr\"{o}dinger equation in~\cref{eq:LFSchrodinger} with a $1/2$ factor in front of $P^-$. This is different from the convention $x^\pm = (x^0\pm x^3)/\sqrt{2}$ \emph{(Kogut-Soper)} used in~\cref{app:lightfront}. 

\noindent\textbf{\textit{Quarks in light-front quantization}.} The quark field $\Psi$ in a classical background gluon field $\mathcal{A}_\mu = \mathcal{A}_\mu^a t^a$ is governed by the Dirac Lagrangian $\mathcal{L} = \bar{\Psi}(i\gamma^\mu \mathcal{D}_\mu - m)\Psi$, where the minimal coupling is $\partial_\mu \to \mathcal{D}_\mu = \partial_\mu + ig\mathcal{A}_\mu$. The Fock space of the quark may be expanded as $|q\rangle + |qg\rangle + \ldots$  but here it is limited to the single-quark sector $|q\rangle$, without considering gluon emissions and virtual corrections~\cite{Li:2020uhl, Li:2021zaw, Li:2023jeh, Li:2025wzq}. In the light-front formalism, the Legendre transform gives the light-front Hamiltonian 

\newpage
\begin{equation}
    \begin{aligned}\label{eq:LFHamiltonian}
        P^-(x^+) = &\int \diff x^- \diff^2\vec{x}_\perp \Bigg[
        \frac{1}{2} \bar{\Psi}\gamma^+
        \frac{m^2-\nabla_\perp^2}{i\partial^+-g \mathcal{A}^+} \Psi
        \\
        &+ g \bar{\Psi} (\gamma^+ \mathcal{A}_+ + \gamma^i \mathcal{A}_i) \Psi + \frac{g^2}{2} \bar{\Psi}\gamma^i\mathcal{A}_i
        \frac{\gamma^+}{i\partial^+-g \mathcal{A}^+}
        \gamma^j \mathcal{A}_j \Psi
        \Bigg]\,.
    \end{aligned}
\end{equation}
where the time-dependence of the Hamiltonian is implicit through the classical background field $\mathcal{A}_\mu(x^+)$. If the gluon sector were included, the gauge field would be $A_\mu+\mathcal{A}_\mu(x^+)$, with $A_\mu$ the quantized gluon field. The Hamiltonian would also contain gluon self-interactions, as well as the quark-gluon interaction terms which allows for gluon emissions and absorptions. Only the classical external field $\mathcal{A}_\mu$ is considered in this Fock truncation, and the quantized gluon field is absent. Importantly, it does not need to satisfy the light-cone gauge $\mathcal{A}^+=0$, which is used only for the quantized gluon field $A_\mu^+=0$. In the case of the glasma, the fields are obtained in the temporal gauge $\mathcal{A}_\tau=0$ and then transformed at $\tau=0$ to the transverse Coulomb gauge $\partial_{\underline{i}}\mathcal{A}_{\underline{i}}=0$ for numerical accuracy purposes.

The fermion field is expanded using creation and annihilation operators as
\begin{align}\label{eq:free_mode_expansion}
    \Psi_c(x) = \sum_{\lambda} \int \frac{\diff p^+ \diff^2\vec{p}_\perp}{\sqrt{2p^+(2\pi)^3}}
    \Big[ u_\lambda(p)\, e^{-ip\cdot x}\, b_{\lambda,c}(p) + v_\lambda(p)\, e^{ip\cdot x}\, d^\dagger_{\lambda,c}(p) \Big]\,,
\end{align}
where $p\cdot x = p^+x^-/2 - \vec{p}_\perp\cdot\vec{x}_\perp$ is the light-front spatial $3$-product, $\lambda$ the helicity, $c$ the color index, and $u_\lambda$, $v_\lambda$ the Dirac spinors. The non-vanishing anti-commutation relations at equal light-front time are $\{b_{\lambda,c}(p),\, b^\dagger_{\lambda',c'}(p')\} = \{d_{\lambda,c}(p),\, d^\dagger_{\lambda',c'}(p')\} = \delta^{(3)}(p-p')\,\delta_{\lambda,\lambda'}\,\delta_{c,c'}$. The single-quark state is created by acting with the creation operator on the vacuum as $|q\rangle = b^\dagger_{\lambda,c}(p) \lvert 0 \rangle$.

For eikonal jets with $p^+ \to \infty$, the denominators in~\cref{eq:LFHamiltonian} are expanded keeping terms of $\mathcal{O}(1/p^+)$ but dropping $\mathcal{O}(\mathcal{A}^+/(p^+)^2)$. Applying the free-field expansion~\cref{eq:free_mode_expansion} and switching to the many-body representation gives
\begin{align}\label{eq:1stQuantHamiltonian}
   P^- = \frac{m^2 + (\vec{p}_\perp - g\vec{\mathcal{A}}_\perp)^2}{p^+} + 2g\mathcal{A}_+ - 2g\frac{\mathcal{B}_x S^x}{p^+}\,,
\end{align}
where $\vec{p}^{\,kin}_\perp \equiv \vec{p}_\perp - g\vec{\mathcal{A}}_\perp$ is the transverse kinetic momentum, $\vec{\mathcal{B}} = \nabla\times\vec{\mathcal{A}}$ the chromo-magnetic field, and $S^x = \sigma^3/2$ the spin projection along the jet axis. In the strict eikonal limit, all terms suppressed by $1/p^+$ vanish, giving the expected eikonal Hamiltonian
\begin{equation}\label{eq:EikonalHamiltonian}
   \lim_{p^+\to\infty} P^- = 2g\mathcal{A}_+\,.
\end{equation}
The full Hamiltonian~\cref{eq:1stQuantHamiltonian} is used to derive quantum operators, while the eikonal Hamiltonian~\cref{eq:EikonalHamiltonian} is used to evolve the quark state. This is important for correctly evaluating the kinetic momentum operator, which contains the transverse gauge potential $\vec{\mathcal{A}}_\perp$ and thus receives contributions from the subeikonal terms in~\cref{eq:1stQuantHamiltonian}.

\newpage
\noindent\textbf{\textit{Time-dependent basis}.} The Hilbert space of the quark is spanned by the eigenstates $|\beta\rangle$ of the free kinetic Hamiltonian
\begin{align}
    P^-_\beta\,|\beta\rangle=\frac{m^2 + \vec{p}_\perp^{\,2}}{p^+}\,|\beta\rangle\,.
\end{align}
Each basis state is labeled by the longitudinal momentum $p^+$, transverse momentum $\vec{p}_\perp$, light-front helicity $\lambda$, and color index $c$. Since the background field for eikonal quarks preserves $p^+$ and $\lambda$, only the transverse momentum and color labels are dynamical, therefore $\beta \equiv \{k^y, k^z, c\}$. The quark state is expanded in this basis as
\begin{equation}\label{eq:QuarkStateExpansion}
    |\psi;\, x^+\rangle = \sum_\beta c_\beta(x^+)\,|\beta\rangle\,,
\end{equation}
where the coefficients $c_\beta(x^+) \equiv \langle\beta|\psi;\,x^+\rangle$ encompass every time dependence into a vector $\mathbf{c}(x^+)$. The evolution is thus reduced to matrix multiplications at every discretized $x^+$.

The state evolves according to~\cref{eq:LFSchrodinger}, whose solution is $|\psi;\, x^+\rangle = U(x^+;\,0)\,|\psi;\,0\rangle$ with the time-evolution operator
\begin{align}\label{eq:SchrodingerSolution}
    U(x^+_f;\,x^+_i) = \mathcal{T}_+\exp\!\left[-\frac{i}{2}\int_{x^+_i}^{x^+_f}\!\diff x^+\,P^-(x^+)\right]\,,
\end{align}
where $\mathcal{T}_+$ is the light-front time ordering. This becomes an expression for the state coefficients in the basis representation as
\begin{equation}\label{eq:CoefficientEvolution}
    \mathbf{c}(x^+) = \mathcal{T}_+\exp\!\left[-\frac{i}{2}\int_0^{x^+}\!\diff s\,\mathcal{M}(s)\right]\mathbf{c}(0)\,,
\end{equation}
where $\mathcal{M}_{\beta\beta'}(x^+) \equiv \langle\beta|P^-|\beta'\rangle$ are the light-front Hamiltonian matrix elements. Observables are evaluated as quantum expectation values
\begin{align}\label{eq:Schrodinger_expval}
    \langle\mathcal{O}(x^+)\rangle_\psi = \sum_{\beta\beta'} c^*_\beta(x^+)\,c_{\beta'}(x^+)\,\mathcal{O}_{\beta\beta'}(x^+)\,,
\end{align}
with $\mathcal{O}_{\beta\beta'} \equiv \langle\beta|\mathcal{O}|\beta'\rangle$. Alternatively, one may use the Heisenberg picture, where operators evolve as $\mathcal{O}_H(x^+) = U^\dagger(x^+;\,0) \mathcal{O}(x^+) U(x^+;\,0)$. 

\subsection{Quantum momentum broadening}

\noindent\textbf{\textit{Glasma fields on the light-cone}.} The boost-invariant glasma fields are evolved in discretized coordinates $(\tau, x, y)$, while the quark wavefunction is solved in light-cone coordinates $(x^+, y, z)$. For a high-energy quark jet propagating on an eikonal trajectory along $x^+$, the proper time is $\tau(x^+,z)=\sqrt{(x^+/2)^2-z^2}$. The exact transformation from the glasma lattice to the light-front discretization is computationally expensive since it requires storing the entire Milne proper-time history. The proper time is reduced to $\tau(x^+) \approx x(x^+) \approx x^+/2$ for eikonal jets propagating along $x^+$ near midrapidity $\eta\approx 0$. A first-order Taylor expansion in $z$ of the fields is sufficient in this situation. With the shorthand $\mathcal{A}_\mu(x^+) \equiv \mathcal{A}_\mu(\tau(x^+), x(x^+), y)$, the approximate transformation is
\begin{align}\label{eq:ApproxJetTransformation}
    &\mathcal{A}_+(x^+,z) \approx \frac{1}{2}\!\left[\mathcal{A}_x(x^+) - \frac{4z}{(x^+)^2}\,\mathcal{A}_\eta(x^+)\right], \quad \mathcal{A}_z(x^+)   \approx \frac{2}{x^+}\,\mathcal{A}_\eta(x^+)\,.
\end{align}
The only $z$ dependence enters through the $\mathcal{A}_\eta$ term of $\mathcal{A}_+$. Recall that on the glasma lattice, the transverse gauge potentials $\mathcal{A}_{\underline{i}}$ with $\underline{i} \in (x,y)$, are extracted from the gauge links via $iga_\perp \mathcal{A}_{\underline{i}} = \ln U_{\underline{i}}$. The resulting classical glasma gauge fields $\mathcal{A}_+$ and $\mathcal{A}_z$ are then used as input for the quantum evolution. 

\noindent\textbf{\textit{Direct calculation}.} The transverse kinetic momentum operator is obtained from the Heisenberg equation of motion applied to the coordinate operator, giving $\vec{p}^{\,kin}_\perp = \vec{p}_\perp - g\vec{\mathcal{A}}_\perp$, where $\vec{p}_\perp$ is the canonical momentum. Even though the full Hamiltonian is required to derive this operator, to perform the time evolution one only needs the eikonal Hamiltonian. The momentum broadening operator is defined as
\begin{align}\label{eq:DirectBroadOperatorDefiniiton}
    (\delta p^{kin}_{i}(x^+))^2 \equiv (p^{kin}_{i,H}(x^+) - p^{kin}_{i,H}(0))^2\,,
\end{align}
and subtracts the initial kinetic momentum. Expanding the square and evaluating the quantum expectation value yields the contributions
\begin{equation}
    \begin{aligned}
        \label{eq:DirectBroadOperatorExpval}
        \langle(\delta p^{kin}_i(x^+))^2\rangle_\psi =& \langle\psi;x^+|(p_i - g\mathcal{A}_i(x^+))^2|\psi;x^+\rangle + \langle\psi;0|(p_i - g\mathcal{A}_i(0))^2|\psi;0\rangle\\
        &- 2\,\mathrm{Re}\big[\langle\psi;x^+|\,p_i - g\mathcal{A}_i(x^+)\,|\tilde{\psi}_i;x^+\rangle\big]\,,
    \end{aligned}
\end{equation}
where computing the two squared terms is trivial, and the cross term requires an auxiliary state $|\tilde{\psi}_i;0\rangle \equiv p^{kin}_i(0)|\psi;0\rangle$ evolved simultaneously with the same Hamiltonian. This is the same as the color rotation used in the classical calculations. The time evolution of the state is performed using the time-evolution operator $U(x^+;0)$, which in the eikonal limit reduces to the path-ordered exponential
\begin{align}\label{eq:TimeEvolutionOperator}
    U_+(x^+;0) = \mathcal{T}_+\exp\!\left[-ig\int_0^{x^+}\!\diff s^+\,\mathcal{A}_+(s^+,y,z)\right]\,,
\end{align}
with $\mathcal{A}_+$ given by the midrapidity approximation from~\cref{eq:ApproxJetTransformation}. The final result is averaged over the quantum state and glasma field configurations $\langle\,\cdots\,\rangle \equiv \langle\langle\,\cdots\,\rangle_\psi\rangle_{\rho}$.

\noindent\textbf{\textit{Lorentz force calculation}.} A complementary method, which resembles the classical calculation, is to express the momentum broadening from the quantum Lorentz force. Starting from $\diff p^{kin}_{i,H}/\diff x^+ = g\,f_{i,H}(x^+)$, the broadening operator may be expressed as
\begin{align}\label{eq:LFTimeKineticBroadening}
    \langle(\delta p^{kin}_i(x^+))^2\rangle_\psi = g^2\int_0^{x^+}\!\diff s^+\!\int_0^{x^+}\!\diff\bar{s}^+\,\langle\psi;0|\,f_{i,H}(s^+)\,f_{i,H}(\bar{s}^+)\,|\psi;0\rangle\,.
\end{align}
By applying the Heisenberg equation of motion to the kinetic momentum and using the eikonal Hamiltonian operator, the Lorentz force operator in the Schrödinger picture may be written as $f_i(x^+) = \mathcal{F}_{i+}(x^+)$. Here, $\mathcal{F}_{\mu\nu}$ represents the classical background field strength tensor. This is the same structure as the classical non-Abelian Lorentz force. In the Heisenberg picture, the Lorentz force operator is
\begin{equation}
    f_{i,H}(x^+) = U^\dagger(x^+;0)\,\mathcal{F}_{i+}(x^+)\,U(x^+;0)\,,
\end{equation}
where $U$ performs the time evolution of the force along the jet trajectory, similar to~\cref{eq:TimeEvolutionOperator}. The expressions hold for any $(x^+, y, z)$. For a midrapidity eikonal quark at $\eta= 0$ along the $x$ direction, the Lorentz force components simplify to $f^i = \mathcal{E}_i - \epsilon_{ij} \mathcal{B}_j$ with $i\in(y,z)$. The color rotation operator reduces to $U(x^+;0) = \mathcal{T}_+\exp\!\big[-(ig/2)\int_0^{x^+}\!\diff s^+\,[\mathcal{A}_x(s^+)]_{y=z=0}\big]$. This quantum calculation is expected to correctly recover the classical momentum broadening result in the boost-invariant glasma fields.

\begin{figure}[t]
    \centering
    \includegraphics[width=0.5\textwidth]{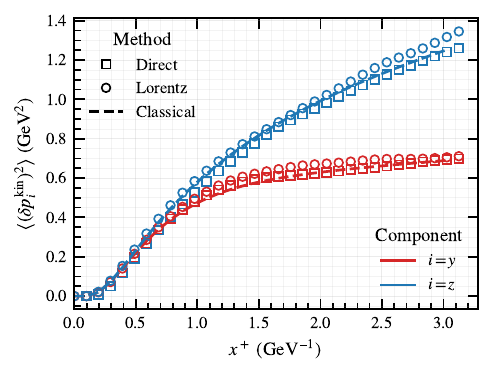}
    \caption{Kinetic momentum broadening $\langle(\delta p_i^{kin})^2\rangle$ as a function of the light-cone time $x^+$ for a well localized quark jet. The \textit{squares} correspond to the direct calculation using the gauge field, the \textit{circles} to the quantum Lorentz force calculation, and the \textit{dashed line} to the classical result. Shown in \textit{red} is the $y$ component and in \textit{blue} the $z$ component. Figure reproduced from Ref.~\cite{Avramescu:2026fgl} under license CC BY 4.0.}
    \label{fig:broadening-comparison}
\end{figure}

These distinct ways to perform the momentum broadening calculations are compared in~\cref{fig:broadening-comparison}. The direct and Lorentz force quantum results agree with each other, and with the classical results, confirming the numerical robustness of the calculations. Without the Coulomb gauge fixing, the deviations between the classical and quantum results would be significantly larger, as argued in the gauge dependence discussion. The momentum broadening is anisotropic, with the $z$ component systematically larger than the $y$ component. This is consistent with previous classical glasma studies~\cite{Ipp:2020mjc,Avramescu:2023qvv}.

\noindent\textbf{\textit{Delocalization effects}.} Let us take a quark defined by a spatially extended Gaussian wave packet with transverse widths $\sigma_y$ and $\sigma_z$, specifically
\begin{align}\label{eq:GaussianWavefunction}
    |\psi\rangle = \frac{1}{(2\pi\sigma_y^2)^{1/4}}\frac{1}{(2\pi\sigma_z^2)^{1/4}}\int\!\diff y\,\int\!\diff z\;\exp\!\left[-\frac{y^2}{4\sigma_y^2}-\frac{z^2}{4\sigma_z^2}\right]|y,z\rangle\,.
\end{align}
The averaged momentum broadening may potentially receive corrections from the finite spatial extent. Working in the Abelian limit, the classical momentum broadening becomes
\begin{align}\label{eq:FiniteSizeEffects}
    \langle(\delta p_y^{kin})^2\rangle &= \frac{g^2}{4}\int_0^{x^+}\!\diff s^+\!\int_0^{x^+}\!\diff\bar{s}^+\left\langle\!\left[f_y f_y\right]_{y=z=0} + \sigma_z^2\left[\frac{4}{(s^+)^2}\partial_y\mathcal{A}_\eta\,\frac{4}{(\bar{s}^+)^2}\partial_y\mathcal{A}_\eta\right]_{y=0}\!\right\rangle,\notag\\
    \langle(\delta p_z^{kin})^2\rangle &= \frac{g^2}{4}\int_0^{x^+}\!\diff s^+\!\int_0^{x^+}\!\diff\bar{s}^+\left\langle\!\left[f_z f_z\right]_{y=z=0}\right\rangle\,.
\end{align}
The resulting broadening in both directions is independent of $\sigma_y$ due to translational invariance after averaging over the transverse coordinates and field configurations. Broadening in $z$ is also independent of $\sigma_z$, since the linear $z$ dependence of $\mathcal{A}_+$ is removed when taking the derivative to obtain $f_z$. However, the broadening in $y$ acquires a $\sigma_z^2$ correction through $\mathcal{A}_\eta$. This shows sensitivity to the longitudinal extent of the quark wavefunction. This correction is not intrinsically quantum and it is also present within a classical framework for probes with the same Gaussian distribution. However, it becomes particularly relevant in the quantum setting where finite wave packet widths are inherent.

\begin{figure}[t]
    \centering
    \begin{subfigure}[b]{0.48\textwidth}
        \centering
        \includegraphics[width=\textwidth]{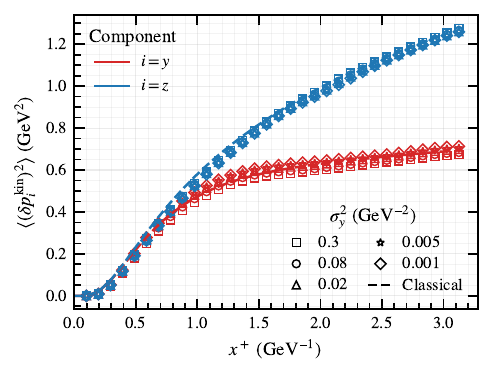}
        \caption{Varying $\sigma_y^2$ at fixed $\sigma_z^2$.}
        \label{fig:y-width}
    \end{subfigure}
    \hfill
    \begin{subfigure}[b]{0.48\textwidth}
        \centering
        \includegraphics[width=\textwidth]{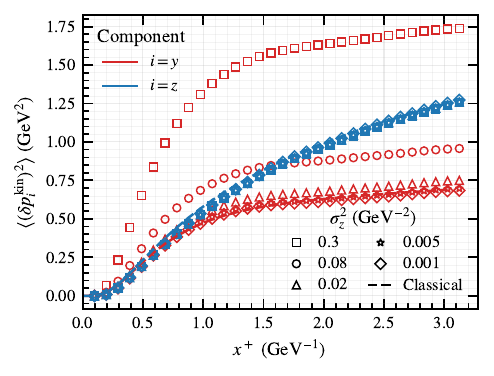}
        \caption{Varying $\sigma_z^2$ at fixed $\sigma_y^2$.}
        \label{fig:z-width}
    \end{subfigure}
    \caption{Kinetic momentum broadening $\langle(\delta p_i^{kin})^2\rangle$ as a function of the light-cone time $x^+$ for a quantum jet initialized as a Gaussian wave packet. The $y$ and $z$ components are shown in \textit{red} and \textit{blue} respectively. The \textit{dashed line} corresponds to the classical result, while the \textit{markers} show the quantum result for different values of $\sigma_{y,z}^2$ of the Gaussian wave packet. Figure reproduced from Ref.~\cite{Avramescu:2026fgl} under license CC BY 4.0.}
    \label{fig:width-dep}
\end{figure}

\cref{fig:width-dep} shows the numerical results which confirm the analytical prediction of~\cref{eq:FiniteSizeEffects}. As expected, varying $\sigma_y^2$ at fixed $\sigma_z^2$ does not change the broadening in either direction, and all values coincide with the localized classical result. In contrast, varying $\sigma_z^2$ at fixed $\sigma_y^2$ leaves the $z$ broadening unchanged, but the $y$ broadening grows substantially and departs from the classical limit. The deviation scales with $\sigma_z^2$, in agreement with the analytic estimate.  Note that these results rely on the first-order Taylor expansion of the fields near $z=0$ used in~\cref{eq:ApproxJetTransformation}, but for larger $\sigma_z$ the coordinate matching breaks down and a more comprehensive description would require the exact transformation between the Milne and light-cone coordinates. To conclude, the dependence of the momentum broadening on the width of the Gaussian wave packet is not a quantum effect, but rather a finite size effect that arises from the coordinate transformation. The expectation is that investigating the momentum broadening in the setup with the $|qg\rangle$ Fock sector included would capture genuine quantum effects.

\newpage

\begin{emphbox}{gray!55!black}{Future directions}
The quantum probes framework presented in this chapter opens several future directions~\cite{Avramescu:2026fgv,Avramescu:2026fgl}:
\begin{itemize}[label=$\circ$, leftmargin=*]
    \item \textbf{Medium-induced gluon radiation.} The tBLFQ calculation is currently truncated to the single-quark $|q\rangle$ sector, which reproduces the classical results for quark momentum broadening without radiation. Extending to the $|qg\rangle$ sector will enable explicit quantum calculations of in-medium gluon radiation and radiative energy loss in the glasma for the first time.
    \item \textbf{Connection to jet quenching.} The relation between the light-front Hamiltonian approach and the BDMPS-Z formalism for medium-induced gluon radiation has not been established, for any background field. Such a connection would bridge the real-time quantum evolution to the standard energy loss calculations used in jet quenching phenomenology.
    \item \textbf{Beyond eikonal approximation.} The light-front evolution currently gets contributions only from $V_I = 2g\mathcal{A}_+$. Going beyond the eikonal limit would incorporate contributions from the transverse gauge fields $\mathcal{A}_\perp$, in both the light-front Hamiltonian evolution and the quantum operators.  
\end{itemize}
\end{emphbox}

\chapter{Conclusion}
\label{chap:conclusion}

This thesis studied the dynamics of hard probes, heavy quarks and jets, propagating in the glasma, the pre-equilibrium classical gluon fields produced immediately after a relativistic heavy-ion collision. The results presented here establish both the classical and quantum frameworks for describing the interaction of probes with the glasma, and characterize the effect of the early-time dynamics on transport coefficients and phenomenologically relevant observables. The structure of the thesis follows:

\noindent\textbf{\textit{Glasma fields.}} The foundation of this thesis is the classical gluon field produced immediately after the collision, the glasma, derived using the Color Glass Condensate effective theory. The McLerran-Venugopalan model provides the color charge distributions of the colliding nuclei, which are combined with the pre-collision analytical gluon fields to construct the boost-invariant glasma initial conditions. These are then solved numerically on the lattice using classical Yang-Mills equations. The resulting fields consist of strong longitudinal electric and magnetic flux tubes, are characterized by the saturation momentum, and form the background medium for the propagation of hard probes studied in this thesis.

\noindent\textbf{\textit{Classical probes.}} The first work included in this thesis develops the classical transport framework for hard probes in the glasma, using the colored particle-in-cell method, Article~\cite{ar1}. Hard quarks and jets are treated as classical colored test particles obeying Wong's equations and coupled to the classical Yang-Mills fields on the lattice. The particle solver includes SU($3$) classical color charges, constructed to reproduce the correct quantum limit. The main findings obtained with this solver are that both heavy quarks and jets accumulate significant anisotropic momentum broadening during the earliest stage.

The study is extended beyond single-particle dynamics to heavy-flavor $Q\bar{Q}$ pairs, which are strongly decorrelated in azimuthal angle as they propagate through the glasma, Articles~\cite{ar2,ar3}. This decorrelation is robust for both charm and beauty quarks, and scales with $Q_s$. For $c\overline{c}$ pairs, the corresponding $D\overline{D}$ correlation is potentially sensitive to the glasma stage. In addition, the glasma imprints a characteristic shape on single-inclusive heavy-flavor momentum spectra but the effect is modest compared to the nuclear parton distribution functions.

\noindent\textbf{\textit{Quantum probes.}} The second body of work goes beyond the classical probe description. The first step is the distinction between kinetic and canonical momentum broadening in the glasma, Article~\cite{ar4}. The canonical momentum is gauge-dependent and numerically large, while the kinetic momentum is the physical gauge-invariant quantity. Imposing Coulomb gauge at the initial time is shown to substantially reduce gauge artifacts. This is needed to correctly match the classical and quantum descriptions in numerical simulations.

Article~\cite{ar5} presents the real-time quantum evolution of a high-energy quark in the classical glasma background fields. This is done using the time-dependent Basis Light-Front Quantization technique. In this work, the Fock space is restricted to the single-quark sector. The resulting quantum momentum broadening and transport coefficient agree with the classical results, providing a validation of the framework. In its most general form, the light-front approach incorporates quantum radiation and subeikonal corrections, and will enable the evaluation of medium-induced gluon radiation in the glasma.

\begin{figure}[H]
    \centering
    \includegraphics[width=0.7\textwidth]{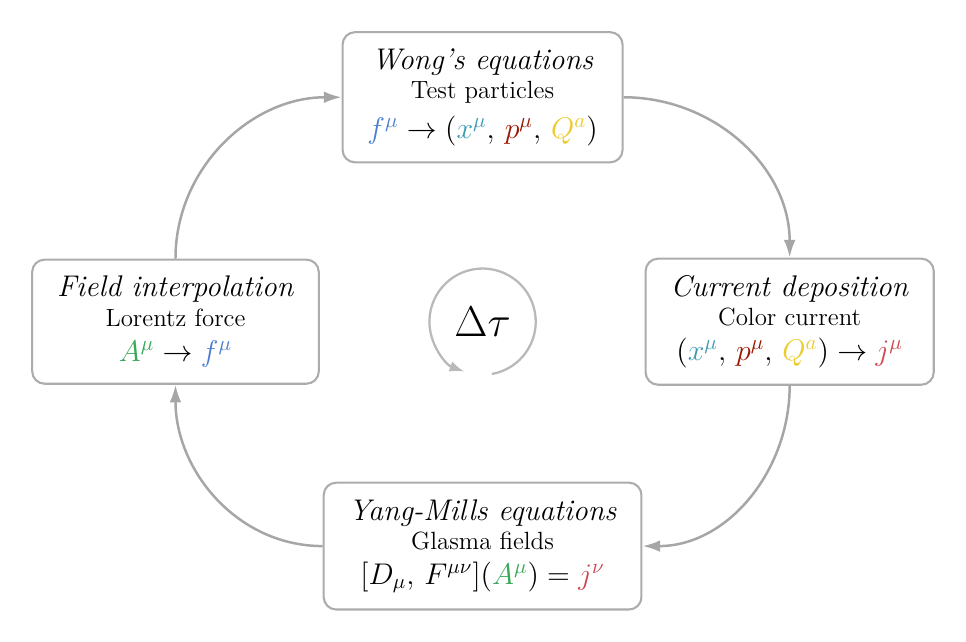}
    \caption{The CPIC simulation cycle. At each timestep, the particle trajectories and color charges are updated using Wong's equations. The resulting color currents are used as sources in the Yang-Mills equations, and the updated fields are interpolated back to the particle positions.}
    \label{fig:cpic-cycle}
\end{figure}

\newpage

\begin{emphbox}{gray!50!black}{Open questions}
\begin{itemize}[label=$\circ$, leftmargin=*]
    \item \textbf{Radiation in the glasma.} Assessing how much energy loss is experienced by hard probes in the glasma is not an easy task. Going beyond momentum broadening to include medium-induced radiation may be done using several different approaches: classical radiation in the CPIC framework, the BDMPS-Z formalism adapted for background fields with correlation domains, solving the Dirac equation, or alternatively the light-front Hamiltonian approach, for quantum evolution in the glasma.
    \item \textbf{Coupling to kinetic theory.} Hard probes need to be tracked continuously from the glasma through the EKT pre-equilibrium stage into the QGP. Matching the CPIC framework to a Boltzmann transport description in a framework that treats the particles and the fields in a self-consistent way remains an open problem. First steps towards such an approach, based on the CPIC simulation cycle, are shown schematically in Fig.~\ref{fig:cpic-cycle}.
    \item \textbf{Unified framework across stages.} The long-term goal is a single framework that simulates hard probes from production through the glasma, EKT, viscous QGP, and hadronization.
    This would enable disentangling the contribution of each stage to the experimental measurements, and potentially finding observables genuinely sensitive to the glasma. 
\end{itemize}
\end{emphbox}

%

\tailmatter

\addtocontents{toc}{\protect\newpage}
\addcontentsline{toc}{chapter}{APPENDICES}
\appendices
\numberwithin{equation}{chapter}
\crefalias{chapter}{appendix}
\makeatletter
\renewcommand\section{\@startsection{section}{1}{\z@}%
                                   {-28pt}{28pt}%
                                   {\normalfont\large\bfseries}}
\renewcommand\subsection{\@startsection{subsection}{2}{\z@}%
                                     {-14pt}{14pt}%
                                     {\normalfont\normalsize\bfseries}}
\renewcommand\subsubsection{\@startsection{subsubsection}{3}{\z@}%
                                     {-14pt}{14pt}%
                                     {\normalfont\normalsize\bfseries}}
\makeatother

\appendix{Bookkeeping}
\label{app:qcd}

\section{Basics of QCD}

\noindent\textbf{\textit{Lagrangian \& field equations}.} The QCD Lagrangian decomposes into Dirac and gluon parts~\cite{Peskin:1995ev,Schwartz:2013pla,Maggiore:2005qv,Greiner:2002ui,Tong:2018,Gelis:2019yfm}, $\mathcal{L}_\mathrm{QCD} = \mathcal{L}_\mathrm{Dirac} + \mathcal{L}_\mathrm{gluon}$. The Dirac Lagrangian for quarks in the presence of a gauge field is
\begin{equation}
    \label{eq:dirac-lagrangian}
    \mathcal{L}_\mathrm{Dirac} = \bar{\psi}(i \slashed{D}-M) \psi\,,
\end{equation}
where $M=\mathrm{diag}(m_1,\ldots,m_{N_f})$ is the quark mass matrix. The quark and anti-quark fields $\psi_{\alpha,i,f}(x)$ and $\bar{\psi}_{\alpha,i,f}(x)$ carry a color index $\alpha=1,\ldots,N_c$, a spin index $i=1,\ldots,4$, and flavor index $f=1,\ldots,N_f$.

The covariant derivative $D_\mu = \partial_\mu - i g A_\mu(x)$ acts as $[D_\mu,\chi]=\partial_\mu\chi-i g [A_\mu,\chi]$ on a field $\chi$, where $\chi = \chi^a t^a$. The gluon field $A_\mu(x)=A_\mu^a(x) t^a$ carries a color index $a=1,\ldots,N_c^2-1$, with $t^a$ the generators of the fundamental representation of $\mathrm{SU}(N_c)$. The interaction term $\mathcal{L}_\mathrm{int} = g J^{\mu,a} A_\mu^a$ allows the Dirac Lagrangian to also be written as $\mathcal{L}_\mathrm{Dirac} = \bar{\psi}(i \slashed{\partial}-M) \psi + \mathcal{L}_\mathrm{int}$, with color current $J^{\mu,a} = \bar{\psi} \gamma^\mu t^a \psi$.

The gluon Lagrangian is
\begin{equation}
    \label{eq:gluon-lagrangian}
    \mathcal{L}_\mathrm{gluon} = -\frac{1}{2}\mathrm{tr}\left[ F_{\mu\nu} F^{\mu\nu} \right]\,,\quad F_{\mu\nu} = \partial_\mu A_\nu - \partial_\nu A_\mu - i g [A_\mu, A_\nu]\,,
\end{equation}
where the field strength tensor is
equivalently expressed as $F_{\mu\nu} = i[D_\mu, D_\nu]/g$. In components, $F_{\mu\nu}=F_{\mu\nu}^a t^a$ with $F_{\mu\nu}^a=\partial_\mu A_\nu^a - \partial_\nu A_\mu^a + g f^{abc} A_\mu^b A_\nu^c$.

The fermion equations of motion obtained from the Dirac Lagrangian are
\begin{equation}
    \label{eq:fermion-eom}
    (i \overrightarrow{\slashed{D}} - M) \psi = 0\,,\quad \bar{\psi}(i \overleftarrow{\slashed{D}} - M) = 0\,.
\end{equation}
The classical Yang-Mills equations of motion~\cite{Yang:1954ek} obtained from the gluon Lagrangian are given by
\begin{equation}
    \label{eq:yang-mills-eom}
    [D_\mu, F^{\mu\nu}] = gJ^\nu\,,
\end{equation}
with $J^{\mu}=J^{\mu,a} t^a$ the color current. 

\noindent\textbf{\textit{Gauge transformation}.}
Under a local gauge transformation $G(x)=\exp\!\left[i \alpha^a(x)t^a\right]$, the fermion field transforms as $\psi(x) \to \psi'(x) = G(x) \psi(x)$ and $\bar{\psi}(x) \to \bar{\psi}'(x) = \bar{\psi}(x) G^\dagger(x)$. The gluon field transforms as
\begin{equation}
    \label{eq:gauge-field-gauge-transf}
    A_\mu(x) \to A'_\mu(x) = G(x) A_\mu(x) G^\dagger(x) + \frac{i}{g} G(x)\left[ \partial_\mu G^\dagger(x)\right]\,,
\end{equation}
and the field strength tensor as $F_{\mu\nu}(x) \to F'_{\mu\nu}(x) = G(x) F_{\mu\nu}(x) G^\dagger(x)$. A pure gauge field $A_\mu^\mathrm{pure}(x) = i G(x) \left[\partial_\mu G^\dagger(x)\right]/g$ can be gauge transformed to zero, $A^\mathrm{pure}_\mu(x) \to 0$, and has vanishing field strength $F_{\mu\nu}^\mathrm{pure} = 0$.

\section{Color algebra}
\label{app:su3}

\noindent\textbf{\textit{Representations}.}
Let $t^a$ with $a\in (1,\dots,N_c^2-1)$ be the generators of $\mathrm{SU}(N_c)$. Common choices in the fundamental (quark) representation are $t^a=\sigma^a/2$ for $\mathrm{SU}(2)$ and $t^a=\lambda^a/2$ for $\mathrm{SU}(3)$, where $\sigma^a$ are the Pauli matrices and $\lambda^a$ the Gell-Mann matrices. Regardless of representation, the generators satisfy the Lie algebra $[t^a,t^b] = if^{abc}t^c$, where $f^{abc}$ are the totally anti-symmetric structure constants. The adjoint (gluon) representation has generators $(t^a)_{bc}=-if^{abc}$. The dimensions of the fundamental and adjoint representations are $D_F = N_c$ and $D_A = N_c^2-1$.

The generators are normalized as $\tr{t^at^b}_R = T_R\,\delta^{ab}$, where the Dynkin index $T_R$ takes the values $T_F=1/2$ and $T_A=N_c$.
The quadratic and cubic Casimir operators are defined by
\begin{equation}
    t^at^a \equiv \mathbbm{1}_{D_R}\,C_2(R)\,,\quad d^{abc}t^at^bt^c \equiv \mathbbm{1}_{D_R}\,C_3(R)\,,
\end{equation}
where summation over the color indices is implied, and $d^{abc}$ are the totally symmetric structure constants, vanishing for $\mathrm{SU}(2)$. Their values are
\begin{subequations}
    \label{eq:casimirvals}
    \begin{align}
        C_2(F) &= \frac{N_c^2-1}{2N_c}\,, & C_2(A) &= N_c\,,\\
        C_3(F) &= \frac{(N_c^2-4)(N_c^2-1)}{4N_c^2}\,, & C_3(A) &= 0\,.
    \end{align}
\end{subequations}
Casimir invariants are used to label representations, since they commute with all generators and are thus constant within a representation.

\noindent\textbf{\textit{Darboux variables}.}
The classical color charges $Q^a$ form a representation of the gauge group Lie algebra obeying the Poisson bracket
\begin{equation}
    \label{eq:liepb}
    \left\{Q^a,Q^b\right\}_\mathrm{PB}=f^{abc}Q^c\,.
\end{equation}

For SU($2$), whose generators $Q^a$ with $a\in(1,2,3)$ obey Eq.~\eqref{eq:liepb} with $f^{abc}=\epsilon^{abc}$, the Darboux variables $(\phi_i,\xi_i)$ satisfy the canonical Poisson bracket $\left\{\phi_i,\xi_j\right\}_\mathrm{PB}=\delta_{ij}$. To parametrize it, a single pair $(\phi,\xi)$ suffices since the phase space evolution preserves the quadratic Casimir $Q^aQ^a=q_2$.

Distributing the color charges uniformly on a sphere of fixed radius $J^2$ satisfies both the Poisson bracket algebra and the Casimir constraint $q_2=J^2$. The SU($2$) color charges are parametrized as
\begin{equation}
    Q^1=\cos\phi\sqrt{J^{2}-\xi^{2}}\,,\quad
    Q^2=\sin\phi\sqrt{J^{2}-\xi^{2}}\,,\quad
    Q^3=\xi\,,
\end{equation}
with $\phi\in[0,2\pi)$ and $\xi\in[-J,J]$ sampled uniformly and randomly.

For SU($3$), the generators $Q^a$ with $a\in(1,\ldots,8)$ obey Eq.~\eqref{eq:liepb} with $f^{abc}$ the anti-symmetric structure constants of SU($3$). The phase space of the SU($3$) color charges is a six-dimensional submanifold of $\mathbbm{R}^8$, constrained by the quadratic Casimir $Q^aQ^a=q_2$ and the cubic Casimir $d^{abc}Q^aQ^bQ^c=q_3$.

The parametrization is constructed starting from an SU($2$) subgroup,
\begin{equation}\label{eq:darboux-su3-1}
    Q^1=\cos\phi_1\,\pi_+\pi_-\,,\quad Q^2=\sin\phi_1\,\pi_+\pi_-\,,\quad Q^3=\pi_1\,,
\end{equation}
with $\pi_\pm\equiv\sqrt{\pi_3\pm\pi_1}$, for which $(Q^1)^2+(Q^2)^2+(Q^3)^2=\pi_3^2$. Setting
\begin{equation}\label{eq:darboux-su3-2}
    Q^8=\pi_2\,,
\end{equation}
the remaining charges are chosen to satisfy Eq.~\eqref{eq:liepb} as
\begin{equation}\label{eq:darboux-su3-3}
    \begin{aligned}
        Q^4&=C_{++}\pi_+A+C_{+-}\pi_-B\,, &\quad Q^5&=S_{++}\pi_+A+S_{+-}\pi_-B\,,\\
        Q^6&=C_{-+}\pi_-A-C_{--}\pi_+B\,, &\quad Q^7&=S_{-+}\pi_-A-S_{--}\pi_+B\,,
    \end{aligned}
\end{equation}
with
\begin{equation*}
    C_{\pm\pm}=\cos\left[\frac{1}{2}\left(\pm\phi_1+\sqrt{3}\,\phi_2\pm\phi_3\right)\right]\,,\quad S_{\pm\pm}=\sin\left[\frac{1}{2}\left(\pm\phi_1+\sqrt{3}\,\phi_2\pm\phi_3\right)\right]\,,
\end{equation*}
and with integration constants
\begin{equation*}
    \begin{aligned}
        A&=\frac{1}{2\pi_3}\bigg[\left(\frac{J_1-J_2}{3}+\pi_3+\frac{\pi_2}{\sqrt{3}}\right)\left(\frac{J_1+2J_2}{3}+\pi_3+\frac{\pi_2}{\sqrt{3}}\right)\\
        &\hspace{4em}\times\left(\frac{2J_1+J_2}{3}-\pi_3-\frac{\pi_2}{\sqrt{3}}\right)\bigg]^{1/2}\,,\\
        B&=\frac{1}{2\pi_3}\bigg[\left(\frac{J_2-J_1}{3}+\pi_3-\frac{\pi_2}{\sqrt{3}}\right)\left(\frac{J_1+2J_2}{3}-\pi_3+\frac{\pi_2}{\sqrt{3}}\right)\\
        &\hspace{4em}\times\left(\frac{2J_1+J_2}{3}+\pi_3-\frac{\pi_2}{\sqrt{3}}\right)\bigg]^{1/2}\,,
    \end{aligned}
\end{equation*}
where $(J_1,J_2)$ label the representation. The Casimirs are then given by
\begin{equation}\label{eq:darboux-casimirs}
    q_2=\frac{1}{3}(J_1^2+J_1J_2+J_2^2)\,,\quad q_3=\frac{1}{18}(J_1-J_2)(J_1+2J_2)(2J_1+J_2)\,.
\end{equation}
Equivalently, $A$ and $B$ can be fixed directly by $(q_2,q_3)$, bypassing $(J_1,J_2)$
\begin{equation*}
    \begin{aligned}
        A&=\frac{1}{6\pi_3}\left[-\sqrt{3}\left(\pi_2+\sqrt{3}\pi_3\right)^3+3\sqrt{3}\left(\pi_2+\sqrt{3}\pi_3\right)q_2+6q_3\right]^{1/2}\,,\\
        B&=\frac{1}{6\pi_3}\left[\sqrt{3}\left(\pi_2-\sqrt{3}\pi_3\right)^3-3\sqrt{3}\left(\pi_2-\sqrt{3}\pi_3\right)q_2-6q_3\right]^{1/2}\,.
    \end{aligned}
\end{equation*}
The parametrization in Eqs.~\eqref{eq:darboux-su3-1}--\eqref{eq:darboux-su3-3} satisfies Eq.~\eqref{eq:liepb} for arbitrary $A$ and $B$, and the Casimir constraints fix their values for a given representation.

\section{Coordinates}
\label{app:milne}

\begin{figure}[t]
    \centering
    \includegraphics[width=0.7\textwidth]{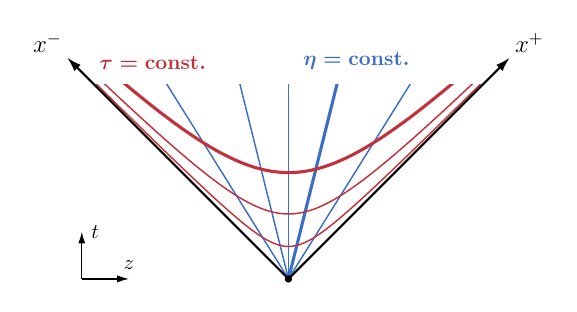}
    \caption{Milne coordinates $(\tau,\eta)$ in the $(t,z)$ plane. Lines of constant proper time $\tau=\sqrt{t^2-z^2}$ are hyperbolae \emph{(red)}, and lines of constant spacetime rapidity $\eta=\ln[(t+z)/(t-z)]/2$ are straight rays from the origin \emph{(blue)}.}
    \label{fig:tau-eta-coords}
\end{figure}

\noindent\textbf{\textit{Milne coordinates}.}
The Milne coordinates $(\tau,\eta)$, illustrated in~\cref{fig:tau-eta-coords}, are defined in terms of the laboratory coordinates $(t,z)$ by
\begin{equation}
    \label{eq:milne-coordinates}
    \tau = \sqrt{t^2 - z^2}\,, \quad \eta = \frac{1}{2} \ln\left(\frac{t+z}{t-z}\right)\,,
\end{equation}
where $\tau$ is the proper time and $\eta$ the space-time rapidity. The Milne metric is $g_{\mu\nu}^M=\mathrm{diag}(1,-1,-1,-\tau^2)$ and $g^{\mu\nu}_M=\mathrm{diag}(1,-1,-1,-1/\tau^2)$. The infinitesimal volume element is $\diff^4 x = \tau \diff\tau\diff\eta\diff^2 \vec{x}_\perp$, with $\sqrt{-\mathrm{det}\, g_M} = \tau$.

The inverse transformations are $t=\tau\cosh\eta$ and $z=\tau\sinh\eta$. The non-vanishing Christoffel symbols for the Milne metric are
\begin{align}
    \Gamma^\tau_{\eta\eta}=\tau\,, \quad \Gamma^\eta_{\tau\eta}=\Gamma^\eta_{\eta\tau}=\frac{1}{\tau}\,.
\end{align}
These are used to express the covariant derivatives of any vector in the Milne coordinates. Using $v^\tau\diff/\diff\tau = m\diff/\diff \lambda$ to express proper time derivatives in terms of Milne proper time $\tau$, where $\lambda$ is the particle proper time, the covariant momentum derivative decomposes as 
\begin{equation}
    \label{eq:covder-milne}
        \frac{\mathrm{D} v_\tau}{\diff\tau} = \frac{\diff v^\tau}{\diff\tau} + \frac{\tau(v^\eta)^2}{v^\tau}\,,\quad
        \frac{\mathrm{D} v_i}{\diff\tau} = -\frac{\diff v^i}{\diff\tau}\,,\quad
        \frac{\mathrm{D} v_\eta}{\diff\tau} = -\tau^2\frac{\diff v^\eta}{\diff\tau} - 2\tau v^\eta\,,
\end{equation}
for $i\in(x,y)$.

The Yang-Mills action in Milne coordinates is
\begin{equation}
    \label{eq:ym-action-milne-app}
    S = \int \diff\tau\, \diff^2\vec{x}_\perp\, \diff\eta\, \mathrm{tr}\left[\tau F_{\tau i} F_{\tau i}+ \frac{1}{\tau} F_{\tau\eta} F_{\tau\eta}- \frac{\tau}{2} F_{ij} F_{ij}- \frac{1}{\tau} F_{\eta i} F_{\eta i}\right]\,.
\end{equation}
The conjugate field momenta are
\begin{equation}
    P^\eta \equiv \frac{\delta S}{\delta(\partial_\tau A_\eta)} = \frac{1}{\tau} F_{\tau\eta}\,, \quad
P^i \equiv \frac{\delta S}{\delta(\partial_\tau A_i)} = \tau F_{\tau i}\,.
\end{equation}
We denote $\delta_\mu X\equiv \delta X/\delta A_\mu$. The variation of the field strength tensor is
\begin{equation}
    \label{eq:field-strength-variation}
    \delta_\mu F_{\mu\nu}=-[D_\nu, \delta A_\mu]\,.
\end{equation}
Varying the action up to surface terms gives
\begin{equation}
    \label{eq:action-variation}
    \delta_\mu S = 2\int \diff\tau\, \diff^2\vec{x}_\perp\, \diff\eta\, \tau\,\mathrm{tr}\big[ [D_\nu,F^{\mu\nu}] \delta A_\mu \big]\,.
\end{equation}
The classical Yang-Mills equations of motion in Milne coordinates are $[D_\nu,F^{\mu\nu}] = 0$, as expected.
The field equation for $\mu=\tau$ is the Gauss law constraint $[D_i, P^i] + [D_\eta, P^\eta] = 0$,
or equivalently $\tau [D_i, F_{\tau i}]+[D_\eta, F_{\tau\eta}]/\tau = 0$.

The field strength tensor $F_{\mu\nu}$ contains electric and magnetic fields. In Minkowski coordinates, these are
\begin{equation}
    E_i\equiv F_{ti}\,, \quad B_i\equiv -\frac{1}{2}\varepsilon_{ijk}F_{jk}\,,
\end{equation}
with $\varepsilon_{ijk}$ the Levi-Civita symbol, or in matrix form
\begin{equation}
    F_{\mu\nu}=\begin{pmatrix}
    0 & E_x & E_y & E_z\\
    -E_x & 0 & -B_z & B_y \\
    -E_y & B_z & 0 & - B_x \\
    -E_z & -B_y & B_x & 0
    \end{pmatrix}\,.
\end{equation}
In Milne coordinates, this becomes
\begin{equation}
    F^M_{\mu\nu}=\begin{pmatrix}
    0 & E^M_x & E^M_y & \tau E^M_\eta\\
    -E^M_x & 0 & -B^M_\eta & \tau B^M_y \\
    -E^M_y & B^M_\eta & 0 & - \tau B^M_x \\
    -\tau E^M_\eta & -\tau B^M_y & \tau B^M_x & 0
    \end{pmatrix}\,.
\end{equation}
The $\tau$ factors ensure correct physical dimensions. The contravariant tensors $F^{\mu\nu}$ and $F^{\mu\nu}_M$ are obtained by raising indices with the Minkowski and Milne metrics, respectively.

Applying the tensor transformation $F^{\mu\nu}_M=(\partial x_M^\mu/\partial x^\alpha)(\partial x_M^\nu/\partial x^\beta)F^{\alpha\beta}$ with the non-vanishing Jacobian components $\partial\tau/\partial t= \cosh\eta$, $\partial\tau/\partial z = -\sinh\eta$, $\partial\eta/\partial t = -\sinh\eta/\tau$, and $\partial\eta/\partial z = \cosh\eta/\tau$ gives the field transformation relations
\begin{equation}
    \begin{alignedat}{2}
        E^M_i &= \cosh\eta\, E_i - \sinh\eta\,\epsilon_{ij}B_j\,, &\qquad E^M_\eta &= E_z\,,\\
        B^M_i &= \cosh\eta\, B_i + \sinh\eta\,\epsilon_{ij}E_j\,, &\qquad B^M_\eta &= B_z\,.
    \end{alignedat}
\end{equation}

\phantomsection\label{app:light-cone-coords}
\noindent\textbf{\textit{Light-cone coordinates}.}
Minkowski coordinates $(x^0,x^1,x^2,x^3)$ map to light-cone (LC) coordinates $(x^+,x^1,x^2,x^-)$ via
\begin{equation}
    \label{eq:light-cone-coordinates}
    x^\pm \equiv \frac{1}{\sqrt{2}}\left(x^0 \pm x^3\right)\,,
\end{equation}
with $\vec{x}_\perp \equiv (x^1, x^2)$. The inverse relations are $x^0=(x^++x^-)/\sqrt{2}$ and $x^3=(x^+-x^-)/\sqrt{2}$.

The Minkowski metric $g_{\mu\nu}=\mathrm{diag}(1,-1,-1,-1)$ becomes off-diagonal under this change of variables,
\begin{equation}
    \label{eq:lc-metric}
    g_{\mu\nu}^{LC} = g^{\mu\nu}_{LC} =
    \begin{pmatrix}
        0 & 0  & 0  & 1 \\
        0 & -1 & 0  & 0 \\
        0 & 0  & -1 & 0 \\
        1 & 0  & 0  & 0
    \end{pmatrix}\,,
\end{equation}
where the non-zero components are $g_{LC}^{+-}=g_{LC}^{-+}=1$ and $g_{LC}^{11}=g_{LC}^{22}=-1$. The scalar product becomes
\begin{equation}
    \label{eq:lc-scalar-product}
    x \cdot y = x^+y^- + x^-y^+ - \vec{x}_\perp\cdot\vec{y}_\perp\,.
\end{equation}
The volume element has unit Jacobian, that is $\diff^4 x = \diff x^+\,\diff x^-\,\diff^2\vec{x}_\perp$. With $\partial_\alpha \equiv \partial/\partial x^\alpha$ and $\partial^\alpha \equiv \partial/\partial x_\alpha$ for $\alpha=(+,-,\perp)$, the partial derivatives satisfy
\begin{equation}
    \label{eq:lc-derivatives}
    \partial_+ = \partial^-\,, \qquad
    \partial_- = \partial^+\,, \qquad
    \partial_i = -\partial^i\,.
\end{equation}

\noindent\textbf{\textit{Transverse light-cone coordinates}.}
Transverse light-cone (TLC) coordinates $(\tilde{x}^+,\tilde{x}^-,y,z)$ are defined by
\begin{equation}
    \label{eq:tlc-coordinates}
    \tilde{x}^\pm \equiv C\left(t \pm x\right)\,,
\end{equation}
with $\tilde{x}^+$ the TLC time. The inverse relations are $t=(\tilde{x}^++\tilde{x}^-)/(2C)$ and $x=(\tilde{x}^+-\tilde{x}^-)/(2C)$. We choose $C=1$ for simplicity, but the relations scale accordingly if $C=1/\sqrt{2}$ is used instead.

The relations between all coordinate systems are collected in Table~\ref{tab:coords}.

\begin{table}[!ht]
\centering
{\renewcommand{\arraystretch}{2.5}
\resizebox{\textwidth}{!}{%
\begin{tabular}{lcccc}
\toprule
 & \textit{Laboratory (L)} & \textit{Light-cone (LC)} & \textit{Milne (M)} & \makecell[c]{\textit{Transverse}\\\textit{light-cone (TLC)}} \\
\midrule
\textit{Notation}
    & $(t,x,y,z)$
    & $(x^+,x^-,x,y)$
    & $(\tau,x,y,\eta)$
    & $(\tilde{x}^+,\tilde{x}^-,y,z)$ \\
\midrule
\makecell[l]{\textit{Definition}\\\textit{(in $L$)}}
    &
    & $x^\pm=C(t\pm z)$
    & \shortstack{$\tau=\sqrt{t^2-z^2}$\\[6pt]$\eta=\dfrac{1}{2}\ln\dfrac{t+z}{t-z}$}
    & $\tilde{x}^\pm=C(t\pm x)$ \\
\midrule
\makecell[l]{\textit{Relation}\\\textit{(to LC)}}
    & \shortstack{$t=\dfrac{x^++x^-}{2C}$\\[6pt]$z=\dfrac{x^+-x^-}{2C}$}
    &
    & \shortstack{$\tau=\dfrac{1}{C}\sqrt{x^+x^-}$\\[6pt]$\eta=\dfrac{1}{2}\ln\dfrac{x^+}{x^-}$}
    & $\tilde{x}^\pm=\dfrac{x^++x^-}{2}\pm Cx$ \\
\midrule
\makecell[l]{\textit{Relation}\\\textit{(to M)}}
    & \shortstack{$t=\tau\cosh\eta$\\[6pt]$z=\tau\sinh\eta$}
    & $x^\pm=C\tau e^{\pm\eta}$
    &
    & $\tilde{x}^\pm=C(\tau\cosh\eta\pm x)$ \\
\midrule
\makecell[l]{\textit{Relation}\\\textit{(to TLC)}}
    & \shortstack{$t=\dfrac{\tilde{x}^++\tilde{x}^-}{2C}$\\[6pt]$x=\dfrac{\tilde{x}^+-\tilde{x}^-}{2C}$}
    & $x^\pm=\dfrac{\tilde{x}^++\tilde{x}^-}{2}\pm Cz$
    & \shortstack{$\tau=\sqrt{\left(\dfrac{\tilde{x}^++\tilde{x}^-}{2C}\right)^{\!2}-z^2}$\\[8pt]$\eta=\dfrac{1}{2}\ln\dfrac{\tilde{x}^++\tilde{x}^-+2Cz}{\tilde{x}^++\tilde{x}^--2Cz}$}
    & \\
\bottomrule
\end{tabular}
}%
}
\caption{Summary of coordinate systems and their mutual relations. The constant $C$ equals $1/\sqrt{2}$ in the standard convention \emph{(Kogut-Soper)}~\cite{Kogut:1969xa}, used throughout this thesis, or $1$ in the other \emph{(Brodsky-Lepage)}, used for the light-front Hamiltonian evolution in~\cref{chap:quantum}.}
\label{tab:coords}
\end{table}

\section{Lattice gauge theory}

\noindent\textbf{\textit{Wilson lines \& loops}.}
The Wilson line in the fundamental representation is
\begin{equation}
    \label{eq:wilson-line-def}
    U(x,y) = P \exp\left[ i g \int_y^x \mathrm{d} z^\mu A_\mu(z) \right]\,,
\end{equation}
with $P$ the path-ordering operator and $A_\mu(z) = A_\mu^a(z) t^a$ the gauge field in the fundamental representation.

An equivalent representation of the Wilson line is the ordered series
\begin{equation}
    U(x,y) = \mathbbm{1} + \sum_{n=1}^\infty \frac{(ig)^n}{n!} \int_0^1 \mathrm{d}\lambda_1 \cdots \int_0^1 \mathrm{d}\lambda_n\; P\left[A(\lambda_1)\cdots A(\lambda_n)\right]\,.
\end{equation}

Expanding the Wilson line, the color structure of the first two terms is
\begin{equation}
    \begin{aligned}
        U(x,y) =&\, \mathbbm{1} + ig \int_0^1 \mathrm{d}\lambda\,\frac{\diff z^\mu_\lambda}{\diff\lambda} A^a_\mu(z_\lambda) t^a\,  - \frac{1}{2}g^2 \int_0^1 \mathrm{d}\lambda \int_0^1 \mathrm{d}\tau\,\frac{\diff z^\mu_\lambda}{\diff\lambda} \frac{\diff z^\nu_\tau}{\diff\tau}  \\
        &\times A^a_\mu(z_\lambda) A^b_\nu(z_\tau)\left[ t^a t^b \theta(\lambda - \tau) + t^b t^a \theta(\tau - \lambda) \right]+ \ldots\,,
    \end{aligned}
\end{equation}
where $z(0)=y$ and $z(1)=x$ for a path parametrized as $z_\lambda\equiv z(\lambda)$ with $\lambda\in[0,1]$ and similarly for $z_\tau$. Defining the shorthand $A(\lambda) \equiv (\diff z^\mu_\lambda/\diff\lambda) A^a_\mu(z_\lambda) t^a$, the path-ordering operator $P$ orders products of fields according to the path parameter, with the convention that \textit{left means later}. Its action on a product of two fields is
\begin{equation}
    P\left[A(\lambda)\,A(\tau)\right] \equiv \theta(\lambda-\tau)\,A(\lambda)\,A(\tau) + \theta(\tau-\lambda)\,A(\tau)\,A(\lambda)\,.
\end{equation}

Parallel transport of a field $\chi$ from $y$ to $x$ is given by $\chi(x) = U(x,y) \chi(y)$, and the covariant derivative can be expressed as
\begin{equation}
    D_\mu \chi(x) = \lim_{\epsilon \to 0} \frac{1}{\epsilon}
    \left[ U(x, x+\epsilon \hat{\mu}) \chi(x+\epsilon \hat{\mu}) - \chi(x) \right]\,,
\end{equation}
with $\hat{\mu}$ a unit vector in the $\hat{\mu}$ direction.

The Wilson line satisfies the differential equation $\partial_\mu U(x,y) = i g A_\mu(x) U(x,y)$, which can be verified by differentiating the path-ordered exponential and using the definition of $P$.

In the adjoint representation the Wilson line is
\begin{equation}
    \label{eq:wilson-line-adjoint}
    U_A(x,y) = P \exp\left[ i g \int_y^x \mathrm{d} z^\mu A_\mu^a(z) T^a \right]\,,
\end{equation}
with $T^a$ the generators of the adjoint representation of $\mathrm{SU}(N_c)$. It is related to the Wilson line in the fundamental representation by
\begin{equation}
    \label{eq:wilson-line-adjoint-fundamental}
    U_A^{ab} = 2\, \mathrm{tr}\left[ t^a U t^b U^\dagger \right]\,,
\end{equation}
or equivalently $U_A^{ab}\, t^b = U t^a U^\dagger$.

The Wilson loop along a closed path $\gamma$ is
\begin{equation}
    \label{eq:wilson-loop}
    U(\gamma) = P\,\mathrm{exp}\left[ i g \oint_\gamma \mathrm{d} z^\mu A_\mu(z)  \right]\,,
\end{equation}
and it is gauge invariant. One may use Stokes' theorem to express it in terms of the field strength tensor. For an Abelian gauge field, the classical Stokes' theorem relates the line integral of the connection, that is the gauge potential $A_\mu$, around $\gamma = \partial\Sigma$, to the flux of the curvature $F_{\mu\nu}$ through $\Sigma$
\begin{equation}
    \label{eq:abelian-stokes}
    \oint_{\gamma} A_\mu\, \mathrm{d}z^\mu
    = \frac{1}{2}\int_\Sigma F_{\mu\nu}\, \mathrm{d}z^\mu \wedge \mathrm{d}z^\nu \,,
\end{equation}
where $\mathrm{d}z^\mu \wedge \mathrm{d}z^\nu$ is an oriented area element on $\Sigma$.

For a non-Abelian gauge field, a direct replacement $F_{\mu\nu} \to F_{\mu\nu}^a t^a$ fails because the curvature at different surface points does not commute. In that case, the integrand is not well defined as a product. The non-Abelian Stokes theorem resolves this by parallel transporting $F_{\mu\nu}$ with $U(x, x_0) = P\exp[ig\int_{x_0}^x \mathrm{d}z^\mu A_\mu(z)]$ from a fixed base point $x_0 \in \gamma$ to the surface point $x$, giving the transported curvature $\mathcal{F}_{\mu\nu}(x) \equiv U^{-1}(x,x_0)\,F_{\mu\nu}(x)\,U(x,x_0)$. This brings the curvature at every point back to the fiber over $x_0$, and the theorem reads~\cite{Broda:2001}
\begin{equation}
    \label{eq:stokes-theorem}
    P\exp\!\left[ ig \oint_{\gamma} \mathrm{d}z^\mu A_\mu(z) \right]
    = \mathcal{P}\exp\!\left[ \frac{ig}{2} \int_\Sigma \mathcal{F}_{\mu\nu}(x)\, \mathrm{d}z^\mu \wedge \mathrm{d}z^\nu \right]\,,
\end{equation}
where $\mathcal{P}$ is now surface ordering, the two-parameter analogue of path ordering. Taking the trace of both sides gives a gauge-invariant relation between the Wilson loop and the field strength.

\noindent\textbf{\textit{Lattice discretization}.}
\label{app:lattice}
Minkowski spacetime is discretized on a lattice with spacing $a^\mu$ and lattice sites labeled by $n_\mu \in \mathds{Z}$. Each lattice point is given by $x = \sum_{\mu} n_\mu \hat{a}^\mu$ with $\hat{a}^\mu=a^\mu\hat{\mu}$, where $\hat{\mu}$ is a unit vector in the $\mu$ direction.

\begin{figure}[t]
    \centering
    \begin{subfigure}[b]{0.48\textwidth}
        \centering
        \includegraphics[width=0.8\textwidth]{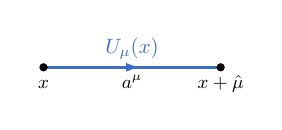}
        \captionsetup{justification=centering}
        \caption{Gauge link $U_{\boldsymbol{x},i}$}
        \label{fig:gauge-link}
    \end{subfigure}
    \hfill
    \begin{subfigure}[b]{0.48\textwidth}
        \centering
        \includegraphics[width=\textwidth]{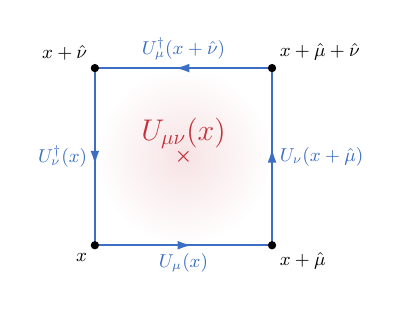}
        \captionsetup{justification=centering}
        \caption{Plaquette $U_{\boldsymbol{x},ij}$}
        \label{fig:plaquette}
    \end{subfigure}
    \caption{Lattice degrees of freedom on the transverse lattice, namely the gauge link $U_{\boldsymbol{x},i}$, between neighboring sites with spacing $a_\perp$ and the plaquette $U_{\boldsymbol{x},ij}$, the minimal Wilson loop around an elementary lattice square.}
    \label{fig:lattice}
\end{figure}

The gauge link $U_{\mu}(x)\equiv U(x+\hat{\mu},x)$, illustrated in~\cref{fig:gauge-link}, is the Wilson line along a lattice direction $\mu$ connecting two neighboring lattice sites $x$ and $x+\hat{\mu}$,
\begin{equation}
    \label{eq:gauge-link}
    \begin{aligned}
        U_{\mu}(x) &= P \exp\left[ i g \int_{x}^{x+\hat{\mu}} \mathrm{d} z^\mu A_\mu(z)  \right]\\
        &\approx \exp\left[ i g a^\mu A_\mu\left(x+\frac{a^\mu}{2}\hat{\mu}\right) \right]+\mathcal{O}(a^2)\,.
    \end{aligned}
\end{equation}
A gauge link is evaluated using the gauge field at the midpoint between the two lattice sites it connects. Under a gauge transformation, the link transforms as $U_\mu(x)\rightarrow U_\mu^\prime(x)=G(x+\hat{\mu})U_\mu(x)G^\dagger(x)$.

The plaquette $U_{\mu\nu}(x)\equiv U_\gamma$, illustrated in~\cref{fig:plaquette}, is the Wilson loop where the path $\gamma$ goes around an elementary square in the $(\mu,\nu)$ plane with one corner at $x$,
\begin{equation}
    \label{eq:plaquette}
    \begin{aligned}
        U_{\mu\nu}(x) &= U_\mu(x) U_\nu(x+\hat{\mu}) U_\mu^\dagger(x+\hat{\nu}) U_\nu^\dagger(x)\\
        & \approx \exp\left[ i g a^\mu a^\nu F_{\mu\nu}\left(x+\frac{a^\mu}{2}\hat{\mu}+\frac{a^\nu}{2}\hat{\nu}\right) \right]+\mathcal{O}(a^3)\,.
    \end{aligned}
\end{equation}
A plaquette is evaluated using the field strength at the center of the square it spans. This relation is obtained by expanding in powers of the lattice spacing the gauge link $U_\mu \approx \mathbbm{1} + i g a^\mu A_\mu - g^2 a^\mu a^\nu [A_\mu, A_\nu]/2 + \mathcal{O}(a^3)$ and the gauge field $A_\nu(x+\hat{\mu})\approx A_\nu(x) + a^\mu \partial_\mu A_\nu(x)+\mathcal{O}(a^2)$, and using the Baker-Campbell-Hausdorff formula $\exp[A]\exp[B]\approx \exp[A+B+[A,B]/2+\ldots]$. Under a gauge transformation, the plaquette transforms as $U_{\mu\nu}(x)\rightarrow U_{\mu\nu}^\prime(x)=G(x)U_{\mu\nu}(x)G^\dagger(x)$.

We denote by $U_\mu^\dagger(x)\equiv U_{-\mu}(x+\hat{\mu})$ the Wilson line in the negative $\mu$ direction. The identity $U_\mu(x)U_{-\mu}(x+\hat{\mu})=\mathbbm{1}$ follows directly.

Using the approximation $\mathrm{tr}\left[\mathbbm{2}-U_{\mu\nu}(x)-U_{\mu\nu}^\dagger(x)\right]\approx (ga^\mu a^\nu)^2\mathrm{tr}\left[F_{\mu\nu}^2(x)\right]+\mathcal{O}(a^6)$, the Yang-Mills action on the lattice becomes
\begin{equation}
    \label{eq:ym-action-lattice}
    S_\mathrm{YM} = -V \sum_{x} \sum_{\mu\nu} \frac{1}{2(g a^\mu a^\nu)^2} \mathrm{tr}\left[\mathbbm{2} - U_{\mu\nu}(x) - U_{\mu\nu}^\dagger(x) \right]\,,
\end{equation}
where $V=\prod_\mu a^\mu$ is the volume of an elementary lattice cell. Gauge invariance of the action follows from the gauge transformation of a plaquette.

This allows us to formulate a lattice gauge invariant action in Minkowski space. One starts by splitting the action into its electric and magnetic part $S_\mathrm{YM} = S_E - S_B$ with
\begin{align}
    S_E = \int \diff^4x \sum_i \mathrm{tr}\left[F_{0i}(x)^2\right]\,, \quad S_B = \int \diff^4x \sum_{ij} \frac{1}{2}\mathrm{tr}\left[F_{ij}(x)^2\right]\,,
\end{align}
where $i,j \in (1,2,3)$. One further uses the gauge invariant combination of plaquettes from Eq.~\eqref{eq:plaquette} to express
\begin{align}
    S_E &= V\sum_x\sum_i \frac{1}{(ga^0a^i)^2}\mathrm{tr}\left[\mathbbm{2} - U_{0i}(x) - U_{0i}^\dagger(x)\right]\,, \\
    S_B & = V\sum_x\sum_{i,j} \frac{1}{2(ga^ia^j)^2}\mathrm{tr}\left[\mathbbm{2} - U_{ij}(x) - U_{ij}^\dagger(x)\right]\,.
\end{align}
These expressions are manifestly gauge invariant and reduce to the continuum action in the limit of vanishing lattice spacing.

\appendix{Light-front Quantization}
\label{app:lightfront}

\section{Conventions}

\noindent\textbf{\textit{Instant and light-front forms}.}
There are three forms of relativistic dynamics: the instant form, the point form, and the light-front form~\cite{Dirac:1949cp,Dirac:1950pj,Susskind:1967rg,Heinzl:2000ht,Burkardt:1995ct}. They differ in the choice of initial hypersurface on which initial conditions are imposed. The starting point is that a relativistic quantum system must possess invariance under the Poincaré group and have a Hamiltonian formulation. Different choices of initial hypersurface yield different forms of the dynamics, and these choices are independent, meaning there is no Lorentz transformation that maps one into another. Each form has the property that a subgroup of the Poincaré generators leaves the hypersurface invariant.

\begin{figure}[t]
    \centering
    \begin{subfigure}[b]{0.48\textwidth}
        \centering
        \includegraphics[width=\textwidth]{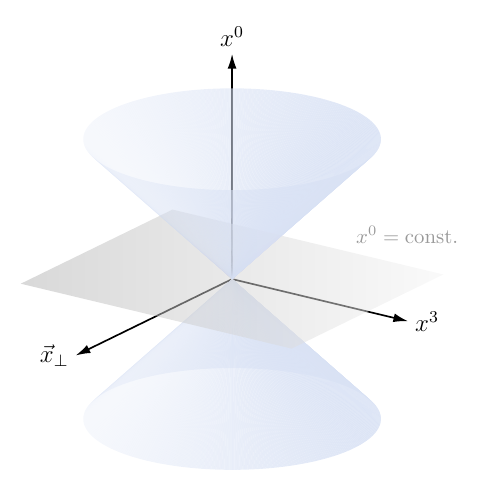}
        \captionsetup{justification=centering}
        \caption{Instant form, initialized at $x^0=0$.}
        \label{fig:lc1}
    \end{subfigure}
    \hfill
    \begin{subfigure}[b]{0.48\textwidth}
        \centering
        \includegraphics[width=\textwidth]{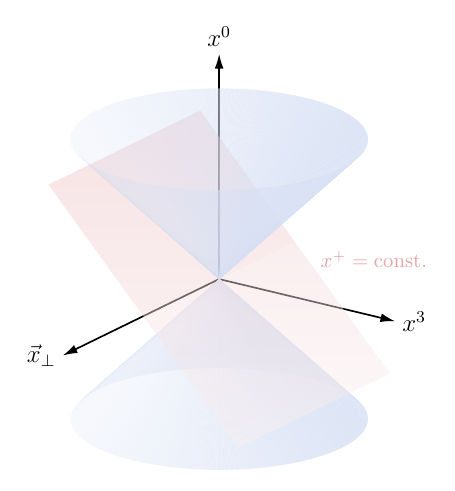}
        \captionsetup{justification=centering}
        \caption{Front form, initialized at $x^+=0$.}
        \label{fig:lc2}
    \end{subfigure}
    \caption{The initial hypersurfaces for the instant and front forms of relativistic dynamics.}
    \label{fig:lc-pair}
\end{figure}

In the instant form, states evolve from initial conditions on the space-like hypersurface at $x^0=0$. It is described by Minkowski coordinates $x^\mu=(x^0,\vec{x})$ with $\vec{x}\equiv(x^1,x^2,x^3)$, metric $g_{\mu\nu}=\mathrm{diag}(1,-1,-1,-1)$, and scalar product $x\cdot y = x^0y^0-\vec{x}\cdot\vec{y}$. The Hamiltonian $H_0$ generates translations in $x^0$ and the state evolves as $H_0\psi(t,\vec{x})=i(\partial/\partial t)\psi(t,\vec{x})$.

In the light-front form, states evolve from initial conditions on the light-like hypersurface at $x^+=0$. It is described by light-cone coordinates $x^\pm\equiv(x^0\pm x^3)/\sqrt{2}$, off-diagonal metric $g_{LC}^{+-}=g_{LC}^{-+}=1$, $g_{LC}^{11}=g_{LC}^{22}=-1$, and scalar product $x\cdot y = x^+y^-+x^-y^+-\vec{x}_\perp\cdot\vec{y}_\perp$. Derivatives satisfy $\partial_+=\partial^-$, $\partial_-=\partial^+$, $\partial_i=-\partial^i$. Full coordinate conventions are collected in~\cref{app:light-cone-coords}.

\noindent\textbf{\textit{Poincaré algebra on the light cone}.}
Any form of relativistic quantum mechanics is constructed starting from symmetries of the Poincaré group $\mathbbm{R}^{1,3}\ltimes\mathcal{SO}(1,3)$. The associated Lie algebra is given by the commutation relations
\begin{equation}
    \label{eq:poincare-algebra}
    \begin{aligned}
        &[P^\mu,\,P^\nu] = 0\,,\quad [M^{\mu\nu},\,P^\rho] = i\left(g^{\nu\rho}P^\mu - g^{\mu\rho}P^\nu\right) \\
        &[M^{\mu\nu},\,M^{\rho\sigma}] = i\left(g^{\mu\sigma}M^{\nu\rho} - g^{\mu\rho}M^{\nu\sigma} + g^{\nu\rho}M^{\mu\sigma} - g^{\nu\sigma}M^{\mu\rho}\right)\,.
    \end{aligned}
\end{equation}
where $P^\mu$ are the generators of translations and $M^{\mu\nu}$ the generators of Lorentz transformations. Introducing notation for angular momentum $J^i \equiv \tfrac{1}{2}\epsilon^{ijk}M^{jk}$ and Lorentz boost $K^i \equiv M^{0i}$, the commutation relations become
\begin{alignat}{3}
    \label{eq:instant-poincare-jk}
    &[J^i,P^j] = i\epsilon^{ijk}P^k\,,  &\quad &[J^i,J^j] = i\epsilon^{ijk}J^k\,,   &\quad &[J^i,P^0] = 0\,, \\
    &[K^i,P^j] = -i\delta^{ij}P^0\,,    &\quad &[K^i,K^j] = -i\epsilon^{ijk}J^k\,,  &\quad &[K^i,P^0] = -iP^i\,, \quad [J^i,K^j] = i\epsilon^{ijk}K^k\,. \notag
\end{alignat}

The Poincaré generators transform as a four-vector and a second-rank tensor under the change to light-cone coordinates. The light-front momentum vector and angular momentum tensor are
\begin{equation}
    \label{eq:lc-poincare-generators}
    \tilde{P}^\mu = (P^+,\,P^1,\,P^2,\,P^-)\,, \qquad
    \tilde{M}^{\mu\nu} = \begin{pmatrix}
        0    & -S^1 & -S^2 & K^3 \\
        S^1  &    0 &  J^3 & B^1 \\
        S^2  & -J^3 &    0 & B^2 \\
        -K^3 & -B^1 & -B^2 &   0
    \end{pmatrix}\,,
\end{equation}
with the light-front momentum, boost and rotation generators defined as
\begin{equation}
    \label{eq:lc-generator-defs}
    P^\pm \equiv \frac{P^0\pm P^3}{\sqrt{2}}\,, \quad B^i \equiv \frac{K^i+\epsilon^{ij}J^j}{\sqrt{2}}\,, \quad S^i \equiv \frac{K^i-\epsilon^{ij}J^j}{\sqrt{2}}\,,
\end{equation}
where $i,j\in(1,2)$ and $\epsilon^{ij}$ is the antisymmetric symbol in the transverse plane, with $\epsilon^{12}=-\epsilon^{21}=1$.

The full set of light-front Poincaré generators is $\{P^+, P^-, \vec{P}_\perp,
    \vec{B}_\perp, \vec{S}_\perp,J^3,K^3\}$, where $\vec{X}_\perp\equiv (X^1,X^2)$.

One important feature of the light-front parametrization is that the Poincaré group possesses a subgroup isomorphic to the two-dimensional Galilean group. The commutation relations for the light-front Poincaré generators are
\begin{equation}
\label{eq:lc-poincare-commutators}
\begin{alignedat}{3}
    &[P^+,P^-]=0\,,                          &\quad &[P^+,P_\perp^i]=0\,,                        &\quad &[P^+,B_\perp^i]=0\,, \quad [P^+,J^3]=0\,,\\
    &[P^-,P_\perp^i]=0\,,                    &\quad &[P^-,B_\perp^i]=iP_\perp^i\,,               &\quad &[P^-,J^3]=0\,, \\
    &[P_\perp^i,B_\perp^j]=i\delta^{ij}P^+\,, &\quad &[J^3,P_\perp^i]=i\epsilon^{ij}P_\perp^j\,,  &\quad &[J^3,B_\perp^i]=i\epsilon^{ij}B_\perp^j\,.
\end{alignedat}
\end{equation}

For comparison, the two-dimensional Galilean algebra is generated by the rotation $\mathcal{J}$, boosts $\mathcal{B}^i$, spatial translations $\mathcal{P}^i$, the Hamiltonian $\mathcal{H}$, and the mass $\mathcal{M}$, with commutation relations
\begin{equation}
\label{eq:galilean-algebra}
\begin{alignedat}{3}
    &[\mathcal{M},\mathcal{H}]=0\,,                        &\quad &[\mathcal{M},\mathcal{P}^i]=0\,,                        &\quad &[\mathcal{M},\mathcal{B}^i]=0\,, \quad [\mathcal{M},\mathcal{J}]=0\,, \\
    &[\mathcal{H},\mathcal{P}^i]=0\,,                      &\quad &[\mathcal{H},\mathcal{B}^i]=i\mathcal{P}^i\,,           &\quad &[\mathcal{H},\mathcal{J}]=0\,, \\
    &[\mathcal{P}^i,\mathcal{B}^j]=i\delta^{ij}\mathcal{M}\,, &\quad &[\mathcal{J},\mathcal{P}^i]=i\epsilon^{ij}\mathcal{P}^j\,, &\quad &[\mathcal{J},\mathcal{B}^i]=i\epsilon^{ij}\mathcal{B}^j\,.
\end{alignedat}
\end{equation}

Comparing~\cref{eq:lc-poincare-commutators} with~\cref{eq:galilean-algebra}, the subalgebra $\{P^+,P^-,\vec{P}_\perp,\vec{B}_\perp,J^3\}$ is isomorphic to the two-dimensional Galilean algebra and gives the identifications $P^+ \leftrightarrow \mathcal{M}$, $P^- \leftrightarrow \mathcal{H}$, $\vec{P}_\perp^i \leftrightarrow \mathcal{P}^i$, $\vec{B}_\perp^i \leftrightarrow \mathcal{B}^i$, $J^3 \leftrightarrow \mathcal{J}$. This allows physical interpretations of the light-front generators
\begin{equation}
    \label{eq:galilean-correspondence}
    \begin{alignedat}{2}
        &P^+\leftrightarrow\text{mass}\,,             &\quad &P^-\leftrightarrow\text{time translation}\,,\quad \vec{P}_\perp\leftrightarrow\text{space translation}\\
        &\vec{B}_\perp\leftrightarrow\text{boosts}, &\quad &J^3\leftrightarrow\text{rotation}\,.
    \end{alignedat}
\end{equation}
The generator $P^-$ is the light-front Hamiltonian and $x^+$ plays the role of light-front time. The isomorphism implies that the light-front Hamiltonian has a non-relativistic structure. This is why the Dirac equation on the light front takes the non-relativistic form of a Schr\"{o}dinger equation.

\section{Light-cone QCD}

\noindent\textbf{\textit{Projection operators}.}
The anti-commutation and conjugation relations for Dirac gamma matrices in light-cone coordinates are
\begin{equation}
    \label{eq:lc-dirac-algebra}
    \{\gamma^\mu,\gamma^\nu\} = 2\tilde{g}^{\mu\nu}\mathbbm{1}\,, \qquad
    (\gamma^\pm)^\dagger = \gamma^\mp\,.
\end{equation}
The light-cone Dirac matrices are defined by
\begin{equation}
    \label{eq:lc-dirac-matrices}
    \gamma^\pm = \frac{1}{\sqrt{2}}\left(\gamma^0 \pm \gamma^3\right)\,,
\end{equation}
with conjugate $(\gamma^\pm)^\dagger = \gamma^\mp$ and the identity $\gamma^\pm\gamma^\mp\gamma^\pm = 2\gamma^\pm$. The light-cone projection operators are
\begin{equation}
    \label{eq:light-cone-projection-operators}
    \Lambda_\pm \equiv \frac{1}{2}\gamma^\mp\gamma^\pm\,,
\end{equation}
or equivalently $\Lambda_\pm= \gamma^0\gamma^\pm/\sqrt{2}$. The projectors are orthogonal, complete, Hermitian, and idempotent:
\begin{subequations}
    \label{eq:projector-properties}
    \begin{align}
        \Lambda_\pm\Lambda_\mp = 0\,, \quad \Lambda_+ + \Lambda_- = \mathbbm{1}\,, \quad \left(\Lambda_\pm\right)^\dagger = \Lambda_\pm\,, \quad \left(\Lambda_\pm\right)^2 = \Lambda_\pm\,. \label{eq:projector-hermitian-idem}
    \end{align}
\end{subequations}

In the chiral representation, the Dirac matrices take the explicit form
\begin{equation}
    \label{eq:chiral-rep}
    \begin{aligned}
    \gamma^0=\begin{pmatrix}0&\mathbbm{1}\\\mathbbm{1}&0\end{pmatrix}\,, \quad
    \gamma^i=\begin{pmatrix}0&-\sigma^i\\\sigma^i&0\end{pmatrix}\,, \quad \gamma^5\equiv i\gamma^0\gamma^1\gamma^2\gamma^3=\begin{pmatrix}\mathbbm{1}&0\\0&-\mathbbm{1}\end{pmatrix}\,,
    \end{aligned}
\end{equation}
The projectors become diagonal: $\Lambda_-=\mathrm{diag}(0,1,1,0)$ and $\Lambda_+=\mathrm{diag}(1,0,0,1)$.
Defining the projected quark fields $\psi_\pm \equiv \Lambda_\pm\psi$, the projector $\Lambda_+$ selects spinor components $(\Psi_1,\Psi_4)$ and $\Lambda_-$ selects $(\Psi_2,\Psi_3)$. The conjugate relations are $\psi_\pm^\dagger = \psi^\dagger\Lambda_\pm$.

\noindent\textbf{\textit{Lagrangian density}.}
In light-cone gauge $A^+=0$, the field strength tensor components are $F^{+-} = \partial^+A^-$, $F^{+i} = \partial^+A^i$, $F^{-i} = \partial^-A^i - \partial^iA^- - ig[A^-,A^i]$, and $F^{ij} = \partial^iA^j - \partial^jA^i - ig[A^i,A^j]$ for $i,j=1,2$. From these, the gluon Lagrangian obtained from~\cref{eq:gluon-lagrangian} becomes
\begin{equation}
    \label{eq:lc-gluon-lagrangian}
    \mathcal{L}_\mathrm{gluons} = -\frac{1}{4}F_a^{ij}F_{a,ij} - \frac{1}{2}F_a^{+-}F_a^{-+} + F_a^{+i}F_a^{-i}\,.
\end{equation}

The quark Lagrangian obtained from~\cref{eq:dirac-lagrangian}, expressed in light-cone coordinates using $\bar\psi(i\tilde{g}^{\mu\nu}\gamma^\nu D_\mu - m)\psi = \bar\psi[i(\gamma^+D^-+\gamma^-D^+-\vec{\gamma}_\perp\cdot\vec{D}_\perp)-m]\psi$ and projected onto $\psi_\pm$, is
\begin{equation}
    \label{eq:lc-quark-lagrangian}
    \begin{split}
    \mathcal{L}_\mathrm{quarks} =\; &i\sqrt{2}\!\left(\psi_+^\dagger D^-\psi_+ + \psi_-^\dagger\partial^+\psi_-\right) \\
    &- \frac{1}{\sqrt{2}}\left[\psi_+^\dagger\gamma^-\!\big(m + i\vec{\gamma}_\perp\cdot\vec{D}_\perp\big)\psi_- + \psi_-^\dagger\gamma^+\!\big(m + i\vec{\gamma}_\perp\cdot\vec{D}_\perp\big)\psi_+\right]\,,
    \end{split}
\end{equation}
where $\vec{\gamma}_\perp \equiv (\gamma^1,\gamma^2)$, $\vec{D}_\perp \equiv (D^1,D^2)$, and $D^+=\partial^+-igA^+=\partial^+$ in light-cone gauge. The combined light-front QCD Lagrangian is
\begin{equation}
    \label{eq:lightfront-lagrangian}
    \begin{split}
    \mathcal{L} =\; &-\frac{1}{4}F_a^{ij}F_{a,ij} + \frac{1}{2}\left(F_a^{+-}\right)^2 + F_a^{+i}F_a^{-i} + i\sqrt{2}\!\left(\psi_+^\dagger D^-\psi_+ + \psi_-^\dagger\partial^+\psi_-\right) \\
    &- \frac{1}{\sqrt{2}}\left[\psi_+^\dagger\gamma^-\!\big(m + i\vec{\gamma}_\perp\cdot\vec{D}_\perp\big)\psi_- + \psi_-^\dagger\gamma^+\!\big(m + i\vec{\gamma}_\perp\cdot\vec{D}_\perp\big)\psi_+\right]\,.
    \end{split}
\end{equation}

\noindent\textbf{\textit{Field equations}.}
The colored Dirac equation from~\cref{eq:fermion-eom}, written in light-cone gauge as $[i (\gamma^+D^-+\gamma^-\partial^+-\vec{\gamma}_\perp\cdot\vec{D}_\perp)-m]\psi = 0$, decouples into equations for the projected components $\psi_\pm$ as
\begin{subequations}
    \label{eq:fermion-eom-lc}
    \begin{align}
        &\partial^+\psi_- = -\frac{i}{2}\!\big(-i\vec{\gamma}_\perp\cdot\vec{D}_\perp + m\big)\gamma^+\psi_+\,, \label{eq:fermion-eom-lc-a} \\
        &D^-\psi_+        = -\frac{i}{2}\!\big(-i\vec{\gamma}_\perp\cdot\vec{D}_\perp + m\big)\gamma^-\psi_-\,. \label{eq:fermion-eom-lc-b}
    \end{align}
\end{subequations}
Equation~\cref{eq:fermion-eom-lc-a} is a constraint for $\psi_-$ while~\cref{eq:fermion-eom-lc-b} gives the time evolution of $\psi_+$.

The Yang-Mills equations from~\cref{eq:yang-mills-eom} in light-cone gauge, for the $\nu=i$ and $\nu=+$ components, give
\begin{subequations}
    \label{eq:gauge-eom-lc}
    \begin{align}
        &D_a^-\!\left(\partial^+A_a^i\right) + \partial^+F_a^{-i} + \left(D_j F^{ji}\right)_a = gJ_a^i\,, \label{eq:gauge-eom-lc-a} \\
        &\left(\partial^+\right)^2\!A_a^- + \left[D_i\!\left(\partial^+A^i\right)\right]_a = gJ_a^+\,. \label{eq:gauge-eom-lc-b}
    \end{align}
\end{subequations}
Equation~\cref{eq:gauge-eom-lc-b} is a constraint for $A_a^-$ while~\cref{eq:gauge-eom-lc-a} gives the time evolution of $A_a^i$.

\noindent\textbf{\textit{Hamiltonian density}.}
The conjugate momenta are obtained as $\Pi_\phi = \delta\mathcal{L}/\delta(\partial^-\phi)$ for each field $\phi$. For the fields in light-front QCD, the conjugate momenta are
\begin{subequations}
    \label{eq:conjugate-momenta-lc}
    \begin{align}
        \Pi_{A_a^i}  = \partial^+A_a^i\,,\quad \Pi_{\psi_+} = i\sqrt{2}\,\psi_+^\dagger\,, \quad \Pi_{A_a^-} =  \Pi_{\psi_-} = 0\,.
    \end{align}
\end{subequations}
The vanishing conjugate momenta for $A_a^-$ and $\psi_-$ signal Dirac constraints. Only the fields $A_a^i$ and $\psi_+$ are dynamical degrees of freedom. Formally inverting~\cref{eq:fermion-eom-lc-a,eq:gauge-eom-lc-b} gives
\begin{subequations}
    \label{eq:constrained-fields-lc}
    \begin{align}
        \psi_- &= -\frac{i}{2\partial^+}\!\big(m - i\vec{\gamma}_\perp\cdot\vec{D}_\perp\big)\gamma^+\psi_+\,, \label{eq:constrained-fields-lc-a} \\
        A_a^- &= \frac{1}{\left(\partial^+\right)^2}\!\left\{gJ_a^+ - \left[D_i\!\left(\partial^+A^i\right)\right]_a\right\}\,. \label{eq:constrained-fields-lc-b}
    \end{align}
\end{subequations}
It is important to note that the choice of light-cone gauge $A^+=0$ is essential here. Without it, terms containing $A^+$ in the denominator would appear and not allow for this simple inversion.

The longitudinal color current in light-cone gauge simplifies to $J_a^+ \equiv \bar\psi\gamma^+\psi\, t^a = \sqrt{2}\,\psi_+^\dagger t^a\psi_+$.

Substituting~\cref{eq:constrained-fields-lc} back into~\cref{eq:lightfront-lagrangian} replaces all $\psi_-$ and $A_a^-$ dependence with expressions in terms of the dynamical fields $\psi_+$ and $A_a^i$. The terms from $\psi_-$ elimination contribute as
\begin{equation}
    \label{eq:lc-lag-psi-minus}
    \mathcal{L}_{\psi_-} = \frac{i}{\sqrt{2}}\,\psi_+^\dagger\!\big(m - i\vec{\gamma}_\perp\cdot\vec{D}_\perp\big)\frac{1}{\partial^+}\!\big(m + i\vec{\gamma}_\perp\cdot\vec{D}_\perp\big)\psi_+\,,
\end{equation}
and the terms from $A_a^-$ elimination contribute as
\begin{equation}
    \label{eq:lc-lag-a-minus}
    \begin{aligned}
    \mathcal{L}_{A_a^-} =\; &\left(\partial^+A_a^i\right)\left(\partial^-A_a^i\right) + i\sqrt{2}\,\psi_+^\dagger\partial^-\psi_+ \\
    &+ \frac{1}{2}\!\left\{gJ_a^+ - \left[D_i\!\left(\partial^+A^i\right)\right]_a\right\}
    \frac{1}{\left(\partial^+\right)^2}\!\left\{gJ_a^+ - \left[D_i\!\left(\partial^+A^i\right)\right]_a\right\}\,.
    \end{aligned}
\end{equation}

The Legendre transformation $\mathcal{H} \equiv \Pi_{\psi_+}(\partial^-\psi_+) + \Pi_{A_a^i}(\partial^-A_a^i) - \mathcal{L}$ then gives the light-front QCD Hamiltonian density
\begin{equation}
    \label{eq:lightfront-hamiltonian}
    \begin{split}
    \mathcal{H} =\; &-\frac{1}{2}\!\left\{gJ_a^+ - \left[D_i\!\left(\partial^+A^i\right)\right]_a\right\}\frac{1}{\left(\partial^+\right)^2}\!\left\{gJ_a^+ - \left[D_i\!\left(\partial^+A^i\right)\right]_a\right\} \\
    &- \frac{i}{\sqrt{2}}\,\psi_+^\dagger\!\big(m - i\vec{\gamma}_\perp\cdot\vec{D}_\perp\big)\frac{1}{\partial^+}\!\big(m + i\vec{\gamma}_\perp\cdot\vec{D}_\perp\big)\psi_+ + \frac{1}{4}F_a^{ij}F_{ij}^a\,.
    \end{split}
\end{equation}

\noindent\textbf{\textit{Quantized gluon field}.}
The mode expansion for the dynamical gluon field is
\begin{equation}
    \label{eq:gauge-field-mode-expansion}
    A_a^\mu(x^+, \vec{x}) = \int\limits_{k^+>0} \frac{\diff k^+\,\diff^2\vec{k}_\perp}{\sqrt{(2\pi)^3\,2k^+}} \sum_\lambda \left[ a_a^\lambda(x^+, \vec{k})\,\epsilon_\lambda^\mu(\vec{k})\,e^{-i\vec{k}\cdot\vec{x}} + a_a^{\lambda\dagger}(x^+, \vec{k})\,\epsilon_\lambda^{\mu\dagger}(\vec{k})\,e^{i\vec{k}\cdot\vec{x}} \right]\,,
\end{equation}
where $\vec{k}=(k^+,\vec{k}_\perp)$, the index $\lambda=\pm 1$ labels the two physical polarizations, and the polarization vectors in light-cone gauge satisfy $\epsilon_\lambda^+=0$ and are given by
\begin{equation}
    \label{eq:polarization-vectors-lc}
    \epsilon_\lambda^\mu = \left(0,\,\frac{\vec{\epsilon}_\lambda\cdot\vec{k}_\perp}{k^+},\,\vec{\epsilon}_\lambda\right)\,,
\end{equation}
with $\vec{\epsilon}_\lambda$ the two-dimensional transverse unit vectors, and satisfy the orthogonality and completeness relations.

The equal-light-front-time commutation relations for $A_a^i$ are
\begin{subequations}
    \label{eq:field-commutators-lc}
    \begin{align}
        &\left[A_a^i(x^+,\vec{x}),\,A_b^j(y^+,\vec{y})\right]\Big|_{x^+=y^+} = 0, \quad
        \left[\Pi_{A_a^i}(x^+,\vec{x}),\,\Pi_{A_b^j}(y^+,\vec{y})\right]\Big|_{x^+=y^+} = 0, \label{eq:field-commutators-lc-a} \\
        &\left[A_a^i(x^+,\vec{x}),\,\Pi_{A_b^j}(y^+,\vec{y})\right]\Big|_{x^+=y^+} = i\delta_{ab}\delta^{ij}\delta(x^- - y^-)\delta^{(2)}(\vec{x}_\perp - \vec{y}_\perp)\,. \label{eq:field-commutators-lc-b}
    \end{align}
\end{subequations}

The commutation relations for the creation and annihilation operators are
\begin{subequations}
    \label{eq:mode-commutators-lc}
    \begin{align}
        &\left[a_a^\lambda(x^+,\vec{k}),\,a_b^{\rho}(y^+,\vec{q})\right]\Big|_{x^+=y^+} = 0, \quad
        \left[a_a^{\lambda\dagger}(x^+,\vec{k}),\,a_b^{\rho\dagger}(y^+,\vec{q})\right]\Big|_{x^+=y^+} = 0, \label{eq:mode-commutators-lc-a} \\
        &\left[a_a^\lambda(x^+,\vec{k}),\,a_b^{\rho\dagger}(y^+,\vec{q})\right]\Big|_{x^+=y^+} = \delta_{ab}\delta^{\lambda\rho}\,\delta(k^+ - q^+)\delta^{(2)}(\vec{k}_\perp - \vec{q}_\perp)\,. \label{eq:mode-commutators-lc-b}
    \end{align}
\end{subequations}

\noindent\textbf{\textit{Quantized quark field}.}
The mode expansion for the dynamical quark field $\psi_+$ is
\begin{equation}
    \label{eq:quark-field-mode-expansion}
    \psi_+(x^+,\vec{x}) = \int\limits_{p^+>0} \frac{\diff p^+\,\diff^2\vec{p}_\perp}{\sqrt{(2\pi)^3\,2p^+}} \sum_s \left[ b_s(x^+,\vec{p})\,u_+(p)\,e^{-i\vec{p}\cdot\vec{x}} + d_s^\dagger(x^+,\vec{p})\,v_+(p)\,e^{i\vec{p}\cdot\vec{x}} \right]\,,
\end{equation}
where $\vec{p}=(p^+,\vec{p}_\perp)$, $s$ labels the helicity, and $u_+(p)\equiv\Lambda_+ u(p)$, $v_+(p)\equiv\Lambda_+ v(p)$ are the light-front projected Dirac spinors satisfying $\Lambda_+ u_+ = u_+$ and $\Lambda_+ v_+ = v_+$.

The equal-light-front-time anti-commutation relation for $\psi_+$, obtained from $\Pi_{\psi_+}=i\sqrt{2}\,\psi_+^\dagger$, is
\begin{equation}
    \label{eq:fermion-commutators-lc}
    \left\{\psi_{+,\alpha}(x^+,\vec{x}),\,\psi_{+,\beta}^\dagger(y^+,\vec{y})\right\}\Big|_{x^+=y^+} = \frac{1}{\sqrt{2}}\left(\Lambda_+\right)_{\alpha\beta}\delta(x^- - y^-)\delta^{(2)}(\vec{x}_\perp - \vec{y}_\perp)\,.
\end{equation}

The anti-commutation relations for the quark and anti-quark operators are
\begin{subequations}
    \label{eq:quark-mode-commutators-lc}
    \begin{align}
        &\left\{b_s(x^+,\vec{p}),\,b_{s'}^\dagger(y^+,\vec{q})\right\}\Big|_{x^+=y^+} = \delta_{ss'}\,\delta(p^+ - q^+)\delta^{(2)}(\vec{p}_\perp - \vec{q}_\perp)\,, \label{eq:quark-mode-commutators-lc-a} \\
        &\left\{d_s(x^+,\vec{p}),\,d_{s'}^\dagger(y^+,\vec{q})\right\}\Big|_{x^+=y^+} = \delta_{ss'}\,\delta(p^+ - q^+)\delta^{(2)}(\vec{p}_\perp - \vec{q}_\perp)\,, \label{eq:quark-mode-commutators-lc-b}
    \end{align}
\end{subequations}
with all other anti-commutators vanishing.

\section{Quantum state evolution}

\noindent\textbf{\textit{Dispersion relation}.}
In the instant form, the free Hamiltonian is $H=\sqrt{\vec{P}^2+M^2}$. The momentum has components $(P^0,\vec{P})$, with the longitudinal momentum $P_z\in(-\infty,+\infty)$. In quantized form, the free Hamiltonian is
\begin{equation}
    \label{eq:instant-hamiltonian-quantized}
    P^0 = \sum_{\vec{k}} a^\dagger_{\vec{k}} a_{\vec{k}}\,\sqrt{M^2+\vec{k}^2}\,,
\end{equation}
with the vacuum condition $a_{\vec{k}}|0\rangle=0$. As the vacuum state is the lowest energy state, many $\vec{P}=0$ excited states contribute to the interacting vacuum. Consequently, the interacting vacuum is a complicated superposition of infinitely many free Fock states.

In the light-front form, the mass-shell condition $M^2 = P_\mu P^\mu = 2P^+P^- - \vec{P}_\perp^2$ gives the dispersion relation
\begin{equation}
    \label{eq:lc-dispersion}
    P^- = \frac{M_\perp^2}{2P^+}\,,
\end{equation}
where $M_\perp^2 \equiv M^2+\vec{P}_\perp^2$. This has a manifestly non-relativistic form where $P^-$ plays the role of energy, $P^+$ of mass, and $M_\perp^2/2$ of the potential. The momentum has components $(P^-,P^+,\vec{P}_\perp)$, with the longitudinal momentum $P^+\in(0,+\infty)$. The quantized free Hamiltonian is
\begin{equation}
    \label{eq:lc-hamiltonian-quantized}
    P^- = \sum_{k^+,\vec{k}_\perp} a^\dagger_{k^+,\vec{k}_\perp} a_{k^+,\vec{k}_\perp}\,\frac{M^2+\vec{k}_\perp^2}{2k^+}\,,
\end{equation}
with the vacuum condition $a_{k^+,\vec{k}_\perp}|0\rangle=0$. Since $P^+>0$ for all physical states, while the vacuum has $P^+=0$, no physical state can mix with it and the light-front vacuum is trivial. This is unlike the instant form, where many $\vec{P}=0$ excited states contribute to the interacting vacuum. Consequently, any interacting state may be constructed by simply applying creation operators on the trivial free vacuum.

\noindent\textbf{\textit{Schr\"{o}dinger equation}.}
The quantum state $|\psi; x^+\rangle$ is defined at fixed light-front time $x^+$, and its dynamical evolution is governed by the light-front Schrödinger equation
\begin{equation}
    \label{eq:lf-schrodinger}
    i\frac{\partial}{\partial x^+}|\psi; x^+\rangle = P^-(x^+)\,|\psi; x^+\rangle \,.
\end{equation}
Here $P^-$ is the light-front Hamiltonian, which generates translations in the light-cone time $x^+$. The state $|\psi\rangle$ is an expansion in multi-particle Fock states built from the quantized fields of~\cref{eq:quark-field-mode-expansion,eq:gauge-field-mode-expansion}. The $x^+$ dependence may arise from interaction with an external background gauge field. The formal solution of~\cref{eq:lf-schrodinger} is
\begin{equation}
    \label{eq:lf-time-evolution}
    |\psi; x^+\rangle = \mathcal{T}_+\exp\!\left[-i\int_0^{x^+}\!\diff z^+\,P^-(z^+)\right]|\psi;\,0\rangle\,,
\end{equation}
where $\mathcal{T}_+$ denotes light-front time ordering. For bound-state problems, in which the Hamiltonian is time-independent,~\cref{eq:lf-schrodinger} reduces to the eigenvalue equation
\begin{equation}
    \label{eq:lf-eigenvalue}
    P^-|\psi\rangle = p^-|\psi\rangle \,,
\end{equation}
where the eigenvalue $p^- = M_\perp^2/(2P^+)$ follows from the dispersion relation~\cref{eq:lc-dispersion}, relating $p^-$ to the invariant mass $M$ of the state. Solving~\cref{eq:lf-eigenvalue} directly yields the mass eigenvalues $M$ and the light-front wavefunctions. The eigenstate $|\psi\rangle = |\psi;\,M,\,P^+,\,\vec{P}_\perp,\,j,\,m_j\rangle$ is labeled by six quantum numbers consisting of the invariant mass $M$, the longitudinal momentum $P^+$, the transverse momentum $\vec{P}_\perp$, the total angular momentum $j$, and its projection $m_j$.

\noindent\textbf{\textit{Basis representation}.}
For numerical computation, both the Fock space and the single-particle basis must be truncated. The framework of \emph{Discretized Light-Cone Quantization} (DLCQ) places the system in a finite box with periodic (bosons) or anti-periodic (fermions) boundary conditions. More precisely, the longitudinal $x^-\in[-L^+,L^+]$ and transverse $\vec{x}_\perp \in[-L_\perp,L_\perp]^2$ coordinates give the volume $V = 2L^+\,(2L_\perp)^2$. The discretized momenta are then
\begin{equation}
    \label{eq:dlcq-momenta}
    p^+ \;\to\; \frac{n\pi}{L^+}\,, \qquad
    \vec{p}_\perp \;\to\; \left(\frac{n_1\pi}{L_\perp},\,\frac{n_2\pi}{L_\perp}\right) \,,
\end{equation}
and the mode expansions in~\cref{eq:quark-field-mode-expansion,eq:gauge-field-mode-expansion} become discrete sums
\begin{subequations}
    \label{eq:dlcq-mode-expansions}
    \begin{align}
        \psi_+(x) &= \frac{1}{\sqrt{V}}\sum_q \frac{1}{\sqrt{2p^+}}
            \sum_s \Bigl[b_s(\vec{p})\,u_+(p)\,e^{-ip\cdot x}
                       + d_s^\dagger(\vec{p})\,v_+(p)\,e^{ip\cdot x}\Bigr]\,, \label{eq:dlcq-quark}\\[4pt]
        A_a^\mu(x) &= \frac{1}{\sqrt{V}}\sum_q \frac{1}{\sqrt{2k^+}}
            \sum_\lambda \Bigl[a_a^\lambda(\vec{k})\,\epsilon_\lambda^\mu(\vec{k})\,e^{-ik\cdot x}
                             + a_a^{\lambda\dagger}(\vec{k})\,\epsilon_\lambda^{\mu\dagger}(\vec{k})\,e^{ik\cdot x}\Bigr]\,, \label{eq:dlcq-gluon}
    \end{align}
\end{subequations}
where $q \equiv \{n,n_1,n_2,\lambda,c,f\}$ collects the longitudinal index $n$, transverse indices $(n_1,n_2)$, helicity $\lambda$, color $c$, and flavor $f$ of each single-particle state. The Hamiltonian then becomes a finite matrix, and the eigenvalue problem~\cref{eq:lf-eigenvalue} is solved by numerical diagonalization.

The \emph{Basis Light-Front Quantization} (BLFQ) generalizes DLCQ to any complete orthogonal basis. If $f_i(p^+)$ and $g_j(\vec{p}_\perp)$ are the chosen longitudinal and transverse basis functions, the mode operators are expanded as
\begin{equation}
    \label{eq:blfq-expansion}
    b_q(p^+,\vec{p}_\perp) = \sum_{i,j} b_{\bar{q},ij}\,f_i(p^+)\,g_j(\vec{p}_\perp)\,,
\end{equation}
where $\bar{q}$ denotes the quantum numbers at fixed $i$ and $j$. The optimal basis choice approximates the eigenstates by using the symmetries of the Hamiltonian. In practice, this accelerates numerical convergence.

\noindent\textbf{\textit{Time-dependent basis light-front quantization}.}
For time-dependent problems, the Hamiltonian splits as
\begin{equation}
    \label{eq:time-dependent-hamiltonian}
     P^-(x^+) = P^- + V(x^+)\,,
\end{equation}
where $P^-$ is the time-independent QCD Hamiltonian of~\cref{eq:lightfront-hamiltonian} and $V(x^+)$ contains interaction terms from the background field. Moving to the interaction picture via $|\psi; x^+\rangle_I \equiv e^{iP^-\,x^+}|\psi; x^+\rangle$, the Schrödinger equation~\cref{eq:lf-schrodinger} becomes
\begin{equation}
    \label{eq:tblfq-interaction-schrodinger}
    i\frac{\partial}{\partial x^+}|\psi; x^+\rangle_I = V_I(x^+)\,|\psi; x^+\rangle_I \,,
\end{equation}
with the interaction-picture potential $V_I(x^+) \equiv e^{iP^-\,x^+}\,V(x^+)\,e^{-iP^-\,x^+}$. The formal solution is
\begin{equation}
    \label{eq:tblfq-formal-solution}
    |\psi; x^+\rangle_I = \mathcal{T}_+\exp\!\left[-i\int_0^{x^+}\!\diff z^+\,V_I(z^+)\right]|\psi;\,0\rangle_I \,.
\end{equation}

For non-perturbative evolution, the time-ordered exponential is decomposed into $n$ small steps of size $\delta x^+ = x^+/n$ as
\begin{equation}
    \label{eq:tblfq-product-expansion}
    \mathcal{T}_+\exp\!\left[-i\int_0^{x^+}\!\diff z^+\,V_I(z^+)\right]
    = \mathcal{T}_+\lim_{n\to\infty}\prod_{k=1}^{n}\!\left[1 - iV_I(x_k^+)\,\delta x^+\right]\,,
\end{equation}
with intermediate times $x_k^+ = k\,\delta x^+$. This product form is exact in the continuum limit $n\to\infty$, and in practice $n$ is chosen large enough to achieve the desired convergence.

The physical state is expanded as
\begin{equation}
    |\psi; x^+\rangle_I = \sum_\beta c_\beta(x^+)|\beta\rangle\,,
\end{equation}
where $|\beta\rangle$ are the BLFQ eigenstates of $P^-$, satisfying
\begin{equation}
    \label{eq:tblfq-eigenstate-expansion}
    P^-|\beta\rangle = p^-_\beta|\beta\rangle\,,
\end{equation}
with initial conditions specified by the column vector $\boldsymbol{c}(0)$. The formal solution given in~\cref{eq:tblfq-formal-solution} then takes the matrix form
\begin{equation}
    \label{eq:tblfq-matrix-solution}
    \boldsymbol{c}(x^+) = \mathcal{T}_+\exp\!\left[-i\int_0^{x^+}\!\diff z^+\,\mathcal{M}(z^+)\right]\boldsymbol{c}(0) \,,
\end{equation}
where the matrix elements are $\mathcal{M}_{\beta\beta'}(x^+) = \langle\beta|\,V_I(x^+)\,|\beta'\rangle$. The state $\boldsymbol{c}(x^+)$ gives direct access to all observables at any intermediate time during the evolution.

The step-by-step evolution is implemented numerically using a finite-difference approximation to~\cref{eq:tblfq-interaction-schrodinger}. The second-order explicit midpoint method gives the update formula $|\psi;\,x^++\delta x^+\rangle_I = -2i\,V_I(x^+)\,\delta x^+\,|\psi;\,x^+\rangle_I + |\psi;\,x^+-\delta x^+\rangle_I +\mathcal{O}[(\delta x^+)^{3}]$.

The \emph{time-dependent Basis Light-Front Quantization} (tBLFQ) provides a fully non-perturbative, step-by-step evolution of a quantum state in the light-front frame. More details of the method are given in Ref.~\cite{Li:2019ijx}.

\appendix{Wong's Equations}
\label{app:wong}

\section{Color Dirac equation}

One starts with the color Dirac Hamiltonian for a quark in the presence of a non-Abelian field $A_\mu=(A_0,\boldsymbol{A})$ with $\boldsymbol{A}\equiv (A_x, A_y, A_z)$, given by
\begin{equation}
    \label{eq:coldir}
    H=\boldsymbol{\alpha}\cdot\big(\boldsymbol{p}-g\boldsymbol{A}^a T^a\big)+gA_0^a T^a+m\beta\,.
\end{equation}
The Dirac matrices are $\alpha^i=\gamma^0\gamma^i$ and $\beta=\gamma^0$. The Yang-Mills coupling constant is $g$ and the SU($N_c$) generators are $T^a$, which satisfy the commutation relations $[T^a, T^b]=i f^{abc}T^c$. All involved objects are quantum operators, unless otherwise stated. In the Heisenberg picture, these operators are time-dependent and their time evolution is expressed as
\begin{equation}
    \label{eq:heis}
    \frac{\dd}{\dd t}\mathcal{O}(t)=i [H,\mathcal{O}(t)]+\frac{\partial}{\partial t}\mathcal{O}(t)\,.
\end{equation}

\noindent\textbf{\textit{Coordinates}.}
Using Eq.~\eqref{eq:heis} for the position operator $\boldsymbol{x}(t)$ yields
\begin{equation}
    \frac{\dd \boldsymbol{x}}{\dd t}=i [H,\boldsymbol{x}]\,.
\end{equation}
The involved commutator simplifies to
\begin{equation}
    [H,x_i]=[\alpha_jp_j, x_i]=\alpha_j[p_j,x_i]=-i \alpha_i\,,
\end{equation}
using the canonical commutation relation $[x_i, p_j]=i \delta_{ij}$. This yields the time evolution of the position operator
\begin{equation}
    \label{eq:heisx}
    \frac{\dd \boldsymbol{x}}{\dd t}=\boldsymbol{\alpha}\,.
\end{equation}
The operator $\boldsymbol{\alpha}$ can be interpreted as the particle's velocity in the presence of the background field.

\noindent\textbf{\textit{Momenta}.}
The kinetic momentum operator $\boldsymbol{p}_{kin}$ is defined in terms of the canonical momentum $\boldsymbol{p}$ coupled to a gauge field $\boldsymbol{A}$ as
\begin{equation}
    \label{eq:kinmom}
    \boldsymbol{p}_{kin}\equiv \boldsymbol{p}-g\boldsymbol{A}^a T^a\,.
\end{equation}
Its time evolution is given by Eq.~\eqref{eq:heis} as
\begin{equation}
    \frac{\dd \boldsymbol{p}_{kin}}{\dd t}=i [H,\boldsymbol{p}_{kin}]+\frac{\partial \boldsymbol{p}_{kin}}{\partial t}\,.
\end{equation}
The commutator term can be split into
\begin{equation}
    \label{eq:comhpi}
    [H,p^{kin}_i]=[H, p_i]-g[H, A_i^a]T^a-gA_i^c[H,T^c]\,.
\end{equation}
Each of the involved commutators is computed in turn. The first is
\begin{equation}
    \label{eq:comhp}
    [H, p_i]=-g\alpha_j[A_j^a,p_i]T^a+g[A^a_0,p_i]T^a=-ig(\boldsymbol{\alpha}\cdot\partial_i\boldsymbol{A}^a-\partial_i A_0^a)T^a\,,
\end{equation}
using $[f(\boldsymbol{x}),p_i]=i \partial_i f(\boldsymbol{x})$. The second commutator is
\begin{equation}
    \label{eq:comha}
    [H, A_i^a]=\alpha_j[p_j, A_i^a]=-i\alpha_j \partial_j A_i^a\,.
\end{equation}
The third commutator involves the SU($N_c$) generator algebra $[T^b, T^c]=if^{abc}T^a$,
\begin{equation}
    \label{eq:comht}
    [H, T^c]=(-g\alpha_i A_i^b+gA_0^b)[T^b, T^c]=-ig f^{abc}(\boldsymbol{\alpha}\cdot \boldsymbol{A}^b-A_0^b)T^a\,.
\end{equation}
The explicit partial time derivative of the kinetic momentum is
\begin{equation}
    \frac{\partial p^{kin}_i}{\partial t}=-g\frac{\partial A_i^a}{\partial t}T^a=g\partial_0 A_i^a T^a\,,
\end{equation}
using $A^i=-A_i$ for the spatial components of $\boldsymbol{A}$ in the metric $g^{\mu\nu}=\mathrm{diag}(+,-,-,-)$. Collecting all the intermediate commutators from \cref{eq:comhp,eq:comha,eq:comht} back in Eq.~\eqref{eq:comhpi} yields
\begin{equation}
    \label{eq:pieb}
    \begin{aligned}
        \frac{\dd p^{kin}_i}{\dd t}=&g\left(\partial_0A_i^a-\partial_i A_0^a+g f^{abc}A_0^b A_i^c\right)T^a\\
        &-g\alpha_j\left(\partial_i A_j^a-\partial_j A_i^a+g f^{abc}A_i^b A_j^c\right)T^a\,,
    \end{aligned}
\end{equation}
where $\alpha^i=\alpha_i$ in the chosen metric.

The color components of the field strength tensor for a Yang-Mills field $F_{\mu\nu}^a=\partial_\mu A_\nu^a-\partial_\nu A_\mu^a+g f^{abc}A_\mu^bA_\nu^c$ are introduced next. The electric field is defined as $E_i^a\equiv F_{0i}^a$, giving explicitly
\begin{equation}
    \label{eq:elfield}
    E_i^a=\partial_0 A_i^a-\partial_i A_0^a+g f^{abc}A_0^bA_i^c\,,
\end{equation}
and the magnetic field is defined via $B_k^a\equiv -\epsilon_{ijk}F_{ij}^a/2$, giving
\begin{equation}
    \label{eq:magfield}
    B_k^a=-\frac{1}{2}\epsilon_{ijk}\left(\partial_i A_j^a-\partial_j A_i^a+g f^{abc}A_i^bA_j^c\right)\,.
\end{equation}
Additionally, it is useful to note that
\begin{equation}
    \label{eq:alphafij}
    (\boldsymbol{\alpha}\times\boldsymbol{B}^a)_i=\epsilon_{ijk}\alpha_j B_k^a=-\alpha_j F_{ij}^a\,,
\end{equation}
since $\epsilon_{ijk}B_k^a=-\epsilon_{ijk}\epsilon_{klm}F_{lm}^a/2=-(F_{ij}^a-F_{ji}^a)/2=-F_{ij}^a$, using the contraction $\epsilon_{ijk}\epsilon_{klm}=\delta_{il}\delta_{jm}-\delta_{im}\delta_{jl}$ along with the antisymmetry $F_{ij}^a=-F_{ji}^a$. Combining \cref{eq:pieb,eq:elfield,eq:magfield,eq:alphafij}, the kinetic momentum equation can be recast as the non-Abelian Lorentz force
\begin{equation}
    \label{eq:heisp}
    \frac{\dd \boldsymbol{p}_{kin}}{\dd t}=g(\boldsymbol{E}^a+\boldsymbol{\alpha}\times\boldsymbol{B}^a)T^a\,.
\end{equation}

\noindent\textbf{\textit{Color charge}.}
The temporal evolution of the color generators from Eq.~\eqref{eq:heis} is
\begin{equation}
    \label{eq:heist}
    \frac{\dd T^a}{\dd t}=i [H,T^a]\,.
\end{equation}
Using the commutator from Eq.~\eqref{eq:comht}, one obtains
\begin{equation}
    \frac{\dd T^a}{\dd t}=-g f^{abc}(\boldsymbol{\alpha}\cdot\boldsymbol{A}^b-A_0^b)T^c\,.
\end{equation}

\noindent\textbf{\textit{Classical limit}.}
\Cref{eq:heisx,eq:heisp,eq:heist} describe the time evolution of the quantum operators $\boldsymbol{x}$, $\boldsymbol{p}_{kin}$ and $T^a$. Note that the $\hat{X}$ with $X\in (\boldsymbol{x},\boldsymbol{p}_{kin},T^a)$ notation for quantum operators was dropped for simplicity. The classical limit is taken by working with the classical $\boldsymbol{x}$, $\boldsymbol{p}_{kin}$ and $Q^a$, where the notation is changed from quantum $T^a$ to classical $Q^a$ to emphasize the different nature of the classical color generators of SU($N_c$). All commutators are replaced by Poisson brackets $[A,B]_\mathrm{com}\rightarrow i \left\{A, B\right\}_\mathrm{PB}$.
For example, the canonical commutation relation between position and momentum becomes in the classical limit $\left\{x_i, p_j\right\}_\mathrm{PB}=\delta_{ij}$,
and similarly the color algebra in the classical limit is $\left\{Q^a, Q^b\right\}_\mathrm{PB}=f^{abc}Q^c$.

\noindent\textbf{\textit{Equation of motion for kinetic momentum}.}
Collecting the results for the kinetic momentum defined in Eq.~\eqref{eq:kinmom}
\begin{equation}
    \boldsymbol{p}_{kin}\equiv \boldsymbol{p}-g\boldsymbol{A}^a T^a\,.
\end{equation}
The time evolution of the Heisenberg-picture kinetic momentum operator is
\begin{equation}
    \frac{\dd \boldsymbol{p}_{kin}} {\dd t}=i [H,\boldsymbol{p}_{kin}]+\frac{\partial \boldsymbol{p}_{kin}}{\partial t}=[H, \boldsymbol{p}\,]-g[H, \boldsymbol{A}^a]T^a-g\boldsymbol{A}^c[H, T^c]+\frac{\partial \boldsymbol{p}_{kin}}{\partial t}\,,
\end{equation}
with the individual terms calculated in \cref{eq:comhp,eq:comha,eq:comht} as
\begin{subequations}
    \begin{align}
        &[H, p_i]=-ig(\boldsymbol{\alpha}\cdot\partial_i\boldsymbol{A}^a-\partial_i A_0^a)T^a\,, & &[H, A_i^a]=-i\alpha_j \partial_j A_i^a\,,\\
        &[H, T^c]=-ig f^{abc}(\boldsymbol{\alpha}\cdot \boldsymbol{A}^b-A_0^b)T^a\,, & &\partial p^{kin}_i/\partial t=g\partial_0 A_i^a T^a\,.
    \end{align}
\end{subequations}
Collecting these, the kinetic momentum evolves as
\begin{equation}
    \frac{\dd p^{kin}_i}{\dd t}=g\Big[(\partial_0 A^a_i-\partial_i A_0^a+g f^{abc}A_0^b A_i^c)-\alpha_j(\partial_i A^a_j-\partial_j A_i^a+g f^{abc}A_i^b A_j^c)\Big]T^a\,,
\end{equation}
where the first bracket is $E_i^a$ from Eq.~\eqref{eq:elfield} and the second is $(\boldsymbol{\alpha}\times\boldsymbol{B}^a)_i$ from Eq.~\eqref{eq:alphafij}. Using also $\boldsymbol{\alpha}=\dd\boldsymbol{x}/\dd t=\boldsymbol{p}_{kin}/p_{kin}^0$ from Eq.~\eqref{eq:heisx}, where $p_{kin}^\mu=(p_{kin}^0, \boldsymbol{p}_{kin})$ with $p_{kin}^0$ the energy in the laboratory frame, the kinetic momentum Wong equation can be recast as
\begin{equation}
    \label{eq:pii}
    \frac{\dd p^{kin}_i}{\dd t}=gF^a_{\mu i}\frac{p_{kin}^\mu}{p_{kin}^0} T^a\,.
\end{equation}

\noindent\textbf{\textit{Equation of motion for canonical momentum}.}
Repeating the derivation for the canonical momentum $\boldsymbol{p}$ instead of $\boldsymbol{p}_{kin}$, one finds that only certain commutator contributions survive. The equation of motion is
\begin{equation}
    \frac{\dd p_i}{\dd t}=g\Big(-\partial_i A_0^a-\alpha_j\partial_i A^a_j\Big)T^a\,,
\end{equation}
where $\boldsymbol{\alpha}=\boldsymbol{p}/p^0$ and only partial contributions to the full electric and magnetic fields are present $-\partial_i A_0^a \sim E_i^a$ and $-\alpha_j\partial_i A_j^a \sim (\boldsymbol{\alpha}\times\boldsymbol{B}^a)_i$.

\noindent\textbf{\textit{Eikonal limit for kinetic momentum}.}
The light-cone coordinates $(x^-, x^+, y, z)$ are related to the Cartesian ones $(t,x,y,z)$ through $x^\pm=t\pm x$, that is
\begin{equation}
    A^\pm=A^0\pm A^x\,,\quad p_{kin}^\pm=p_{kin}^0\pm p_{kin}^x\,.
\end{equation}
The eikonal limit $p_{kin}^+\rightarrow\infty$ is therefore equivalent to $p_{kin}^0, p_{kin}^x\rightarrow\infty$. In this limit, the kinetic momentum evolution from Eq.~\eqref{eq:pii} becomes
\begin{equation}
    \label{eq:eikpii}
    \begin{aligned}
        \lim_{p_{kin}^+\rightarrow\infty}\frac{\dd p^{kin}_i}{\dd t}=&g\Big((\partial_0 A_i - \partial_i A_0 - ig[A_0, A_i])^a \\
        &- \frac{p_{kin}^x}{p_{kin}^0}(\partial_i A_x -\partial_x A_i - ig[A_i, A_x])^a\Big)T^a\,,
    \end{aligned}
\end{equation}
for $i=y,z$, where the first group is $E_i^a$, the ratio $p_{kin}^x/p_{kin}^0\to 1$ in the eikonal limit, and the second group is $\epsilon_{xij}B_j^a$. This gives the expected classical eikonal limit in terms of the electric and magnetic fields
\begin{equation}
    \lim_{p_{kin}^+\rightarrow\infty}\frac{\dd p^{kin}_y}{\dd t}=g(E_y^a-B_z^a)T^a\,,\quad \lim_{p_{kin}^+\rightarrow\infty}\frac{\dd p^{kin}_z}{\dd t}=g(E_z^a+B_y^a)T^a\,.
\end{equation}

\noindent\textbf{\textit{Eikonal limit for canonical momentum}.}
The eikonal limit $p^+\rightarrow\infty$ is equivalent to $p^0, p^x\rightarrow\infty$, which leads to, for $i=(y,z)$,
\begin{equation}
    \label{eq:eikpi}
    \lim_{p^+\rightarrow\infty}\frac{\dd p_i}{\dd t}=g\left(-\partial_i A_0^a - \frac{p^x}{p^0}\partial_i A_x^a\right)T^a\,,
\end{equation}
where $p^x/p^0\to 1$ in the eikonal limit. This gives the classical eikonal limit for $A_y=A_z=0$
\begin{equation}
    \lim_{p^+\rightarrow\infty}\frac{\dd p_y}{\dd t}=g(E^a_{p,y}-B^a_{p,z})T^a\,,\quad \lim_{p^+\rightarrow\infty}\frac{\dd p_z}{\dd t}=g(E^a_{p,z}+B^a_{p,y})T^a\,,
\end{equation}
where the electric and magnetic fields contain only the $\partial_i A_+$ contribution with $A_+=A_0+A_x$, namely
\begin{equation}
    E^p_i=-\partial_i A_0\,,\quad B^p_i=\epsilon_{xji}\partial_j A_x\,,
\end{equation}
or on $i=y,z$ components $E^p_y=-\partial_y A_0$, $B^p_z=\partial_y A_x$, $E^p_z=-\partial_z A_0$ and $B^p_y=-\partial_z A_x$. Since $\partial_z=-\sinh\eta\,\partial_\tau+\cosh\eta/\tau\,\partial_\eta$, which simplifies to $\partial_z=1/\tau\,\partial_\eta$ at midrapidity $\eta=0$, the $y$ magnetic field component vanishes: $B^p_y=0$. On the other hand, using $A_0=-\sinh\eta/\tau\,A_\eta$ in the temporal gauge $A_\tau=0$, the $z$ electric field component at midrapidity is $E^p_z=A_\eta/\tau^2$.

\section{Boltzmann-Vlasov equation}

Wong's equations arise as equations of motion for test particles parametrizing the distribution function of the non-Abelian Boltzmann-Vlasov equation~\cite{Schenke:2008hw}.

The phase-space distribution $f(t,\tilde{\mathbf{x}},\tilde{\mathbf{p}},\tilde{\boldsymbol{Q}})$ depends on continuous variables, namely position $\tilde{\mathbf{x}}$, momentum $\tilde{\mathbf{p}}$, and the color charge vector $\tilde{\boldsymbol{Q}}=(\tilde{Q}^1,\ldots,\tilde{Q}^{N_c^2-1})$. In the collisionless regime, it satisfies the manifestly covariant non-Abelian Vlasov equation
\begin{equation}
    \label{eq:vlasov_cov}
    \tilde{p}^\mu\partial_\mu f + g\tilde{Q}^a F^a_{\mu\nu}\,\tilde{p}^\mu\,\partial^{\nu}_{(\tilde{p})}f + g f^{abc}\tilde{Q}^cA^b_\mu\,\tilde{p}^\mu\,\partial_{\tilde{Q}^a}f = 0\,,
\end{equation}
where $\partial^\nu_{\tilde{p}}\equiv\partial/\partial\tilde{p}_\nu$. Dividing by $\tilde{p}^0$, and expressing the time derivative explicitly, the Vlasov equation can be recast as
\begin{equation}
    \label{eq:vlasov_3plus1}
    \partial_t f + \frac{\tilde{p}^k}{\tilde{p}^0}\,\partial_{\tilde{x}^k} f + \frac{g}{\tilde{p}^0}\tilde{Q}^a F^{k\nu,a}\tilde{p}_\nu\,\partial_{\tilde{p}^k} f + \frac{g}{\tilde{p}^0}f^{abc}\tilde{Q}^cA^b_\mu\tilde{p}^\mu\,\partial_{\tilde{Q}^a} f = 0\,.
\end{equation}

The distribution function is represented by a collection of $N_{tp}$ classical point particles. Denote the position, momentum and color charge vector of the $i$-th test particle by $\mathbf{x}_i(t)$, $\mathbf{p}_i(t)$ and $\boldsymbol{Q}_i(t)$, with color charge components $Q^a_i(t)$. Introducing the shorthand notations
\begin{equation}
    \delta^{(i)}_x\equiv\delta^{(3)}\!\bigl(\tilde{\mathbf{x}}-\mathbf{x}_i(t)\bigr)\,,\quad \delta^{(i)}_p\equiv(2\pi)^3\delta^{(3)}\!\bigl(\tilde{\mathbf{p}}-\mathbf{p}_i(t)\bigr)\,,\quad \delta^{(i)}_Q\equiv\delta^{(N_c^2-1)}\!\bigl(\tilde{\boldsymbol{Q}}-\boldsymbol{Q}_i(t)\bigr)\,,
\end{equation}
the test particle approximation for the distribution function reads
\begin{equation}
    \label{eq:testparticle}
    f\bigl(t,\tilde{\mathbf{x}},\tilde{\mathbf{p}},\tilde{\boldsymbol{Q}}\bigr) = \frac{1}{N_{tp}}\sum_{i=1}^{N_{tp}} \delta^{(i)}_x\,\delta^{(i)}_p\,\delta^{(i)}_Q\,.
\end{equation}

Substituting the ansatz into Eq.~\eqref{eq:vlasov_3plus1}, the time derivative acts via the chain rule, giving
\begin{equation}
    \label{eq:dt_ansatz}
    \begin{aligned}
        \partial_t f = -\frac{1}{N_{tp}}\sum_i\biggl[&\frac{\dd x^k_i}{\dd t}\,\partial_{\tilde{x}^k}\delta^{(i)}_x\,\delta^{(i)}_p\delta^{(i)}_Q \\
         &+ \frac{\dd p^k_i}{\dd t}\,\delta^{(i)}_x\,\partial_{\tilde{p}^k}\delta^{(i)}_p\,\delta^{(i)}_Q + \frac{\dd Q^a_i}{\dd t}\,\delta^{(i)}_x\delta^{(i)}_p\,\partial_{\tilde{Q}^a}\delta^{(i)}_Q\biggr]\,.
    \end{aligned}
\end{equation}
The remaining derivative terms act on the continuous variables of the delta functions. The delta functions $\delta^{(i)}_p$ and $\delta^{(i)}_Q$ localize the continuous variables to the particle values $\tilde{p}^\mu\to p^\mu_i$ and $\tilde{Q}^a\to Q^a_i$, while $\delta^{(i)}_x$ localizes the gauge field arguments as $F^{k\nu,a}(\tilde{\mathbf{x}})\to F^{k\nu,a}(\mathbf{x}_i)$ and $A^b_\mu(\tilde{\mathbf{x}})\to A^b_\mu(\mathbf{x}_i)$. Collecting all contributions, the Vlasov equation requires
\begin{align}
    \label{eq:vlasov_collected}
    0 = \frac{1}{N_{tp}}\sum_i\Bigg[&\left(\frac{p^k_i}{p^0_i}-\frac{\dd x^k_i}{\dd t}\right)\partial_{\tilde{x}^k}\delta^{(i)}_x\,\delta^{(i)}_p\delta^{(i)}_Q \notag\\
    &+\left(\frac{g}{p^0_i}Q^a_i F^{k\nu,a}(\mathbf{x}_i)p_{i,\nu}-\frac{\dd p^k_i}{\dd t}\right)\delta^{(i)}_x\,\partial_{\tilde{p}^k}\delta^{(i)}_p\,\delta^{(i)}_Q \notag\\
    &+\left(\frac{g}{p^0_i}f^{abc}Q^c_i A^b_\mu(\mathbf{x}_i)p^\mu_i-\frac{\dd Q^a_i}{\dd t}\right)\delta^{(i)}_x\delta^{(i)}_p\,\partial_{\tilde{Q}^a}\delta^{(i)}_Q\Bigg]\,.
\end{align}
Since test particles occupy distinct phase-space points, contributions from different $i$ are independent. For each $i$, the three distributions $\partial_{\tilde{x}}\delta^{(i)}_x$, $\partial_{\tilde{p}}\delta^{(i)}_p$ and $\partial_{\tilde{Q}}\delta^{(i)}_Q$ are linearly independent, therefore each bracket in Eq.~\eqref{eq:vlasov_collected} must vanish separately, giving
\begin{equation}
    \label{eq:wong_derived}
    \frac{\dd x^k_i}{\dd t} = \frac{p^k_i}{p^0_i}\,,\qquad\frac{\dd p^k_i}{\dd t} = \frac{g}{p^0_i}Q^a_i F^{k\nu,a}(\mathbf{x}_i)p_{i,\nu}\,,\qquad\frac{\dd Q^a_i}{\dd t} = \frac{g}{p^0_i}f^{abc}Q^c_i A^b_\mu(\mathbf{x}_i)p^\mu_i\,.
\end{equation}
These are Wong's equations for the $i$-th test particle, reparametrized from the worldline parameter $\lambda$ to coordinate time $t$ via $\dd t/\dd\lambda = p^0_i/m$. The Boltzmann-Vlasov equation therefore ensures that each particle in the test particle representation follows Wong's equations of motion.

The converse also holds, namely, if each test particle obeys Wong's equations, the ansatz satisfies the Boltzmann-Vlasov equation. Requiring $\dd f/\dd t = 0$ and applying the chain rule to the ansatz gives $\sum_i[(\dd x^k_i/\dd t)\partial_{\tilde{x}^k}\delta^{(i)}_x\,\delta^{(i)}_p\delta^{(i)}_Q +(\dd p^k_i/\dd t)\delta^{(i)}_x\,\partial_{\tilde{p}^k}\delta^{(i)}_p\,\delta^{(i)}_Q +(\dd Q^a_i/\dd t)\delta^{(i)}_x\delta^{(i)}_p\,\partial_{\tilde{Q}^a}\delta^{(i)}_Q]/N_{tp}=0$.

Further, one substitutes Wong's equations from~\cref{eq:wong_derived} for the three trajectory derivatives, and uses the delta functions to localize $p^\mu_i\to\tilde{p}^\mu$, $Q^a_i\to\tilde{Q}^a$, $F^{k\nu,a}(\mathbf{x}_i)\to F^{k\nu,a}(\tilde{\mathbf{x}})$ and $A^b_\mu(\mathbf{x}_i)\to A^b_\mu(\tilde{\mathbf{x}})$. The spatial drift term $(p^k_i/p^0_i)\,\partial_{\tilde{x}^k}$ can be written as $(\tilde{p}^\mu/\tilde{p}^0)\,\partial_\mu$ because $\delta^{(i)}_x$ does not depend on $t$ as a phase-space label, so $\partial_t\delta^{(i)}_x = 0$ and the $\mu=0$ component vanishes. Likewise, $\delta^{(i)}_p$ is a three-momentum delta function, so $\partial^0_{\tilde{p}}\delta^{(i)}_p = 0$ and the force term $\tilde{p}^\mu F^a_{\mu\nu}\partial^\nu_{\tilde{p}}$ reduces to its spatial components only. With these two observations, each particle $i$ contributes
\begin{equation}
    \begin{aligned}
        \label{eq:vlasov_i}
        \frac{\tilde{p}^\mu}{N_{tp}}\Big[&\partial_\mu\delta^{(i)}_x\,\delta^{(i)}_p\delta^{(i)}_Q \\
        &+g\tilde{Q}^aF^a_{\mu\nu}\,\partial^{\nu}_{(\tilde{p})}\delta^{(i)}_p\,\delta^{(i)}_x\delta^{(i)}_Q +gf^{abc}\tilde{Q}^cA^b_\mu\,\partial_{\tilde{Q}^a}\delta^{(i)}_Q\,\delta^{(i)}_x\delta^{(i)}_p\Big]=0\,,
    \end{aligned}
\end{equation}
Summing over all $i$ and identifying $\sum_i\delta^{(i)}_x\delta^{(i)}_p\delta^{(i)}_Q/N_{tp} = f$ recovers the non-Abelian Vlasov equation from Eq.~\eqref{eq:vlasov_cov}. Together, the two directions establish the equivalence between Wong's equations and the non-Abelian Boltzmann-Vlasov equation.

\makeatletter\diss@chapter@allowedtrue\makeatother
\references

\let\origcleardoublepage\cleardoublepage
\let\cleardoublepage\clearpage
\backmatter
\let\cleardoublepage\origcleardoublepage

\printindex

%

\end{document}